\pdfoutput=1

\documentclass[11pt, twoside]{report}  

\usepackage{fancyheadings}

\usepackage{a4,fancyhdr}
\usepackage[intlimits]{amsmath}
\usepackage{graphicx}
\usepackage{graphics}
\usepackage{amsfonts}
\usepackage{amssymb}
\usepackage{dsfont}
\usepackage{units}

\usepackage[usenames]{color}
\usepackage{colortbl}
\usepackage{titlesec}
\usepackage{framed}

\usepackage{cite}

\definecolor{webgreen}{rgb}{0,0.75,0}
\definecolor{webred}{rgb}{0.75,0,0}
\definecolor{webblue}{rgb}{0,0,0.75}
\definecolor{darkblue}{rgb}{0,0,0.6}
\definecolor{dunkelgrau}{rgb}{0.8,0.8,0.8}
\definecolor{lgray}{rgb}{0.95,0.95,0.95}
\definecolor{lgreen}{rgb}{0.95,1.00,0.90}
\definecolor{lblue}{rgb}{0.9,0.95,1.00}
\definecolor{lred}{rgb}{1.00,0.90,0.80}

\definecolor{shadecolor}{rgb}{1.00,0.92,0.82}

\usepackage[colorlinks=true,linkcolor=darkblue,citecolor=darkblue,urlcolor=darkblue]{hyperref}

\newcommand{\backdef}{\color{black}\mdseries\rmfamily\upshape }

\newenvironment{fshaded}  { \begin{shaded}\vspace{-0.4cm} }{ \vspace{-0.4cm}\end{shaded}  \vspace{-0.5cm} \noindent }
\newenvironment{fshaded2} { \begin{shaded}                }{ \vspace{-0.3cm}\end{shaded}  \vspace{-0.5cm} \noindent }

\newenvironment{fshaded1} { \begin{shaded}\vspace{-0.1cm} }{ \vspace{-0.3cm}\end{shaded}  \vspace{-0.5cm} \noindent }
\newenvironment{fshaded3} { \begin{shaded}\vspace{-0.1cm} }{ \vspace{-0.5cm}\end{shaded}  \vspace{-0.5cm} \noindent }
\newenvironment{fshaded4} { \begin{shaded}\vspace{-0.5cm} }{ \vspace{-0.4cm}\end{shaded}  \vspace{-0.5cm} \noindent }
\newenvironment{fshaded5} { \begin{shaded}\vspace{-0.3cm} }{ \vspace{-0.3cm}\end{shaded}  \vspace{-0.5cm} \noindent }

\newcommand{\vect}[1]{{\mbox{\boldmath $#1$}}}
\newcommand{\be}{\begin{equation}}
\newcommand{\ee}{\end{equation}}
\newcommand{\beqa}{\begin{eqnarray}}
\newcommand{\eeqa}{\end{eqnarray}}

\newcommand{\fatcol}[1]{\noindent\textcolor{webred}{\textbf{#1}}}

\newcommand{\conjg}[1]%
{\ensuremath{\hspace{1pt}\overline{\hspace{-1pt}#1\hspace{-1pt}}}\hspace{1pt}}

\def\Slash#1{\setbox0=\hbox{$#1$} 
\dimen0=\wd0 
\setbox1=\hbox{/} \dimen1=\wd1 
\ifdim\dimen0>\dimen1 
\rlap{\hbox to \dimen0{\hfil/\hfil}} 
#1 
\else 
\rlap{\hbox to \dimen1{\hfil$#1$\hfil}} 
/ 
\fi}

\def\longlongrightarrow{
\relbar\joinrel\relbar\joinrel\relbar\joinrel\relbar\joinrel\rightarrow}
\def\longlonglongrightarrow{
\relbar\joinrel\relbar\joinrel\relbar\joinrel\relbar\joinrel\relbar\joinrel\relbar\joinrel\rightarrow}
\def\longlonglonglongrightarrow{
\relbar\joinrel\relbar\joinrel\relbar\joinrel\relbar\joinrel\relbar\joinrel\relbar\joinrel\relbar\joinrel\relbar\joinrel\relbar\joinrel\relbar\joinrel\relbar\joinrel\rightarrow}

\begin{document}

\begin{titlepage}

\vspace*{3cm}

\begin{center} \LARGE \bfseries Gernot Eichmann
\end{center}

\vspace{1cm}

\begin{center}
\huge \bfseries Hadron properties from \\ QCD bound-state equations
\end{center}

\vspace{8cm}

\begin{center}
\LARGE Dissertation \\ \large
zur Erlangung des Doktorgrades der Naturwissenschaften \\
verfasst am Institut f{\"u}r Physik \\
an der Karl-Franzens-Universit{\"a}t Graz \\
Betreuer: Univ.-Prof. Dr. R. Alkofer \\
Graz, 2009
\end{center}

\end{titlepage}

\newpage
\section*{}
\newpage

\tableofcontents

\titleformat{\chapter}[display]
{\color{webred}\normalfont\huge\bfseries}{\chaptertitlename\
\thechapter}{20pt}{\Huge}

\titleformat{\section}
{\color{webred}\normalfont\Large\bfseries}{\thesection}{1em}{}

\titleformat{\subsection}
{\color{webred}\normalfont\large\bfseries}{\thesubsection}{1em}{}


\chapter[Outline]{Outline}

     Quantum chromodynamics (QCD) is the quantum field theory of the strong interaction,
     with quarks and gluons being its elementary degrees of freedom.
     Nevertheless only color-neutral hadrons as bound states of quarks and gluons
     appear as observable particles in detector facilities, and
     a detailed study of hadron properties is of fundamental importance
     for an understanding of the quark-gluon dynamics of QCD.
     A wealth of information has been collected in the past decades' high-energy scattering experiments
     but many issues are still unresolved.
     Present and future experimental programs at JLAB, SLAC, COMPASS/CERN and FAIR/GSI
     are devoted to charmed meson and baryon spectroscopy,
     the search for exotic mesons, or the investigation of the nucleon's electromagnetic
     and spin structure in terms of form factors and generalized parton distributions.

     The interplay between QCD's elementary and observable degrees of freedom
     addresses two phenomena whose origin is not yet fully understood:
     color confinement and dynamical chiral symmetry breaking.
     The latter is the mass generation mechanism that equips light quarks with large dynamical constituent-quark masses 
     whereas it retains the light pseudoscalar mesons as would-be Goldstone bosons of spontaneously broken chiral symmetry.
     While QCD is well under control in the high-energy region whose weak coupling strength enables the application of perturbative methods,
     these phenomena characterize the large-distance or low-energy structure of QCD where the coupling is strong and demands a non-perturbative treatment.

     Many approaches which describe particular aspects of QCD have emerged in the past decades. For instance,
     quark models have pioneered our understanding of hadron structure and successfully described many of their properties, see e.g.~\cite{Isgur:1979be,Loring:2001kx,Faessler:2005gd,Melde:2008yr,Metsch:2008zz};
     effective field theories provide rigorous results in certain limits, e.g.~\cite{Gasser:1983yg,Ecker:1994gg,Hemmert:2002uh,Procura:2006bj,Bernard:2007zu,Brambilla:2004jw};
     experimentally extracted generalized parton distributions (GPDs) combine electromagnetic form factors
     and parton distribution functions and establish a three-dimensional tomography of hadrons, see \cite{Belitsky:2005qn,Boffi:2007yc,Feldmann:2007zz} for recent reviews.
     Each has its own strengths but also weaknesses, e.g., parameter dependence,
     limited range of applicability, or the reliance upon a factorization scale.

     \newpage

     The long-term goal of rigorously solving QCD necessitates a quantum-field theoretical,
     non-perturbative description starting directly from the QCD Lagrangian as a prerequisite.
     Lattice QCD is constantly pushing forward towards the physical quark mass
     in its investigation of hadron structure~\cite{Zanotti:2008zm,WalkerLoud:2008pj}. 
     On the other hand, the increase of computational power in the past decade has also played a vital role
     in the development of functional continuum methods:
     the Dyson-Schwinger equations (DSEs) of QCD provide a tool
     to map out the infrared structure of QCD where confinement and dynamical chiral symmetry breaking occur;
     non-perturbative bound-state equations have been utilized to perform hadron spectroscopy
     and thereby determine the meson and baryon amplitudes that are needed to calculate hadronic observables
     such as electromagnetic form factors and decay properties. For reviews on Dyson-Schwinger equations and their application
     to hadron physics, see e.g. Refs.~\cite{Roberts:1994dr,Alkofer:2000wg,Maris:2003vk,Holl:2006ni,Fischer:2006ub}.

     Functional methods are in some sense complementary to lattice QCD.
     While access to the full momentum region and quark-mass range is readily available without any need for extrapolations,
     one faces an infinite system of integral equations which needs to be truncated with regard to a practical numerical computation.
     Such truncations, while subject to specific constraints, induce a model dependence which is not easily quantified.
     Lattice calculations  and their chiral extrapolations are not only useful for comparing results
     but can also serve as a guideline to construct truncations which capture the important physics. 
     Much of the current effort is focused on further resolving the dynamics of quarks and gluons from their DSEs 
     and elevate simplistic truncations towards a more complete picture of the involved physical content.
     An exploration of the full potential contained in such functional approaches has only begun, and the future holds many exciting perspectives:
     from a phenomenologist's point of view, some of the intermediate goals are access to charm and bottom physics, the investigation of excited hadrons,
     electromagnetic form factors in the timelike region, or the  medium- and large-$Q^2$ domain of spacelike form factors.

     The aim of this thesis is to report progress that has been achieved in the framework of meson and baryon bound-state equations
     and thereby contributes to an understanding of the dynamics of hadrons from their underlying constituents. 
     After recapitulating basic concepts in QCD and the bound-state formalism in Chapter~\ref{chapter:QCD},
     we provide a systematic study of quark-antiquark (Chapter~\ref{sec:mesons}), three-quark (Chapter~\ref{chapter:faddeev}), diquark and quark-diquark bound states (Chapter~\ref{chapter:quark-diquark}). 
     The common parenthesis is provided by a rainbow-ladder (RL) truncation, i.e. an iterated vector-vector gluon exchange in the $q\bar{q}$ kernel, which guarantees
     the correct implementation of chiral symmetry and its spontaneous breaking and
     is elaborated in Chapter~\ref{sec:mesons}.
     In the $q\bar{q}$ sector we report results for $\pi$ and $\rho$-meson observables;
     the three-body equation is solved for the nucleon, and the quark-diquark approach is applied to $N$ and $\Delta$ baryons.
     In Chapter~\ref{chap:ffs} we present results for the nucleon's electromagnetic form factors in the quark-diquark framework.
     In Chapter~\ref{sec:conclusions} we summarize and conclude.

     Calculational details, as well as details on the structure of Green functions and bound-state amplitudes that appear throughout the text,
     are collected in Appendices~\ref{app:vertices} and~\ref{app:utilities}.
     We work in the Euclidean formulation (corresponding relations are given in App.\,\ref{app:conventions})
     and restrict ourselves to two-flavor QCD with isospin symmetry.

  \chapter{Bound states in QCD}\label{chapter:QCD}


            The dynamical content of QCD as a local quantum field theory of quarks and gluons is described by its Lagrangian or, equivalently, its action.
            It is a difficult task to unfold the physical content of the QCD Lagrangian toward a quantum-field theoretical description of hadrons.
            At large energies and small distances (the 'ultraviolet' region), the interaction between quarks and gluons is weak and can be described 
            by perturbative methods. At low energies and large distances (the 'infrared'), the coupling becomes strong
            and perturbation theory can no longer explain the most striking features of QCD in this regime: the dynamical breaking of chiral symmetry (D$\chi$SB), the generation of large constituent-quark masses
            from almost massless quarks, the formation of hadrons --- mesons as quark-antiquark and baryons as three-quark  bound states ---\,, and the confinement of quarks and gluons inside hadrons.

            A rigorous starting point to describe the non-perturbative dynamics of quarks and gluons is the path-integral approach.
            Dyson-Schwinger equations (DSEs) are the quantum equations of motion. They relate QCD's Green functions --- the basic
            propagators and vertices of the theory, e.g. the dressed quark and gluon propagators, and the quark-gluon vertex --- to each other and lead
            to a self-consistent and infinitely coupled system of integral equations.

             Hadrons appear as free-particle poles in the respective $n$-particle Green functions. For instance, the quark-antiquark and
             three-quark Green functions exhibit meson and baryon poles, respectively. Hadron properties can be extracted upon
             solving bound-state equations which are valid at these poles and need the Green-function content of QCD as an input.
             In combination they provide a powerful tool to calculate experimentally accessible hadron observables, e.g. meson and baryon mass spectra,
             decay constants, scattering processes, and electromagnetic properties such as form factors, magnetic moments and charge radii.

             In the present chapter we briefly introduce the theoretical foundation and phenomenological aspects of QCD.
             Starting from its generating functional, we sketch the derivation of Dyson-Schwinger equations,
             discuss confinement and D$\chi$SB and their manifestation in the bound-state approach, and derive the general form
             of a bound-state equation which will be referred to in the following chapters.

\newpage
     \section{Basic concepts in QCD}\label{sec:qcd:phenomenology}

    \bigskip
    \fatcol{QCD action.}
            Based upon the principle of local gauge invariance, the QCD action which describes the interaction of quark and gluon fields $\psi$, $\conjg{\psi}$, $A_\mu^a$ is written as
            \begin{equation}\label{QCD:action}
               S_\text{QCD}[A,\psi,\conjg{\psi}\,] = \int \! d^4 x \; \Big[ \conjg{\psi} \left( -\Slash{D} + m \right) \psi + \textstyle\frac{1}{4}\,  F^a_{\mu\nu} \,F^a_{\mu\nu} \Big]\,,
            \end{equation}
            where $D_\mu = \partial_\mu + i g A_\mu$ is the covariant derivative and the gluon field-strength tensor reads $F_{\mu\nu} = \partial_\mu A_\nu-\partial_\nu A_\mu + ig \left[ A_\mu,A_\nu\right]$.
            The action is per construction invariant along the gauge orbit
            \begin{equation}
                \psi^U = U \psi\,, \qquad A_\mu^U = U A_\mu U^\dag - \frac{i}{g}\,U (\partial_\mu U^\dag)\,, \qquad F_{\mu\nu}^U = U F_{\mu\nu} U^\dag\,,
            \end{equation}
            where $U(x) \in SU(3)_C$ is a local gauge transformation and $A_\mu = A_\mu^a\, t^a$, $F_{\mu\nu} = F_{\mu\nu}^a\, t^a$ are elements of the
            corresponding color algebra whose basis elements $t^a$, $a=1\dots 8$ satisfy the commutator relation $[t^a,t^b] = i f^{abc}t^c$.
            In the fundamental representation they are given by the Gell-Mann matrices $t^a = \lambda^a/2$.

            The non-Abelian nature of the color group $SU(3)_C$ induces gluonic self-interaction terms $\sim A^3$ and $\sim A^4$
            encoded in $F_{\mu\nu}$ which lead to significant complications compared to the Abelian gauge theory QED.
            They are commonly believed to be the origin of phenomena such as confinement and dynamical chiral symmetry breaking in QCD.

    \bigskip
    \fatcol{Generating functional.}
            In an Euclidean path-integral approach, the quantum field theory which corresponds to the action \eqref{QCD:action} is defined by the generating functional
            \begin{equation}
                Z[J,\eta,\bar{\eta},\sigma,\bar{\sigma}] = \int \!\mathcal{D}[A,\psi,\bar{\psi},c,\bar{c}] \, e^{\,-\,S_\text{QCD}[A,\psi,\bar{\psi}] \,-\, S_\text{GF}[A,c,\bar{c}] \,+\, S_\text{C}}
            \end{equation}
            from which all physical quantities can be derived.
            An integration over infinitely many physically equivalent gauge field configurations $A^U$,
            and the emergence of zero eigenvalues of the perturbative inverse gluon propagator as obtained from \eqref{QCD:action}, are avoided by adding the gauge-fixing term
            \begin{equation}\label{QCD:GaugeFixingAction}
               S_\text{GF}[A,c,\bar{c}\,] = \int \! d^4 x \;\bigg[ \frac{(\partial_\mu A_\mu^a)^2}{2\,\xi} +(\partial_\mu\bar{c}^a)(D^{ab}_\mu c^b) \bigg]
            \end{equation}
            to the classical action\footnote{Gauge fixing is not necessary for a direct calculation of gauge-independent quantities via Eq.\,\eqref{QCD:GreenFunctions} as is for instance routinely done in lattice QCD.}.
            It introduces unphysical auxiliary fields, the scalar anticommuting ghost fields $c$ and $\bar{c}$, where
            $D^{ab}= \delta^{ab}\partial_\mu + g f^{abc} A_\mu^c$ is the covariant derivative in the adjoint representation.
            The expression \eqref{QCD:GaugeFixingAction} arises from evaluating the gauge-fixing condition $\delta(\partial_\mu A_\mu^a-\xi B)\, \text{det} \,M[A]$
            in the path integral which involves the determinant of the Faddeev-Popov operator
            \begin{equation}
                M[A]^{ab}(x,y) = -\partial_\mu D_\mu^{ab}\,\delta^4(x-y)\,.  
            \end{equation}
                        The source term
            \begin{equation}
               S_\text{C} = \int \! d^4 x \; \Big[ A_\mu^a J_\mu^a + \conjg{\eta}\,\psi + \conjg{\psi}\,\eta + \conjg{\sigma}\,c + \conjg{c}\,\sigma \Big]
            \end{equation}
            contains the external currents which are auxiliary quantities to enable the derivation of Green functions in terms of functional derivatives of
            $Z$.

    \bigskip
    \fatcol{Green functions.}
            All physical properties can be extracted from the Green functions of a quantum field theory, and
            the theory is solved once all of them are determined.
            Abbreviating the fields 
            generically by $\varphi(x)$ and the associated sources by $J(x)$, the Green functions are
            defined as the time-ordered vacuum expectation values of products of fields, denoted by $G[\varphi]$:
            \begin{equation}\label{QCD:GreenFunctions}
                \langle  G[\varphi]  \rangle := \langle 0 \,| \,T\,G[\hat{\varphi}]\, | \,0 \rangle
                = \frac{\int \mathcal{D}\varphi  \, e^{- S[\varphi]}\, G[\varphi] }{\int \mathcal{D}\varphi \, e^{- S[\varphi]}}
                = G\left[\frac{\delta}{\delta J}\right]_{J=0} \frac{Z[J]}{Z[0]}\,.
            \end{equation}
            For instance, a two-point function is associated with $G[\varphi]=\varphi(x) \varphi(y)$.
            In the functional formulation the Green functions correspond to averages over all field configurations with a probability distribution $e^{- S[\varphi]}$.
            As indicated in \eqref{QCD:GreenFunctions}, they are most conveniently obtained as moments of the generating functional $Z[J]$ by taking functional derivatives with respect to the sources $J$.
            The same procedure yields connected Green functions as derivatives of the functional $W[J] := -\ln{Z[J]}$
            and one-particle irreducible (1PI) Green functions from the effective action $\Gamma[\widetilde{\varphi}]$ which is related to $W[J]$ via a Legendre transformation:
            \begin{equation}
                Z[J] = e^{- W[J]} = \int \! \mathcal{D}\varphi\,e^{-S[\varphi]+\int\! \varphi(x)\,J(x) }  =: e^{- \Gamma[\widetilde{\varphi}] +\int\! \widetilde{\varphi}(x)\,J(x)  } \,.
            \end{equation}
            The averaged field $\widetilde{\varphi}$ is the expectation value of $\varphi$ in the presence of the source~$J$:
            \begin{equation}
                -\frac{\delta W[J]}{\delta J(x)}  = \frac{\int\! \mathcal{D}\varphi\, e^{- S[\varphi] + \int \varphi(x)\,J(x) }\,\varphi(x)}{ \int\! \mathcal{D}\varphi\,e^{-S[\varphi] + \int \varphi(x)\,J(x) } }
                                                  = \langle \varphi(x) \rangle_J = \widetilde{\varphi}(x)\,.
            \end{equation}
            With the following shorthand notation for functional derivatives:
            \begin{equation}
                W''[0]_{xy} = \left.\frac{\delta W[J]}{\delta J(x) \,\delta J(y)} \right|_{J=0}\,, \qquad
                \Gamma''[0]_{xy} = \left.\frac{\delta \Gamma[\widetilde{\varphi}]}{\delta \widetilde{\varphi}(x) \,\delta\widetilde{\varphi}(y)} \right|_{\widetilde{\varphi}=0}\,,
            \end{equation}
            the propagator associated with the field $\varphi$ is given by $\Gamma''[0]^{-1}_{xy}$, the respective three-point vertex is $\Gamma'''[0]_{xyz}$, etc.
            1PI vertices involving different types of fields are obtained upon differentiating with respect to each corresponding variable $\widetilde{\varphi}_i(x)$.

    \bigskip
    \fatcol{Dyson-Schwinger equations.}
            Dyson-Schwinger equations \cite{Dyson:1949ha,Schwinger:1951ex} are the quantum equations of motion.
            They follow from an invariance of the generating functional under a variation $\varphi(x) \rightarrow \varphi(x) + \epsilon(x)$ of the fields.
            Assuming that the integral measure is invariant under such a transformation, the condition $Z'[J] = Z[J]$ leads to the relation
            \begin{equation}\label{QCD:GeneratingDSE1}
                \left\langle \frac{\delta S[\varphi]}{\delta \varphi(x) } \right\rangle_J = Z[J]^{-1} \,\frac{\delta S}{\delta \varphi}\left[\frac{\delta}{\delta J(x)}\right]  Z[J] = J(x)
            \end{equation}
            for each field $\varphi$ that appears in the action.
            This is the defining relation for the infinite tower of Dyson-Schwinger equations which
            relate the Green functions of the field theory to each other.
            Inserting $Z[J] = e^{-W[J]}$ yields the corresponding equation for connected Green functions:
            \begin{equation}
                \frac{\delta S}{\delta \varphi}\left[-W'[J]_x + \frac{\delta}{\delta J(x)}\right]\,1 = J(x)\,.
            \end{equation}
            With $W'[J]_x = -\widetilde{\varphi}(x)$, $\Gamma'[\widetilde{\varphi}]_x = J(x)$, and
            \begin{equation}
                \frac{\delta}{\delta J(x)} = \int_y \frac{\delta \widetilde{\varphi}(y)}{\delta J(x)} \,\frac{\delta}{\delta \widetilde{\varphi}(y)}
                                           = -\int_y W''[J]_{xy} \,\frac{\delta}{\delta \widetilde{\varphi}(y)}
                                           = \int_y \Gamma''[\widetilde{\varphi}]^{-1}_{xy} \,\frac{\delta}{\delta \widetilde{\varphi}(y)}\,,
            \end{equation}
            one arrives at
            the generating DSE for 1PI Green functions:
            \begin{fshaded2}
            \begin{equation}\label{QCD:GeneratingDSE3}
                \Gamma'[\widetilde{\varphi}]_x = \frac{\delta S}{\delta \varphi} \left[ \, \widetilde{\varphi}(x) +  \int_y \! \Gamma''[ \widetilde{\varphi}]_{xy}^{-1} \,\,\frac{\delta}{\delta \widetilde{\varphi}(y)}\right] 1\,.
            \end{equation}
            \end{fshaded2}

            \noindent
            Further $(n-1)$-fold differentiation and finally setting $\widetilde{\varphi}=0$
            yields the system of DSEs\footnote{A Mathematica package which enables an automated derivation of DSEs is described in Ref.\,\cite{Alkofer:2008nt}.}
            for the 1PI $n$-point functions $\Gamma^{(n)}[0]_{x_1, \dots x_n}$.

\begin{figure}[t]
            \begin{center}
            \includegraphics[scale=0.54]{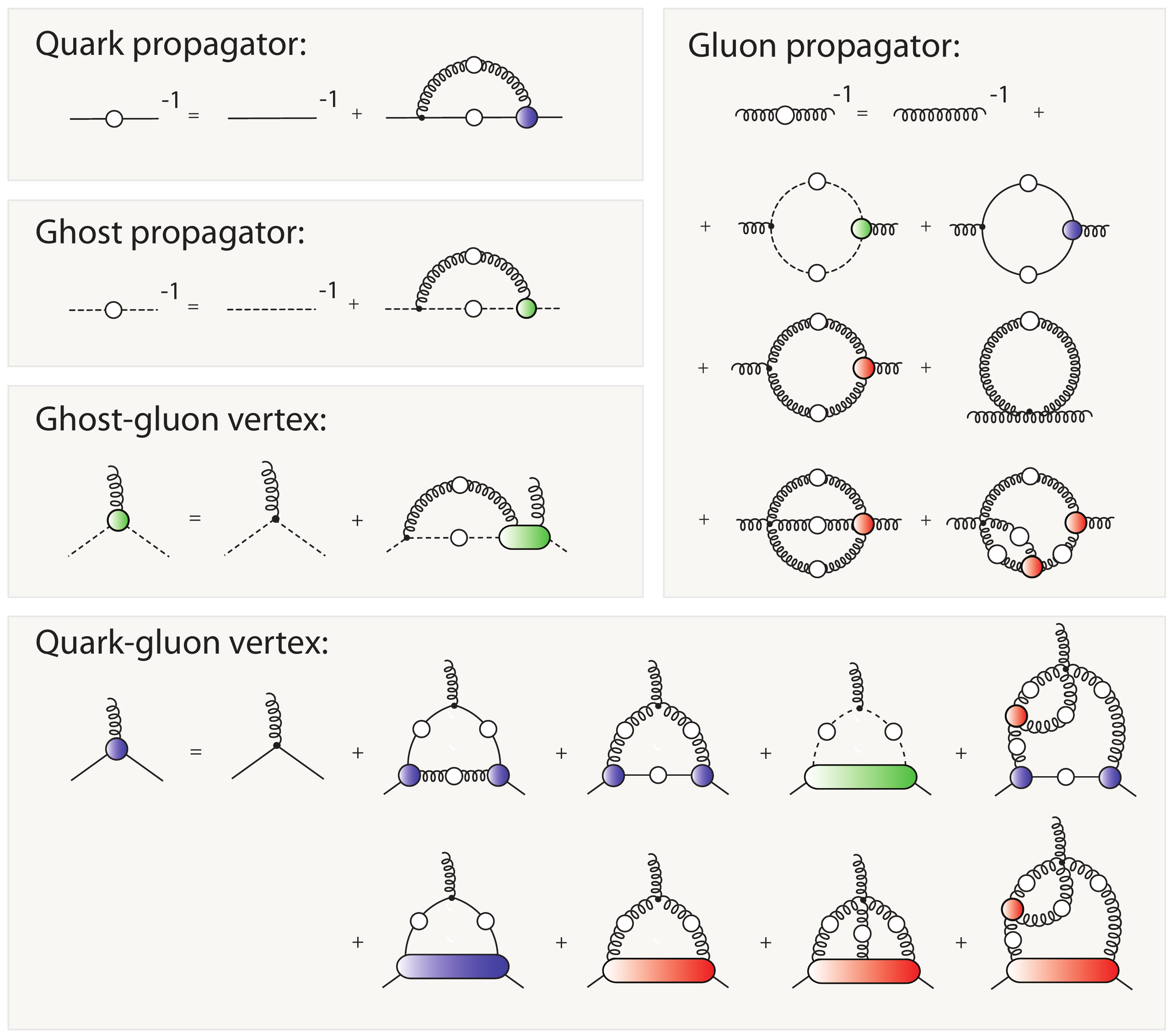}
            \caption[$c(m)$]{\backdef DSEs for quark, gluon and ghost propagators and the ghost-gluon and quark-gluon vertices.
                                      White blobs denote dressed propagators, filled blobs represent 1PI Green functions.}
            \label{fig:QCD-DSEs}
            \end{center}
            \end{figure}

    \bigskip
    \fatcol{Dyson-Schwinger equations in QCD.}
            In the context of QCD, the action whose functional derivative appears in Eqs.\,(\ref{QCD:GeneratingDSE1}--\ref{QCD:GeneratingDSE3})
            is the sum of the classical action and the gauge-fixing contribution \eqref{QCD:GaugeFixingAction}.
            Dyson-Schwinger equations for 1PI Green functions are obtained via functional derivatives of Eq.\,\eqref{QCD:GeneratingDSE3}
            with respect to the gluon, quark or ghost fields $A$, $\psi$, $\bar{\psi}$, $c$ and $\bar{c}$,
            where the latter four are anticommuting Grassmann variables.
            A detailed derivation of these relations can be found in Refs.\,\cite{Itzykson:1980rh,Roberts:1994dr,Alkofer:2000wg}.
            The DSEs for quark, gluon and ghost propagators and quark-gluon and ghost-gluon vertices are illustrated in Fig.\,\ref{fig:QCD-DSEs}.

            The Dyson-Schwinger approach provides an appealing tool for several reasons.
            DSEs operate in fully relativistic quantum field theory;
            they provide access to both perturbative and non-perturbative regimes of QCD; and they represent a continuum approach which is able to cover the full
            quark-mass range between chiral limit and the heavy-quark domain.
            Of course the main caveat concerns the complexity of this framework: for numerical studies one relies upon a truncation of the infinite
            systems of equations to a subset that captures the physical content and is solved explicitly, combined with the use of
            ans\"atze for those Green functions that enter the equations but are not solved for.
            These ans\"atze are constrained by symmetry properties, multiplicative renormalizability, perturbative limits, etc.
            During the past years a cross-fertilization between functional methods and lattice QCD has provided further insight
            into the non-perturbative structure of Green functions.

    \bigskip
    \fatcol{Gauge invariance.}
            While Green functions depend on the gauge, physical observables must be gauge independent.
            Similarly as the classical equations of motion are implemented in the functional description
            through equations for the 1PI vertices, the requirement of gauge invariance of the generating functional
            leads to Ward-Takahashi (WTIs) and Slavnov-Taylor identities (STIs) which interrelate the Green functions of a gauge theory.
            Although the gauge-fixed action of QCD is no longer invariant with respect to local gauge transformations,
            it still satisfies global gauge symmetry and BRST symmetry \cite{Becchi:1975nq,Iofa:1976je}.
            The latter is formally equivalent to a gauge transformation by a ghost field and can be used to derive the STIs. 
            Moreover it can be shown that requiring BRST invariance of a gauge theory generates both the ghosts and the gauge fixing
            while ensuring gauge independence of physical observables \cite{Llewellyn-Smith1980}.


     \section{Phenomenological aspects of QCD}\label{sec:qcd:phenomenology}

    \bigskip
    \fatcol{Confinement.}
            The absence of colored states in the physical spectrum is one of the most characteristic non-perturbative phenomena in QCD,   
            and its underlying mechanism is still not fully understood.
            Several different ideas have emerged which are likely to explain different aspects of confinement; for reviews, see Refs.\,\cite{DiGiacomo:1995ck,Greensite:2003bk,Alkofer:2000wg,Alkofer:2006fu,Fischer:2006ub,Diakonov:2009jq}.
            A common line of thought builds upon the idea of certain topological gauge field configurations dominating the path integral near
            the Gribov horizon which corresponds to a vanishing eigenvalue of the Faddeev-Popov operator \cite{Gribov:1977wm}.
            This entails a linearly rising quark-antiquark potential in Coulomb gauge \cite{Zwanziger:1998ez,Szczepaniak:2001rg,Zwanziger:2002sh,Cucchieri:2002su,Greensite:2003xf,Feuchter:2004mk}
            and a strongly infrared-divergent ghost propagator and infrared-vanishing gluon propagator in Landau gauge 
            which trigger the infrared behavior of other Landau-gauge Green functions \cite{Zwanziger:1991gz,Lerche:2002ep,Alkofer:2008tt}.

            A somewhat different point of view concerns the identification of physical subspaces from the asymptotic state space of QCD.
            The Kugo-Ojima scenario \cite{Kugo:1979gm} which relies on BRST symmetry describes the cancellation of longitudinal and timelike gluons with ghost and antighost fields
            similar to the Gupta-Bleuler mechanism in QED. On the other hand, the Osterwalder-Schrader axiom of reflection positivity \cite{Osterwalder:1973dx}
            implies that a certain degree of freedom whose propagator violates positivity and thereby prevents a Lehmann representation
            cannot describe an asymptotic physical state.

            The manifestation of confinement in the bound-state framework is elusive.
            In the particular approach employed in this work, ans\"atze are employed for the gluon propagator and
            quark-gluon vertex such that, out of the coupled system of DSEs, the quark propagator is the only Green function which is explicitly solved for.
            While a rainbow-ladder truncation indeed induces complex conjugate poles which ensure positivity violation and describe a confined quark,
            this result is truncation dependent and sensitive to the structure of the quark-gluon vertex \cite{Alkofer:2003jj,Fischer:2005nf}.

            On the other hand, the solution of the quark DSE is insensitive to the precise shape of the interaction
            in the deep infrared region since the self-energy integral is dominated by momenta larger than the quark mass.
            As a consequence, the impact of infrared physics upon hadronic ground states is at best modest.
            It should certainly become important in the context of highly excited states with a large spatial extension or small-$x$ physics.

    \bigskip
    \fatcol{Dynamical chiral symmetry breaking and bound states.}
            A phenomenon which has a direct impact upon the spectra of light hadrons
            is the spontaneous breaking of chiral symmetry.
            An unbroken chiral $SU_L(N_f) \times SU_R(N_f)$ symmetry, realized in the massless QCD Lagrangian,
            would imply mass-degenerate meson parity doublets in the chiral limit whose remnants
            should be visible in nature. 
            The surprisingly small mass of the pion 
            compared to its parity partners 
            indicates that chiral symmetry is broken spontaneously, and
            the pion is identified as the massless Goldstone boson of the two-flavor case in the chiral limit of massless quarks.
            It acquires a small mass due to the explicit breaking of chiral symmetry at small non-zero current-quark masses,
            a behavior which is described by the Gell-Mann-Oakes-Renner (GMOR) relation \cite{GellMann:1968rz}.

            An order parameter of chiral symmetry breaking is the scalar quark condensate.
            It modifies the structure of the QCD vacuum and, through its interaction with the quarks,
            equips them with a large dynamical constituent mass.
            This behavior is reflected in the quark mass function $M(p^2)$ that appears in the dressed quark propagator:
            it connects the perturbative ultraviolet momentum region with the nonperturbative infrared and
            thereby communicates the transition from a current quark to a dynamically generated 'constituent quark'.
            Since quarks are the building blocks of hadrons, the masses of light mesons and baryons are generated dynamically as well.
            D$\chi$SB therefore offers an explanation for the quark mass-generation mechanism,
            and it illustrates why the simple constituent-quark reckoning $m_\text{Meson}\sim 2\,M(0)$, $m_\text{Baryon} \sim 3\,M(0)$
            works reasonably well except for the light pseudoscalar mesons.

            D$\chi$SB becomes manifest not only in the quark propagator but also in the quark-gluon vertex whose scalar Dirac structures
            are dynamically generated together with the quark mass function \cite{Alkofer:2008tt}.
            Such an effect is naturally missing upon employing the rainbow-ladder truncation.
            Here the quark-gluon coupling $\alpha(k^2)$ associated with a bare vertex enters the quark DSE as a parametrization
            including a scale $\Lambda_\text{IR}$ as its input (Sections~\ref{sec:coupling-ansaetze} and \ref{sec:MESON:results}).
            A non-zero quark mass function in the chiral limit only occurs
            if the coupling exhibits enough strength in the infrared. 
            Above a critical threshold, $\Lambda_\text{IR}$ is directly proportional to all mass-dimensionful quantities in the chiral limit, hence all of them represent a 'scale of D$\chi$SB'.

    \bigskip
    \fatcol{Pion cloud.}
            It is a longstanding prediction that, induced by D$\chi$SB, the low-energy and low quark-mass behavior of hadrons
            is modified by their interaction with pseudoscalar mesons, i.e. the long-range part of $q\bar{q}$ correlations.
            Established in the cloudy bag model \cite{Theberge:1980ye,Thomas:1982kv,Lu:1997sd}, where the pion field is coupled to a
	        constituent-quark bag \cite{Chodos:1974je}, meson-cloud effects have been studied in a number of quark models, e.g. \cite{Faessler:2005gd,Weigel:1995cz,Diakonov:1997sj,Glozman:1997fs,Miller:2002ig,Melde:2008dg}.
            They are systematically implemented in chiral effective field theories \cite{Gasser:1983yg,Bernard:1995dp} which,
            in combination with lattice simulations, provide an efficient tool for describing masses and electromagnetic properties of hadrons \cite{Hemmert:2002uh,Young:2002cj,Gockeler:2003ay}.

            In these frameworks hadrons consist of a 'quark core' that is augmented by a pseudoscalar meson cloud
            which mediates a stronger binding, decreases the hadron's mass and increases its size.
            Chiral effective field theories combined with lattice techniques typically
            predict a reduction of $20- 30\%$ for dynamically generated hadron masses by chiral corrections, 
            an effect which is suppressed  with increasing distance from the chiral limit.

            In the covariant bound-state approach used herein, the freedom of choosing a current-mass dependent rainbow-ladder coupling strength can be exploited
            to construct a hadronic quark core which subsequently needs to be dressed by meson-cloud effects~\cite{Eichmann:2008ae}.
            Following up on previous quark-diquark model investigations \cite{Hecht:2002ej,Alkofer:2004yf,Cloet:2008re}, we will frequently
            present results in such a 'core model' and compare to those of chirally extra\-polated lattice calculations.
            A natural extension is the explicit implementation of pionic effects in the covariant bound-state equations;
            corresponding results have been recently reported \cite{Fischer:2008wy}.


\newpage

     \section{Bound-state equations}\label{sec:qcd:bses}

      \bigskip
      \fatcol{T-matrix and scattering kernel.}
              As bound states of valence quarks, hadrons correspond to poles in the quark 4- or 6-point
            functions $G^{(2)}$ and $G^{(3)}$
            or, equivalently, in their amputated connected parts, the scattering
            matrices $T^{(2)}$ and $T^{(3)}$ defined via
            \begin{equation}\label{bs:tmatrix}
                G^{(n)} = G_0^{(n)} + G_0^{(n)} T^{(n)} G_0^{(n)},
            \end{equation}
            where $G_0^{(n)}$ is the product of $n$ dressed quark propagators $S_i$.
            We dropped all Dirac, color and flavor indices of each quark leg in the above quantities for brevity.
            The product in Eq.\,\eqref{bs:tmatrix} is understood
            as a summation over all occurring indices as well as integration over all internal momenta.
            The full notation is given at the end of this section.

            The $n$-quark T-matrix is related to the $n$-quark scattering kernel $K^{(n)}$ ($n\geq 2$) by a non-perturbative Dyson sum.
            In the context of bound-state equations one frequently encounters renormalization-group invariant combinations
            of $T^{(n)}$ or $K^{(n)}$ with $n$ quark propagator legs attached on their right-hand sides.
            Denoting them by
            \begin{equation} \label{BS:tilde}
                \widetilde{T}^{(n)} := T^{(n)} G_0^{(n)}, \quad \widetilde{K}^{(n)} := K^{(n)} G_0^{(n)},
            \end{equation}
            the defining relations for the scattering kernels $\widetilde{K}^{(n)}$ read
            \begin{equation}\label{BS:dysonsum}
                \widetilde{T}^{(n)} = \widetilde{K}^{(n)} + \widetilde{K}^{(n)} \widetilde{K}^{(n)} + \widetilde{K}^{(n)} \widetilde{K}^{(n)} \widetilde{K}^{(n)} + \dots  
            \end{equation}
            which may be rephrased as an integral equation, namely Dyson's equation (also referred to as inhomogeneous BSE) for the T-matrix:
            \begin{fshaded2}
            \begin{equation} \label{BS:dyson}
                \widetilde{T}^{(n)} = \widetilde{K}^{(n)} \left( 1 +  \widetilde{T}^{(n)} \right)   = \left( 1 +  \widetilde{T}^{(n)} \right)  \widetilde{K}^{(n)}\,.
            \end{equation}
            \end{fshaded2}

\begin{figure}[tbp]
            \begin{center}
            \includegraphics[scale=0.115]{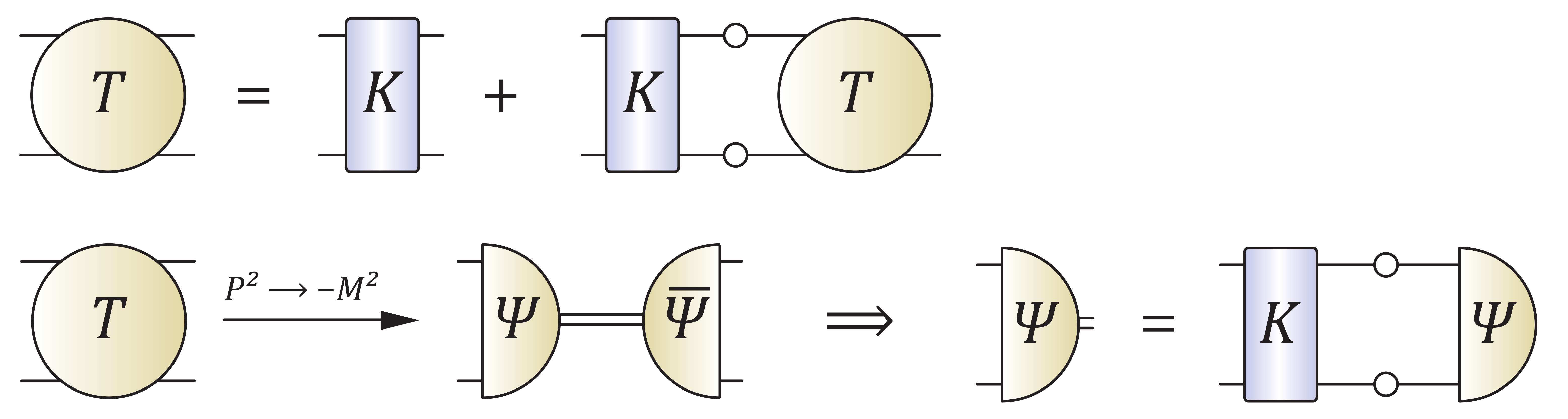}
            \caption[Bound-state equations]{\backdef Schematic derivation of a two-body bound-state equation.
                                                     The first row illustrates Dyson's equation \eqref{BS:dyson}.
                                                     The behavior at the mass pole defines the bound-state amplitude
                                                     and leads to the corresponding bound-state equation (second row).} \label{fig:BS:eq}
            \end{center}
            \end{figure}

            \noindent
            This equation provides the central foundation of the approach and is depicted in the upper part of Fig.~\ref{fig:BS:eq}.
            It allows for a derivation of bound-state equations for $q\bar{q}$, $qq$ and $qqq$ systems together with their canonical normalization conditions, cf. Eq.\,\eqref{bs:boundstate-eq}.
            Moreover, it is of virtue when constructing an off-shell ansatz for the 2-quark T-matrix (App.~\ref{app:mesondiquark-dqprop}),
            and we will resort to Eq.\,\eqref{BS:dyson} for deriving an electromagnetic current operator (Section~\ref{sec:em}).
            Schematically, its inverse form reads:
            \begin{equation}
                \left(T^{(n)}\right)^{-1} = \left(K^{(n)}\right)^{-1} - G_0^{(n)} \quad \Longleftrightarrow\quad \left( \widetilde{T}^{(n)} \right)^{-1} = \left( \widetilde{K}^{(n)} \right)^{-1} - 1\,.
            \end{equation}

\begin{figure}[tbp]
            \begin{center}
            \includegraphics[scale=0.14]{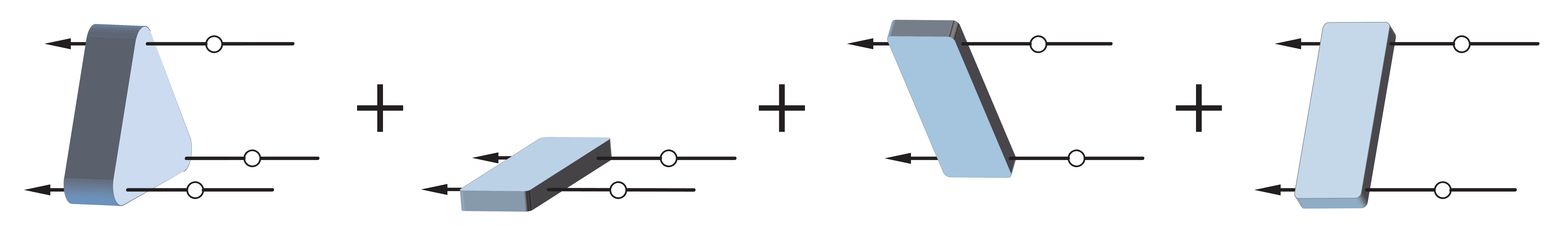}
            \caption[Three-body kernel]{\backdef Three-quark kernel $\widetilde{K}^{(3)}$ of Eq.\,\eqref{FADDEEV:3body-kernel}.} \label{fig:BS:3body-kernel}
            \end{center}
            \end{figure}

            The scattering kernels $K^{(n)}$ consist of $l$-quark irreducible components, with $l=2\dots n$.
            For instance, the three-body kernel $K^{(3)}$ which appears in the bound-state equation of a baryon
            is the sum of a 3-quark irreducible contribution $K^{(3)}_\text{irr}$ and three permuted 2-body kernels $K_i^{(2)} \otimes \,S_i^{-1}$ \cite{Taylor:1966zza,Cahill:1988dx,Loring:2001kv}.
            With the notation of \eqref{BS:tilde}, the kernel $\widetilde{K}^{(3)}$ reads
            \begin{equation}\label{FADDEEV:3body-kernel}
            \widetilde{K}^{(3)} = \widetilde{K}^{(3)}_\text{irr} + \sum_{i=1}^3 \widetilde{K}_i^{(2)},
            \end{equation}
            where the subscript $i$ identifies the spectator quark. $\widetilde{K}^{(3)}$ is illustrated in Fig.~\ref{fig:BS:3body-kernel}.

      \bigskip
      \fatcol{Bound-state equations.}
            At the pole corresponding to the bound-state mass $M$, bound-state amplitudes $\Psi$ are
            introduced as the residues of the scattering matrix via
            \begin{equation}\label{BS:Tmatrixpole}
            T^{(n)} \; \stackrel{P^2\rightarrow -M^2}{\longlonglongrightarrow} \; \mathcal{N} \frac{\Psi\,\conjg{\Psi}}{P^2+M^2}\;,
            \end{equation}
            where $P$ is the total momentum of the $n$ quarks.
            The possibly dimensionful constant $\mathcal{N}$ accounts for the dimensionality of $T^{(n)}$ and depends on the spin of the resulting particle.
            For instance, the propagators of free spin-$0$ and spin-$\nicefrac{1}{2}$ particles are given by:
            \begin{equation}
                J=0: \; \frac{1}{P^2+M^2}\,, \quad
                J=\nicefrac{1}{2}: \; \frac{-i \Slash{P}+M}{\;\; P^2+M^2} = 2M\,\frac{\Lambda_+(P)}{P^2+M^2}\,.
            \end{equation}
            For a scalar or pseudoscalar particle: $\mathcal{N}=1$.
            In the spin-$1/2$ case, the matrix-valued amplitude $\Psi$ includes the positive-energy projector $\Lambda_+(P) = (\mathds{1}+\hat{\Slash{P}})/2$ (cf. Section~\ref{chapter:faddeev}),
            where $\hat{P}$ denotes the normalized total momentum; this yields $\mathcal{N}=2M$.

            Inserting the pole condition \eqref{BS:Tmatrixpole} into Dyson's equation and comparing the residues of the most singular terms
            leads to a bound-state equation at the pole $P^2=-M^2$, cf. Fig.~\ref{fig:BS:eq}. An examination of the relation $T' = -T \,(T^{-1})' \,T$ at the bound-state pole, where $'$ denotes
            \pagebreak[4]
            the derivative $d/dP^2$, yields the associated canonical normalization condition.
            Their combination completely determines the amplitudes $\Psi$ on the mass shell:
            \begin{fshaded5}
                \parbox{4.5cm}{\begin{equation} \label{bs:boundstate-eq}
                             \Psi = \widetilde{K}^{(n)}\Psi,
                             \end{equation}
                             } \quad\quad
                \parbox{7cm}{\begin{equation} \label{bs:normalization}
                              \conjg{\Psi} \left[\,\mathcal{N} \frac{d}{dP^2}\,\left(T^{(n)}\right)^{-1}\right]\Psi  = 1.
                              \end{equation} }
            \end{fshaded5}

            \noindent
            Eq.\,\eqref{bs:boundstate-eq} is a fully relativistic linear homogeneous integral equation.
            It is the Bethe-Salpeter equation in the two-body case ($n=2$) and its quantum-field theoretical analogue for a three-body system ($n=3$).
            Its solution necessitates knowledge of the quark propagator $S_i$ and the kernel $K^{(n)}$.
            Both ingredients can in principle be obtained from the infinite coupled set of
	        Dyson-Schwinger equations, cf. Section~\ref{sec:qcd:phenomenology}.
            Feasible present-day numerical DSE solutions include 2- and 3-point functions within certain truncations,
            but the complexity of DSEs for 4-point functions has so far prevented a sophisticated numerical treatment.

            The construction of appropriate kernels is restricted by the underlying symmetries of the theory.
            Symmetries in quantum-field theory are implemented by Ward-Takahashi (WTIs) and Slavnov-Taylor identities (STIs)
            which relate different $n$-point functions to each other.
            A prominent example is the axial-vector Ward-Takahashi identity (AVWTI)~\cite{Maris:1997hd} which
	        relates the two-quark kernel $K^{(2)}$ to the kernel of the quark DSE. It is imperative to satisfy these
	        identities in the truncation used in a numerical study, and it will be the guiding principle which
            motivates an application of the rainbow-ladder truncation in Section~\ref{sec:qcdgreenfunctions}.

      \bigskip
      \fatcol{The bound-state approach as an eigenvalue problem.}
              The linear homogeneous integral equation \eqref{bs:boundstate-eq} can be viewed as an eigenvalue problem for the kernel $\widetilde{K}^{(n)}$:
              \begin{equation}\label{BOUND-STATE:Eigenvalue-Eq}
                  \widetilde{K}^{(n)}(P^2)\,\Psi_i = \lambda_i(P^2) \,\Psi_i\,,
              \end{equation}
              where $P$ is the total momentum of the $n$-quark bound state and enters the equation as an external parameter.
              Upon projection onto given quantum numbers, the eigenvalues of $\widetilde{K}^{(n)}$  constitute the trajectories $\lambda_i(P^2)$.
              An intersection $\lambda_i(P^2)=1$ at some value $P^2=-M_i^2$ reproduces Eq.\,\eqref{bs:boundstate-eq} and corresponds to a potential physical state\footnote{
              In this context one has to keep in mind the possibility of anomalous states in the excitation spectra of BSE solutions, see e.g.~\cite{Ahlig:1998qf}.}
               with mass $M_i$.

              To enable an iterative solution, the spectrum of $\widetilde{K}^{(n)}$ must be bounded.
              The largest eigenvalue $\lambda_0$ of $\widetilde{K}^{(n)}$ represents the ground state
              of the quantum numbers under consideration and the remaining ones $\lambda_{i\geq 1}$ its excitations;
              the associated eigenvectors $\Psi_i$ are the bound-state amplitudes.
              A repeated multiplication by itself is equivalent to a projection onto the largest eigenvalue $\lambda_0$, i.e., the ground state.
              Solving for the first excited state involves a subtraction of the ground-state contribution from the kernel.
              Hence the $\Psi_i$ are obtained upon applying the projectors
              \begin{equation}
                  \mathcal{P}_0 := \lim_{m\rightarrow\infty} \left[ \frac{\widetilde{K}^{(n)}}{\lambda_0}\right]^m , \quad
                  \mathcal{P}_1 := \lim_{m\rightarrow\infty} \left[ \frac{\left(\widetilde{K}^{(n)}-\lambda_0 |\Psi_0\rangle \langle \Psi_0 | \right)}{\lambda_1}\right]^m , \quad \dots
              \end{equation}
              onto a general amplitude $\Psi$.
              As an alternative, one may identify the ground state and excited states from their poles in the off-shell
              vertex whose quantum numbers coincide with those of the bound-state amplitude. 
              This is realized by solving an inhomogeneous Bethe-Salpeter equation for the respective vertex \cite{Maris:1997hd,Holl:2004fr,Bhagwat:2007rj}.

      \bigskip
      \fatcol{Solution strategy.}
              In this thesis we will merely be concerned with ground states. 
              To solve Eq.\,\eqref{bs:boundstate-eq}, one must specify the color, flavor and spin structure of the bound-state amplitude under investigation.
              The latter is constructed to be Poincar\'e covariant and expressed through a certain number of matrix-valued Dirac amplitudes $\tau_i$ which incorporate all involved momenta.
              The respective dressing functions depend on the Lorentz-invariant scalar products of these momenta.
              For instance, a two-body amplitude can be characterized by a relative momentum $p$ and a total on-shell momentum~$P$:
              \begin{equation}
                 \Psi(p,P) = \sum_i f_i(p^2, p\cdot P, P^2)\,\tau_i(p,P) \; \otimes \; \text{Color} \;\otimes \;\text{Flavor}\,.
              \end{equation}
              Projecting the bound-state equation onto its color and flavor quantum numbers and the orthogonal Dirac basis
              results in coupled integral equations for the components $f_i$.

              According to Eq.~\eqref{BOUND-STATE:Eigenvalue-Eq}, the equation can be solved via iteration within a 'guess range' $P^2 \in \{ -M_\text{min}^2,\,-M_\text{max}^2 \}$,
              where $M_\text{max}$ is determined from the singularity
              structure of the equation's ingredients, for instance the quark propagator (see App.~\ref{app:singularities}).
              The eigenvalue $\lambda(-M^2)=1$ determines the bound-state mass $M$. These procedures are explained in detail, e.g., in \cite{Oettel:2001kd}
              in the context of a quark-diquark Bethe-Salpeter equation.

              The current-quark mass is an input to the quark DSE, cf. Section~\ref{sec:qcdgreenfunctions},
              and can be mapped onto the pion mass upon solving the pseudoscalar meson BSE.
              This allows for a determination of all subsequent results as a function of $m_\pi^2$, where the physical point is characterized by $m_\pi=138$ MeV.
              Varying the current mass, and thus the pion mass, enables a direct comparison to lattice data and their chiral extrapolations.

      \bigskip

      \bigskip
      \fatcol{Remarks on the notation.}
             In the current section we have dropped almost all potential occurrences of notational inconvenience,
             a strategy which we will continue to pursue when deriving formal relations.
             In this formal notation, internal loop integrals, summations or four-dimensional delta functions are not retained;
             possible color-flavor prefactors or symmetrization factors are not considered either.
             They will be stated in subsequent chapters when the expressions are explicitly evaluated.

             Consider for example the quark 6-point function $G^{(3)}$: is defined as the vacuum expectation
             value of time-ordered quark and antiquark field operators,
            \begin{equation}\label{6-p-func}
                G^{(3)}(x_1,x_2,x_3;y_1,y_2,y_3)_{\alpha_1 \alpha_2\alpha_3;\,\beta_1\beta_2\beta_3} = \langle 0 |\, T \prod_{i=1}^3 \,q_{\alpha_i}(x_i) \,\conjg{q}_{\beta_i}(y_i) \,| 0 \rangle\,,
            \end{equation}
            where the Euclidean momentum-space representation is given by
            \begin{equation}
            \begin{split}
                (2\pi)^4 \delta^4\left(\textstyle\sum_i (k_i-p_i)\right) &\, G^{(3)}(k_1,\dots,p_3)_{\alpha_1 \dots \beta_3} := \\ = &
                \displaystyle\prod_{i=1}^3 \int d^4 x_i \int d^4 y_i \,e^{i \left(
                k_i \cdot x_i - p_i \cdot y_i\right)} \,G^{(3)}(x_1,\dots,y_3)_{\alpha_1 \dots \beta_3}\,.
            \end{split}
            \end{equation}
            Greek subscripts collect Dirac, color and flavor indices of the quark legs.
            A $\delta$-function was extracted in the Fourier transform due to translation invariance and thus total momentum conservation.
            The dressed propagator of a single quark $i$ is given by
            \begin{equation}
                (2\pi)^4 \delta^4(k_i-p_i)\,S(k_i)_{\alpha\beta} := \int d^4 x_i \int d^4 y_i
                \,e^{i \left( k_i \cdot x_i - p_i \cdot y_i\right)} \,G^{(1)}(x_i,y_i)_{\alpha\beta}
            \end{equation}
            such that $G_0^{(3)}$, the disconnected product of three single propagators, is written as
            \begin{equation}
                (2\pi)^4 \delta^4\left(\textstyle\sum_i (k_i-p_i)\right) \,
                G_0^{(3)}(k_1,\dots,p_3)_{\alpha_1 \dots \beta_3}=\displaystyle\prod_{i=1}^3 (2\pi)^4
                \delta^4(k_i-p_i)\, S(k_i)_{\alpha_i\beta_i}\,.
            \end{equation}
            By virtue of the delta functions, the 6-point function effectively depends on 5 momenta, the quark propagator on 1 momentum and the
            three-propagator product on 3 momenta.

            In the three-body case, the pole assumption \eqref{BS:Tmatrixpole} is written as:
            \begin{equation}
            T^{(3)}(k_1,\dots,p_3)_{\alpha_1 \dots \beta_3} \;\; \stackrel{P^2\rightarrow -M^2}{\longlongrightarrow} \;\; 2M\,\frac{\Psi(k_1,k_2,k_3)_{\alpha_1 \alpha_2 \alpha_3} \conjg{\Psi}(p_1,p_2,p_3)_{\beta_1 \beta_2 \beta_3}}{P^2+M^2}\,,
            \end{equation}
            where the three-quark 'wave function' $\Phi:= G_0^{(3)} \Psi$ corresponding to the three-quark amplitude $\Psi$ is
            defined to be the transition matrix element between the vacuum and
            the bound state with momentum $P=\sum_i p_i = \sum_i k_i$:
            \begin{equation}\label{BS:Three_body_amplitude_proper_notation}
            \begin{split}
            (2\pi)^4 \delta^4(\textstyle\sum_i p_i-P) &\,\Phi(p_1,p_2,p_3)_{\alpha_1 \alpha_2 \alpha_3} := \\ & =
            \displaystyle\prod_{i=1}^3 \int d^4 x_i \,e^{i  p_i \cdot x_i }
            \,\langle 0 | \, q(x_1)_{\alpha_1}\,q(x_2)_{\alpha_2}\,q(x_3)_{\alpha_3} \,|P\rangle\,.
            \end{split}
            \end{equation}

            The products in, e.g., Eq.\,\eqref{bs:tmatrix} are understood
            as summations over all occurring indices as well as integrations over all internal momenta.
            For instance, the expression appearing in Eq.\,\eqref{BS:dysonsum},
            \begin{equation}
                \widetilde{K}^{(3)} \widetilde{K}^{(3)} = K^{(3)} G_0^{(3)} K^{(3)} G_0^{(3)}\,,
            \end{equation}
            involves nine 4-dimensional integrations in the first place.
            Since aside from total momentum conservation also the total momenta of $K^{(3)}$ and $G_0^{(3)}$ are conserved on their own,
            the original nine integrals are reduced to six.
            Four of those are canceled by the two delta functions in every occurrence of $G_0^{(3)}$,
            such that two integrals (thus integration over 8 real variables) according to the two internal loop momenta remain in the final expression.


\chapter{Mesons}\label{sec:mesons}

            The simplest bound states in QCD are those composed of a quark and an antiquark.
            The corresponding Bethe-Salpeter equation has been formulated in \cite{Salpeter:1951sz} and
            relies upon non-perturbative expressions for the involved quark propagator and $q\bar{q}$ kernel.
            First reliable numerical solutions were obtained in the context of a rainbow-ladder truncation \cite{Cahill:1987qr,Jain:1991pk,Munczek:1991jb,Burden:1996nh,Maris:1997tm},
            i.e. an iterated vector-vector gluon exchange between quark and antiquark,
            which has become a standard approach since then.

            A variety of results related to meson spectroscopy \cite{Maris:1997tm,Maris:1999nt,Alkofer:2002bp,Holl:2004fr,Krassnigg:2004if},
            electromagnetic properties \cite{Maris:1998hc,Maris:1999bh,Maris:2000sk,Holl:2005vu,Bhagwat:2006pu}, decays \cite{Maris:1998hc,Maris:2002mz,Jarecke:2002xd} and scattering processes
            \cite{Bicudo:2001jq,Cotanch:2002vj} have been reported in such a setup;
            see Refs.~\cite{Maris:2003vk,Maris:2005tt} for an overview. The overall tendencies may be summarized as:
            (isovector) pseudoscalar and vector-meson ground state properties agree well with experiment starting from the chiral limit up to the bottom quark;
            radial excitations are sensitive to the model parameters in the interaction;
            axial-vector mesons are too light compared to experimental data.

            The apparent reason for the popularity of rainbow-ladder is tied to its nature of being the simplest truncation of the $q\bar{q}$ kernel
            that implements the correct scheme of chiral symmetry and its spontaneous breaking. Quark-antiquark interactions beyond a simple gluon exchange
            must be consistent with the truncation of the quark propagator to maintain the GMOR relation for the pion.
            Efforts to go beyond rainbow-ladder have been made, and are underway, but typically suffer from a drastic amplification in complexity.
            Phenomenologically important corrections beyond RL involve pseudoscalar meson-cloud effects: the attractive nature of the pion cloud
            should induce a sizeable decrease in the vector-meson mass toward the chiral limit and similarly affect related observables.

            In the current chapter we introduce the basic relations which will frequently be referred to in subsequent parts of this thesis.
            Those are: the quark DSE, meson BSE, and the effective coupling $\alpha(k^2)$ which provides the common link in our studies of $q\bar{q}$, $qq$, $qqq$ and $q(qq)$ systems.
            We will investigate the properties of pseudoscalar and vector mesons using different inputs for the effective coupling, compare their current-mass evolution to lattice data,
            discuss pionic corrections, and extract simple relations which can be used to describe the results.

 \newpage


\section{Quark DSE and meson BSE}\label{sec:qcdgreenfunctions}

                  \bigskip
                  \fatcol{Quark propagator.}
                          The basic quantity which appears in any of the bound-state equations \eqref{bs:boundstate-eq} is
              	          the renormalized dressed quark propagator $S(p,\mu)$. It consists of two dressing
                          functions $\sigma_v$ and $\sigma_s$ which correspond to a general fermion propagator's
              	          vector and scalar Lorentz structures. They can be expressed via
                          the quark renormalization function $Z_f$ and the quark mass function $M$,
                          where the latter is independent of the renormalization point $\mu$:
                          \begin{equation}\label{dse:qprop}
                              S(p,\mu) =  -i \Slash{p} \,\sigma_v(p^2,\mu^2) + \sigma_s(p^2,\mu^2) = \frac{Z_f(p^2,\mu^2)}{p^2+M(p^2)^2} \left( -i \Slash{p} + M(p^2) \right)\,.
                          \end{equation}
                          Another frequently used notation involves
                          the functions $A(p^2,\mu^2) = 1/Z_f(p^2,\mu^2)$ and $B(p^2,\mu^2) = M(p^2)/Z_f(p^2,\mu^2)$.

\begin{figure}[tbp]
                          \begin{center}
                          \includegraphics[scale=0.6]{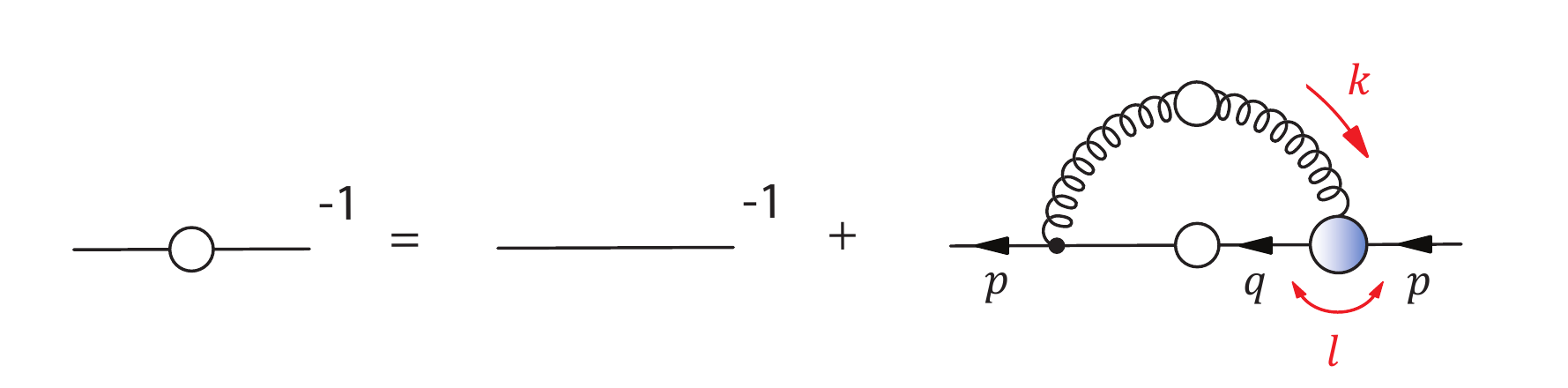}
                          \caption[Quark DSE]{\backdef The quark DSE \eqref{dse:qdse} in pictorial form. } \label{fig:DSE}
                          \end{center}
                          \end{figure}

                          These dressing functions represent the solution of the quark Dyson-Schwinger equation, also known
              	          as the QCD gap equation (see Fig.~\ref{fig:DSE})
                          \begin{equation}\label{dse:qdse}
                              S^{-1}(p,\mu) =  Z_2(\mu^2,\Lambda^2) \left( i \Slash{p} + m_0(\Lambda^2) \right) + \Sigma(p,\mu,\Lambda) \; ,
                          \end{equation}
                          where $Z_2(\mu^2,\Lambda^2)=A(p^2=\Lambda^2,\mu^2)$ %
              	          is the quark renormalization constant and $\Lambda$ an ultra-violet regularization scale.
                          $m_0$ is the cutoff-dependent bare current-quark mass.
                          The quark self-energy $\Sigma$ is defined via
                          \begin{equation}\label{dse:qselfenergy}
                              \Sigma(p,\mu,\Lambda) = - \frac{4 g^2}{3}\,Z_{1F}(\mu^2,\Lambda^2) \int_q^\Lambda
                                                       i\gamma^\mu \, S(q,\mu) \, D^{\mu\nu}(k,\mu)\, \Gamma^\nu(l,k,\mu)\;,
                          \end{equation}
                          where the prefactor $(N_C^2-1)/(2 N_C)=4/3$ stems from the color trace.
                          $\Sigma$ involves the gluon propagator $D^{\mu\nu}$ with gluon momentum $k=q-p$, and
                          a bare ($g\, Z_{1F} \,i\gamma^\mu$) and dressed ($g \, \Gamma^\mu$) quark-gluon vertex with renormalization constant $Z_{1F}$, where
                          we introduced the average momentum $l=(q+p)/2$.
                          These Green functions either need to be  known a priori
                          or determined in the course of a self-consistent solution of the DSEs of the quark  and gluon propagators
              	          together with the quark-gluon vertex.

                          Technical details of the solution of Eq.\,\eqref{dse:qdse} are sketched in App.\,\ref{sec:quarkpropagator}.
                          We note that the current-quark mass dependence which appears in Eq.\,\eqref{dse:qdse} in terms of the bare mass $m_0(\Lambda)$
                          can alternatively be expressed via the mass function $M(\mu^2)$ at the renormalization point through Eq.\,\eqref{QUARK:Mmu+Z2}.
                          The asymptotic form of $M(p^2)$ is given in Eq.\,\eqref{dse:asymptoticmassf} and defines the renormalization-point independent current mass $\hat{m}$.
                          Given a sufficiently large renormalization point $\mu$, $\hat{m}$ can be determined from $M(\mu^2)$ via one-loop evolution, cf. Eq.\,\eqref{QUARK:M+Cond},
                          whereas in the chiral limit: $\hat{m}=0$.

                  \bigskip

                  \fatcol{Gluon propagator and quark-gluon vertex.}
                          The dressed gluon propagator, characterized by a dressing function $Z$,
                          is in Landau gauge transverse to the gluon momentum $k=q-p$:
                          \begin{equation}
                              D^{\mu\nu}(k,\mu)=\frac{Z(k^2,\mu^2)}{k^2}\,T^{\mu\nu}_k, \quad  T^{\mu\nu}_k := \delta^{\mu\nu}-\frac{k^\mu k^\nu}{k^2} \;.
                          \end{equation}
                          The dressed quark-gluon vertex consists of 12 tensor structures and can be written as
                          \begin{equation}\label{qgv:structure}
                              \Gamma^{\mu}(l,k,\mu) = \sum_{i=1}^{4} \left( f_i^{(1)} i\gamma^\mu + f_i^{(2)} l^\mu + f_i^{(3)} k^\mu\right)\,\tau_i(l,k)\;,
                          \end{equation}
                          where the $f_i^{(j)}(l^2,l\!\cdot\! k,k^2,\mu^2)$ are Lorentz-invariant dressing functions.
                          A possible representation of the Dirac basis elements is given by
                          \begin{equation}
                              \tau_i(l,k) = \left\{\mathds{1}, \Slash{k}, \, \Slash{l}, \, [\Slash{l},\Slash{k}] \right\}.
                          \end{equation}
                          The four longitudinal basis elements $\sim k^\mu$ do not survive in the quark-DSE integral because of the transversality of the gluon propagator.
                          Likewise, only the transverse projections of the remaining ones provide a non-vanishing contribution.
                          In accordance with the notation of the quark propagator's dressing functions,
                          the two covariants $i\gamma^\mu$ and $l^\mu$ are referred to as the \textit{vector} and \textit{scalar} components, respectively.

                          Using the STIs in Landau gauge, $Z_{1F} = Z_2/\tilde{Z}_3$ and $Z_g \, \tilde{Z}_3\, Z_3^{1/2} = 1$,
                          where $\tilde{Z}_3$, $Z_3$ and $Z_g$ are ghost, gluon and charge renormalization constants,
                          the quark self-energy integral of Eq.\,\eqref{dse:qselfenergy} becomes
                          \begin{equation}\label{eq:mesons:alpha-i1}
                              \Sigma(p,\mu,\Lambda) = - \frac{16}{3}\, Z_2^2 \int_q^\Lambda i\gamma^\mu   S(q,\mu) \, \frac{T^{\mu\nu}_k}{k^2} \,
                                                        \sum_{i=1}^4 \left( \alpha_i^\text{(1)} i\gamma^\nu + \alpha_i^\text{(2)} l^\nu\right) \,\tau_i(l,k),
                          \end{equation}
                          where we defined the coefficients $\alpha_i^{(j)}$ as combinations of the gluon dressing function and the vertex dressings:
                          \begin{equation}\label{eq:mesons:alpha-i2}
                              \alpha_i^{(j)}(l^2,l\!\cdot\! k,k^2)  = \frac{g^2}{4\pi} \frac{1}{Z_2 \tilde{Z}_3} \, Z(k^2,\mu^2) \, f_i^{(j)}(l^2,l\!\cdot\! k,k^2,\mu^2).
                          \end{equation}
                          They are independent of the renormalization point, as can be inferred
                          from $Z_g \, \tilde{Z}_3\, Z_3^{1/2} = 1$  and the renormalization-scale dependence
                          of the quantities $g \sim 1/Z_g$, $Z \sim 1/Z_3$ and $f_i \sim Z_2/\tilde{Z}_3$.

                  \bigskip

                  \fatcol{Solution of a coupled DSE system.}
                          Both gluon propagator and quark-gluon vertex satisfy their own DSEs.
                          Progress on a consistent solution of this system of DSEs has been made both analytically
                          in terms of infrared exponents of the Green functions, e.g. \cite{Zwanziger:2001kw,Lerche:2002ep,Cucchieri:2004sq,Alkofer:2006gz,Fischer:2008uz,Fischer:2009tn},
                          as well as numerically for general momenta in certain truncations \cite{Fischer:2002hna,Fischer:2003rp,Alkofer:2008tt}.
                  	      The implementation of such a scheme is however beyond the scope of the present
                  	      study. In the following we will motivate the use of underlying symmetry properties of QCD to arrive at a consistent truncation
                  	      whose numerical implementation is feasible in our bound-state approach.

 \renewcommand{\arraystretch}{1.5}

            \begin{figure}[tbp]
            \begin{center}
            \includegraphics[scale=0.2]{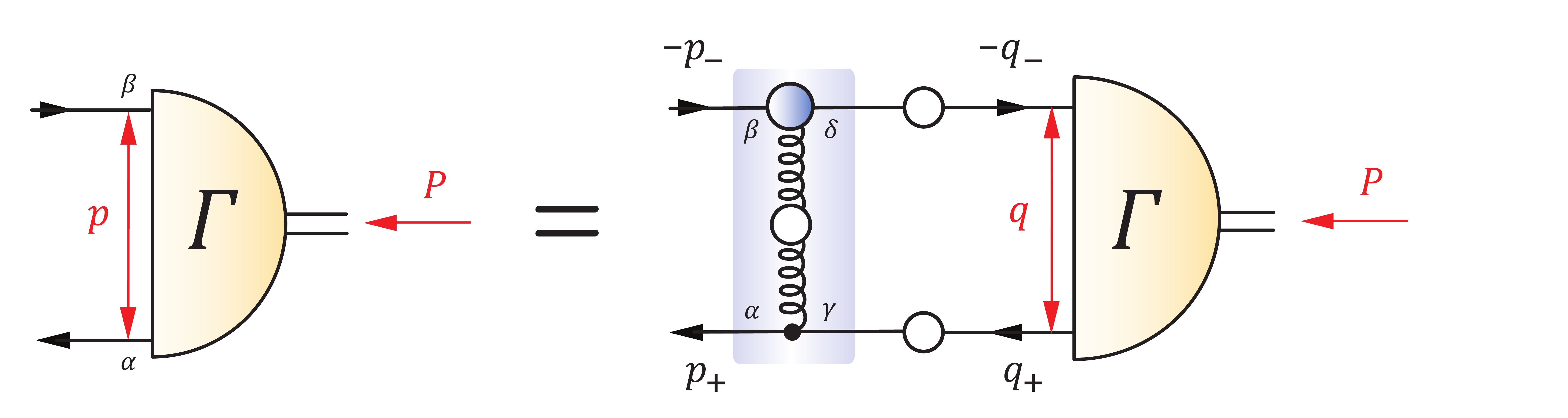}
            \caption[Meson BSE]{\backdef The meson BSE \eqref{bse:bse} in RL truncation, Eq.\,\eqref{bse:rlkernel}.} \label{fig:mesonBSE}
            \end{center}
            \end{figure}

            \bigskip
            \fatcol{Meson BSE.}
             The quark-antiquark bound state amplitude $\Gamma(p,P)$ with relative momentum $p$, total momentum $P$, and mass $M$ (at $P^2=-M^2$)
             is the solution of a homogeneous Bethe-Salpeter equation (see Fig.~\ref{fig:mesonBSE})
                 \begin{equation}\label{bse:bse}
                     \Gamma_{\alpha\beta}(p,P) = \int^\Lambda_q K_{\alpha\gamma,\delta\beta}(p,q,P)
     		                                      \big\{ S(q_+) \Gamma(q,P) S(-q_-) \big\}_{\gamma\delta}\,,
                 \end{equation}
             where Greek indices represent Dirac, color and flavor indices, and the quark and
   	         antiquark loop momenta are $q_+= q+\sigma P$ and $q_-= -q+(1-\sigma) P$.
             The conventions for the momenta were chosen to comply with the diquark case, cf. Eq.\,\eqref{dq:bse}.
             The momentum partitioning parameter $\sigma \in [0,1]$ is arbitrary since
             in a covariant
             description there is no frame-independent definition of a relative momentum.
             Translation invariance implies that for each BSE solution $\Gamma(p,P;\sigma)$ a family of solutions
             of the form $\Gamma(p+(\sigma-\sigma')P,P;\sigma')$ exists \cite{Oettel:2002wf}.
   	         For equal quark and antiquark masses a value $\sigma=1/2$ maximizes the calculable meson mass (see App.~\ref{app:singularities})
             and simplifies its bound-state amplitude.

             The arguments $q_\pm^2$ of the quark propagator's dressing functions in (\ref{bse:bse}) are complex for timelike $P^2$;
 	         methods to evaluate the quark propagator in the complex plane of Euclidean four-momentum-squared are discussed in App.\,\ref{sec:quarkpropagator}.
             The amplitude's dependence on $p,P$ can be formulated in terms of the three Lorentz invariants $p^2$, $P^2$, and $z=\hat{p}\cdot \hat{P}$.
             Calculations are simplified if the dependence on the angular variable $z$ is expanded
 	         in Chebyshev polynomials (see App.~\ref{appendixchebyshev}).
             The spin structure of the amplitudes for different sets of quantum numbers and the method for
 	         solving the BSE are presented in App.\,\ref{app:mesondiquark}.

             The kernel $K$ is the amputated quark-antiquark scattering kernel which is irreducible with respect to a pair of $q\bar{q}$ lines.
             Along with the quark propagator, it provides the physical input to the meson BSE and must be known in advance
             to obtain a solution for the meson's amplitude and mass.

            \begin{figure}[tbp]
            \begin{center}
            \includegraphics[scale=0.50]{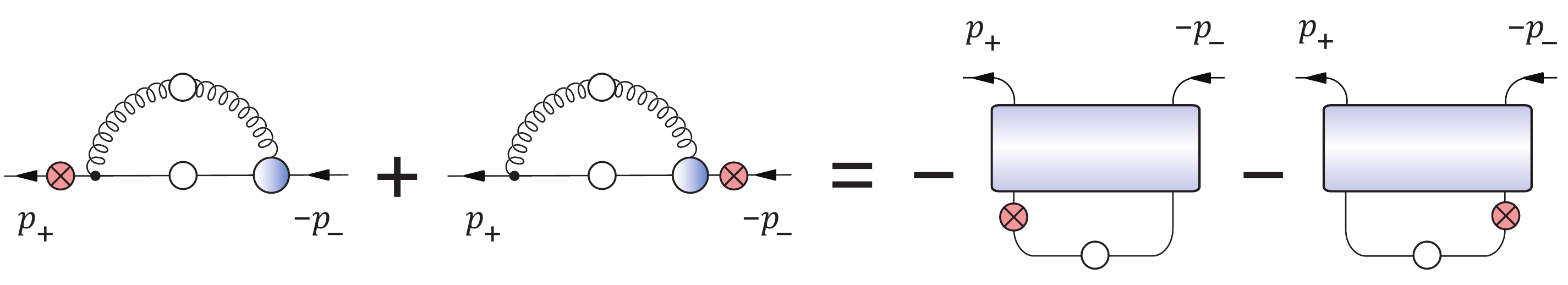}
            \caption[AVWTI]{\backdef The axial-vector Ward-Takahashi  identity \eqref{eq:mesons:axwti} relates quark self-energy and quark-antiquark kernel.
                            Crossed circles denote a $\gamma^5$ insertion. } \label{fig:mesons:axwti}
            \end{center}
            \end{figure}

            \bigskip
            \fatcol{Rainbow-ladder truncation.}
            The central identity which ensures the correct implementation of chiral symmetry and its dynamical breaking
            in a bound-state approach is the AVWTI \cite{Maris:1997hd}. It provides a relation between the quark self-energy and the quark-antiquark kernel,
            the latter of which appears in the meson's bound-state equation. The identity can be expressed as
            \begin{equation}\label{eq:mesons:axwti}
                \left\{\gamma^5 \,\Sigma(-p_-)+\Sigma(p_+) \,\gamma^5 \right\}_{\alpha\beta} = -\int K_{\alpha\gamma,\delta\beta}(p,q,P) \left\{ \gamma^5 \,S(-q_-) + S(q_+) \,\gamma^5 \right\}_{\gamma\delta}
            \end{equation}
            and is sketched in Fig.\,\ref{fig:mesons:axwti}.
            A $q\bar{q}$ kernel which preserves the AVWTI ensures a massless pion in the chiral limit as the Goldstone boson related to dynamical chiral symmetry breaking.
            In addition, Eq.\,\eqref{eq:mesons:axwti} leads to a generalization of the Gell-Mann-Oakes-Renner relation for all pseudoscalar mesons and current-quark masses \cite{Maris:1997hd}.
            In this respect it is imperative for any meaningful truncation of the system of DSEs and BSEs to satisfy this identity.

            A systematic procedure to formulate a $q\bar{q}$ kernel which preserves the AVWTI through functional derivatives of the quark self-energy
            has been introduced in \cite{Munczek:1994zz}.
            Following this prescription, several such constructions have been devised in the literature
            \cite{Bender:1996bb,Bender:2002as,Watson:2004kd,Bhagwat:2004kj,Bhagwat:2004hn,Watson:2004jq,Matevosyan:2006bk,Fischer:2007ze,Alkofer:2008tt,Chang:2009zb,Fischer:2009jm}.

            The simplest setup
            which corresponds to the lowest order in such a symmetry-preserving truncation scheme
            is the rainbow-ladder (RL) truncation.
            In this framework
            the $q\bar{q}$ kernel is expressed by a gluon ladder exchange, including the gluon propagator, one bare and one 'dressed' quark-gluon vertex.
            To satisfy Eq.\,\eqref{eq:mesons:axwti}, the dressed vertex may however only involve Dirac basis tensors with an odd number of gamma matrices, and it can only depend on the gluon momentum $k$.
            This leaves a vector part $\gamma^\mu$ with a purely $k^2$-dependent vertex dressing the only option in both quark DSE and meson BSE.
            The resulting ladder kernel is written as
            \begin{equation}\label{bse:rlkernel}
                K_{\alpha\gamma,\beta\delta} (p,q,P) = Z_2^2 \,\frac{4\pi\alpha(k^2)}{k^2}
                                                      \left(\frac{\lambda^i}{2}\right)_{\!AC} \! \left(\frac{\lambda^i}{2}\right)_{\!BD}
                                                      (i\gamma^\mu)_{\alpha\gamma} \, T^{\mu\nu}_k \, (i\gamma^\nu)_{\beta\delta}\;,
            \end{equation}
            where the $SU(3)_C$ Gell-Mann matrices $\lambda_i$ are explicitly stated.
            The flavor structure in the BSE gives no contribution in an equal-mass system with isospin symmetry.

            In the same way as earlier, the gluon propagator and quark-gluon vertex dressings have been combined into an effective coupling $\alpha(k^2)$.
            By virtue of the RL truncation, it is related to the $\alpha_i^{(j)}$ of Eq.\,\eqref{eq:mesons:alpha-i2} via
            \begin{equation}
                \alpha_1^{(1)}(l^2,l\cdot k,k^2) = \alpha(k^2)\,, \quad  \text{and all other} \quad  \alpha_i^{(j)} = 0 \,.
            \end{equation}
            It poses the single unknown function in the rainbow-ladder approach.


\section{Effective quark-gluon interaction}\label{sec:coupling-ansaetze}

            By virtue of the RL truncation, the entire framework rests upon a choice for the effective coupling $\alpha(k^2)$.
            Rainbow-ladder represents the perturbative remainder of both quark-gluon vertex and the $q\bar{q}$ kernel.
            To satisfy the one-loop relations of perturbative QCD, $\alpha(k^2)$ must approach the asymptotic behavior of QCD's running coupling:
            \begin{equation}\label{dse:asympcoupling}
                \alpha(k^2) \stackrel{k^2\rightarrow\infty}{\longlongrightarrow} \frac{\pi\gamma_m}{\ln{k^2/\Lambda_{QCD}^2}}\;,
            \end{equation}
            where $\gamma_m=12/(11 N_C-2 N_f)$ is the anomalous dimension of the quark propagator (in our calculation we use $\gamma_m=12/25$ which corresponds to $N_f=4$).
            On the other hand, the interaction should exhibit sufficient strength at small gluon momenta
            to enable dynamical chiral symmetry breaking and the generation of a constituent-mass scale for the quark ---
            a feature which would be the result of a combined DSE solution.
            This translates into strong non-perturbative enhancements of the quark dressing functions
	        $A(p^2)$ and $M(p^2)$ at infrared momenta, see, e.\,g., \cite{Roberts:1994dr}.

            Several models for $\alpha(k^2)$ combining the UV limit with
	        an ansatz in the infrared have been employed in the past and applied to detailed studies of meson physics
	        \cite{Jain:1991pk,Munczek:1991jb,Frank:1995uk,Alkofer:1995jx,Maris:1997tm,Maris:1999nt,Alkofer:2002bp}.
            In the present study we implement the interaction of Maris and Tandy \cite{Maris:1999nt} which reads
            \begin{fshaded5}
            \begin{equation}\label{dse:maristandy}
                \alpha(k^2) = \frac{c\,\pi}{\omega^7} \,\left(\frac{k^2}{\Lambda_0^2}\right)^2 e^{-k^2/(\omega^2 \,\Lambda_0^2)} +
                              \frac{\pi \gamma_m \left(1-e^{-k^2/\Lambda_0^2}\right)}{\ln \sqrt{ e^2 - 1 + \left(1+k^2/\Lambda_{QCD}^2\right)^2 } }\,,
            \end{equation}
            \end{fshaded5}

            \noindent
            where we use $\Lambda_{QCD} = 0.234\,\text{GeV}$ and $\Lambda_0 = 1\,\text{GeV}$.
            The first term of \eqref{dse:maristandy} characterizes the infrared properties,
            expressed through the two parameters $c$ and $\omega$ which will be discussed in detail below.
            It provides the characteristic infrared strength which is crucial for a dynamical quark mass generation.
            The second term accounts for the ultraviolet behavior of \eqref{dse:asympcoupling} and is thus constrained by perturbative QCD.

	        We note that the ansatz \eqref{dse:maristandy} behaves as $\alpha(k^2) \rightarrow k^2$
            for $k^2\rightarrow 0$. This facilitates the numerics as the self-energy integral is of lesser divergence in the infrared, see App.~\ref{sec:quarkpropagator}.
            Nevertheless, since the quark DSE is dominated by the interaction at intermediate momenta,
            the explicit behavior in the deep infrared does not play an important role for hadronic ground states.
            This has been explicitly verified in \cite{Fischer:2007ze} with a parametrization different from \eqref{dse:maristandy}.

    \bigskip
    \fatcol{Physical input.}
            The two parameters $c$ and $\omega$ which appear in the ansatz \eqref{dse:maristandy} 
            modulate the coupling's strength and width in the infrared. 
            Studies using this parametrization have demonstrated that pseudoscalar and vector-meson ground state properties are insensitive to  a variation of the coupling width $\omega$
            in a certain range \cite{Maris:1999nt}, i.e. independent of the detailed shape of the interaction in the infrared.
            Such an insensitivity has been found for the ground state of the nucleon in the quark-diquark model as well \cite{Eichmann:2008ef}.
            To highlight this property we will present subsequent results in terms of \linebreak[4] '$\omega$--bands' which denote a variation $\omega=\conjg{\omega}\pm\Delta\omega$.

            In the context of the aforementioned hadron masses, the infrared coupling strength $c$ is accordingly the only active physical parameter in the entire setup.
            In contemporary meson studies it has usually been fixed to reproduce the experimental value of $f_\pi=131$ MeV at the
            $u/d$ current-quark mass.
            Guided by results from experiments and lattice QCD, one may choose a current-mass dependent coupling strength $c(\hat{m})$,
            where $\hat{m}$ is the renormalization-point independent current-quark mass,
            to implement phenomenologically reasonable assumptions that comply with the heavy-quark sector.
            In the following we will give a short overview of the strategies that have been undertaken in the literature
            and will be used as an input to our calculations.

                  \bigskip

                  \fatcol{(C1) Current-mass independent coupling strength.}\hypertarget{coupling:c1}{}
                  A setup which has been extensively used in the literature \cite{Maris:1997tm,Maris:1999nt,Maris:1999bh,%
            Maris:2000sk,Ji:2001pj,Maris:2002mz,Jarecke:2002xd,Holl:2004fr,Krassnigg:2004if,Holl:2005vu,Maris:2005tt,%
            Bhagwat:2006pu,Maris:2006ea,Bhagwat:2006xi,Bhagwat:2007rj} involves a coupling strength that is
                  independent of the quark mass, chosen to reproduce the phenomenological quark condensate
                  and experimental pion decay constant $f_\pi=131$ MeV
                  at the $u/d$ current-quark mass associated with $m_\pi=138$ MeV.
                  The corresponding value of the coupling strength is $c=0.37$.
                  It enables a reasonable description of masses, decay constants and electromagnetic properties of ground-state pseudoscalar
                  and vector mesons up to bottomonium \cite{Maris:2006ea}.

                  \bigskip

                  \fatcol{(C2) Fit to the lattice mass function.}\hypertarget{coupling:c2}{}
                  The availability of lattice data for the quark propagator, gluon propagator and quark-gluon vertex, e.g.
                  \cite{Bowman:2002bm,Skullerud:2003qu,Sternbeck:2005tk,Bowman:2005vx,Kizilersu:2006et,Bowman:2007du,Kamleh:2007ud,Cucchieri:2007md},
                  has provided a means to test the properties of Dyson-Schwinger solutions, in particular with regard to their larger quark-mass behavior.
                  Several parametrizations for the quark-gluon interaction have been employed in the literature and compared to lattice results,
                  either in a rainbow-ladder-like context \cite{Bhagwat:2003vw,Bhagwat:2006tu,Fischer:2005nf,Fischer:2007ea}
                  or more general setups \cite{Bhagwat:2004kj,Alkofer:2003jj,Fischer:2007ze,Alkofer:2008tt,Fischer:2008sp}.

                  A strategy explored in Ref.~\cite{Eichmann:2007nn} was to adopt the ansatz \eqref{dse:maristandy} in the rainbow-ladder approach,
                  where its strength $c(\hat{m})$ is kept intact in the chiral region to recover $f_\pi$ at the physical pion mass,
                  but diminished for larger masses such that the resulting quark mass function $M(p^2)$ agree with quenched
	              lattice results of Ref.~\cite{Bowman:2002bm}.
                  A current-mass dependent coupling therefore emulates to some extent a quark-mass dependent structure in the quark-gluon vertex beyond rainbow-ladder.
                  Upon solving the respective meson and quark-diquark BSEs, corresponding results underestimate $m_\rho$ and $M_N$ obtained from lattice calculations at larger quark masses,
                  e.g., by $\sim 10\%$ at the strange-quark mass \cite{Eichmann:2007nn}.
                  This result either suggests inconsistencies between different lattice techniques or an inaccuracy of rainbow-ladder well beyond the strange-quark mass.

            \begin{figure}[p]
            \begin{center}
            \label{fig:c(m)}
            \includegraphics[scale=0.45]{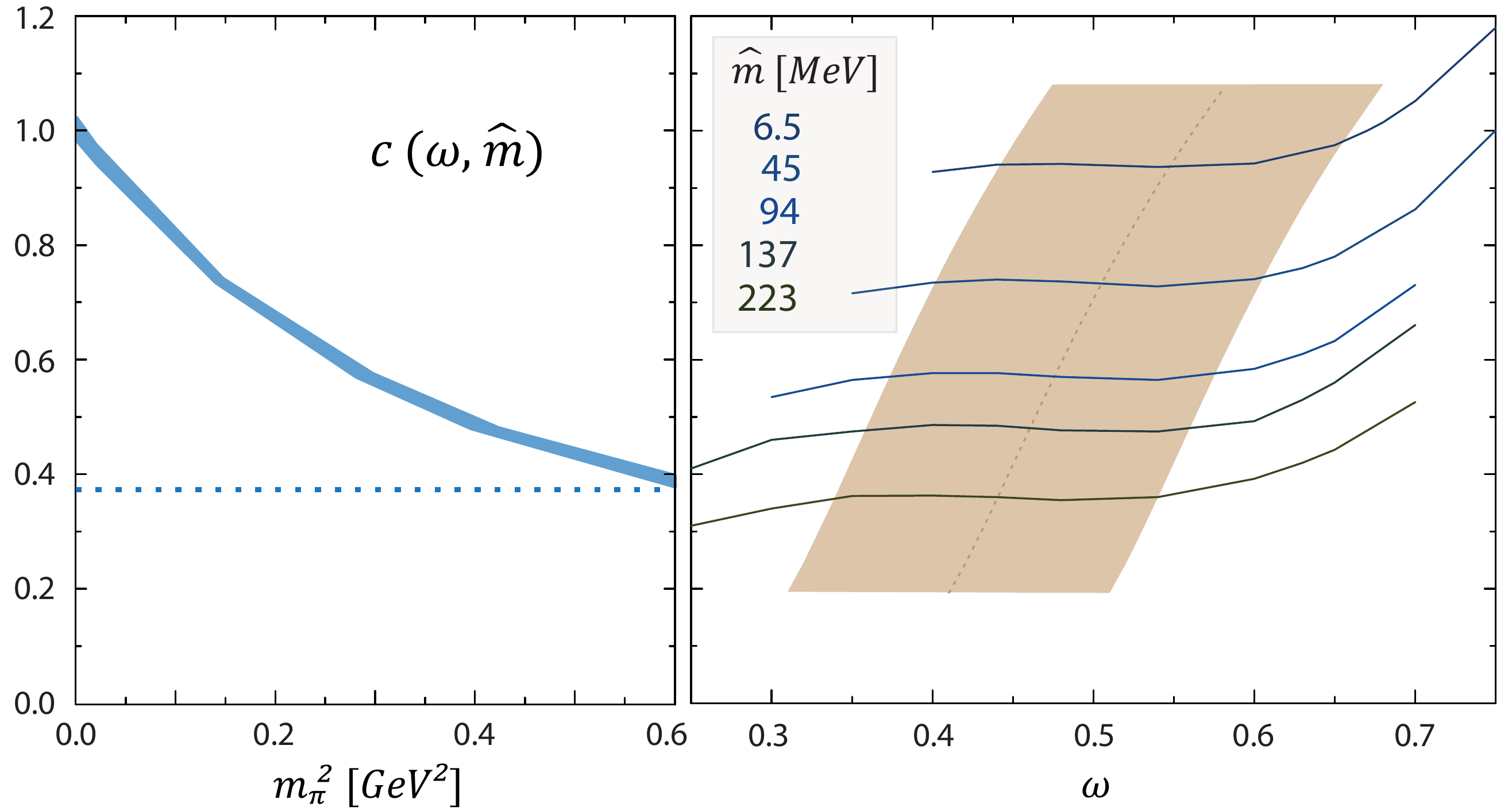}
            \caption[$c(m)$]{\backdef \textit{Left panel}: Coupling strength $c(\omega,\hat{m})$ in setup \hyperlink{coupling:c1}{(C1)} (\textit{dashed line})
                                      and \hyperlink{coupling:c3}{(C3)} (\textit{solid band}), cf. Eq.\,\eqref{core:C}.
                                      The current-mass evolution is expressed in terms of the squared pion mass obtained from the pseudoscalar-meson BSE.
                                      \textit{Right panel}: Coupling strength in \hyperlink{coupling:c3}{(C3)} as a function of the width parameter $\omega$
                                      for five different current-quark masses. The current-mass dependent $\omega$ plateaus are reflected in \eqref{core:omegarange}.
                                      The dotted vertical line corresponds to $\Delta\omega=0$, the shaded region to $|\Delta\omega| < 0.1$.}
            \label{fig:c(m)}
            \end{center}
            \end{figure}

            \begin{figure}[p]
            \begin{center}
            \includegraphics[scale=0.9]{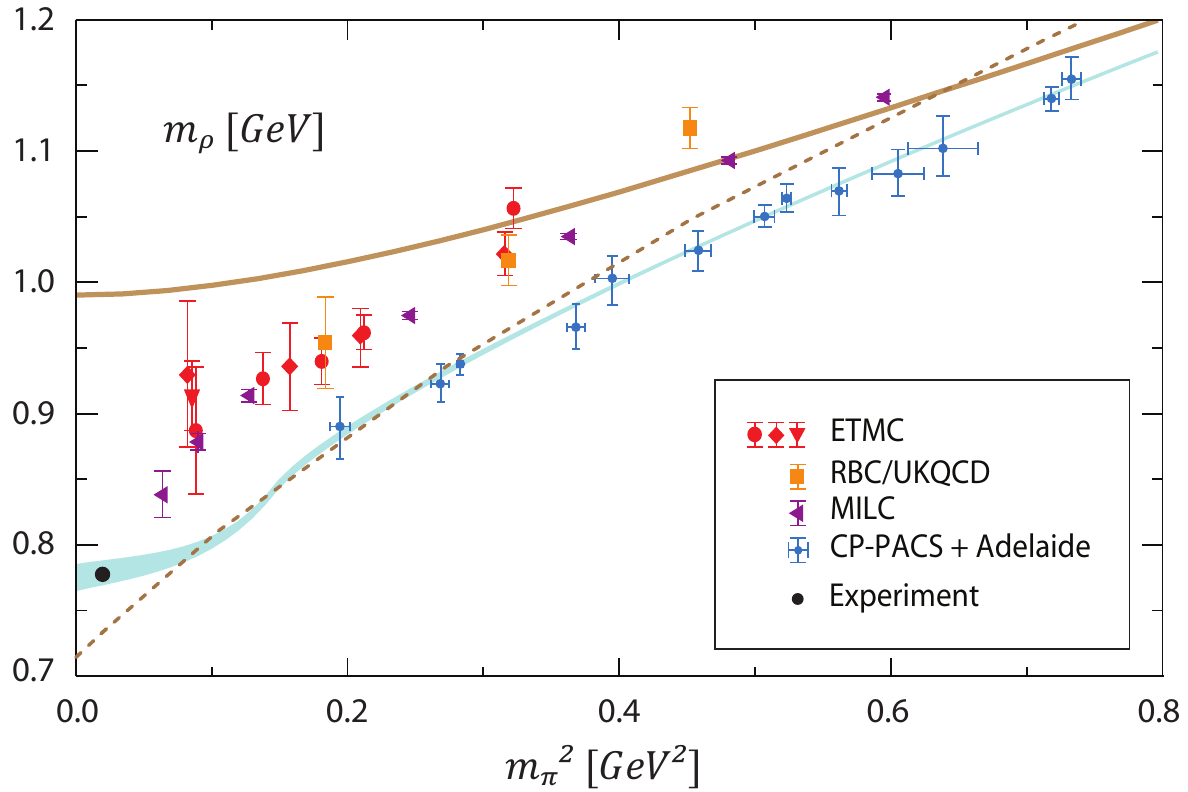}
            \caption[]{\backdef   Solution for $m_\rho(m_\pi^2)$ in setup \hyperlink{coupling:c1}{(C1)} (\textit{dashed line}) and \hyperlink{coupling:c3}{(C3)} (\textit{solid line}),
                                where the latter is identical to the parametrization of Eq.\,\eqref{core:mrho}.
                                The curves may be regarded as the input which defines the coupling strength $c(\hat{m})$ in both setups.
                                We compare to a selection of lattice data \cite{Boucaud:2007uk,Allton:2007hx,Bernard:2001av} extracted from Ref.\,\cite{McNeile:2007fu},
                                and the results of \cite{AliKhan:2001tx} together with a chiral extrapolation~\cite{Allton:2005fb} (\textit{blue band}).} \label{fig:mrho}
            \end{center}
            \end{figure}

                  \bigskip

                  \fatcol{(C3) Fit to the $\rho$-meson "quark core".}\hypertarget{coupling:c3}{}
                  A third practical strategy is to implement phenomenological assumptions about the nature and possible impact of corrections beyond RL upon \textit{hadron properties},
                  and to adjust the current-mass dependence of the effective coupling in a way where these corrections are explicitly \textit{missing}.

                  One important effect in the chiral and low-energy regime of QCD is imposed by pseudoscalar meson-cloud contributions, cf. Section~\ref{sec:qcd:phenomenology}.
                  Such corrections provide a substantial attractive contribution to the 'quark core' of hadronic observables in the chiral regime
                  whereas they vanish with increasing current-quark mass.
                  Their impact on the chiral structure of the quark mass function and condensate, $f_\pi$, $m_\rho$, and nucleon and $\Delta$ observables has been demonstrated
                  in the NJL-model \cite{Oertel:2000jp,Cloet:2008fw}, DSE studies \cite{Pichowsky:1999mu,Hecht:2002ej,Fischer:2007ze,Fischer:2008wy},
                  and chiral extrapolations of lattice results \cite{Young:2002cj}.
                  A sizeable reduction of $\rho$-meson, nucleon and $\Delta$ masses is paired with an increase of hadronic charge radii towards a logarithmic divergence in the chiral limit.
                  The latter effect is clearly missing in a RL truncation as can be inferred, e.g., from the BSE result for the pion charge radius displayed in Fig.~\ref{fig:fpi+rpi}.

                  A further resummation of non-resonant Abelian diagrams in the quark-gluon vertex and $q\bar{q}$ kernel provides additional attraction in the vector-meson channel
                  which decreases with increasing current-quark mass \cite{Watson:2004kd,Bhagwat:2004hn,Matevosyan:2006bk}.
                  On the other hand, an inclusion of the three-gluon vertex exhibits a substantial amount of repulsion \cite{Fischer:2009jm}
                  which suggests a non-perturbative cancellation mechanism beyond RL \cite{Bender:2002as}.
                  Nonetheless one may construct a rainbow-ladder 'quark core' which overestimates the experimental $\rho$-meson mass, most noticeably towards the chiral limit,
                  and resembles the hadronic quark core of chiral effective field theories which is subsequently dressed by chiral corrections.
                  Such an inflated quark core mass for $m_\rho$ has been used in \cite{Eichmann:2008ae} via (see Fig.~\ref{fig:mrho})
                      \begin{equation}\label{core:mrho}
                          x_\rho^2 = 1 + x_\pi^4/(0.6+x_\pi^2)\,, \quad x_\rho := m_\rho/m_\rho^0\,, \quad x_\pi := m_\pi/m_\rho^0
                      \end{equation}
                  with the chiral-limit value $m_\rho^0 = 0.99$ GeV. The sum of corrections beyond RL would then
                  reduce $m_\rho$ in the chiral limit by $\sim 25\%$ whereas the quark core contribution to $m_\rho$ approaches lattice results above the $s$-quark mass.
                  To reproduce Eq.\,\eqref{core:mrho} upon solving the $\rho$-meson BSE, the coupling strength $c$
                  of Eq.\,\eqref{dse:maristandy} must be equipped with the following current-mass dependence (see Fig.~\ref{fig:c(m)}):
                      \begin{equation}\label{core:C}
                           c(\omega, \hat{m}) = 0.11 +  \frac{ 0.86 \,b(\Delta\omega) }{ 1 + 0.885\,x_q + (0.474\,x_q)^2}, \quad x_q:=\hat{m}/(0.12\text{GeV}).
                      \end{equation}
                  At each value of the current-quark mass $\hat{m}$, the parametrization
                      \begin{equation}\label{core:B}
                           b(\Delta\omega) =  1 - 0.15\,\Delta\omega + (1.50\,\Delta\omega)^2 +(2.95\,\Delta\omega)^3
                      \end{equation}
                  eliminates the residual $\omega$ dependence of the BSE result for $m_\rho$ and ensures the validity of Eq.~\eqref{core:mrho}
                  in the range $\omega = \bar{\omega}(\hat{m}) \pm |\Delta\omega|$,
                  with the central value given by
                  \begin{equation}\label{core:omegarange}
                      \bar{\omega}(\hat{m}) = 0.38 + 0.17/(1+x_q)\,, \qquad  |\Delta\omega| \lesssim 0.1\,.
                  \end{equation}
                  As demonstrated in Refs. \cite{Eichmann:2008ae,Eichmann:2008ef} and this thesis,
                  the procedure induces consistent overestimated values for a range of $\pi$, $\rho$, nucleon and $\Delta$ observables as
                  obtained from their respective meson BSEs, quark-diquark BSEs and Faddeev equation.
                  The results are only weakly $\omega$-independent within the range \eqref{core:omegarange}.

            \bigskip

             By implementing either of the discussed models for the current-mass dependent coupling strength,
             the combination of quark DSE \eqref{dse:qdse}, meson BSE \eqref{bse:bse}, rainbow-ladder kernel \eqref{bse:rlkernel}, and
             effective coupling \eqref{dse:maristandy}
             completely determines the subsequent results.
                  In the course of this thesis, these results will be primarily presented
                  in the 'quark core' model \hyperlink{coupling:c3}{(C3)} and occasionally compared
                  to those obtained with a fixed coupling strength \hyperlink{coupling:c1}{(C1)}.
                  Shaded bands denote the sensitivity to a variation of $\omega$ which we consider for the
                  \hyperlink{coupling:c3}{(C3)} setup only; all results corresponding to \hyperlink{coupling:c1}{(C1)} are
                  plotted with a current-mass independent value $\conjg{\omega}=0.4$.
                  A connection between the two setups will be established in the next section.


\section{Analysis of rainbow-ladder meson results}\label{sec:MESON:results}

\begin{figure}[p]
            \begin{center}
            \includegraphics[scale=0.4]{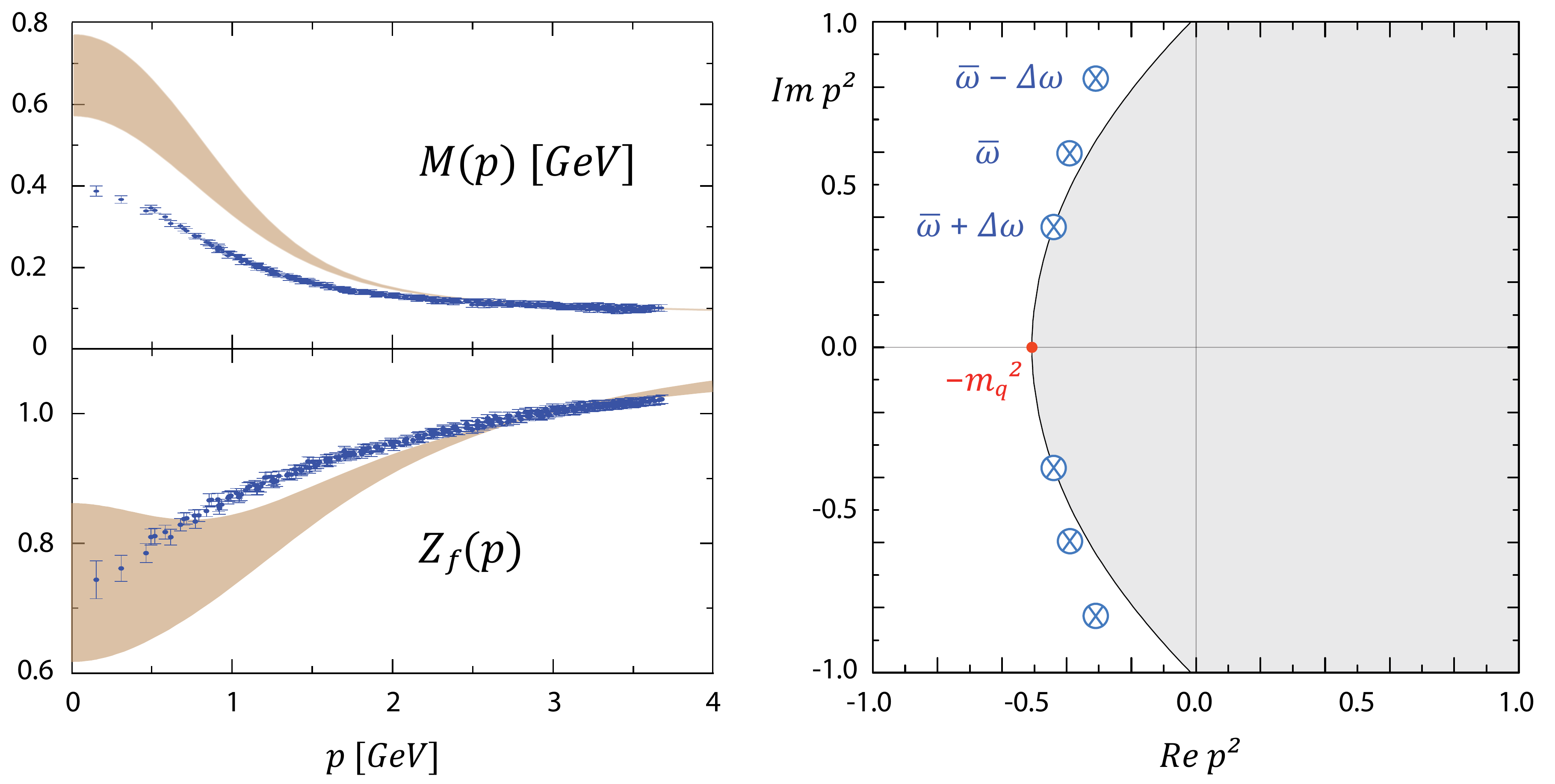}
            \caption[DSE in the complex plane]{\backdef
                                               \textit{Left panel}: Quark mass function $M(p^2)$ and wave-function renormalization $Z_f(p^2)$
                                               in setup \hyperlink{coupling:c3}{(C3)}, at a current-mass value $\hat{m}$ which corresponds to $m_\pi=0.63$ GeV,
                                               compared to lattice results \cite{Bowman:2005vx}.
                                                A renormalization point $\mu=2.9$ GeV  was chosen.
                                               The lower (upper) edge of the bands is related to the largest (smallest) value of $\omega$ for $M(p^2)$,
                                               and vice versa for $Z_f(p^2)$.
                                               \textit{Right panel}: Complex conjugate poles of the quark propagator in the complex $p^2$ plane for different values of $\omega$.
                                               The apex of the corresponding parabola \eqref{APP:SING:polemass} defines the $\omega$-dependent 'pole mass' $m_q$.} \label{fig:MZ}
            \end{center}
            \end{figure}

    \bigskip
    \fatcol{Quark propagator.}
            A characteristic feature of QCD is the dynamically generated enhancement of the quark mass function $M(p^2)$  at small momenta, visible in  Fig.~\ref{fig:MZ}.
            The resulting constituent-mass scale, e.g. $M(p^2=0)$, is typically several hundred MeV larger than the current-quark mass which
            is the input of the DSE. In addition, Fig.~\ref{fig:MZ} illustrates the impact of the overestimated quark-core model \hyperlink{coupling:c3}{(C3)} on the quark propagator
            dressing functions $M(p^2)$ and $Z_f(p^2)$. The resulting mass function at infrared momenta is considerably larger than the respective lattice results.  
            An implementation of pionic effects in the $q\bar{q}$ kernel reduces this difference \cite{Fischer:2007ze,Fischer:2008sp}.

            \begin{table}[p]
                    \begin{center}
                        \renewcommand{\arraystretch}{1.0}
                        \begin{tabular}{l||c|c||c|c|c|c|c}
                                                        & $\Lambda_\textrm{IR}$  &  $\eta$      &  $\langle\bar{q}q\rangle_\text{1 GeV}^{1/3}$      &  $f_\pi$      &  $f_\rho$    &  $m_\rho$    &  $r_\pi$   \\[0.1cm] \hline
                                                        &                        &              &                                                            &               &              &              &            \\[-0.4cm]
                         Phen./Exp.                     &                        &              &  $0.236$                                                   &  $0.131$      &  $0.216$     &  $0.77$      &  $0.67$     \\ \hline
                                                        &                        &              &                                                            &               &              &              &            \\[-0.4cm]
                         \hyperlink{coupling:c1}{(C1)}  & $0.72$                 &  $1.8$       &  $0.235$                                                   &  $0.131$      &  $0.208$     &  $0.73$      &  $0.66$     \\
                         \hyperlink{coupling:c3}{(C3)}  & $0.98$                 &  $1.8(2)$    &  $0.319$                                                   &  $0.176$      &  $0.280(6)$  &  $0.99$      &  $0.49$  \\[0.1cm] \hline
                                                        &                        &              &                                                            &               &              &              &            \\[-0.4cm]
                         Ratio                          & $0.73$                 &              &  $0.74$                                                    &  $0.74$       &  $0.74$      &  $0.74$      &  $0.74$     \\ \hline
                        \end{tabular}
                         \caption[]{\backdef
                             Comparison of the quark condensate, $\pi$ and $\rho$ decay constants, $\rho$-meson mass, and pion charge radius
                             in setups \hyperlink{coupling:c1}{(C1)} and \hyperlink{coupling:c3}{(C3)}, characterized by the parameters $\Lambda_\textrm{IR}$ and $\eta$ defined in \eqref{MESON:c+eta}.
                             A variation of $\eta = 1.8 \pm 0.2$ in \hyperlink{coupling:c3}{(C3)} coincides with $\omega \approx \conjg{\omega} \pm 0.06$.
                             The results correspond to a current mass $\hat{m}=6.1$ MeV which is related to the
   			              physical pion mass $m_\pi=138$ MeV. Experimental or phenomenological values are quoted in the first row.
                             In the first three rows, $\eta$ is dimensionless and $r_\pi$ is given in fm, all other units are GeV.
                             The last row plots the ratios of sets \hyperlink{coupling:c1}{(C1)} and \hyperlink{coupling:c3}{(C3)}.  }  \label{tab:meson:masses}
                    \end{center}
            \end{table}

            \begin{figure}[tbp]
            \begin{center}
            \includegraphics[scale=0.43]{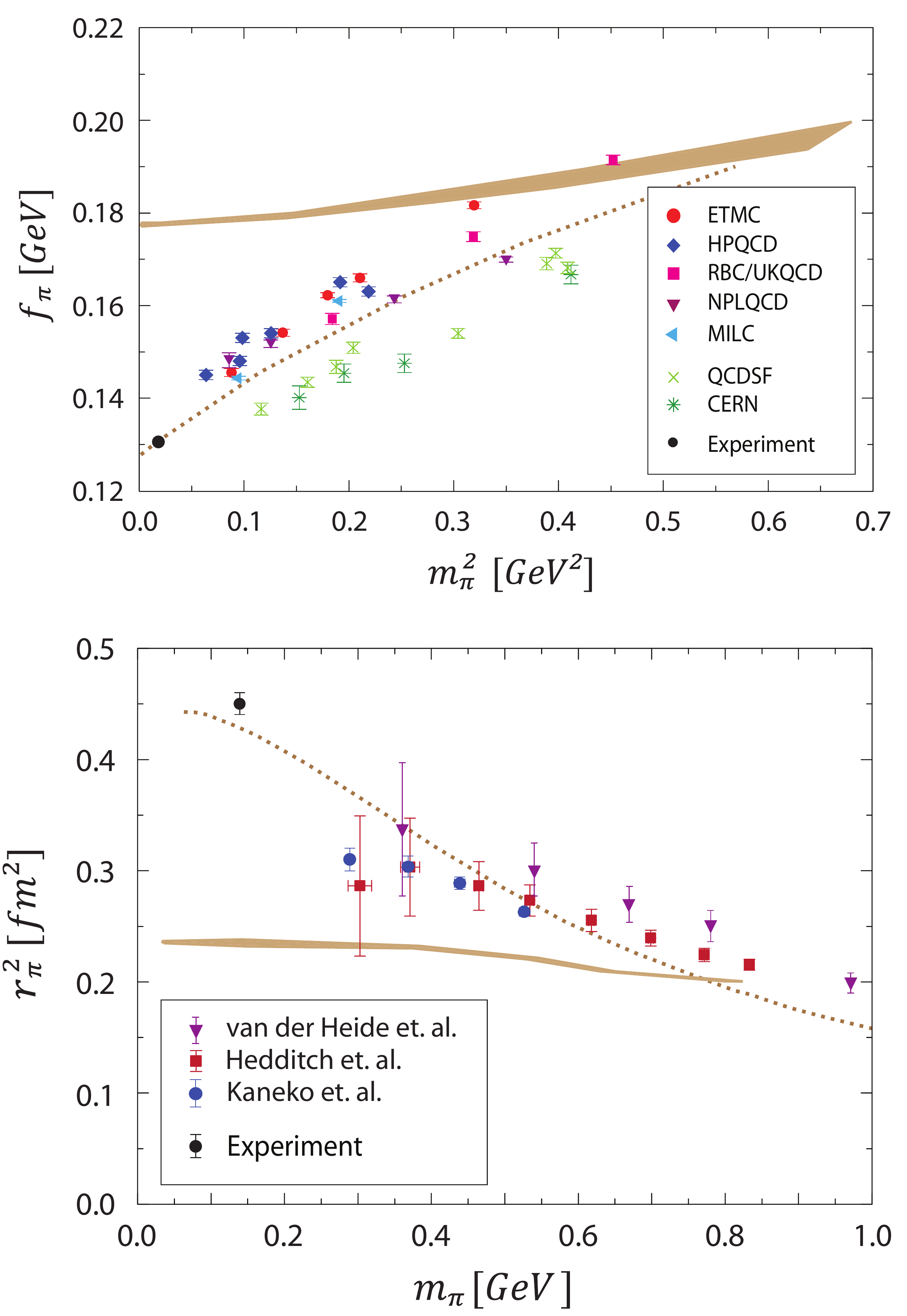}
            \caption[]{\backdef Pion-mass evolution of pion decay constant (\textit{upper panel}) and squared pion charge radius (\textit{lower panel})
                              in setup \hyperlink{coupling:c1}{(C1)} (\textit{dashed lines}) and \hyperlink{coupling:c3}{(C3)} (\textit{shaded bands}).
                              The compilation of lattice results for $f_\pi$ \cite{Boucaud:2007uk,Allton:2007hx,Follana:2007uv,Beane:2006kx,Bernard:2007ps,Gockeler:2006ns,DelDebbio:2006cn,DelDebbio:2007pz}
                              was extracted from Ref.~\cite{McNeile:2007fu}. Lattice data for $r_\pi^2$ are from Refs.~\cite{vanderHeide:2003kh,Hedditch:2007ex,Kaneko:2007nf}
                              .} \label{fig:fpi+rpi}
            \end{center}
            \end{figure}

           The singularity structure of the resulting quark propagator, i.e., of its denominator $1/\left(p^2+M^2(p^2)\right)$, is that of complex conjugate poles,
           a feature which may be an artifact of the rainbow truncation but has been found in more general truncations beyond rainbow-ladder as well \cite{Fischer:2008sp,Alkofer:2008tt}.
           The pole positions depend on the infrared width $\omega$, see Fig.~\ref{fig:MZ}: a larger value $\omega$ forces the poles closer to the timelike axis while
           a smaller value shifts them further into the complex plane.
           The resulting trajectory is similar to the boundary of the parabolic integration domain which is needed in the subsequent bound-state equations.
           It constitutes the singularity limitations encountered in the calculation of form factor diagrams, cf. App.~\ref{app:singularities}.

           The vacuum quark condensate can be calculated from Eqs.~(\ref{dse:asymptoticmassf}--\ref{QUARK:M+Cond}).
           The value shown in Table~\ref{tab:meson:masses} is obtained from the perturbative tail of the chiral-limit mass function and evolved to the scale $\mu=1$ GeV via
           the one-loop formula \eqref{QUARK:M+Cond}.
           As discussed in connection with Eq.~\eqref{MESON:IRscale1}, the quark-core setup \hyperlink{coupling:c3}{(C3)} operates with an inflated scale of dynamical chiral symmetry breaking.
           This overestimation compared to the phenomenological value is visible in the quark condensate and consistent with the solution for the quark mass function.
           Pionic contributions to the condensate are attractive, cf. Refs.~\cite{Oertel:2000jp,Fischer:2007ze}.

    \bigskip
    \fatcol{Meson properties.}
            A solution of the pseudoscalar and vector-meson BSEs provides the respective bound-state amplitudes which are subsequently used to calculate
            physical meson properties. Table~\ref{tab:meson:masses} collects results for $\pi$ and $\rho$ decay constants, obtained from Eqs.\,\eqref{bse:pdc},
            and the pion charge radius, Eq.\,\eqref{MESON:pion-charge-radius}. One observes a uniform response to the inflated quark core for $m_\rho$:
            expressed in units of mass, the tabulated quantities related to the input \hyperlink{coupling:c3}{(C3)}
            consistently overestimate their experimental values by $\sim 30-35\%$.
	        Moreover, $f_\pi$ and $r_\pi$ tend to approach lattice results for heavier quarks (see Fig.~\ref{fig:fpi+rpi}).
            This validates the notion of a pseudoscalar meson cloud which increases a hadron's charge distribution towards the chiral limit where
            the charge radius would diverge.

        \bigskip
        \fatcol{Relation between the models.}
                Simple relations between the setups \hyperlink{coupling:c1}{(C1)}, \hyperlink{coupling:c2}{(C2)} and \hyperlink{coupling:c3}{(C3)} can be established in the chiral limit.
                To this end it is beneficial to rewrite the infrared part of the coupling \eqref{dse:maristandy} in a more suggestive way.
                Upon replacing the infrared parameters $c$, $\omega$ and $\Lambda_0$ by two new parameters $\Lambda_\text{IR}$ and $\eta$,
                defined via
                \begin{equation}\label{MESON:c+eta}
                    c =: \left(\frac{\Lambda_\text{IR}}{\Lambda_0}\right)^3, \quad \omega =: \frac{1}{\eta}\left(\frac{\Lambda_\text{IR}}{\Lambda_0}\right),
                \end{equation}
                the infrared contribution to the effective coupling $\alpha(k^2)$ is expressed through
                    \begin{equation}
                        \alpha_\textrm{IR}(k^2) = \pi \,\eta^7 x^2 e^{-\eta^2 x}, \quad x = k^2/\Lambda_\textrm{IR}^2.
                    \end{equation}
                $\Lambda_\text{IR}$ only appears in the denominator of $x$, hence it is the only dimensionful scale of the rainbow-ladder truncated DSE-BSE system
                in the chiral limit as long as the UV part of the coupling is not taken into account.
                It represents the scale of dynamical chiral symmetry breaking which, in a coupled solution of quark, gluon and ghost DSEs, would be generated self-consistently.
                As a consequence, mass-dimensionful quantities which are sensitive to the infrared properties
                scale with $\Lambda_\text{IR}$.
                The chiral-limit values of $\Lambda_\text{IR}$ are
                \begin{equation} \label{MESON:IRscale1}
                    \text{\hyperlink{coupling:c1}{(C1):}} \quad  \Lambda_\text{IR} = 0.72\,\text{GeV}, \quad  \text{\hyperlink{coupling:c3}{(C3):}} \quad  \Lambda_\text{IR} = 0.98\,\text{GeV},
                \end{equation}
                which makes clear that the model \hyperlink{coupling:c1}{(C1)} for a given set of observables
                will, upon entering its 'core version' \hyperlink{coupling:c3}{(C3)}, produce results which are overestimated by the same percentage.
                Dimensionless chiral-limit ratios between the two setups will be equal, namely $(0.72)/(0.98) = 0.73$.
                Table~\ref{tab:meson:masses} demonstrates that this scaling property still persists at the physical $u/d$ mass
                where the disturbance from a non-vanishing current-quark mass is negligible.

                The second parameter $\eta$ replaces the coupling width $\omega$. An insensitivity of observables with respect to a variation of the width $\omega$ at a certain coupling
        	    strength $c$ translates into an invariance with respect to $\eta$ at a fixed scale $\Lambda_\text{IR}$.
                The combined change of $c$ and $\omega$
                from \hyperlink{coupling:c1}{(C1)} to \hyperlink{coupling:c3}{(C3)} according to Eqs.~(\ref{core:C}--\ref{core:B}) is equivalent to a rescaling of $\Lambda_\text{IR}$ where $\eta$ is essentially unchanged.

\begin{figure}[p]
                \begin{center}
                \includegraphics[scale=0.35]{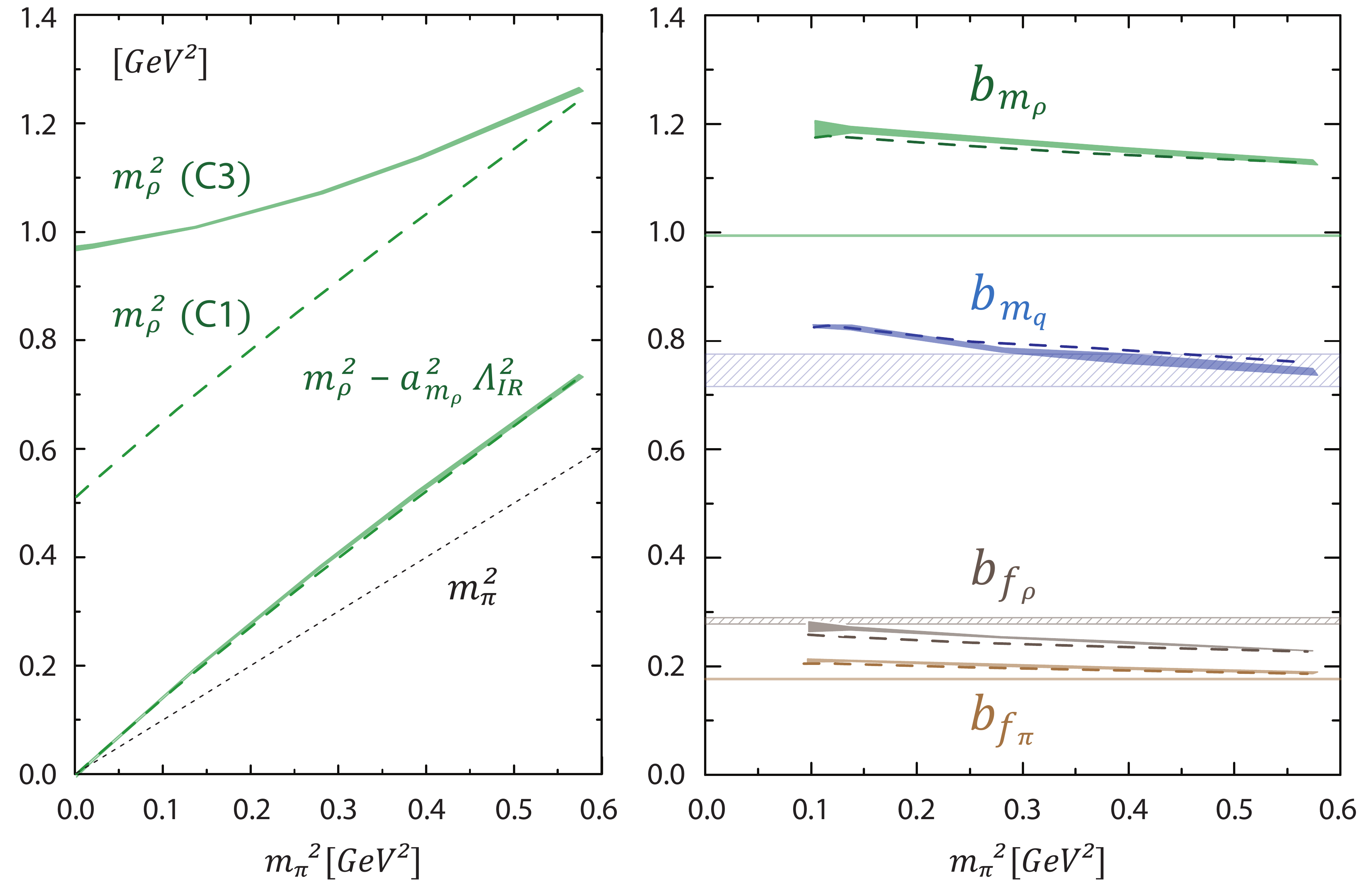}
                \caption[]{\backdef \textit{Left panel:} $m_\rho^2$ and the difference $m_\rho^2 - a_\rho^2 \,\Lambda_\text{IR}^2$
                           for both model versions \hyperlink{coupling:c1}{(C1)} and \hyperlink{coupling:c3}{(C3)}, compared to $m_\pi^2$.
                           \textit{Right panel:} The values of $b_U$ extracted from Eq.\,\eqref{eq:mesons:massformula}
                           for the four calculated quantities $f_\pi$, $f_\rho$, $m_q$ and $m_\rho$.
                           The meaning of the quark pole mass $m_q = \text{Im}\sqrt{p^2}_\text{pole}$ is clarified in Fig.~\ref{fig:MZ}.
                           Shaded bands: \hyperlink{coupling:c3}{(C3)}; dashed curves: \hyperlink{coupling:c1}{(C1)} for fixed $\eta=1.8$.
                           The horizontal lines (or shaded areas, respectively) depict the chiral-limit values $a_U(\eta)$ for comparison.
                           Qualitatively similar features are found for diquark and baryon masses. } \label{fig:modelindep}
                \end{center}
                \end{figure}

        \bigskip
        \fatcol{A rainbow-ladder 'mass formula'.}
                The analysis can be extended to finite current-quark masses where the additional scale $\hat{m}$ or, equivalently, the pion mass $m_\pi$ is introduced.
                An investigation of the numerical results exhibits that quantities with the dimension of a squared mass, generically denoted by $U$,
                allow for a scale separation in the following way  (see Fig.~\ref{fig:modelindep}):
                \begin{equation}\label{eq:mesons:massformula}
                    M_U^2(m_\pi^2) \approx a_U^2(\eta) \,\Lambda_\text{IR}^2 + b_U^2(m_\pi^2) \, m_\pi^2\,.
                \end{equation}

                \noindent
                For a vanishing pion mass, the first term in \eqref{eq:mesons:massformula} reproduces the observation from above:
                masses scale with $\Lambda_\text{IR}$, and the dimensionless chiral-limit values $a_U$ may depend on the infrared
                width parameter $\eta$. For a finite current-quark mass, the second term is found to be insensitive with regard to the infrared parameters $\Lambda_\text{IR}$ and $\eta$:
                it only depends on $m_\pi$ and the remaining scales $\Lambda_{QCD}$ and $\Lambda_0$ in the coupling.

                The dimensionless coefficients $b_U$ are generic functions of $m_\pi$ which account for any additional pion-mass dependence of the results.
                In the heavy-quark limit they would represent the dimensionless values of $M_U$ in units of $m_\pi$.
                A BSE analysis of heavy-meson observables \cite{Ivanov:1998ms,Bhagwat:2006xi,Krassnigg:priv} confirms that meson masses become proportional to $m_\pi$,
                such that e.g. $b_{m_\rho}(\infty) \rightarrow 1$, whereas decay constants behave as $\sim \sqrt{m_\pi}$ which implies that $b_{f_\pi}$ and $b_{f_\rho}$ vanish with an inverse square root of the pion mass.

                We tested Eq.~\eqref{eq:mesons:massformula} up to $\hat{m} \approx 200$ MeV for a range of quantities
                \begin{equation}
                    U = M(0),\, m_q,\, m_\rho,\, f_\pi,\, f_\rho,\, 1/r_\pi,\, m_\text{sc},\, m_\text{av},\, M_N,\, M_\Delta,
                \end{equation}
                where the diquark and baryon masses are determined in Chapter~\ref{chapter:quark-diquark}.
                The relation breaks down below $\Lambda_\text{IR}\lesssim 0.5$ GeV which is related to the threshold of dynamical chiral symmetry breaking in the quark DSE.
                In the domain of its validity,
                a rescaling of $\Lambda_\text{IR} \longrightarrow \Lambda_\text{IR}'$
                in the infrared part of the effective coupling \eqref{dse:maristandy}
                produces a current-mass independent \textit{additive} contribution to \textit{squared} masses.
                This infrared component may however depend on the parameter $\eta$.
                In the 'core model' \hyperlink{coupling:c3}{(C3)}, $\Lambda'_\text{IR}$ was chosen pion-mass dependent;
                nonetheless Eq.\,\eqref{eq:mesons:massformula} provides an estimate for its results:
                \begin{equation}
                    {M'_U}^2(m_\pi^2) \approx M_U^2(m_\pi^2) + a_U^2(\eta) \left( {\Lambda'}^2_\text{IR}(m_\pi^2) - \Lambda_\text{IR}^2 \right) .
                \end{equation}

\begin{figure}[p]
                \begin{center}
                \includegraphics[scale=0.35]{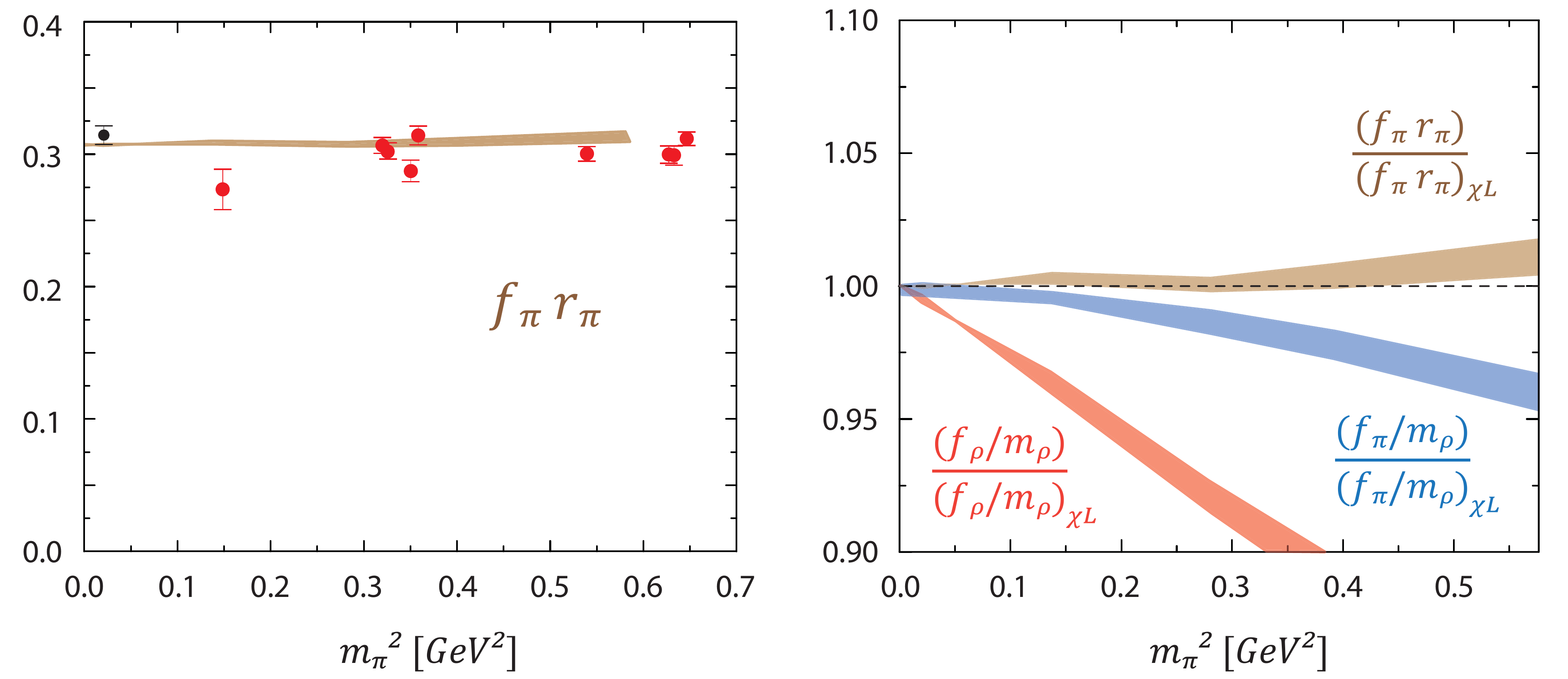}
                \caption[]{\backdef \textit{Left panel:} The dimensionless product $f_\pi r_\pi$ as a function of $m_\pi^2$, compared to the experimental value
                                                         and lattice results; figure taken from \cite{Cloet:2008fw}.
                                    \textit{Right panel:} Pion-mass evolution of several ratios, normalized by their chiral-limit value.} \label{fig:ratios1}
                \end{center}
                \end{figure}

                Moreover, if two quantities $U$ and $V$ independently satisfy $a_U \approx b_U$ and $a_V \approx b_V$, then
                their ratio $M_U/M_V \approx a_U/a_V$ will only weakly depend on $m_\pi^2$ and $\Lambda_\text{IR}$.
                This is most remarkably realized for the dimensionless product $f_\pi r_\pi$ (Fig.\,\ref{fig:ratios1}):
                for the pion-mass range under consideration, it is constant in $m_\pi^2$,
                identical in \hyperlink{coupling:c1}{(C1)} and \hyperlink{coupling:c3}{(C3)} (i.e.,~independent of $\Lambda_\text{IR}$),
                insensitive to $\eta$ (since both $f_\pi$ and $r_\pi$ are very weakly $\eta$-dependent, see Fig.~\ref{fig:fpi+rpi}),
                and it agrees with the experimental value and lattice-QCD results \cite{Cloet:2008fw}.

  \chapter{Baryons: Three-body equation} \label{chapter:faddeev}

            The three-body bound-state problem has a longstanding history which dates back to the original work by Faddeev \cite{Faddeev:1960su}.
            Non-relativistic Faddeev equations have found widespread application in the description of three-nucleon systems, see Ref.~\cite{Gloeckle:1995jg} for an overview.
            The covariant generalization of the Faddeev equation to the three-body analogue of a Bethe-Salpeter equation was formulated in Refs. \cite{Taylor:1966zza,Boehm:1976ya}; a
            comprehensive introduction can be found in \cite{Loring:2001kv}.
            Within the framework of Section \eqref{sec:qcd:bses}, the equation describes the baryon as a bound state of three spin-$\nicefrac{1}{2}$ valence quarks
            where the interaction kernel comprises two- and three-quark contributions.

            A solution of the covariant three-body equation requires knowledge of the dressed quark propagator and the three-quark kernel;
            and a specification of the Poincar\'e-covariant baryon amplitude.
            The relativistic spin structure of the latter has been explored in \cite{Machida:1974xw,Henriques:1975uh}
            and described in the light-front formalism in \cite{Weber:1986qw,Beyer:1998xy,Karmanov:1998jp,Sun:2001ir}.
            A complete classification according to the Lorentz group and the permutation group $\mathbb{S}_3$ was derived in \cite{Carimalo:1992ia} in terms of covariant three-spinors.
            Their analogues in the form of Dirac tensors, kindred to the decomposition of Green functions and meson amplitudes encountered in previous chapters, will be stated
            below (cf. Table~\ref{tab:faddeev:basis}).

            The complexity of the three-body bound state equation has so far prevented a direct numerical solution.
            Upon implementing perturbative quark propagators 
            it has been studied, for instance, in the works of Refs.\,\cite{Kielanowski:1979eb,Falkensteiner:1981ab}, in the context of
            a three-body spectator approximation \cite{Stadler:1997iu}, or a Salpeter equation with instantaneous forces \cite{Loring:2001kv}.
            The corresponding equation of a scalar three-particle system with scalar two-body exchange based on the Wick-Cutkosky model
            \cite{Wick:1954eu,Cutkosky:1954ru} was recently investigated and compared to the light-front approach \cite{Karmanov:2008bx}.
            An appealing strategy to simplify the three-quark problem while maintaining full Poincar\'e covariance is provided by the quark-diquark model
            which will be discussed in Chapter~\ref{chapter:quark-diquark}.

            The present chapter is devoted to a novel solution of the three-quark bound-state equation of the nucleon
            where the full Dirac structure of the covariant amplitude is taken into account.
            The numerical computation is performed upon truncating the kernel to a rainbow-ladder gluon exchange 
            which allows for a direct implementation of the effective quark-gluon coupling~$\alpha(k^2)$ introduced in Chapter~\ref{sec:mesons}.
            The resulting current-mass evolution of the nucleon mass, when compared to lattice data, exhibits a  behavior which is qualitatively similar to that of the vector meson in the same approach.

            A future extension to more sophisticated interaction kernels is certainly possible. This involves
            an inclusion of irreducible three-body forces, supposedly dominated by a three-gluon coupling to any of the three quark lines, 
            and a generalization towards a quark-quark interaction beyond rainbow-ladder.
            In the context of meson studies, the latter has attracted a considerable amount of attention in recent years, and
            an implementation of the findings in the three-body approach will certainly provide further insights into the physics of baryons.


\section{Faddeev amplitude and equation}

            Baryons appear as poles in the three-quark scattering matrix.
            The derivation pursued in Section~\ref{sec:qcd:bses} leads to a relativistic three-body bound-state equation
            which is the direct analogue of the Bethe-Salpeter equation in the quark-antiquark channel:
            \begin{equation}
                \Psi = \widetilde{K}^\text{(3)}\,\Psi\,, \qquad \widetilde{K}^{(3)} = \widetilde{K}^\text{(3)}_\text{irr} + \sum_{i=1}^3 \widetilde{K}_i^{(2)}\,,
            \end{equation}
            where $\Psi$ is the baryon's bound-state amplitude.
            The respective 3-body kernel $K^{(3)}$, stated in Eq.\,\eqref{FADDEEV:3body-kernel},
            comprises a 3-quark irreducible contribution and the sum of permuted two-quark kernels, where the subscript $i$ denotes the respective accompanying spectator quark.
            The quark-antiquark analogue of the two-body kernel $\widetilde{K}^{(2)}$
            appears in the meson BSE \eqref{bse:bse} where we employed a ladder truncation together with a rainbow-truncated quark propagator.

            It has been the guiding assumption for the quark-diquark model that correlations of two quarks
            provide the dominant binding structure in baryons.
            This was inspired by noticing that colored two-quark states
            can appear in an $SU(3)_C$ anti-triplet or sextet configuration
            where the former, in combination with a color-triplet quark, allows for the formation of a color-singlet nucleon.
             The observation motivates
             the omission of the three-body irreducible contribution from the full 3-body kernel.
             The resulting relativistic Faddeev equation includes a permuted sum of two-body $qq$ kernels: 
            \begin{fshaded2}
            \begin{equation}\label{bs:faddeevtruncated}
                \Psi = \sum_{i=1}^3 \widetilde{K}^{(2)}_i \Psi \,,
            \end{equation}
            \end{fshaded2}

            \noindent
            which enables an adoption of the formalism established in Chapter~\ref{sec:mesons} in the three-body problem.
            Eq.\,\eqref{bs:faddeevtruncated} represents a technical simplification as well since the  equation
            merely includes one momentum loop and becomes computationally tractable.

 \renewcommand{\arraystretch}{1.7}

\begin{figure}[tbp]
                    \begin{center}
                    \includegraphics[scale=0.152]{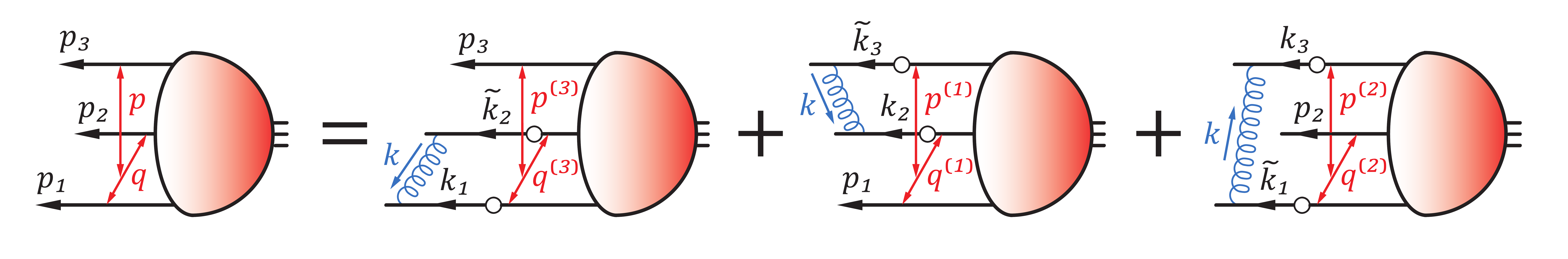}
                    \caption{\backdef Faddeev equation \eqref{faddeev:eq} in rainbow-ladder truncation.}\label{fig:faddeev}
                    \end{center}
            \end{figure}

    \bigskip
    \fatcol{Nucleon amplitude.}
            The bound-state amplitude $\Psi_{\alpha\beta\gamma\delta}$ of a nucleon carries 3 spinor indices $\{\alpha,\beta,\gamma\}$ for the involved valence quarks
            and one index $\delta$ for the spin-$1/2$ nucleon. It depends on three quark momenta $p_1$, $p_2$, $p_3$ which may be reexpressed in terms of
            the total momentum $P$ and two relative Jacobi momenta $p$ and $q$. They are related via:
                \begin{equation}
                \begin{array}{l}  p = (1-\eta)\,p_3 - \eta\,p_d\,,  \\[0.2cm]
                                  q = \dfrac{p_2-p_1}{2}\,,  \\
                                  P = p_1+p_2+p_3\,,
                \end{array} \qquad\quad
                \begin{array}{l}  p_1 = -q + \dfrac{p_d}{2} = -q -\dfrac{p}{2} + \dfrac{1-\eta}{2}\, P\,, \\[0.2cm]
                                  p_2 =  q + \dfrac{p_d}{2} =  q -\dfrac{p}{2} + \dfrac{1-\eta}{2}\, P\,, \\
                                  p_3 =  p + \eta\, P\,,
                \end{array}
                \end{equation}
            where we abbreviated $p_d:=p_1+p_2$.
            The relations between the momenta can also be inferred from Fig.\,\ref{fig:faddeev}.
            We have chosen equal momentum partitioning $1/2$ for the relative momentum $q$ 
            and use the value $\eta=1/3$ in connection with the momentum $p$
            which maximizes the upper boundary for the nucleon mass with respect to singularity restrictions (see App.~\ref{app:singularities}).

            The nucleon amplitude can be decomposed into a certain number of Dirac structures:
                \begin{equation}\label{faddeev:amp}
                    \Psi_{\alpha\beta\gamma\delta}(p,q,P) = \sum_{i=1}^{64} f_i(p^2, q^2, \{z\}) \,\tau_i(p,q,P)_{\alpha\beta\gamma\delta}.
                \end{equation}
            The amplitude dressing functions $f_i$ depend on the five Lorentz-invariant combinations
            \begin{equation}
                p^2\,, \quad q^2\,,\quad   z_0=\widehat{p_T}\cdot\widehat{q_T} \,,\quad z_1 = \hat{p}\cdot\hat{P} \,,\quad  z_2 = \hat{q}\cdot\hat{P}\,,
            \end{equation}
            where a hat denotes a normalized 4-vector and $p_T^\mu = T^{\mu\nu}_P p^\nu$ a transverse projection ($T^{\mu\nu} = \delta^{\mu\nu} - \hat{P}^\mu \hat{P}^\nu$).
            We abbreviated the angular variables by the shorthand notation $\{z\}=\{z_0,z_1,z_2\}$.
            The total momentum-squared $P^2 = -M^2$ is fixed since the nucleon is on its mass shell.
            The Dirac structures $\tau_i(p,q,P)$ will be explained in Section~\ref{sec:faddeev:basis}.

 \renewcommand{\arraystretch}{1.2}

     \bigskip
     \fatcol{Faddeev equation.}
             Using the above kinematics, the full Dirac and momentum dependence of the Faddeev equation \eqref{bs:faddeevtruncated}
             reads (see also Fig.\,\ref{fig:faddeev}):
                \begin{equation}\label{faddeev:eq}
                \begin{split}
                    \Psi_{\alpha\beta\gamma\delta}(p,q,P) = \int_k \Big[  \,\, &  K_{\alpha\alpha'\beta\beta'}(k) \, S_{\alpha'\alpha''}(k_1)  \, S_{\beta'\beta''} (\widetilde{k}_2)\,
                                                                             \Psi_{\alpha''\beta''\gamma\delta}(p^{(3)},q^{(3)},P) \\[-0.35cm]
                                                                        +\, &  K_{\beta\beta'\gamma\gamma'}(k) \, S_{\beta'\beta''}(k_2)  \, S_{\gamma'\gamma''} (\widetilde{k}_3)\,
                                                                             \Psi_{\alpha\beta''\gamma''\delta}(p^{(1)},q^{(1)},P) \\[0.1cm]
                                                                        +\, &  K_{\gamma\gamma'\alpha\alpha'}(k) \, S_{\gamma'\gamma''}(k_3)  \, S_{\alpha'\alpha''} (\widetilde{k}_1)\,
                                                                             \Psi_{\alpha''\beta\gamma''\delta}(p^{(2)},q^{(2)},P) \, \Big]\,,
                \end{split}
                \end{equation}
            where we already anticipated a two-quark kernel $K$ that only depends on the gluon momentum $k$ 
            which is also used as the loop-integration variable.
            The quark propagators $S$ depend on the internal quark momenta
            $k_i=p_i-k$ and $\widetilde{k}_j=p_j+k$,
            and the internal relative momenta are given by
                \begin{equation}
                \begin{array}{l}  p^{(1)} = p+k\,, \\
                                  p^{(2)} = p-k\,, \\
                                  p^{(3)} = p\,,
                \end{array} \qquad\quad
                \begin{array}{l}  q^{(1)} = q-k/2\,, \\
                                  q^{(2)} = q-k/2\,, \\
                                  q^{(3)} = q+k\,.
                \end{array}
                \end{equation}
            Specifically, we use the rainbow-ladder kernel of Eq.\,\eqref{bse:rlkernel}:
                \begin{equation}
                    K_{\alpha\alpha'\beta\beta'}(k) = Z_2^2 \,\frac{4\pi\alpha(k^2)}{k^2} \, T^{\mu\nu}_k \,\gamma^\mu_{\alpha\alpha'} \,\gamma^\nu_{\beta\beta'}\,,
                \end{equation}
            which involves two bare quark-gluon vertices and a gluon propagator, subsumed into the effective coupling $\alpha(k^2)$ of Eq.\,\eqref{dse:maristandy}.
            Tracing the color structure induced by the quark-gluon vertices leads to a color factor $2/3$ in front of the integral \eqref{faddeev:eq}.


 \renewcommand{\arraystretch}{1.4}

     \section{Dirac basis covariants} \label{sec:faddeev:basis}

            A general Green function with 4 fermion legs which depends on 3 independent momenta involves 256 independent components.
            The subspaces corresponding to positive and negative parity consist of 128 Dirac structures each.
            A possible linearly independent basis for the positive-parity and positive-energy nucleon is given by the 64 components
             \begin{equation}\label{faddeev:basisSA}
             \begin{array}{lrlrl}
                \mathsf{S}_{ij}^{\pm}(p,q,P) :=   &               & (\Gamma_i \,\Lambda_\pm) \,  (\gamma_5 C) \, \otimes &               & (\Gamma_j \,\Lambda_+)\,,    \\
                \mathsf{P}_{ij}^{\pm}(p,q,P) :=   & \gamma_5  \!\!& (\Gamma_i \,\Lambda_\pm) \,  (\gamma_5 C) \, \otimes & \gamma_5 \!\! & (\Gamma_j \,\Lambda_+)\,,
             \end{array}
             \end{equation}
             where the $\Gamma_i(p,q,P)$ with $i=1\dots 4$, given in Eq.\,\eqref{FADEEV:Gamma_i}, carry the relative-momentum dependence,
             $\Lambda_+(P) = (\mathds{1}+\hat{\Slash{P}})/2$ is the nucleon's positive-energy projector
             and $C=\gamma^4\,\gamma^2$ the charge-conjugation matrix.
             The notation, e.g. $(\mathsf{S}_{11}^{+})_{\alpha\beta,\gamma\delta}= (\Lambda_+ \gamma_5 C)_{\alpha\beta} (\Lambda_+)_{\gamma\delta}$,
             refers to two outgoing quark legs with indices $\alpha,\beta$ on the left-hand side of the tensor product (hence the $(\gamma^5 C)$ insertion, cf.
             Eq.\,\eqref{dq:amps}) and an outgoing quark and incoming nucleon leg (indices $\gamma$, $\delta$) on the right-hand side, where the positive-energy projector is attached to the latter.
             The ordering of indices in the tensor product is insignificant since, for a complete basis, any permuted version can be expressed
             by a sum of basis elements in the above "canonical" ordering using Fierz identities, see e.g.~\cite{Karmanov:1998jp}.

             Each subspace of definite parity and sign of energy, corresponding to the nucleon's $(1/2)^+$, $\overline{(1/2)^+}$, $(1/2)^-$, $\overline{(1/2)^-}$ states,
             includes 64 covariants. The negative-parity basis elements are obtained by attaching a factor $\gamma^5$ to either left or right-hand side of Eq.\,\eqref{faddeev:basisSA};
             the negative-energy (antibaryon) structures by replacing $\Lambda_+$ on the r.h.s with the negative-energy projector $\Lambda_- = (\mathds{1}-\hat{\Slash{P}})/2$.
             The symbols $\mathsf{S}$ and $\mathsf{P}$ in Eqs.\,\eqref{faddeev:basisSA} were chosen to reflect the combination of two scalar or pseudoscalar covariants
             whose product again exhibits positive parity. 
             All possible further basis elements, e.g.
             \begin{equation}
             \begin{array}{lrlrl}\label{faddeev:basisAVT}
                \mathsf{A}_{ij}^{\pm}(p,q,P) :=   & \gamma^\mu_T   \gamma_5 \,  & (\Gamma_i \,\Lambda_\pm) \,  (\gamma_5 C) \, \otimes & \gamma^\mu_T   \gamma_5\,   & (\Gamma_j \,\Lambda_+)\,,    \\
                \mathsf{V}_{ij}^{\pm}(p,q,P) :=   & \gamma^\mu_T            \,  & (\Gamma_i \,\Lambda_\pm) \,  (\gamma_5 C) \, \otimes & \gamma^\mu_T              \,& (\Gamma_j \,\Lambda_+)\,,    \\
                \mathsf{T}_{ij}^{\pm}(p,q,P) :=   & \sigma^{\mu\nu}_T\,         & (\Gamma_i \,\Lambda_\pm) \,  (\gamma_5 C) \, \otimes & \sigma^{\mu\nu}_T\,         & (\Gamma_j \,\Lambda_+)\,,
             \end{array}
             \end{equation}
             linearly depend on the ones of Eq.\,\eqref{faddeev:basisSA}; respective relations are given in Eq.\,\eqref{faddeev:basis-tf}.
             In the actual calculation we use an orthogonalized set of covariants constructed from the basic structures ($j=1\dots 4$)
             \begin{fshaded2}
             \begin{equation}\label{faddeev:cov1}
                    \mathsf{S}_{1j},  \mathsf{P}_{1j}, \quad  \mathsf{A}_{1j},  \mathsf{V}_{1j},   \quad
                    \mathsf{S}_{3j},  \mathsf{P}_{3j}, \quad  \mathsf{S}_{4j},  \mathsf{P}_{4j},
             \end{equation}
             \end{fshaded2}

             \noindent
             where each carries a superscript $(\pm)$.
             Here we have exchanged the elements $\{\mathsf{S}_{2j}, \mathsf{P}_{2j}\}$ by $\{\mathsf{A}_{1j},\mathsf{V}_{1j}\}$
             to facilitate a comparison with the multispinor notation used in the literature, e.g. Ref.\,\cite{Carimalo:1992ia}.

 \renewcommand{\arraystretch}{1}

      \bigskip
     \fatcol{Momentum-dependent covariants.}
             The momentum-dependent Dirac structures which appear in Eqs.\,(\ref{faddeev:basisSA}--\ref{faddeev:basisAVT}) involve the following 8 basis elements:
             \begin{equation}
                 \Gamma_i(p,q,P)\,\Lambda_\pm(P) \; \in \; \left\{ \,\mathds{1},\,[ \Slash{p}, \Slash{q} ],\,\Slash{p},\,\Slash{q} \,\right\}  \, \times \, \left\{ \,\mathds{1},\, \Slash{P} \,\right\}.
             \end{equation}
             It is convenient to choose a set of momenta $\{\widehat{p_T}, \widehat{q_{t}}, \widehat{P}\}$ which are orthonormal with respect to the Euclidean metric, i.e.
             \begin{equation}
                 \widehat{p_T}^2 = \widehat{q_{t}}^2 = \widehat{P}^2 = 1, \qquad \widehat{p_T}\cdot \widehat{q_{t}} = \widehat{p_T}\cdot\widehat{P} = \widehat{q_{t}}\cdot\widehat{P} = 0.
             \end{equation}
             This is realized via
             \begin{equation}
                 p_T^\mu = T^{\mu\nu}_P  \,p^\nu,  \qquad
                 q_{t}^\mu = T^{\mu\nu}_{{p_T}} \,T^{\nu\lambda}_{P} \,q^\lambda = T^{\mu\nu}_{{p_T}} \,q_T^\nu\,,
             \end{equation}
             where $T^{\mu\nu}_k = \delta^{\mu\nu}-\hat{k}^\mu \hat{k}^\nu$ denotes a transverse projector with respect to any four-momentum $k$. 
             The four covariants $\Gamma_i(p,q,P)$ are those of a fermion-scalar vertex and effectively depend on two momenta:
             \begin{equation}\label{FADEEV:Gamma_i}
                \Gamma_i(p,q,P) = \left\{  \mathds{1},\;
                                                       \textstyle\frac{1}{2}\,\displaystyle[ \widehat{\Slash{p}_T}, \widehat{\Slash{q}_{t}} ],\;
                                                       \widehat{\Slash{p}_T},\;
                                                       \widehat{\Slash{q}_{t}}
                                              \right\}\,.
             \end{equation}
             Without loss of generality one may choose the momentum alignment (cf. Eq.\,\eqref{APP:momentum-coordinates})
                 \begin{equation}\label{FADDEEV:momentumalignment}
                     p^{\mu} =  \sqrt{p^2} \left( \begin{array}{c} 0 \\ 0 \\ \sqrt{1-z_1^2} \\ z_1 \end{array} \right), \quad
                     q^{\mu} =  \sqrt{q^2} \left( \begin{array}{c} 0 \\ \sqrt{1-z_2^2} \,\sqrt{1-z_0^2} \\ \sqrt{1-z_2^2} \,z_0 \\ z_2 \end{array} \right),
                 \end{equation}
             which yields in the baryon's rest frame:
                 \begin{equation}\label{faddeev:orthogonal-unit-vectors}
                     \widehat{P}^{\mu}          = \left( \begin{array}{c} 0 \\ 0 \\ 0 \\ 1 \end{array} \right)\,, \quad
                     \widehat{p_T}^{\mu}    = \left( \begin{array}{c} 0 \\ 0 \\ 1 \\ 0 \end{array} \right)\,, \quad
                     \widehat{q_{T}}^{\mu} = \left( \begin{array}{c} 0 \\ \sqrt{1-z_0^2} \\ z_0 \\ 0 \end{array} \right)\,,\quad
                     \widehat{q_{t}}^{\mu} = \left( \begin{array}{c} 0 \\ 1 \\ 0 \\ 0 \end{array} \right)\,,
                 \end{equation}
             and hence $\Lambda_\pm = \frac{1}{2}(\mathds{1} \pm \gamma^4)$ and $\Gamma_i(p,q,P) = \left\{ \mathds{1},\, \gamma^3 \gamma^2,\, \gamma^3,\, \gamma^2 \right\}$.
             The loop momenta and covariants inside the integral of Eq.\,\eqref{faddeev:eq} will naturally have a more complicated appearance.

     \bigskip
     \fatcol{Angular momentum decomposition.}
             The Dirac basis elements \eqref{faddeev:cov1} can be classified with respect to their quark-spin and orbital angular momentum content in the nucleon's rest frame.
             In general only the total angular momentum $j=1/2$ of the nucleon is Poincar\'e-covariant  while the interpretation in terms of total quark spin
             and orbital angular momentum will differ in every frame.
             The spin is described by the Pauli-Lubanski operator:
             \begin{equation}
                 W^\mu = \frac{1}{2} \,\epsilon^{\mu\nu\alpha\beta} \hat{P}^\nu J^{\alpha\beta}\,,
             \end{equation}
             where we chose the total momentum $P$ to be normalized.
             $J^{\mu\nu}$ and $P^\mu$ are the generators of the Poincar\'e algebra satisfying the usual commutator relations.
             The square of the Pauli-Lubanski operator,
             \begin{equation}
                 W^2 = \frac{1}{2} J^{\mu\nu}J^{\mu\nu} + \hat{P}^\mu \hat{P}^\nu J^{\mu\alpha} J^{\alpha\nu},
             \end{equation}
             is one of the two Casimir operators of the Poincar\'e group. 
             Its eigenvalues are given by $W^2 \longrightarrow j(j+1)$, where $j$ describes the spin of the particle.

             For a system of three particles with total momentum $P$ and relative momenta $p$ and $q$,
             the total angular momentum operator consists of the total quark spin $\vect{S}$ and the relative orbital angular momentum $\vect{L}=\vect{L}_{(p)}+\vect{L}_{(q)}$.
             Upon subsuming them into Lorentz-covariant operators
             \begin{align}
                 S^{\mu} &= \textstyle\frac{1}{4} \displaystyle \epsilon^{\mu\nu\alpha\beta} \hat{P}^\nu
                            \left( \sigma^{\alpha\beta} \otimes \mathds{1} \otimes \mathds{1} +  \mathds{1} \otimes \sigma^{\alpha\beta} \otimes \mathds{1} + \mathds{1} \otimes \mathds{1} \otimes \sigma^{\alpha\beta} \right), \\
                 L_{(p)}^{\mu} &= \textstyle\frac{i}{2} \displaystyle \epsilon^{\mu\nu\alpha\beta} \hat{P}^\nu \left( p^\alpha \partial_p^\beta - p^\beta \partial_p^\alpha \right) \mathds{1} \otimes \mathds{1} \otimes \mathds{1}, \\
                 L_{(q)}^{\mu} &= \textstyle\frac{i}{2} \displaystyle \epsilon^{\mu\nu\alpha\beta} \hat{P}^\nu \left( q^\alpha \partial_q^\beta - q^\beta \partial_q^\alpha \right) \mathds{1} \otimes \mathds{1} \otimes \mathds{1},
             \end{align}
             one may verify that the basis covariants \eqref{faddeev:cov1} are indeed eigenfunctions of the square of the Pauli-Lubanski operator  $W^\mu = S^\mu + L_{(p)}^\mu + L_{(q)}^\mu$
             with $j=1/2$.
             Here the following identities prove to be useful:
             \begin{align*}
                 S^2 &= \textstyle\frac{9}{4} \displaystyle \,\mathds{1} \otimes \mathds{1} \otimes \mathds{1} + \textstyle\frac{1}{4} \displaystyle \left( \sigma_T^{\mu\nu} \otimes \sigma_T^{\mu\nu} \otimes \mathds{1} + \text{perm.} \right)\,, \\[0.1cm]
                 S \!\cdot\! L_{(p)} &= \textstyle\frac{i}{2} \displaystyle \,p_T^\mu\, (\partial_T)_p^\nu \left( \sigma_T^{\mu\nu} \otimes  \mathds{1} \otimes \mathds{1} + \text{perm.} \right)\,, \\[0.15cm]
                 L_{(p)}^2 &= 2  \,p^\mu_T \,(\partial_T)_p^\mu + p^\mu_T  \,p^\nu_T  \,(\partial_T)_p^\mu (\partial_T)_p^\nu - p_T^2  \,(\partial_T)_p \!\cdot\! (\partial_T)_p\,,\\[0.1cm]
                 L_{(p)}\!\cdot\! L_{(q)} &= p_T^\mu\,q_T^\nu\, (\partial_T)_p^\nu \,(\partial_T)_q^\mu -p_T \!\cdot\! q_T \,(\partial_T)_p \!\cdot\! (\partial_T)_q\,.
             \end{align*}

                \renewcommand{\arraystretch}{1.5}

             \begin{table}[tbh]

                \begin{center}
                \begin{tabular}{ | @{\quad} c @{\quad} | @{\quad} c @{\quad} || @{\quad} c @{\quad} | @{\quad} c @{\quad} | @{\quad} c @{\quad} | @{\quad} c @{\quad} || @{\quad} c @{\quad} | } \hline

                 $s$                 &  $l$ &    $\mathsf{X}^\pm_{1j}$ &   $\mathsf{X}^\pm_{2j}$ &   $\sqrt{3}\,\mathsf{X}^\pm_{3j}$                     &   $\sqrt{3}\,\mathsf{X}^\pm_{4j}$                  & \#    \\ [0.1cm]  \hline\hline   

                 $\nicefrac{1}{2}$   &  $0$  &   $\mathsf{S}_{11}$     &   $\mathsf{P}_{11}$     &   $\mathsf{A}_{11}$     &   $\mathsf{V}_{11}$  & 8     \\ \rowcolor{lred}
                 $\nicefrac{1}{2}$   &  $1$  &   $\mathsf{S}_{12}$     &   $\mathsf{P}_{12}$     &   $\mathsf{A}_{12}$     &   $\mathsf{V}_{12}$  & 8     \\ \rowcolor{lred}
                 $\nicefrac{1}{2}$   &  $1$  &   $\mathsf{S}_{13}$     &   $\mathsf{P}_{13}$     &   $\mathsf{A}_{13}$     &   $\mathsf{V}_{13}$  & 8     \\ \rowcolor{lred}
                 $\nicefrac{1}{2}$   &  $1$  &   $\mathsf{S}_{14}$     &   $\mathsf{P}_{14}$     &   $\mathsf{A}_{14}$     &   $\mathsf{V}_{14}$  & 8     \\ [0.1cm]  \hline

                \end{tabular} \vspace{0.2cm}

                \begin{tabular}{ | @{\quad} c @{\quad} | @{\quad} c @{\quad} ||@{\quad} c @{\quad} | @{\quad} c @{\quad} || @{\quad} c @{\quad} |  } \hline

                 $s$                 &  $l$ &    $\sqrt{6}\,\mathsf{X}^\pm_{5j}$                                  &   $\sqrt{6}\,\mathsf{X}^\pm_{6j}$                                        & \#  \\ [0.1cm]  \hline\hline   

                 $\nicefrac{3}{2}$   &  $2$  &   $3\,\mathsf{S}_{33}-\mathsf{V}_{11}$                           &   $3\,\mathsf{P}_{33}-\mathsf{A}_{11}$                        & 4  \\ \rowcolor{lred}
                 $\nicefrac{3}{2}$   &  $1$  &   $3\,\mathsf{S}_{34}-3\,\mathsf{S}_{43}-2\,\mathsf{V}_{12}$     &   $3\,\mathsf{P}_{34}-3\,\mathsf{P}_{43}-2\,\mathsf{A}_{12}$  & 4    \\ \rowcolor{lred}
                 $\nicefrac{3}{2}$   &  $1$  &   $3\,\mathsf{S}_{31}-\mathsf{V}_{13}$                           &   $3\,\mathsf{P}_{31}-\mathsf{A}_{13}$                        & 4  \\ \rowcolor{lred}
                 $\nicefrac{3}{2}$   &  $1$  &   $3\,\mathsf{S}_{41}-\mathsf{V}_{14}$                           &   $3\,\mathsf{P}_{41}-\mathsf{A}_{14}$                        & 4  \\ [0.1cm]  \hline

                \end{tabular} \vspace{0.2cm}

                \begin{tabular}{ | @{\quad} c @{\quad} | @{\quad} c @{\quad} ||@{\quad} c @{\quad} | @{\quad} c @{\quad} || @{\quad} c @{\quad} | } \hline

                 $s$                 &  $l$ &    $\sqrt{2}\,\mathsf{X}^\pm_{7j}$                                                                &   $\sqrt{2}\,\mathsf{X}^\pm_{8j}$                & \#      \\ [0.1cm]  \hline\hline 

                 $\nicefrac{3}{2}$   &  $2$  &   $2\,\mathsf{S}_{44}+\mathsf{S}_{33}-\mathsf{V}_{11}$           &   $2\,\mathsf{P}_{44}+\mathsf{P}_{33}-\mathsf{A}_{11}$        & 4    \\ 
                 $\nicefrac{3}{2}$   &  $2$  &   $\mathsf{S}_{34}+\mathsf{S}_{43}$                              &   $\mathsf{P}_{34}+\mathsf{P}_{43}$                           & 4  \\ 
                 $\nicefrac{3}{2}$   &  $2$  &   $-2\,\mathsf{S}_{42}+\mathsf{S}_{31}-\mathsf{V}_{13}$          &   $-2\,\mathsf{P}_{42}+\mathsf{P}_{31}-\mathsf{A}_{13}$       & 4     \\ 
                 $\nicefrac{3}{2}$   &  $2$  &   $2\,\mathsf{S}_{32}+\mathsf{S}_{41}-\mathsf{V}_{14}$           &   $2\,\mathsf{P}_{32}+\mathsf{P}_{41}-\mathsf{A}_{14}$        & 4     \\ [0.1cm]  \hline

                \end{tabular} \caption{\backdef Orthonormal Dirac basis constructed from
                                       Eq.\,\eqref{faddeev:cov1} via a partial-wave decomposition.
                                       The superscripts $\pm$ are not displayed.}\label{tab:faddeev:basis}
                \end{center}

        \end{table}

             The basis states can furthermore be classified with respect to the eigenvalues of $S^2 \longrightarrow s(s+1)$ and $L^2 \longrightarrow l(l+1)$,
             which, in the nucleon's rest frame, assume the interpretation of total quark spin and intrinsic quark orbital angular momentum.
             Such a partial-wave decomposition allows for an arrangement of
             the 64 basis covariants (32 for total quark spin $s=1/2$ and $s=3/2$, respectively)
             into sets of 8 $s$ waves ($l=0$), 36 $p$ waves ($l=1$), and 20 $d$ waves ($l=2$).
             We  denote the resulting orthonormal basis states by the symbol $\mathsf{X}_{ij}^\pm$, with $i=1\dots 8$ and $j=1\dots 4$, and collect them in Table~\ref{tab:faddeev:basis}.

             \bigskip
             \fatcol{Orthogonality.}
             The combinations $\mathsf{X}_{ij}^\pm$ satisfy the following orthogonality relation:
                 \begin{equation}\label{faddeev:orthogonality}
                     \textstyle\frac{1}{4} \,\displaystyle \text{Tr}\{\conjg{\mathsf{X}}^r_{ij}\,\mathsf{X}^{r'}_{i'j'}\} =
                     \textstyle\frac{1}{4} \,\displaystyle \big(\conjg{\mathsf{X}}^r_{ij}\big)_{\beta\alpha,\delta\gamma}\,\big(\mathsf{X}^{r'}_{i'j'}\big)_{\alpha\beta,\gamma\delta} =
                     \delta_{ii'}\,\delta_{jj'}\,\delta_{rr'}\,,
                 \end{equation}
             where the charge conjugation $\conjg{\mathsf{X}}$ is defined as
                 \begin{equation}
                     \conjg{\mathsf{X}}(p,q,P) = (C\otimes C)\,\mathsf{X}(-p,-q,-P)^T(C^T\!\otimes C^T)\,.
                 \end{equation}
             As an example, consider:
                 \begin{equation*}
                 \begin{split}
                     \conjg{\mathsf{S}}_{ij}^\pm(p,q,P) =\,& C\,\big\{\Gamma_i(-p,-q,-P)\, \Lambda_\mp \, \gamma_5 C\big\}^T C^T  \otimes  C\,\big\{\Gamma_j(-p,-q,-P)\, \Lambda_-\big\}^T C^T = \\
                                                      =\,& (C^T\!\gamma_5)\,(\Lambda_\pm \conjg{\Gamma}_i ) \,\otimes  (\Lambda_+\conjg{\Gamma}_j ),
                 \end{split}
                 \end{equation*}
            where we used $\conjg{\Gamma}_i(p,q,P) = C\,\Gamma_i(-p,-q,-P)^T\,C^T$ and the relations $\Lambda_\pm(-P) = \Lambda_\mp(P)$ and $\conjg{\Lambda}_\pm(P) = C\,\Lambda_\pm(-P)\,C^T=\Lambda_\pm(P)$.

          \renewcommand{\arraystretch}{1.2}

             \bigskip
              \fatcol{Multispinor notation.}
              An alternative way to construct a basis for the nucleon amplitude is to use
              the Dirac spinors $U^\sigma(P)$, $V^\sigma(P) = \gamma_5 U^\sigma(P)$ which satisfy the free Dirac equation for a spin-$\nicefrac{1}{2}$ particle, i.e.
              which are eigenspinors of the positive and negative energy projectors $\Lambda_\pm$.
              Normalized to $\conjg{U}^\rho(P) \,U^\sigma(P) = \delta_{\rho\sigma}$, they are expressed as
              \begin{equation}\label{NUCLEON:DiracSpinors}
                  U^\sigma(P) = \sqrt{\frac{\varepsilon+M}{2\,M}} \left( \begin{array}{c} w^\sigma \\ \textstyle\frac{\vect{\scriptstyle\sigma}\cdot \vect{\scriptstyle P}}{\varepsilon+M}\,w^\sigma \end{array}\right), \;
                  \varepsilon=\sqrt{\vect{P}^2+M^2}, \;
                  \renewcommand{\arraystretch}{1.0}
                  w^\uparrow = \left(\begin{array}{c} 1 \\ 0 \end{array}\right), \;  w^\downarrow = \left(\begin{array}{c} 0 \\ 1 \end{array} \right).
              \end{equation}
              The 64 linearly independent basis states for the nucleon wave function are obtained from the multispinors
              $UUU:=U\otimes U\otimes U$, $VVU$, $VUV$, $UVV$ and their parity-reversed counterparts $VVV$, $UUV$, $UVU$, and $VUU$
              by equipping them with the 8 possible spin-up/down arrangements $\uparrow\uparrow\uparrow$, $\uparrow\uparrow\downarrow$, and so on.
              A corresponding basis has been explicitly constructed in Ref.\,\cite{Carimalo:1992ia}.
              Using identities such as
              \begin{equation}
                  \Lambda_+ = U^\uparrow \conjg{U}^\uparrow + U^\downarrow \conjg{U}^\downarrow\,, \qquad  \Lambda_+ \,(\gamma_5 C) = U^\downarrow U^\uparrow - U^\uparrow U^\downarrow
              \end{equation}
              leads to the following relations for the $s$-wave basis states of Table~\ref{tab:faddeev:basis}:
              \begin{equation*}\label{faddeev:spinor}
              \begin{array}{rl}
                 -\mathsf{S}_{11}^+ \, U^\uparrow &=  (U^\uparrow  U^\downarrow - U^\downarrow  U^\uparrow) \, U^\uparrow\,,  \\
                 -\mathsf{S}_{11}^- \, U^\uparrow &=  (V^\uparrow  V^\downarrow - V^\downarrow  V^\uparrow) \, U^\uparrow\,,  \\
                 -\mathsf{P}_{11}^+ \, U^\uparrow &=  (V^\uparrow  U^\downarrow - V^\downarrow  U^\uparrow) \, V^\uparrow\,,  \\
                 -\mathsf{P}_{11}^- \, U^\uparrow &=  (U^\uparrow  V^\downarrow - U^\downarrow  V^\uparrow) \, V^\uparrow\,,
              \end{array} \qquad
              \begin{array}{rl}
                  \mathsf{A}_{11}^+ \, U^\uparrow &=  (U^\uparrow  U^\downarrow + U^\downarrow  U^\uparrow) \, U^\uparrow - 2 \, U^\uparrow  U^\uparrow  U^\downarrow\,,  \\
                 -\mathsf{A}_{11}^- \, U^\uparrow &=  (V^\uparrow  V^\downarrow + V^\downarrow  V^\uparrow) \, U^\uparrow - 2 \, V^\uparrow  V^\uparrow  U^\downarrow\,,   \\
                  \mathsf{V}_{11}^+ \, U^\uparrow &=  (V^\uparrow  U^\downarrow + V^\downarrow  U^\uparrow) \, V^\uparrow - 2 \, V^\uparrow  U^\uparrow  V^\downarrow\,,   \\
                 -\mathsf{V}_{11}^- \, U^\uparrow &=  (U^\uparrow  V^\downarrow + U^\downarrow  V^\uparrow) \, V^\uparrow - 2 \, U^\uparrow  V^\uparrow  V^\downarrow\,.
              \end{array}
              \end{equation*}
              Due to the Poincar\'e-covariant construction of Eqs.\,(\ref{faddeev:basisSA}--\ref{faddeev:basisAVT}), the remaining 56 covariants depend on the relative momenta.
              For instance, the Dirac structure $\mathsf{S}_{13}^-$ satisfies the relation
              \begin{equation}\label{FE:S13}
                  i\,\mathsf{S}_{13}^- \, U^\uparrow =  (V^\uparrow  V^\downarrow - V^\downarrow  V^\uparrow) \, \left( \left([\widehat{p_T}]_1 + i [\widehat{p_T}]_2 \right) V^\downarrow + [\widehat{p_T}]_3 V^\uparrow  \right)\,.
              \end{equation}
              Using the special momentum alignment \eqref{faddeev:orthogonal-unit-vectors} leads to
              \begin{equation}\label{FADDEEV:S13}
                  i\,\mathsf{S}_{13}^- \, U^\uparrow \stackrel{\eqref{faddeev:orthogonal-unit-vectors}}{\longlongrightarrow}  (V^\uparrow  V^\downarrow - V^\downarrow  V^\uparrow) \, V^\uparrow \,
              \end{equation}
              and similar relations for the remaining basis elements in Table~\ref{tab:faddeev:basis}.
              At the same time the expression \eqref{FADDEEV:S13} is the 'parity-flipped' counterpart of $\mathsf{S}_{11}^+$,
              \begin{equation}
                  -(\gamma_5 \otimes \gamma_5) \,\mathsf{S}_{11}^+  \,(\gamma_5^T \otimes U^\uparrow) = (V^\uparrow  V^\downarrow - V^\downarrow  V^\uparrow) \, V^\uparrow \,,
              \end{equation}
              which appears in the amplitude of a negative-parity nucleon ($1/2^-$) and that of a positive-parity antinucleon ($\overline{1/2^+}$).
              Nevertheless the seemingly odd-parity structure \eqref{FADDEEV:S13} in the multispinor basis still contributes to the ($1/2^+$) state via Eq.\,\eqref{FE:S13},
              where the parity flip induced by the spinor replacement
              $U \rightarrow V$ is saturated by an odd parity introduced by the relative momentum $\widehat{p_T}$.
              As a consequence, indeed all 64 three-spinor combinations contribute to the nucleon's amplitude.


     \section{Quark exchange and Pauli principle}

 \renewcommand{\arraystretch}{1}

             The full Dirac--flavor--color amplitude of the nucleon reads (cf. Eq.\,\eqref{FE:nucleon_amplitude_full}):
              \begin{equation}\label{FE:nucleon_amplitude_full-2}
                  \Psi(p,q,P) = \Big\{ \Psi(p,q,P)_\mathcal{M_A} \otimes \mathsf{T}_\mathcal{M_A} + \Psi(p,q,P)_\mathcal{M_S}  \otimes \mathsf{T}_\mathcal{M_S} \Big\} \otimes \frac{\varepsilon_{ABC}}{\sqrt{6}}\,.
              \end{equation}
             The Pauli principle requires the Faddeev amplitude to be antisymmetric under exchange of any two quarks,
             as can be inferred from Eq.\,\eqref{BS:Three_body_amplitude_proper_notation}.
             The color singlet wave function $\varepsilon_{ABC}$ is totally antisymmetric, hence the Dirac-flavor amplitude must be symmetric.
             Since the two isospin-$1/2$ flavor tensors $\mathsf{T}_\mathcal{M_A}$, $\mathsf{T}_\mathcal{M_S}$, given in Eq.\,\eqref{FAD:flavor},
             are either mixed-symmetric or mixed-antisymmetric, the same feature must hold for the
             Dirac remainders $\Psi_\mathcal{M_A}$ and $\Psi_\mathcal{M_S}$ of the nucleon amplitude.

             The Faddeev kernel $\widetilde{K}^{(3)}$ is per construction invariant
             under the permutation group $\mathbb{S}^3$ \textcolor{webblue}:
             it commutes with any permutation of two quark legs. The Dirac parts of the solutions to the Faddeev equation
             can be arranged into irreducible $\mathbb{S}^3$ multiplets
             \begin{equation}
                 \Psi_\mathcal{S}, \; \Psi_\mathcal{A}, \; \left(\begin{array}{c} \Psi_\mathcal{M_A} \\ \Psi_\mathcal{M_S} \end{array}\right),
             \end{equation}
             of which the first two (totally symmetric or antisymmetric solutions) are unphysical while the mixed-symmetry doublet constitutes the nucleon amplitude
             according to Eq.\,\eqref{FE:nucleon_amplitude_full-2}.
             The Faddeev equation will in general mix the two linearly independent solutions $\Psi_\mathcal{M_A}$ and $\Psi_\mathcal{M_S}$.
             However, since the rainbow-ladder kernel presently employed is flavor-independent
             and the two flavor tensors $\mathsf{T}_\mathcal{M_A}$ and $\mathsf{T}_\mathcal{M_S}$ are orthogonal to each other,
             the equations for the Dirac amplitudes $\Psi_\mathcal{M_A}$, $\Psi_\mathcal{M_S}$ decouple:
             \begin{equation}\label{FADDEEV:MS,MA:Independent}
                 \Psi = \widetilde{K}^{(3)}\,\Psi \quad \longrightarrow \quad \begin{array}{c} \Psi_\mathcal{M_A} = \widetilde{K}^{(3)}\,\Psi_\mathcal{M_A}, \\[0.1cm]
                                                                                    \Psi_\mathcal{M_S} = \widetilde{K}^{(3)}\,\Psi_\mathcal{M_S}.\end{array}
             \end{equation}
             These two states do indeed emerge as independent solutions upon solving the equation.
             The dominant amplitudes in either case are the mixed-antisymmetric and mixed-symmetric covariants defined in Eqs.\,(\ref{faddeev:basisSA}) and (\ref{faddeev:basisAVT}):
             \begin{align}
                 \Psi_\mathcal{M_A} \; \sim \; \mathsf{S}_{11}^+ &= \Lambda_+  (\gamma_5 C)  \otimes \Lambda_+ \,, \\
                 \Psi_\mathcal{M_S} \; \sim \; \mathsf{A}_{11}^+ &= \gamma^\mu_T  \gamma_5 \,\Lambda_+   (\gamma_5 C)  \otimes \gamma^\mu_T   \gamma_5\, \Lambda_+.
             \end{align}

             \bigskip
              \fatcol{s-wave components.}
              Upon restricting oneself to the eight momentum-independent covariants of Table~\ref{tab:faddeev:basis} with $s=\nicefrac{1}{2}$, $l=0$ and
              applying the $\mathbb{S}^3$ symmetrizers/anti\-symmetrizers, these 8 basis elements can be rearranged into three mixed-symmetry doublets and
              two symmetric or antisymmetric singlets:
              \begin{equation}\label{FADDEEV:8multiplets}
              \begin{split}
                  \psi_\mathcal{M_A}^{(1)} &= \mathsf{S}_{11}^+\,, \\
                  \psi_\mathcal{M_S}^{(1)} &= \mathsf{A}_{11}^+ \\[0.3cm]
                  \psi_\mathcal{M_A}^{(2)} &= \mathsf{P}_{11}^+ + \mathsf{P}_{11}^- + \mathsf{S}_{11}^-\,, \\
                  \psi_\mathcal{M_S}^{(2)} &= \mathsf{V}_{11}^+ - \mathsf{V}_{11}^- - \mathsf{A}_{11}^- \,, \\[0.3cm]
                  \psi_\mathcal{M_A}^{(3)} &= 2\,\mathsf{S}_{11}^- + \mathsf{V}_{11}^+ + \mathsf{V}_{11}^- - (\mathsf{P}_{11}^+ + \mathsf{P}_{11}^- ) \,,\\
                  \psi_\mathcal{M_S}^{(3)} &= 2\,\mathsf{A}_{11}^- + \mathsf{V}_{11}^+ - \mathsf{V}_{11}^- +3\, (\mathsf{P}_{11}^+ - \mathsf{P}_{11}^- ) \,,\\[0.3cm]
                  \psi_\mathcal{A} &=   (\mathsf{P}_{11}^+ + \mathsf{P}_{11}^-) + (\mathsf{V}_{11}^+ + \mathsf{V}_{11}^-) - 2\,\mathsf{S}_{11}^-  \,,\\[0.15cm]
                  \psi_\mathcal{S} &= 3\, (\mathsf{P}_{11}^+ - \mathsf{P}_{11}^-) - (\mathsf{V}_{11}^+ - \mathsf{V}_{11}^-) - 2\,\mathsf{A}_{11}^-\,.
              \end{split}
              \end{equation}

             \medskip
             \noindent
             Via exchange of the involved momenta, the dressing functions $f_i(p^2,q^2,\{z\})$ defined in Eq.\,\eqref{faddeev:amp} transform as irreducible representations
             of the permutation group as well. If those coefficients $f_i$ were totally symmetric,
             e.g., by being constant or by depending only on certain symmetric combinations of $p^2$, $q^2$ and $\{z\}$ as derived in \cite{Carimalo:1992ia},
             the nucleon amplitude would be a linear combination of the six $\mathcal{M_A}$ and $\mathcal{M_S}$ basis elements in Eq.\,\eqref{FADDEEV:8multiplets}:
             \begin{equation}
                 \left(\begin{array}{c} \Psi_\mathcal{M_A} \\ \Psi_\mathcal{M_S} \end{array}\right) =
                 \left(\begin{array}{c} \sum_{i=1}^3 f_i\,\Psi^{(i)}_\mathcal{M_A} \\ \sum_{i=1}^3 \tilde{f}_i\,\Psi^{(i)}_\mathcal{M_S} \end{array}\right).
             \end{equation}

             \medskip
             \noindent
             Since the coefficients $f_i$ can appear in all symmetry representations
             the inclusion of the remaining Dirac covariants $\psi_\mathcal{A}$ and $\psi_\mathcal{S}$ (and, in general, all 64 basis elements) is however necessary.


 \renewcommand{\arraystretch}{1.2}


     \section{Results}\label{sec:FADDEEV:results}

                The explicit numerical implementation of the Faddeev equation is described in App.\,\ref{app:faddeevamp}.
                We solve for all 64 dressing functions $f_i(p^2, q^2,0,z_1,z_2)$ but
                omit the dependence on the angular variable $z_0=\widehat{p_T}\cdot\widehat{q_T}$ for the sake of numerical efficiency.
                In the context of a quark-diquark model, the dependence on $z_0$ is excluded a priori due to the separability assumption
                of the amplitude\footnote{
                Note however that this assignment is not strictly valid as the quark-diquark amplitude needs to be symmetrized to obtain the corresponding Faddeev amplitude,
                thereby changing the interpretation of the involved momenta.}, cf. Eq.\,\eqref{nuc:defqdqamp}.

            \begin{figure}[tbp]
                    \begin{center}
                    \includegraphics[scale=0.34]{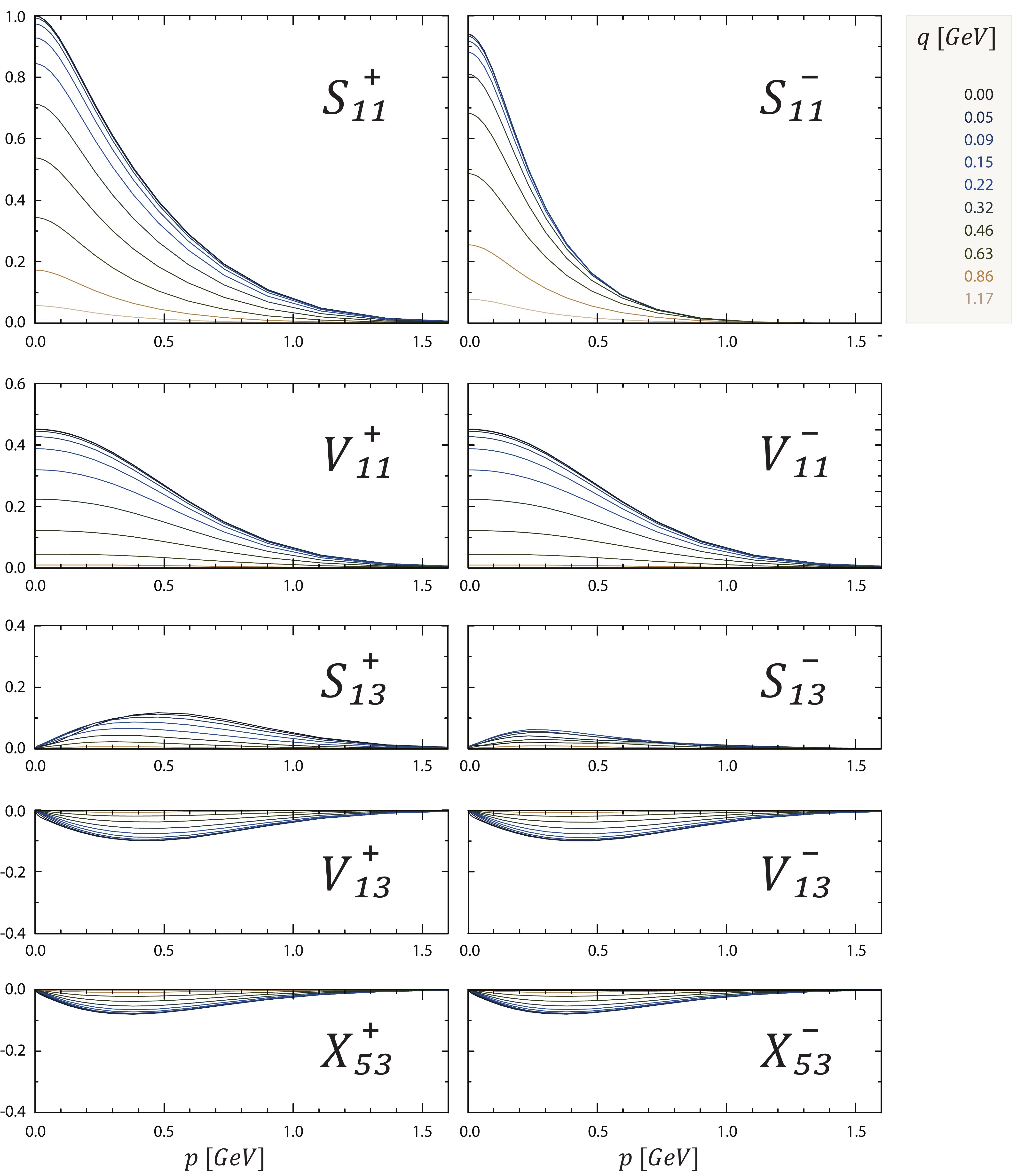}
                    \caption{\backdef $0^\text{th}$ Chebyshev moments of the dressing functions corresponding to the dominant covariants in the Faddeev amplitude $\Psi_\mathcal{M_A}$, plotted as a function of $p$ and $q$.}
                    \label{fig:FADDEEV:amplitudes}
                    \end{center}
            \end{figure}

            \begin{figure}[tbp]
                    \begin{center}
                    \includegraphics[scale=0.34]{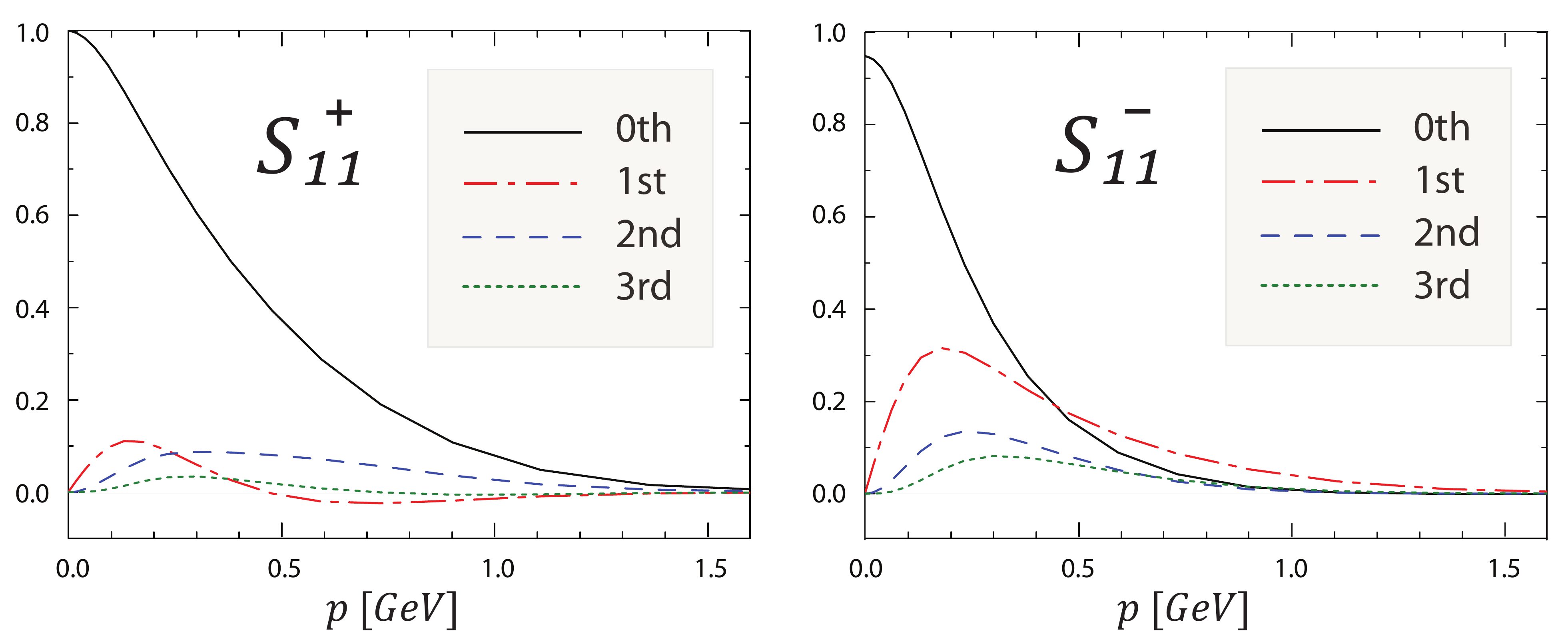}
                    \caption{\backdef First four Chebyshev moments in the variable $z_1$ of the dressing functions at $q^2=0$ associated with the amplitudes $\mathsf{S}_{11}^\pm$.}
                    \label{fig:FADDEEV:amp-chebyshev}
                    \end{center}
            \end{figure}

                      As a consequence of Eq.\,\eqref{FADDEEV:MS,MA:Independent},
                      the two states $\Psi_\mathcal{M_A}$ and $\Psi_\mathcal{M_S}$ emerge as independent solutions of the Faddeev equation,
                      where by virtue of the iterative method
                      the symmetry of the start function determines the symmetry of the resulting amplitude.
                      Both separate equations produce approximately the same nucleon mass,
                      where the deviation of $\sim 2\%$ is presumably a truncation artifact associated with the omission of the angle $z_0$.
                      For each of the two solutions, typically only a small number of covariants are relevant.
                      Comparing the relative strengths of the amplitudes (cf. Fig.~\ref{fig:FADDEEV:amplitudes} for the mixed-antisymmetric solution) allows to identify the dominant contributions:
                      \begin{equation}
                      \begin{array}{rl}
                          \Psi_\mathcal{M_A}: \quad\; & \mathsf{S}_{11}^+, \, \mathsf{S}_{11}^-, \\
                                                    & \mathsf{V}_{11}^+ + \mathsf{V}_{11}^-, \\
                                                    & \mathsf{S}_{13}^+, \, \mathsf{S}_{13}^-, \\
                                                    & \mathsf{V}_{13}^+ + \mathsf{V}_{13}^-, \\
                                                    & \mathsf{X}_{53}^+ + \mathsf{X}_{53}^-   \\
                      \end{array}\qquad
                      \begin{array}{rl}
                          \Psi_\mathcal{M_S}: \quad\; & \mathsf{A}_{11}^+, \, \mathsf{A}_{11}^-,\\
                                                    & \mathsf{V}_{11}^+ - \mathsf{V}_{11}^-, \\
                                                    & \mathsf{P}_{11}^+ - \mathsf{P}_{11}^-, \\
                                                    & \mathsf{V}_{13}^+ - \mathsf{V}_{13}^-, \\
                                                    & \mathsf{X}_{63}^+ + \mathsf{X}_{63}^-

                      \end{array}
                      \end{equation}
                      which indicates a sizeable admixture of $p$ waves to the dominant $s$-wave components (cf. Table~\ref{tab:faddeev:basis}).

                      Fig.~\ref{fig:FADDEEV:amp-chebyshev} displays the angular dependence in the variable $z_1$,
                      in terms of the first few Chebyshev moments,
                       of the amplitudes $\mathsf{S}_{11}^\pm$
                      which contribute to $\Psi_\mathcal{M_A}$.
                      The $z_1$ dependence is much more pronounced than that in the variable $z_2$ where already the
                      zeroth Chebyshev moment provides a satisfactory approximation.
                      This is again kindred
                      to the quark-diquark model, where $q$ is related to the relative momentum between the two quarks in a diquark amplitude
                      and the dependence on the associated angle $z_2$ is small, cf. Eq.\,\eqref{dq:amp-offshell}.

            \begin{figure}[t]
            \begin{center}
            \includegraphics[scale=1.1]{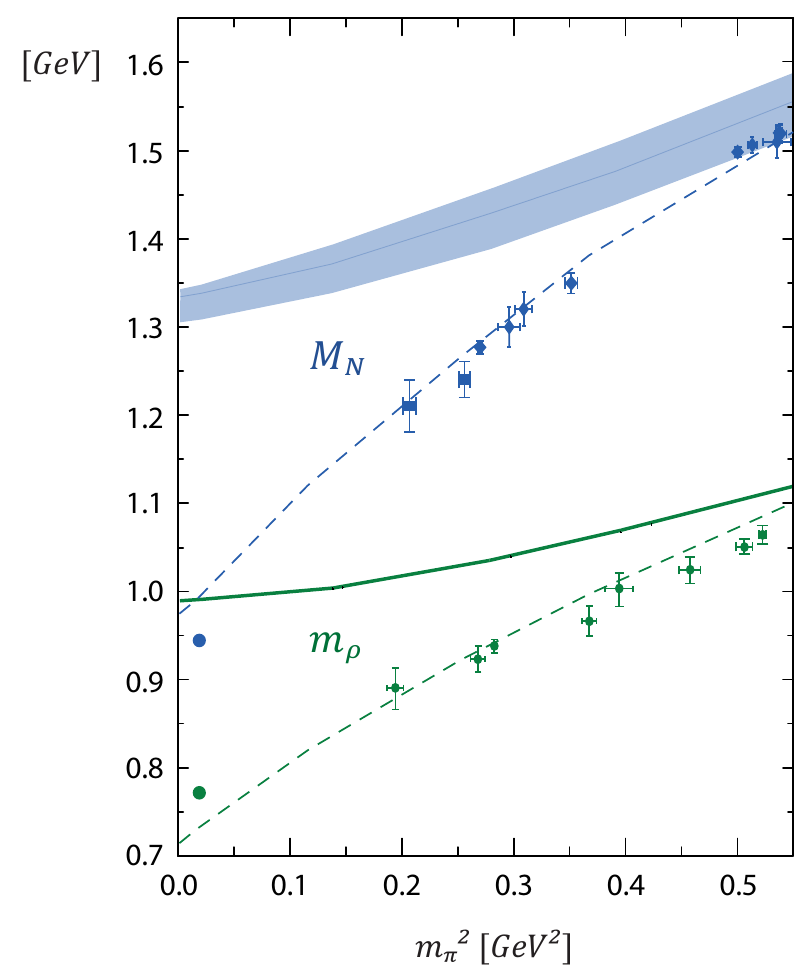}
            \caption[Diquark masses]{\backdef Current-mass evolution of $M_N$ and $m_\rho$ compared to lattice results: \cite{AliKhan:2003cu,Frigori:2007wa,Leinweber:2003dg} for
                                     $M_N$ and \cite{AliKhan:2001tx,Allton:2005fb} for $m_\rho$.
                                     The dashed lines are related to setup \hyperlink{coupling:c1}{(C1)}.
                                     The solid line for $m_\rho$ is the input of Eq.\,\eqref{core:mrho} in setup \hyperlink{coupling:c3}{(C3)};
                                     the band for $M_N$ is the result of the Faddeev equation (mixed-antisymmetric solution).
                                     Both correspond to a variation of $\omega$.
                                     Dots demarcate the experimental values.} \label{fig:FADDEEV:nucleon-mass}
            \end{center}
            \end{figure}

\newpage

                The resulting nucleon masses at the physical $u-d$ quark value in both setups \hyperlink{coupling:c1}{(C1)} and \hyperlink{coupling:c3}{(C3)} of Section~\ref{sec:coupling-ansaetze} are:
                \begin{equation*}
                    \hyperlink{coupling:c1}{\text{(C1):}} \quad \begin{array}{rl} \Psi_\mathcal{M_A}\!: &  \, 0.99\;\text{GeV}, \\
                                                           \Psi_\mathcal{M_S}\!: &  \, 0.97\;\text{GeV},
                                          \end{array}\qquad
                   \hyperlink{coupling:c3}{\text{(C3):}} \quad  \begin{array}{rl} \Psi_\mathcal{M_A}\!: & \,  1.33(2)\;\text{GeV}, \\
                                                            \Psi_\mathcal{M_S}\!: & \,  1.31(2)\;\text{GeV},
                                          \end{array}
                \end{equation*}
                where the $\omega$ dependence is explicitly taken into account in the 'core' setup \hyperlink{coupling:c3}{(C3)}.
                The current-mass evolution of $M_N$ is plotted in Fig.~\eqref{fig:FADDEEV:nucleon-mass} and compared to lattice results.
                The findings are qualitatively similar to those for $m_\rho$: setup \hyperlink{coupling:c1}{(C1)}, where the coupling strength
                is adjusted to the experimental value of $f_\pi$, agrees with the lattice data while the  input of \hyperlink{coupling:c3}{(C3)}
                provides a description of a quark core which consistently overestimates the experimental values while approaching the lattice results at larger quark masses.
                The sensitivity of $M_N$ to the width parameter $\omega$ is more articulate than that of the quark-diquark model result for the nucleon mass, cf. Section~\ref{sec:results:nuclenmass},
                which might also be a side effect of neglecting the angular variable $z_0$.
                For a detailed discussion of the current-mass dependence of $M_N$ we refer to Section~\ref{sec:results:nuclenmass}
                where the quark-diquark model results for $M_N$ and $M_\Delta$
                are compared to estimates from chiral effective field theory.
                In this context it will turn out that the  quark-diquark solution for the nucleon mass provides a quite reasonable approximation
                to the result of the three-body calculation.


            %

  \chapter[Baryons: The quark-diquark picture]{Baryons: \\ The quark-diquark picture} \label{chapter:quark-diquark}

            The underlying assumption of the Faddeev truncation in Chapter\,\ref{chapter:faddeev} 
            was the identification of two-quark correlations as the dominant structure which binds a baryon.
            Upon implementation of a rainbow-ladder truncation --- i.e., an iterated gluon exchange, 
            the same mechanism which has been used to describe $q\bar{q}$ bound states ---
            a direct numerical solution of the relativistic Faddeev equation \eqref{bs:faddeevtruncated} was obtained.

            The same premise can be implemented with less numerical effort in a quark-diquark model.
            It treats such two-quark correlations as a separable pole sum in the $qq$ scattering matrix
            and leads to a description of baryons as bound states of effective quarks and diquarks.
            In the simplest version of the model the nucleon is made of a quark and a scalar ($0^+$) diquark
            whereas the $\Delta$ comprises a quark and an axial-vector ($1^+$) diquark.
            A strong attraction in the color-antitriplet diquark channel has also been proposed to explain
            missing exotic states in the hadron spectrum and the masses of light scalar mesons \cite{Anselmino:1992vg,Jaffe:2004ph}.
            Further support for the diquark concept has recently been drawn from lattice calculations, cf.~Table~\ref{tab:DQmass-splitting}.

            The simplest realization of a quark-diquark description of baryons is the Nambu-Jona-Lasinio (NJL) model \cite{Nambu:1961fr,Nambu:1961tp}
            which allows for a formation of bound states of quarks and point-like diquarks via quark exchange \cite{Cahill:1988zi,Reinhardt:1989rw,Buck:1992wz,Meyer:1994cn,Ishii:1995bu,Keiner:1996at}.
            In this context it was soon realized that axial-vector diquarks provide substantial attraction in the nucleon and should be taken into account as well.
            An extension of the model to include diquarks of finite width led to a series of studies investigating
            nucleon and $\Delta$ masses \cite{Burden:1988dt,Oettel:1998bk,Hecht:2002ej}, nucleon electromagnetic form factors \cite{Hellstern:1997pg,Bloch:1999ke,Oettel:1999gc,Oettel:2000jj,Oettel:2002wf,Bloch:2003vn,Alkofer:2004yf,Cloet:2008re}, and
            the nucleon's pseudoscalar, scalar and axial-vector form factors \cite{Bloch:1999rm,Oettel:2000jj}. 

            The present chapter extends the quark-diquark model insofar as the dynamics of $0^+$ and $1^+$ diquarks
            are determined from their underlying quark and gluon constituents.
            Parametrizations for the diquark amplitudes are removed and replaced by solutions of the corresponding diquark Bethe-Salpeter equations.
            The identification of colored diquarks as poles in the $qq$ scattering matrix becomes possible within a rainbow-ladder truncation:
            it induces timelike $0^+$, $1^+$, $\dots$ diquark poles whose mass scales play an important role in the description of light baryons.
            These poles, however, correspond to unphysical asymptotic states and disappear from the spectrum when going beyond rainbow-ladder.
            The quark-diquark BSE is derived from the relativistic Faddeev equation and will be applied to calculate the masses of $N$ and $\Delta$.
            The current-mass dependence of these masses will be compared to lattice-QCD results, and we will discuss the impact of chiral corrections.


\section{The diquark ansatz for the $qq$ scattering matrix}\label{sec:dq}

\begin{figure}[tbp]
            \begin{center}
            \medskip
            \includegraphics[scale=0.1]{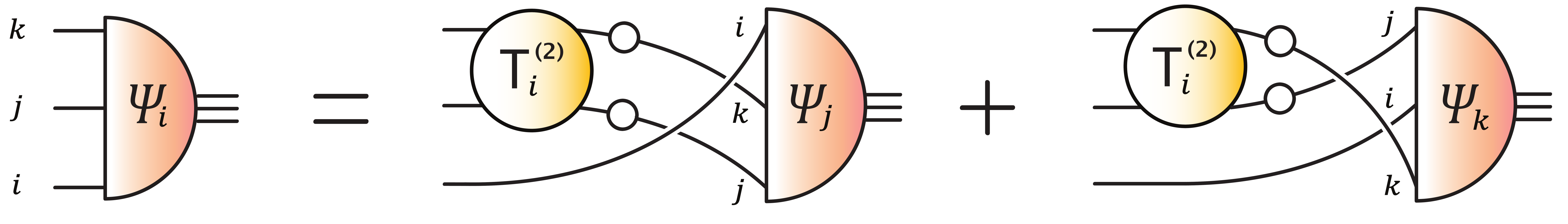}
            \caption[Relativistic Faddeev equation 2]{\backdef
                                         Relativistic Faddeev equation \eqref{QDQ:faddeev2} which involves the 2-quark scattering matrix.} \label{fig:faddeevtruncated2}
            \end{center}
            \medskip
            \end{figure}

            \bigskip
            \fatcol{Faddeev equation revisited.}
            As a first step in deriving the quark-diquark BSE from the relativistic Faddeev equation \eqref{bs:faddeevtruncated},
            one introduces the Faddeev components $\Psi_i$ of the baryon amplitude via
            \begin{equation}\label{QDQ:faddeev1}
                \Psi = \sum_{i=1}^3 \widetilde{K}^{(2)}_i \Psi =: \sum_{i=1}^3 \Psi_i\,.
            \end{equation}
            The $\Psi_i$ are identical to the three graphs on the right-hand side of Fig.~\ref{fig:faddeev}.
            Again, the index $i$ refers to the non-interacting spectator quark which contributes to each diagram.
            The equation is subsequently rewritten in a way where the 2-quark scattering matrices $\widetilde{T}^{(2)}_i$ appear instead of
	        the 2-quark kernels $\widetilde{K}^{(2)}_i$.
            They are related to each other through Dyson's equation \eqref{BS:dyson} which entails
            \begin{equation} \label{QDQ:Dyson-anew}
                \widetilde{T}_i^{(2)} = \left( 1 +  \widetilde{T}_i^{(2)} \right) \widetilde{K}_i^{(2)}   \quad \Longrightarrow \quad
                \widetilde{T}_i^{(2)} \Psi = \left( 1 +  \widetilde{T}_i^{(2)} \right) \Psi_i\,.
            \end{equation}
            The relativistic Faddeev equation is thereby transformed into a set of coupled integral equations for the components $\Psi_i$:
            \begin{fshaded2}
            \begin{equation} \label{QDQ:faddeev2}
                \Psi_i = \widetilde{T}_i^{(2)} \left( \Psi - \Psi_i \right) = \widetilde{T}_i^{(2)} \left( \Psi_j + \Psi_k \right),
            \end{equation}
            \end{fshaded2}

            \noindent
            where $\{i,j,k\}$ is an even permutation of $\{1,2,3\}$. This equation is depicted in Fig.\,\ref{fig:faddeevtruncated2}.

            No new information has been gathered by this transformation.
            Instead of directly implementing an ansatz for the kernel in the original equation \eqref{QDQ:faddeev1},
            the modified version necessitates an expression for the $qq$ scattering matrix $\widetilde{T}^{(2)}$
            which must be determined from the kernel in the intermediate step \eqref{QDQ:Dyson-anew}.
            In particular, the Faddeev components $\Psi_i$ are still three-body amplitudes which
            depend on three independent momenta and carry the same Dirac structure as the amplitude $\Psi$.
            A solution of the equation relies upon the solution techniques  described in Section~\ref{chapter:faddeev}.

            An investigation of the structure of the T-matrix in this setup would certainly be an interesting issue in itself.
            The basic intention of Eq.\,\eqref{QDQ:faddeev2} is however to replace that solution by an ansatz
            which is adequate to reduce the complexity of the equation, i.e., by simplifying the Faddeev approach
            to a  two-body problem. 

\begin{figure}[tbp]
            \begin{center}
            \includegraphics[scale=0.2]{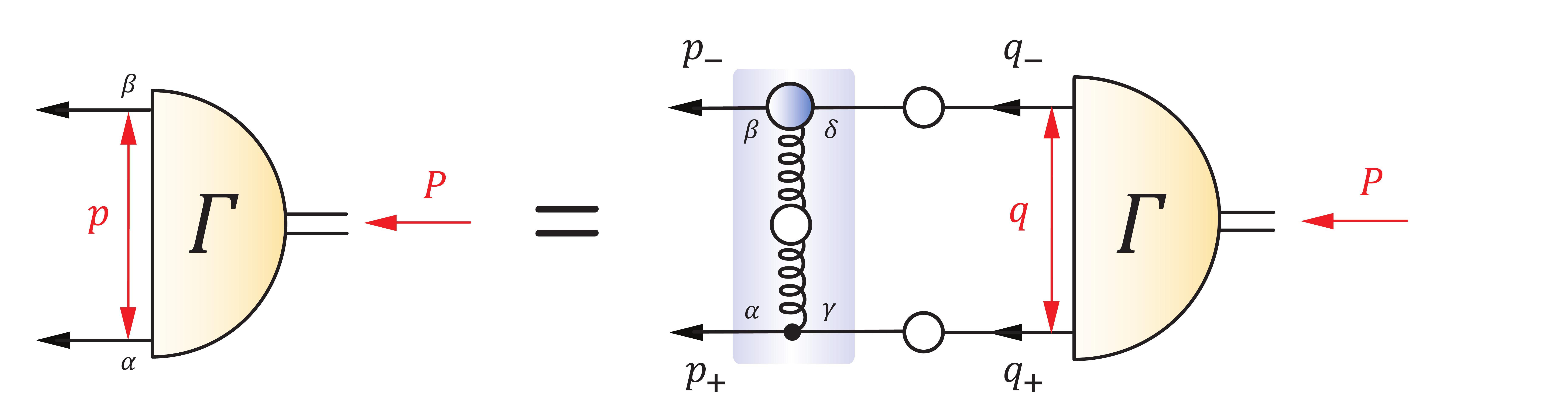}
            \caption[Diquark BSE]{\backdef The Diquark BSE \eqref{dq:bse} in RL truncation.} \label{fig:BSE}
            \end{center}
            \end{figure}

\newpage
            \fatcol{Diquark correlations.}
            Up to this point we have assumed that correlations between two quarks provide the dominant attraction not only in the
            meson but also in the baryon channel; and that these are dominated by an iterated gluon exchange between any pair of two quarks which constitute the baryon.
            In the following discussion we will further exploit this concept and aim at an ansatz for the T-matrix which
            contains \textit{diquark} poles at timelike values of the total quark-quark momentum $P^2$
            but still retains some of the characteristic features of Dyson's equation \eqref{BS:dyson}.
            This is realized through a separable sum over
            diquark correlations. We restrict ourselves to the $0^+$ and $1^+$ channels,
            i.\,e.~to scalar and axial-vector diquarks for reasons explained below.
            The corresponding expression for $T^{(2)}$ reads:
            \begin{fshaded2}
            \begin{equation}\label{dq:tmatrixansatz}
                T^{(2)}_{\alpha\gamma, \beta\delta}(p,q,P) = \sum_{(\mu\nu)}\Gamma^{(\mu)}_{\alpha\beta}(p,P)\,D^{(\mu\nu)}(P^2) \,\conjg{\Gamma}^{(\nu)}_{\delta\gamma}(q,P)\,,
            \end{equation}
            \end{fshaded2}

            \noindent
            where the notation $(\mu\nu)$ implies that the Lorentz indices can be dropped in the scalar-diquark contribution.
            $\Gamma^{(\mu)}$ are the diquark analogues of the respective pseudoscalar and vector meson amplitudes
            and the $D^{(\mu\nu)}$ denote scalar and axial-vector diquark propagators.
            The assumed poles of the T-matrix in the ansatz \eqref{dq:tmatrixansatz} are embedded in the diquark propagators and define the associated diquark masses:
            \begin{equation}
                D(P^2) \stackrel{P^2\rightarrow -M_\text{sc}^2}{\longlonglongrightarrow} \frac{1}{P^2+M_\text{sc}^2}\,, \quad
                D^{\mu\nu}(P^2) \stackrel{P^2\rightarrow -M_\text{av}^2}{\longlonglongrightarrow} \frac{T^{\mu\nu}_P}{P^2+M_\text{av}^2}\,,
            \end{equation}
            where $T^{\mu\nu}_P$ again denotes a transverse projector, now with respect to the diquark momentum $P$.
            In the same manner as described in Section~\ref{sec:qcd:bses},
            this leads to a homogeneous diquark BSE,  $\Gamma = \widetilde{K}^{(2)}\Gamma$,
            for a diquark bound-state on the mass shell $P^2=-M^2$ which resembles the meson BSE given in Eq.\,\eqref{bse:bse}. It was
	        used for detailed studies of diquarks, e.\,g., in \cite{Maris:2002yu,Maris:2004bp}. The equation is shown
	        diagrammatically in Fig.~\ref{fig:BSE} and reads
            \begin{equation}\label{dq:bse}
                \Gamma^{(\mu)}_{\alpha\beta}(p,P) = \int^\Lambda_q K_{\alpha\gamma,\beta\delta}(p,q,P) \left\{ S(q_+) \,\Gamma^{(\mu)}(q,P) \,S^T(q_-) \right\}_{\gamma\delta}\,,
            \end{equation}
            where the momenta have been defined in the discussion of Eq.\,\eqref{bse:bse}.
            The replacements $S(-q_-) \rightarrow S^T(q_-)$ and $K_{\alpha\gamma,\delta\beta} \rightarrow K_{\alpha\gamma,\beta\delta}$
            amount to a substitution of an antiquark- with a quark leg.
            For the sake of consistency the kernel $K$ is identified with the rainbow-ladder kernel \eqref{bse:rlkernel}.

            By working out the color traces of Eqs.\,\eqref{bse:rlkernel} and \eqref{dq:amps} one finds that
            the resulting equation for $\Gamma\,C^\dag$ with quantum numbers $J^P$
            is identical to that of a color-singlet $J^{-P}$ meson except for the diquark's coupling strength
            which is reduced by a factor of $2$. This confirms that the interaction in the color anti-triplet diquark
            channel is strong and attractive. The same analysis entails that the interaction is strong and repulsive in the color
	        sextet channel \cite{Cahill:1987qr,Hecht:2000jh}.
            Comparison with meson phenomenology hence suggests that the lightest diquarks are the
	        scalar diquarks (the parity partners of the pseudoscalar mesons),
            followed by axial-vector, pseudoscalar and vector diquarks.
            This was also observed in Bethe-Salpeter \cite{Maris:2002yu} and lattice studies studies (cf. Table~\ref{tab:dqmasses})
	        and justifies the restriction to the scalar and axial-vector diquark channels for describing light baryons composed of quark and diquark.

            \bigskip
            \fatcol{Diquark poles and diquark confinement.}
            A necessary prerequisite to justify the particular ansatz \eqref{dq:tmatrixansatz} is that  the 
            rainbow-ladder kernel, via \eqref{QDQ:Dyson-anew},
            indeed generates timelike scalar and axial-vector diquark poles in the quark-quark scattering matrix.
            While this has not been explicitly studied in the context of Dyson's equation, 
            the existence of a solution to Eq.\,\eqref{dq:bse} provides a satisfactory indication.

            Asymptotic diquark states correspond to timelike poles in the diquark propagators,
            hence one might suspect a violation of diquark confinement.
            However, the absence of a Lehmann representation of a certain propagator
            is a sufficient but not necessary criterion for confinement of the corresponding state
            due the associated violation of reflection positivity \cite{Osterwalder:1973dx,Krein:1990sf,Haag:1992hx,Alkofer:2000wg}.
            Two-point correlations of colored fields may contain
            real timelike poles in momentum space without contradicting confinement,
            a statement which is also true for the quark propagator \cite{Alkofer:2003jj,Oehme:1994pv,Roberts:2000aa}.
            In addition, free-particle quark and diquark propagators
            can yield quantitatively meaningful results for hadronic observables (see, e.\,g., \cite{Oettel:2000jj}).

	        On the other hand, the introduction of beyond-RL interaction terms in the skeleton
	        expansion of the $qq$ kernel which appears in the diquark BSE removes diquark states from the physical mass spectrum
            due to large repulsive corrections \cite{Bender:2002as,Bender:1996bb,Bhagwat:2004hn,Hellstern:1997nv}.
            In this respect, kernels that do not produce diquark bound states but induce a more complicated singularity structure in the $qq$ T-matrix can still support a physical interpretation
            either in terms of mass scales or inverse correlation lengths of diquark interactions inside a baryon.
            Further motivation for the significance of the
	        diquark concept has come from investigations of diquark confinement in Coulomb-gauge QCD~\cite{Alkofer:2005ug}.

            \begin{figure}[p]
            \begin{center}
            \includegraphics[scale=0.88]{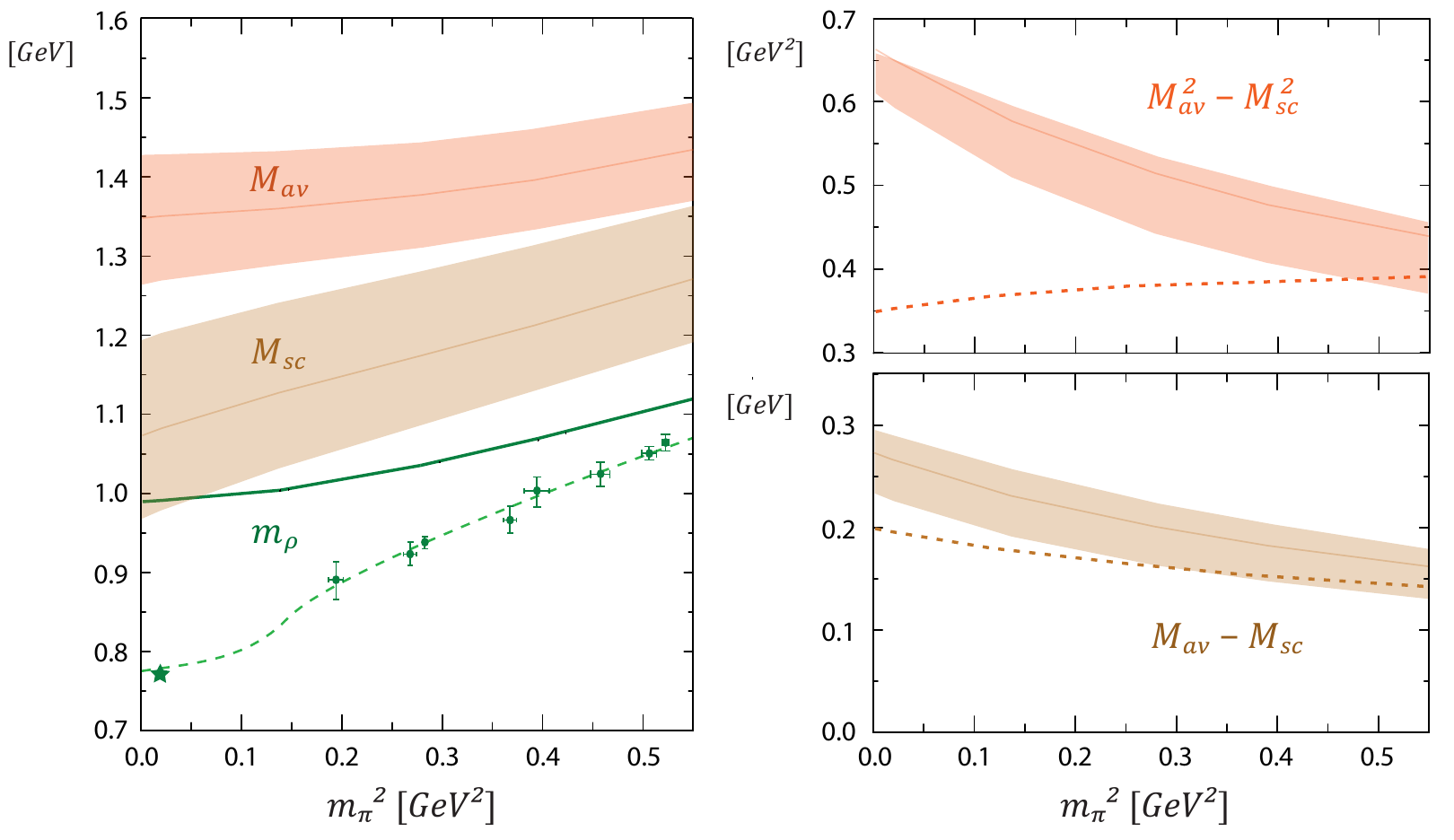}
            \caption[Diquark masses]{\backdef
                                     \textit{Left panel:}
                                     Scalar and axial-vector diquark masses from the diquark BSE vs.~squared pion mass \cite{Eichmann:2008ef}.
                                     The bands denote a variation of $\omega$ in setup \hyperlink{coupling:c3}{(C3)};
                                     the central lines correspond to $\conjg{\omega}$.
                                     For comparison we show the \hyperlink{coupling:c3}{(C3)} input for $m_\rho$ (\textit{solid line}),
                                     together with its experimental value (\textit{star}),
                                     lattice data and a chiral extrapolation (\textit{dashed line}).
                                     Respective references are given in Fig.~\ref{fig:mrho}.
                                     \textit{Right panels:} Scalar--axial-vector mass splitting (\textit{lower panel}) and squared-mass splitting (\textit{upper panel}).
                                     Dashed lines are the results of \hyperlink{coupling:c1}{(C1)}; bands those of \hyperlink{coupling:c3}{(C3)}.
                                     } \label{fig:dqmasses}
            \end{center}
            \end{figure}

 \renewcommand{\arraystretch}{1.0}

            \begin{table*}[p]
                \begin{center}

                \begin{tabular}{l||c|c||c|c}
                                                & $\Lambda_\textrm{IR}$  &  $\eta$      &  $M_\textrm{sc}$  & $M_\textrm{av}$    \\ \hline
                 \hyperlink{coupling:c1}{(C1)}  & $0.72$                 &  $1.8$       &  $0.81$           &  $1.00$              \\
                 \hyperlink{coupling:c3}{(C3)}  & $0.98$                 &  $1.8(2)$    &  $1.08(10)$       &  $1.35(8)$
                \end{tabular}
                 \caption[]{\backdef
                          Comparison of diquark masses $M_\textrm{sc}$, $M_\textrm{av}$ as obtained from the diquark BSE
                          in setups \hyperlink{coupling:c1}{(C1)} and \hyperlink{coupling:c3}{(C3)}, characterized by the parameters $\Lambda_\textrm{IR}$ and $\eta$ defined in \eqref{MESON:c+eta}.
                          A variation of $\eta = 1.8 \pm 0.2$ in \hyperlink{coupling:c3}{(C3)} is equivalent to $\omega \approx \conjg{\omega} \pm 0.06$.
                          The results correspond to a current mass $\hat{m}=6.1$ MeV which is related to the
			              physical pion mass $m_\pi=138$ MeV.
                          The units of $\Lambda_\textrm{IR}$, $M_\textrm{sc}$ and $M_\textrm{av}$ are GeV; $\eta$ is dimensionless. }  \label{tab:dqmasses}

                \end{center}
            \end{table*}

            \begin{table}[p]
                \begin{center}
                \begin{tabular}{cc|cccc} \label{tab:dqmass}
                  \hyperlink{coupling:c1}{(C1)}    & \hyperlink{coupling:c3}{(C3)}  & Ref.\,\cite{Wetzorke:2000ez}  &  Ref.\,\cite{Alexandrou:2006cq}  &   Ref.\,\cite{Babich:2007ah} &   Ref.\,\cite{Orginos:2005vr}   \\ \hline
                                  $0.20$           &               $0.27(3)$        & $0.10(5)$                     &  $0.14(1)$                       &   $0.29(4)$                  &   $0.36(7)$
                \end{tabular}
                \caption{\backdef
                         Scalar-axialvector diquark mass splitting in the chiral limit, cf. Fig.~\ref{fig:dqmasses}. The BSE values are compared
                         to several lattice-QCD results. The units are GeV. } \label{tab:DQmass-splitting}
                \end{center}
            \end{table}

    \bigskip
    \fatcol{Diquark masses.}
            While diquark masses are gauge-dependent, gauge-independent mass differences can be determined from lattice calculations.
            Several such investigations have been performed in different approaches and with various fermion actions
            \cite{Hess:1998sd,Wetzorke:2000ez,Orginos:2005vr,Liu:2006zi,Alexandrou:2006cq,Babich:2007ah}. While they exhibit quite different quantitative
            results, the common qualitative feature is that the mass splitting between scalar and axial-vector diquark
            in the chiral limit is of the size of hundred to several hundred MeV (see Table~\ref{tab:dqmass}) and decreases with increasing
            current quark mass.

            Fig.\,\ref{fig:dqmasses} shows scalar and axial-vector masses together with their mass and squared-mass difference as a result of the diquark BSE \eqref{dq:bse}.
            The diquark masses exhibit large sensitivities to the width parameter $\omega$, a feature which has previously been observed in Ref.~\cite{Maris:2002yu}.
            The scalar--axial-vector mass splitting is comparatively weakly dependent on model details, cf. Table~\ref{tab:DQmass-splitting}.
            The deviation in the squared-mass difference between setups \hyperlink{coupling:c1}{(C1)} and \hyperlink{coupling:c3}{(C3)} has its origin in the current-mass dependence of
            the scale $\Lambda_\text{IR}$ in \hyperlink{coupling:c3}{(C3)} and can be removed by examining the ratio
            \begin{equation}
                \frac{M_\text{av}^2-M_\text{sc}^2}{\Lambda_\text{IR}^2} \stackrel{\eqref{eq:mesons:massformula}}{\approx}
                \left(a_\text{av}^2-a_\text{sc}^2\right) + \left(b_\text{av}^2-b_\text{sc}^2\right) \frac{m_\pi^2}{\Lambda_\text{IR}^2}\,.
            \end{equation}
            Since $\Lambda_\text{IR}$ is constant in setup \hyperlink{coupling:c1}{(C1)}, a weak current-mass dependence (dashed line in the upper right plot of Fig.~\ref{fig:dqmasses})
            implies that the squared-mass difference is predominantly generated by the infrared contribution to the effective coupling \eqref{dse:maristandy}
            which owes to dynamical chiral symmetry breaking.

    \bigskip
    \fatcol{Offshell behavior of the T-matrix.}
            Upon solving the scalar and axial-vector diquark BSEs \eqref{dq:bse},
            the quark-quark scattering matrix is determined at the diquark mass poles
            $P^2= -M_\text{sc}^2$ and $P^2=-M_\text{av}^2$.
            Since the diquarks in the nucleon are off-shell,
            the description of baryons as composites of quark and diquark requires knowledge of the T-matrix
            for general diquark momenta as well.
            The ansatz \eqref{dq:tmatrixansatz} dictates its off-shell behavior
            to be inherited from the separable structure in the vicinity of the poles
            in terms of off-shell diquark amplitudes and propagators.

            Given a certain ansatz for the off-shell amplitudes,
            one may exploit Dyson's equation \eqref{BS:dyson} to obtain an expression for the
            diquark propagators. This procedure is detailed in Appendix~\ref{app:mesondiquark-dqprop} and yields propagators of the form
            \begin{equation}
                D(P^2) = R(P^2) + \frac{1}{P^2+M^2}
            \end{equation}
            where the finite parts $R(P^2)$ emulate off-shell contributions which are suppressed at the mass poles
            but determine the ultraviolet behavior of the T-matrix.


\section{Quark-diquark BSE}\label{sec:qdqbse}

\begin{figure}[tbp]
            \begin{center}
            \includegraphics[scale=0.2]{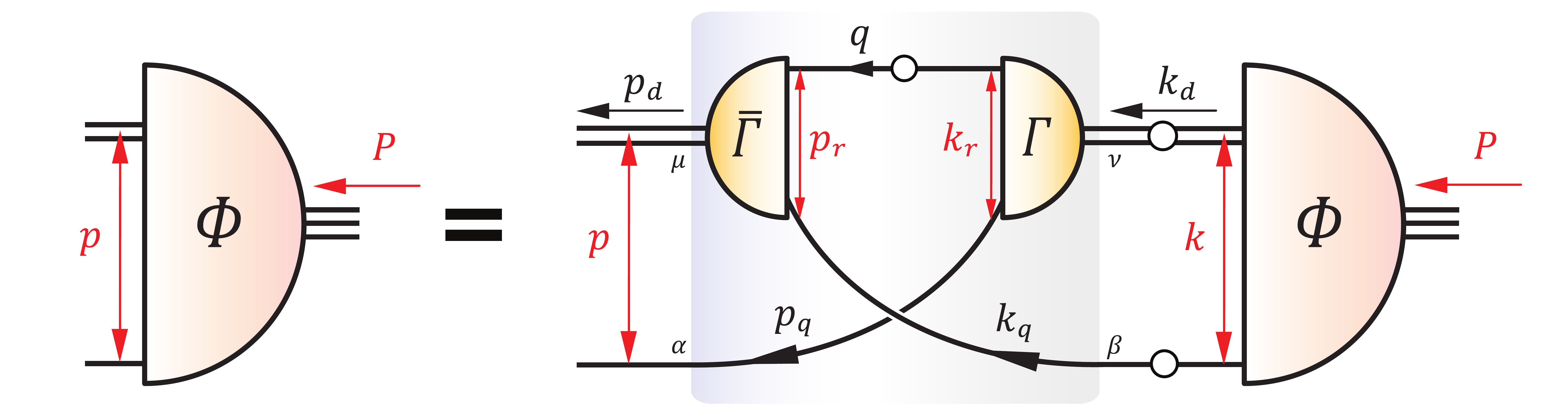}
            \caption[Quark-Diquark BSE]{\backdef The quark-diquark BSE \eqref{nuc:bse} in
            pictorial form. } \label{fig:qdqbse}
            \end{center}
            \end{figure}

        \fatcol{Separable ansatz for the amplitudes.}
            The relativistic Faddeev equations in the form \eqref{QDQ:faddeev2} represent coupled integral equations for either of the Faddeev components $\Psi_i$.
            All elements of the equation, i.\,e., the dressed-quark propagator and the diquark ingredients of the quark-quark scattering matrix,
            have been specified in the previous sections.
            The separability of the diquark ansatz \eqref{dq:tmatrixansatz} for the T-matrix is one of two requirements to reduce the original three-body equation
            for the baryon to a two-body Bethe-Salpeter equation for a quark-diquark bound state.
            The second prerequisite is the separability of the components $\Psi_i$ in terms of the relative momenta
            between quark and diquark ($p$) and within the diquark, where the latter is now denoted by $p_r$, cf. Fig.~\ref{fig:qdqbse}.
            This is realized via the following ansatz for the $\Psi_i$:
            \begin{equation}\label{nuc:defqdqamp}
                \Psi_{\alpha\beta\gamma\delta}(p,p_r,P)   = \sum_{a,b} \Gamma_{\beta\gamma}^a(p_r,p_d) \, D^{ab}(p_d) \, \Phi^{b}_{\alpha\delta}(p,P),
            \end{equation}
            which involves a combination of scalar and axial-vector diquark amplitudes and propagators.
            $p_d$ and $P$ are the total diquark and nucleon momenta, respectively.
            The superscripts $a,b,c$ collect the diquarks' Lorentz indices: $a=5$ denotes scalar and $a=1\dots 4$ axial-vector
            quantities; the diquark propagator $D^{ab}$ is either scalar ($a=b=5$) or axial-vector ($a,b=1\dots 4$).
            The Dirac, color and flavor decomposition of the quark-diquark amplitudes $\Phi^a_{\alpha\beta}(p,P)$ thereby defined
            is described in App.\,\ref{app:qdqamp}.
            The final baryon amplitudes, as stated for the nucleon in Eq.\,\eqref{FE:nucleon_amplitude_full}, are constructed by symmetrizing the Faddeev components in Eq.\,\eqref{nuc:defqdqamp}.


            The spin- and isospin-$\nicefrac{1}{2}$ nucleon is a sum of scalar and axial-vector diquark correlations,
            whereas the $SU(2)$ flavor algebra excludes scalar diquarks from participating in the spin- and isospin-$\nicefrac{3}{2}$ $\Delta$ baryon: only
            an isospin-$1$ axial-vector diquark can be combined with an isospin-$\nicefrac{1}{2}$ quark to obtain $I=\nicefrac{3}{2}$.
            The quark-diquark amplitude $\Phi^a$ thus appears in the following manifestations:
            \begin{equation}\label{QDQ:Amplitudes-N-Delta-1}
            \begin{split}
               N: \quad      & \Phi^{a}(p,P) \quad \longrightarrow \quad \Phi^5(p,P), \;  \Phi^{\mu}(p,P), \\
               \Delta: \quad & \Phi^{a}(p,P) \quad \longrightarrow \quad \Phi^{\mu\nu}(p,P).
            \end{split}
            \end{equation}
            The final $N$ and $\Delta$ quark-diquark spinors  are obtained upon contraction with the
            Dirac and Rarita-Schwinger spinors $u_\delta(P)$, $u^\nu_\delta(P)$
            which describe free spin--$\nicefrac{1}{2}$ or --$\nicefrac{3}{2}$ particles with momentum $P$:
            \begin{equation}
            \begin{split}
                N: \quad & \Phi^5_{\alpha\delta}(p,P) \,u_\delta(P), \; \Phi^{\mu}_{\alpha\delta}(p,P)\,u_\delta(P), \\
                 \Delta: \quad &  \Phi^{\mu\nu}_{\alpha\delta}(p,P) \,u^\nu_\delta(P).
            \end{split}
            \end{equation}

        \bigskip
        \fatcol{Two-body equation.}
            Inserting the ansatz \eqref{dq:tmatrixansatz} for the T-matrix together with \eqref{nuc:defqdqamp}
            into the relativistic Faddeev equation \eqref{QDQ:faddeev2}  yields a quark-diquark Bethe-Salpeter equation on the baryon's mass shell \cite{Oettel:1998bk,Oettel:2000jj}:
            \begin{fshaded2}
            \begin{equation}\label{nuc:bse}
                \Phi_{\alpha\beta}^a(p,P) = \int_k \left\{ K^{ab}(p,k,P) \, S(k_q) \, \Phi^c(k,P) \right\}_{\alpha\beta} \, D^{bc}(k_d),
            \end{equation}
            \end{fshaded2}

            \noindent
            where the kernel of the equation is given by
            \begin{equation}\label{nuc:kernel}
                K^{ab}_{\alpha\beta}(p,k,P) = \left\{ \Gamma^b(k_r,k_d) \, S^T(q) \, \conjg{\Gamma}^a(p_r,-p_d) \right\}_{\alpha\beta}.
            \end{equation}
            It couples scalar and axial-vector diquarks (i.e., $a=1\dots 5$) in the case of the nucleon whereas
            only the axial-vector index $a=1\dots 4$ is required for the $\Delta$.
            The kernel \eqref{nuc:kernel} describes an iterated exchange of roles between the spectator quark and the quarks which constitute the diquark;
            this quark exchange generates the attractive interaction that binds quarks and diquarks to a baryon.

            The momenta in Eqs.\,(\ref{nuc:bse}--\ref{nuc:kernel}) read (cf.\,Fig.\,\ref{fig:qdqbse}):
            \begin{equation} \label{nuc:momenta}
                \begin{aligned}
                    p_q &= p+\eta\,P\,,  \\
                    k_q &= k+\eta\, P\,,
                \end{aligned} \qquad
                \begin{aligned}
                    p_d &= -p+(1-\eta)\,P\,,   \\
                    k_d &= -k+(1-\eta)\,P\,,
                \end{aligned} \qquad
                \begin{aligned}
                    q &= p_d-k_q\,,  \\
                    k_r &= (1-\sigma)\,p_q-\sigma\,q\,, \\
                    p_r &= (1-\sigma)\,k_q-\sigma\,q\,.
                \end{aligned}
            \end{equation}
            $p$ and $p_r$ are the relative momenta between quark and diquark and within the diquark; $p_d$, $P$
            are total diquark and nucleon momenta. The respective momenta $k$ appear inside the loop integral.
            Again, the momentum partitioning parameters $\sigma,\,\eta \in [0,1]$ for diquark and quark-diquark amplitudes
            are arbitrary.
            In contrast to the analogous case in the three-body equation, $\eta=\nicefrac{1}{3}$ is no longer the optimal choice since
            the pole limits now result from the combined singularity structures of quark and diquark propagators
            (and, theoretically, also of the diquark amplitudes).
            We set $\sigma=1/2$ but keep $\eta$ as a variable since it can be used to ease these constraints (cf. App.\,\ref{app:singularities}).

            Upon working out the color and flavor factors of the quark-diquark amplitudes, given in Eqs.\,\eqref{nuc:amplitudes} and \eqref{deltaamplitdecompos},
            and of the diquark amplitudes \eqref{dq:amps}, the BSE kernel picks up a color-flavor factor
            \begin{equation}
            N: \quad \frac{1}{2}
            \left(%
            \begin{array}{r@{\quad}r}
              -1         & \sqrt{3} \\
              \sqrt{3} & 1        \\
            \end{array}%
            \right), \quad\quad\quad
            \Delta: \quad
            \left(%
            \begin{array}{r@{\quad}r}
              0 & 0 \\
              0 & -1        \\
            \end{array}%
            \right),
            \end{equation}
            where the first row (column) represents the scalar part of the
            kernel and the second row (column) the axial-vector part.


\section{Nucleon and $\Delta$ masses}\label{sec:results:nuclenmass}

\begin{figure}[p]
                    \begin{center}
                        \includegraphics[scale=0.9]{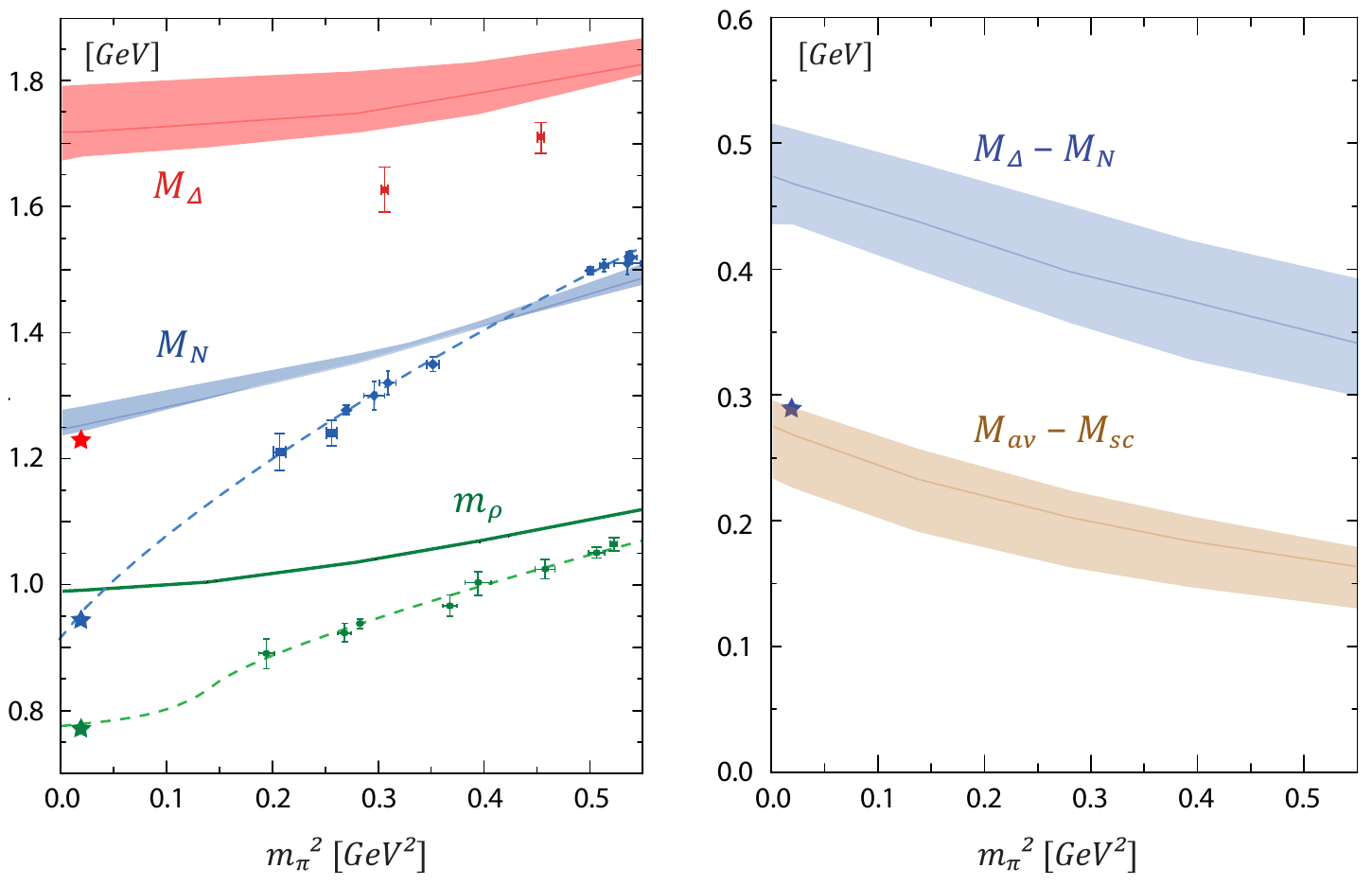}
                        \caption[]{\backdef Evolution of $N$ and $\Delta$ masses (\textit{left panel})
                                 and the mass splittings $M_\Delta-M_N$ and $M_\textrm{av}-M_\textrm{sc}$ (\textit{right panel})
                                 vs.~pion mass squared \cite{Eichmann:2008ef,Nicmorus:2008vb}, obtained in the quark-diquark model using setup  \hyperlink{coupling:c3}{(C3)}.
                                 The bands denote the sensitivity to a variation of $\omega$.
                                 We compare to a selection of dynamical lattice data and their chiral extrapolations (\textit{dashed lines}) for
                                 $M_N$ \cite{AliKhan:2003cu,Frigori:2007wa,Leinweber:2003dg} and $M_\Delta$ \cite{Zanotti:2003fx}.
                                 The depicted data for $m_\rho$ are identical to those of Figs.~\ref{fig:mrho} and \ref{fig:dqmasses}.
                                 Stars denote the experimental values.}\label{fig:2}
                    \end{center}
         \end{figure}

                The results of the quark-diquark BSE \eqref{nuc:bse} for nucleon and
                $\Delta$ masses in the 'quark core' setup \hyperlink{coupling:c3}{(C3)} are shown in Fig.\,\ref{fig:2}.
                The left panel depicts the calculated values for
            	$m_\rho$, $M_N$ and $M_\Delta$, each together with a selection of lattice results and their chiral
            	extrapolations (if available). The corresponding abscissa values $m_\pi^2$ are
            	obtained from the pseudoscalar meson BSE. The solid curve for $m_\rho$ is the input defined in
                Eqs.~(\ref{core:mrho}--\ref{core:B}) and completely fixes the parameters in the interaction.
                The bands represent the sensitivity of the results for $M_N$ and $M_\Delta$ on the width
            	parameter $\omega$.
                At larger quark masses the deviation from the lattice data diminishes, in accordance with the assumption of Eq.~(\ref{core:mrho}),
                namely that beyond-RL corrections to hadronic observables become small in the heavy-quark limit.

                The resulting values at the physical pion mass in both setups \hyperlink{coupling:c1}{(C1)} and \hyperlink{coupling:c3}{(C3)} are displayed in Table~\ref{tab:QDQ:masses}.
                Again, Eq.\,\eqref{eq:mesons:massformula} can be used to relate both models to each other.
	            The intention of the former was to reproduce $\pi$ and $\rho$ properties;
                in addition, the model also yields $N$ and $\Delta$ masses that are close to the experimental values.
                This is quite remarkable since, upon fixing the coupling strength in the effective quark-gluon coupling to meson phenomenology,
                no further parameters have been used as an input of the calculation.

                A comparison with the results of Section~\ref{sec:FADDEEV:results}, obtained through a solution of the relativistic Faddeev equation,
                shows that the corresponding nucleon mass is larger than the result of the quark-diquark approach.
                Fig.~\ref{fig:FADDEEV:nucleon-mass} indicates an approximately uniform shift by $50-100$ MeV
                throughout the examined current-quark mass range.
                This illustrates that the quark-diquark approach is somewhat too attractive in the nucleon channel
                but still a quite reasonable approximation.

         \begin{table}[p]
                 \begin{center}
                     \renewcommand{\arraystretch}{1}
                     \begin{tabular}{l||c|c||c|c}
                                                     & $\Lambda_\textrm{IR}$  &  $\eta$      &  $M_N$        &  $M_\Delta$   \\ \hline
                      Exp.                           &                        &              &  $0.94$       &  $1.23$       \\ \hline
                      \hyperlink{coupling:c1}{(C1)}  & $0.72$                 &  $1.8$       &  $0.94$       &  $1.28$        \\
                      \hyperlink{coupling:c3}{(C3)}  & $0.98$                 &  $1.8(2)$    &  $1.26(2)$    &  $1.73(5)$
                     \end{tabular}
                      \caption[]{\backdef Comparison of $M_N$ and $M_\Delta$ for both model inputs \hyperlink{coupling:c1}{(C1)} and \hyperlink{coupling:c3}{(C3)}
                               at the physical point. The parameter $\eta$ is defined in \eqref{MESON:c+eta}.
                               The first row quotes the experimental values.
                               The units of $\Lambda_\textrm{IR}$, $M_\textrm{sc}$ and $M_\textrm{av}$ are GeV; $\eta$ is dimensionless.
                               }  \label{tab:QDQ:masses}
                 \end{center}
         \end{table}

            \begin{figure}[p]
            \begin{center}
            \includegraphics[scale=0.1]{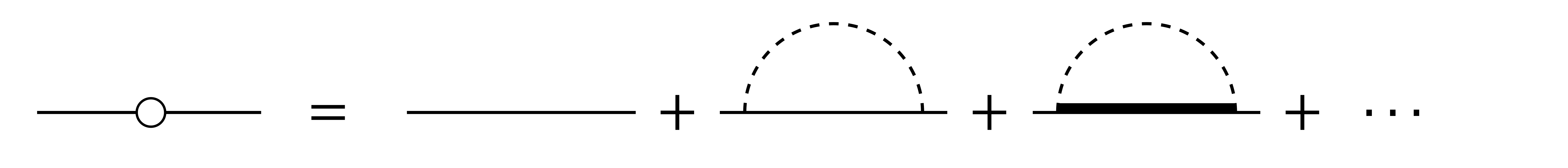}
            \caption[Chiral expansion]{\backdef Expansion of the nucleon propagator in chiral perturbation theory.
                                       Dashed, solid, and thick solid lines correspond to pseudoscalar-meson, nucleon and $\Delta$ degrees of freedom.} \label{fig:chiralexpansion}
            \end{center}
            \end{figure}

      \bigskip
      \fatcol{Pionic corrections.}
            It is instructive to compare our results to baryonic core masses estimated from chiral effective field theory.
            In this framework the mass of a baryon $B$ is obtained from the expression (e.\,g., \cite{Young:2002cj})    
            \begin{equation}\label{BARYON:chpt}
                M_B(m_\pi^2) = M_B^\text{core}(m_\pi^2,\Lambda)  + \Sigma_B(m_\pi^2,\Lambda)\,,
            \end{equation}
            where the "baryon core" includes the bare (and a priori unknown) parameters $a_i$ appearing in the effective Lagrangian:
            \begin{equation}
                M_B^\text{core}(m_\pi^2,\Lambda) = a_B^{(0)}(\Lambda) + a_B^{(2)}(\Lambda) \, m_\pi^2 + a_B^{(4)}(\Lambda) \, m_\pi^4 + \dots
            \end{equation}
            The inherent assumption of a momentum cutoff regularization, expressed through the scale $\Lambda$,
            allows for a physical interpretation in terms of a scale separation
            between quark core and pion cloud.
            The cutoff $\Lambda$ regularizes the short-distance divergences associated
            with pointlike baryons and pions in the effective field theory
            and retains the long-distance or low-energy part of the self-energy integrals.
            A non-pointlike nucleon-pion interaction requires a cutoff of the order of the baryon size ($\Lambda > 0.2$ GeV $\sim r<1$~fm).
            Of course one may alternatively employ a dimensional regularization technique \cite{Donoghue:1998bs}.

            $\Sigma_B$ in Eq.\,\eqref{BARYON:chpt} denotes the sum of meson-loop self-energies, e.g.
            to 1-loop order the sum of intermediate $N\pi$ and $\Delta\pi$ states (cf. Fig.\,\ref{fig:chiralexpansion}).
            In heavy baryon chiral perturbation theory \cite{Jenkins:1990jv,Labrenz:1996jy}, the respective contributions are given by \cite{Thomas:1999mu,Young:2002cj}:
            \begin{equation}\label{BARYON:meson-selfenergy}
            \begin{split}
                \Sigma_N(m_\pi^2, \Lambda) &= -\lambda\,\Lambda^3 \int_0^\infty dx\, \frac{x^4}{\omega^2} \left[ u_{NN}^2(x) + \frac{32}{25}\,\frac{\omega}{\omega+\delta}\,u_{N\Delta}^2(x) \right], \\
                \Sigma_\Delta(m_\pi^2, \Lambda) &= -\lambda\,\Lambda^3 \int_0^\infty dx\, \frac{x^4}{\omega^2} \left[ u_{\Delta\Delta}^2(x) + \frac{8}{25}\,\frac{\omega}{\omega-\delta}\,u_{N\Delta}^2(x) \right],
            \end{split}
            \end{equation}
            where $x$, $\omega(x)$ and $\delta$ are pion momentum, intermediate pion energy and the physical $N$--$\Delta$ mass splitting, normalized by the cutoff $\Lambda$, and
            the prefactor $\lambda$ includes the experimental values for the pion decay constant $f_\pi=131$ MeV and the axial coupling $g_A=1.26$:
            \begin{equation}
                x=\frac{k}{\Lambda}\,, \quad \omega(x) = \sqrt{x^2+\left(\frac{m_\pi}{\Lambda}\right)^2}\,, \quad \delta = \frac{0.29\,\text{GeV}}{\Lambda}\,, \quad \lambda = \frac{3 \,g_A^2}{8\pi^2 f_\pi^2}\,.
            \end{equation}
            The cutoff-dependent $NN\pi$, $\Delta\Delta\pi$ and $N\Delta\pi$ vertex dressings are denoted by $u_{BB'}(x)$.
            A dipole form factor mimics the physical shape of the meson-baryon vertex. 
            Choosing all dressings identical via  $u(x) = ( 1 + x^2 )^{-2}$ with a regulator $\Lambda=0.8$ GeV
        	yields for $m_\pi=140$ MeV:
            \begin{equation}\label{CHPT:sigma}
                \Sigma_N = -0.30 \, \text{GeV}, \quad  \Sigma_\Delta=-0.27 \, \text{GeV},
            \end{equation}
            and hence a similar reduction of both nucleon and $\Delta$ masses. In combination with the experimental
        	numbers for $M_N$ and $M_\Delta$, \eqref{CHPT:sigma} provides the simple estimates
            \begin{equation}\label{CHPT:core}
                M_N^\textrm{core} \sim 1.24 \, \text{GeV}, \quad  M_\Delta^\textrm{core} \sim 1.5 \, \text{GeV}.
            \end{equation}
            These values are roughly consistent with those obtained in, e.\,g.,
            the cloudy-bag model \cite{Pearce:1986du}, NJL model \cite{Ishii:1998tw}, and
        	nucleon-pion Dyson-Schwinger studies \cite{Hecht:2002ej,Oettel:2002cw}
            and make clear that nucleon-pion loops are attractive and the binding energy reduces the baryon's mass.

            The separation into a "core" and a "meson cloud" contribution is cutoff-dependent.
            Due to the approximate $\Lambda^3$--dependence of the self-energy integrals,
            different values of the regulator can vary the above results considerably. 
            Naturally, the chiral extrapolations of the physical $N$ and $\Delta$ masses depicted in Fig.~\ref{fig:2} must be
            independent of the cutoff $\Lambda$.
            Via expansion in $m_\pi^2$, the self-energy contributions $\Sigma_B$ are split into
            cutoff-independent non-analytic terms  $\sim m_\pi^3$, $ \sim m_\pi^4 \ln{m_\pi}$
            and cutoff-dependent terms that are even in $m_\pi^2$ \cite{Procura:2003ig,Bernard:2003rp,Leinweber:2003dg}.
            The latter are combined with the bare parameters $a^{(i)}_B$ to renormalized coefficients
            such that the chiral expansion of $M_N$ and $M_\Delta$ reads \cite{Leinweber:1999ig}
            \begin{align}\label{BARYON:ChiralExpansion}
                M_B(m_\pi^2)  &= \left( M_B(0) + c_B^{(2)} \, m_\pi^2 + c_B^{(4)} \, m_\pi^4 + \dots \right)
                              + \left( c_B^{(3)} \, m_\pi^3 + c_B^{(4L)} \, m_\pi^4 \, \ln{m_\pi} + \dots \right), \nonumber \\
                c_N^{(3)} &= c_\Delta^{(3)} = -\frac{\lambda\pi}{2}\,, \quad
                c_N^{(4L)} = \frac{32}{25} \,\frac{3\lambda}{8(\Delta M)}\,, \quad
                c_\Delta^{(4L)} = -\frac{8}{25} \,\frac{3\lambda}{8(\Delta M)}\,.
            \end{align}
            It is expressed in terms of renormalized low-energy constants $M_B(0)$, $c_B^{(2)}$ and $c_B^{(4)}$ which may be
	        determined by a fit to lattice data \cite{Young:2002cj,Leinweber:2003dg}. Through this renormalization procedure the regulator dependence is removed \cite{Donoghue:1998bs,Young:2002ib}.

      \bigskip
      \fatcol{Discussion.}
            The previous considerations illustrate that both the quark-diquark model and the three-body result
            for $M_N$ in the 'core' setup \hyperlink{coupling:c3}{(C3)} are roughly consistent with
            a pseudoscalar-meson dressing providing the dominant correction to the nucleon's quark core.
            In particular, it complies with the assumptions which motivated the introduction of \eqref{core:mrho}
            in the context of pion and $\rho$-meson observables.

            Nevertheless one has to keep in mind that the identification of the baryonic
        	quark core \eqref{CHPT:core} with the quark-diquark or three-quark 'core' is more
        	complicated than in the meson case. Eq.~(\ref{core:mrho}) assumes that corrections
        	to $m_\rho$ are partly induced by a pseudoscalar-meson cloud and to a similar extent owe to non-resonant corrections to RL truncation.
            In the baryon one has additional lines of improvement:
            an inclusion of irreducible 3-body interactions, which were neglected in the derivation of the Faddeev equation \eqref{bs:faddeevtruncated},
            can still describe a quark core in the sense of Eq.~(\ref{core:mrho}).
            Morever it is conceivable that contributions beyond rainbow-ladder induce different ramifications for $m_\rho$ and $M_N$.

            An interpretation of the quark-diquark result for $M_\Delta$ is more difficult.
            In view of Eq.\,\eqref{CHPT:core}, the solution depicted in Fig.~\ref{fig:2} indicates larger corrections
            beyond the quark-diquark approach than encountered for the nucleon.
            The $M_\Delta$ solution additionally exhibits a sizeable $\omega$ dependence,
            a feature which is less pronounced in $\pi$, $\rho$ and nucleon observables.
            The scalar and axial-vector diquark masses exhibit particularly large sensitivities
        	to $\omega$ (see Fig.~\ref{fig:dqmasses}) which apparently cancel upon constituting the
        	nucleon mass. In the $M_\Delta$ case, the same consideration could suggest that taking into account only
        	an axial-vector diquark may not be sufficient for describing the $\Delta$,
            and that a possible further isospin-$1$ (tensor) diquark component with a mass large enough to be
        	irrelevant for the nucleon could diminish the $\Delta$ 'core' mass.
            In this respect it is highly desirable to extend the three-body study of the nucleon to the $\Delta$.
            Such a framework automatically implements the effect of inserting further diquark channels
            which could affect $N$ and $\Delta$ properties differently.

            The right panel of Fig.~\ref{fig:2} displays the $N$--$\Delta$ mass splitting.
            According to \eqref{CHPT:sigma}, the pseudoscalar meson contribution to
            the experimental value $M_\Delta-M_N=0.29$ GeV is small and positive:
            $\Sigma_\Delta-\Sigma_N=0.03$ GeV, cf.~Ref.~\cite{Young:2002cj}.
            This is not the case in our calculation, where at the $u/d$ mass
    	    $(M_\Delta-M_N)^\textrm{core}=0.48(4)$ GeV and therefore predicts a negative correction to
    	    the full splitting.

            We also compare $M_\Delta-M_N$ with the diquark mass splitting
        	$M_\textrm{av}-M_\textrm{sc}$. Both decrease with increasing current-quark mass;
            nevertheless there is no direct relationship between the two quantities,
            since the axial-vector diquark contribution to the mass of the nucleon does not vanish \cite{Nicmorus:2008vb}.

  \chapter[Nucleon: Electromagnetic form factors]{Nucleon: \\ Electromagnetic form factors}\label{chap:ffs}


              The nucleon's charge and magnetization structure is encoded in its electromagnetic form factors which,
              for space-like values of the photon momentum, are experimentally determined via elastic nucleon-electron scattering.
               Nucleon electromagnetic form factors have been studied in a variety of approaches; 
               an overview on the experimental and theoretical progress can be found in the recent review articles of Refs.\,\cite{Perdrisat:2006hj,Arrington:2006zm}.

              The results presented in this chapter are rooted in a long tradition of nucleon form factor studies within the quark-diquark model
              \cite{Hellstern:1997pg,Bloch:1999ke,Oettel:1999gc,Oettel:2000jj,Oettel:2002wf,Bloch:2003vn}.
              These calculations share some common caveats.  
              First, pionic contributions play an important role in the low-energy and small-quark mass behavior of the nucleon's electromagnetic structure.
              Such effects are not included in a quark-diquark 'core' and must be added on top of it \cite{Hecht:2002ej,Oettel:2002cw,Alkofer:2004yf,Cloet:2008re}.
              Second, access to the large-$Q^2$ region and thereby to the truly perturbative domain is so far only feasible upon implementing pole-free model propagators
              which, in turn, exhibit essential singularities at timelike infinity. The problem is not of fundamental concern; it merely awaits a thorough numerical treatment.
              Third, the quark-mass dependence of magnetic moments and charge radii, while emerging naturally in lattice calculations,
              is practically inaccessible in a quark-diquark model due to the unknown mass dependence of the modeled ingredients.

               The rainbow-ladder based quark-diquark approach, introduced in Chapter~\ref{chapter:quark-diquark}, removes the latter obstacle. 
               Upon resolving the diquarks' substructure, the form factors are immediately related to 
               the parameters in the effective quark-gluon coupling $\alpha(k^2)$, in particular: its quark-mass dependent coupling strength.
               In the following we will restrict ourselves to the quark 'core model' which represents a quark-diquark core
               that needs to be dressed by meson-cloud effects.
               A comparison of the core's static properties with lattice results is appropriate at larger quark masses;
               form factors depending on the photon momentum may be compared to experiment at $Q^2 \gtrsim 2$ GeV$^2$ where pion-cloud effects are diminished.

\section{Electromagnetic current}\label{sec:em}

    \fatcol{Nucleon-photon vertex.}
            The vertex which describes the coupling of the nucleon as a spin-$1/2$ fermion to a photon
            is constructed from 12 Dirac covariants:
            \begin{equation}
                \left\{ \gamma^\mu, P^\mu, Q^\mu \right\} \times \left\{ \,\mathds{1}, \, \Slash{P}, \,\Slash{Q}, \,[ \Slash{P}, \Slash{Q} ] \,\right\},
            \end{equation}
            where the Breit momentum $P$ and photon momentum $Q$ are combinations of incoming and outgoing nucleon momenta:
            \begin{equation}\label{FF:Breit+Photon}
                P = \frac{1}{2} \left( P_i + P_f\right), \quad\quad Q = P_f-P_i\,.
            \end{equation}
            A nucleon on the mass shell satisfies $P_i^2 = P_f^2 = -M^2$ and therefore $P^2 = -M^2 - Q^2/4$ and $P\cdot Q=0$;
            hence the only independent variable is the photon momentum-squared $Q^2$.
            The nucleon's electromagnetic current $J^\mu(Q^2)$ is obtained by sandwiching the on-shell vertex by
            the nucleon spinors $\conjg{U}^\beta(P_f)$, $U^\alpha(P_i)$ of Eq.\,\eqref{NUCLEON:DiracSpinors} which are solutions of the Dirac equation:
            \begin{equation}
                \Lambda_+(P_i) \,U^\alpha(P_i)=U^\alpha(P_i), \quad\quad  \conjg{U}^\beta(P_f) \,\Lambda_+(P_f) = \conjg{U}^\beta(P_f)\,.
            \end{equation}
            Equivalently, one may construct a matrix-valued current by taking the spin sums, i.e., by contracting
            the vertex with the positive-energy projectors
            \begin{equation}
                \Lambda_+(P_{i,f}) = \frac{1}{2}\left( \mathds{1} + \frac{\Slash{P}_{i,f}}{iM}\right) = \sum_{\alpha=1}^2 U^\alpha(P_{i,f})\,\conjg{U}^\alpha(P_{i,f})\,.
            \end{equation}
            They reduce the 12 basis elements to three: $\left\{ \gamma^\mu, P^\mu, Q^\mu \right\}$.
            Current conservation $Q^\mu J^\mu(Q^2) = 0$ eliminates the structure $Q^\mu$ via $\Lambda_+(P_f)\,\Slash{Q} \,\Lambda_+(P_i) = 0$ and
            $Q\cdot P=0$. The most general electromagnetic current of the nucleon is therefore given by
            \begin{equation}\label{FF:current0}
                J^\mu(Q^2) = \Lambda_+(P_f) \left( (F_1+F_2) \,i\gamma^\mu -F_2 \,\frac{P^\mu}{M} \right) \Lambda_+(P_i)\,,
            \end{equation}
            thereby defining the Dirac and Pauli form factors $F_1(Q^2)$ and $F_2(Q^2)$.
            Using the Gordon identity
            \begin{equation}
                \Lambda_+(P_f) \left( i \gamma^\mu + \frac{i \,\sigma^{\mu\nu}Q^\nu}{2M} - \frac{P^\mu}{M}   \right) \Lambda_+(P_i)=0
            \end{equation}
            with $\sigma^{\mu\nu} = -\frac{i}{2}\, [\gamma^\mu, \gamma^\nu ]$ transforms Eq.\,\eqref{FF:current0} into:
            \begin{fshaded1}
            \begin{equation}\label{FF:current1}
                J^\mu(Q^2) = \Lambda_+(P_f) \left( F_1 \,i\gamma^\mu - \frac{F_2}{2M}\, i \,\sigma^{\mu\nu} Q^\nu \right) \Lambda_+(P_i)\,.
            \end{equation}
            \end{fshaded1}

            \noindent
            $F_1$ and $F_2$ are dimensionless; for $Q^2=0$ they reduce to the proton and neutron charges $\lambda_{p,n}=\{1,0\}$
            and anomalous magnetic moments $\kappa_{p,n}$, expressed in nuclear magnetons $e \hbar/(2M)$.
            Charge conservation $F_1(0)=\lambda$ is automatically satisfied if the nucleon amplitudes are
            canonically normalized via Eq.\,\eqref{bs:normalization} \cite{Oettel:1999gc}.

    \bigskip
    \fatcol{Electromagnetic current of a composite system.}
            To relate the electromagnetic current of Eq.\,\eqref{FF:current1} to the underlying description of the nucleon as a composite object,
            the baryon must be resolved into its constituents to each of which the current can couple. 
            The construction of an electromagnetic current operator in the framework of the Bethe-Salpeter equation was first treated by Mandelstam \cite{Mandelstam:1955sd}.
            In the same way as an $n$-particle T-matrix reduces to the form \eqref{BS:Tmatrixpole} at the bound-state pole with mass $M$,
            thereby defining the bound-state amplitude,
            the electromagnetic current matrix $J^\mu$ is the residue of the $(2n+1)-$point function $T^\mu$ which describes the photon's coupling to the T-matrix
            at the bound-state mass:
            \begin{equation} \label{FF:tpole}
                T^\mu \; \stackrel{P^2\rightarrow -(M^2+Q^2/4)}{\longlonglonglongrightarrow} \; -\mathcal{N}^2   \frac{\Psi_{\!f} \, J^\mu \, \conjg{\Psi}_i}{(P^2+M^2+Q^2/4)^2}\,.
            \end{equation}
            The Breit momentum $P$ is defined in \eqref{FF:Breit+Photon} and yields at the hadron pole:
            \begin{equation}
                (P_f^2+M^2)(P_i^2+M^2) = (P^2+M^2+Q^2/4) = 0.
            \end{equation}

            A systematic procedure for the construction
            of a hadron-photon vertex based on electromagnetic gauge invariance is the "gauging of equations" prescription
            \cite{Haberzettl:1997jg,Kvinikhidze:1998xn,Kvinikhidze:1999xp}
            which represents a generalization of the normalization condition in Eq.\,\eqref{bs:normalization} to finite photon momenta. 
            In this context, "gauging", formally denoted by $T \rightarrow T^\mu$,
            is a derivative: it is linear and satisfies Leibniz' rule.                    
            The gauged $n$-quark scattering matrix satisfies
            \begin{equation} \label{FF:tmu}
                T^\mu = -T \left(T^{-1}\right)^\mu T = - T \left(K^{-1}-G_0\right)^\mu  T = T \left(G_0^\mu + K^{-1} K^\mu K^{-1}\right)  T,
            \end{equation}
            where, according to Section~\ref{sec:qcd:bses}, $K$ is the interaction kernel, $G_0$ is the product of $n$ propagators, and
            Dyson's equation has been implemented.
            The gauged T-matrix depends on incoming and outgoing total momenta $P_i$, $P_f$ (or, equivalently, on $P$ and $Q$)
            and two further relative momenta which are not relevant for the following discussion.

            Eq.\,\eqref{FF:tmu} entails in combination with the pole behavior \eqref{BS:Tmatrixpole} and the bound-state equation $\Psi = K G_0 \Psi$:
            \begin{equation}
                T^\mu \; \stackrel{P^2\rightarrow -(M^2+Q^2/4)}{\longlonglonglongrightarrow} \;
                \mathcal{N}^2   \frac{\Psi_{\!f}\conjg{\Psi}_{\!f}}{P_f^2+M^2} \left( G_0^\mu + G_0 \, K^\mu \, G_0 \right) \frac{\Psi_i\conjg{\Psi}_i}{P_i^2+M^2}\,,
            \end{equation}
            and comparison with \eqref{FF:tpole} yields
            \begin{fshaded3}
            \begin{equation}\label{FF:current}
                J^\mu \, = \, \conjg{\Psi}_{\!f} \left( T^{-1} \right)^\mu \Psi_i \; = \;  -\conjg{\Psi}_{\!f} \left( G_0^\mu + G_0 \, K^\mu \, G_0 \right) \Psi_i\;.
            \end{equation}
            \end{fshaded3}

            \noindent
            The impulse approximation $G_0^\mu$ involves, e.g., the quark-photon vertices $(-S^{-1})^\mu$ where\-as $K^\mu$ represents the photon's coupling to the kernel $K$.
            For instance, in the case of a two-body system with a rainbow-ladder kernel, $K$ does not depend on the total momentum,
            hence $K^\mu=0$: the electromagnetic current of a meson consistent with a rainbow-ladder truncation is an impulse-approximation current.
            This is no longer true for a baryon, either described in terms of three valence quarks or a quark-diquark system.

            For vanishing photon momentum, 'gauging' is the derivative with respect to $P^\mu$:
            \begin{equation}
                T^\mu \Big|_{Q^2 \rightarrow 0} = \frac{d\, T}{d P^\mu} = 2 P^\mu \,\frac{d \,T}{d P^2} =: 2 P^\mu \, T'
            \end{equation}
            Comparison with the normalization condition $\mathcal{N} \,\conjg{\Psi} \left(T^{-1}\right)' \Psi = 1$ of Eq.\,\eqref{bs:normalization} yields $\mathcal{N} J^\mu(0) = 2P^\mu$
            (modulo the bound-state wave functions, e.g., the nucleon spinors).
            Examples for this relation are given in \eqref{current:em} for the pion's electromagnetic current ($\mathcal{N}=1$) and \eqref{FF:current1} for that of the nucleon ($\mathcal{N}=2M$).
            At the level of the bound-state constituents which contribute to the total current this relation is ensured by differential Ward identities.

            \begin{figure}[tb]
            \begin{center}
            \includegraphics[scale=0.13]{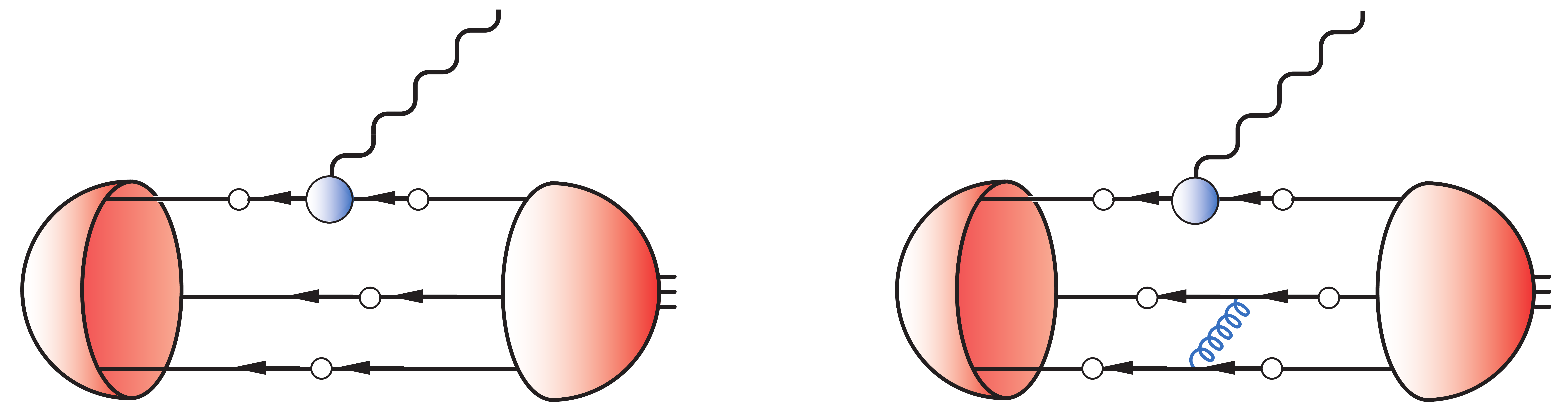}
            \caption[Electromagnetic current diagrams]{\backdef
                     The two types of diagrams which contribute to the nucleon's three-body current in rainbow-ladder truncation, Eq.\,\eqref{FADDEEV:Current}} \label{fig:faddeev-current}
            \end{center}
            \end{figure}

    \bigskip
    \fatcol{Three-body approach.}
            The kernel which appears in the bound-state equation of a baryon described by three valence quarks (cf. Section~\ref{chapter:faddeev}) is given by
            \begin{equation}
                K^{(3)} = K^{(3)}_\text{irr} + \sum_{i=1}^3 K_i^{(2)} \otimes S^{-1}.
            \end{equation}
            The corresponding gauged kernel reads
            \begin{equation}
                \left(K^{(3)}\right)^\mu = \left(K^{(3)}_\text{irr}\right)^\mu + \sum_{i=1}^3 \left(K_i^{(2)}\right)^\mu \otimes S^{-1} + \sum_{i=1}^3 K_i^{(2)} \otimes \left(S_i^{-1}\right)^\mu.
            \end{equation}
            Neglecting the three-body irreducible contribution and using a rainbow-ladder two-body kernel
            therefore leads to the electromagnetic current
            \begin{equation}\label{FADDEEV:Current}
                J^\mu = -\conjg{\Psi}_{\!f} \left[ G_0^\mu + G_0 \sum_{i=1}^3 K_i^{(2)} \otimes \left(S_i^{-1}\right)^\mu G_0 \right] \Psi_i\,
            \end{equation}
            which is depicted in Fig.~\ref{fig:faddeev-current}.
            We will not further investigate this equation in the present work and focus instead
            on the respective current operator in the quark-diquark model.

    \bigskip
    \fatcol{Diagrams in the quark-diquark model.}
            The general discussion of the last section applies to the quark-diquark model as well.
            In this context, the incoming and outgoing baryon states are described by quark-diquark amplitudes $\Phi_i$, $\Phi_f$ introduced in Eq.\,\eqref{nuc:defqdqamp}.
            Upon interaction with the external current, the baryon is resolved into its constituents: quark and diquark
            and the interaction between them to each of which the current can couple \cite{Oettel:1999gc}.

\begin{figure}[tb]
            \begin{center}
            \includegraphics[scale=0.11]{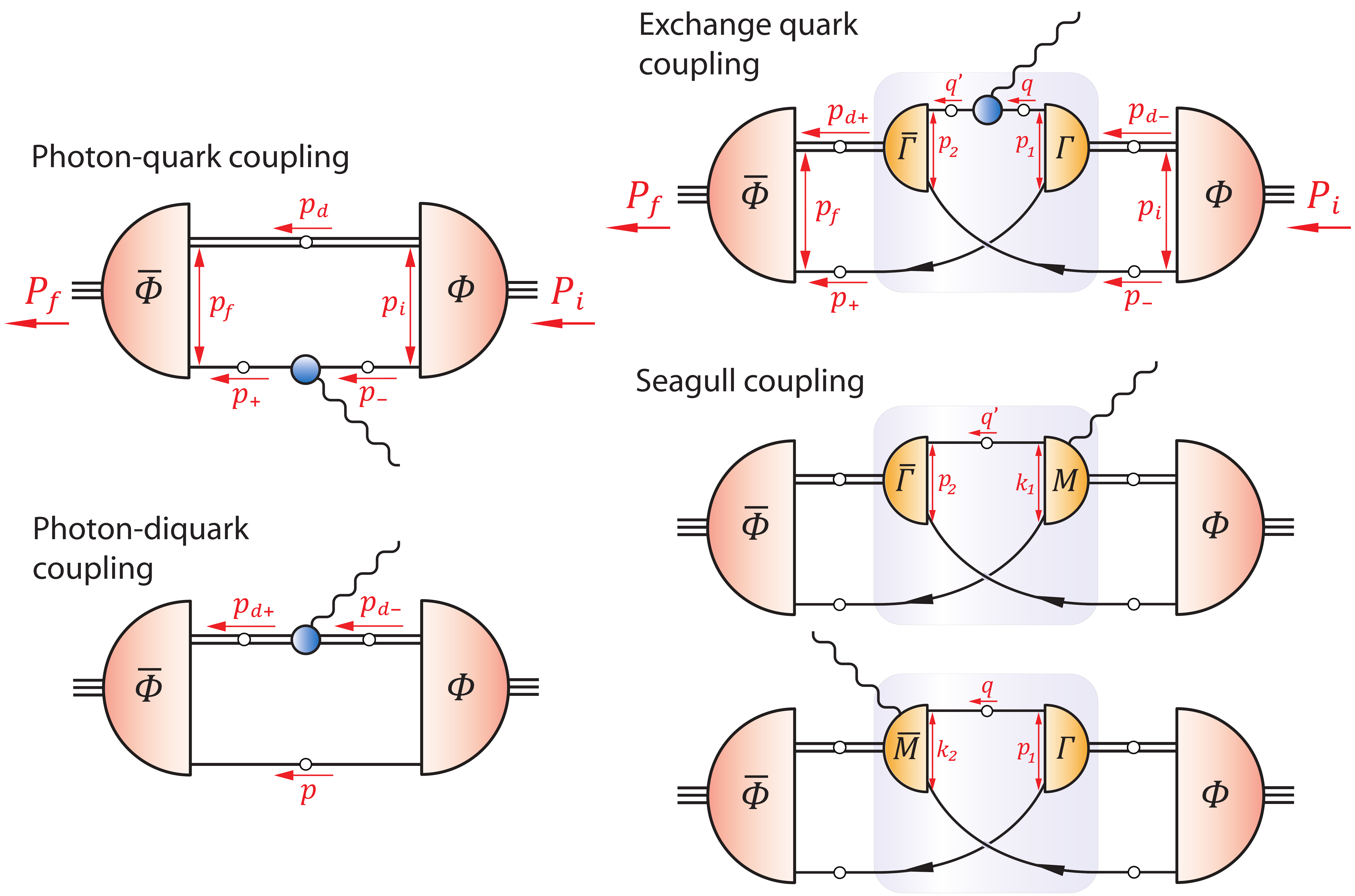}
            \caption[Electromagnetic current diagrams]{\backdef
                     The five diagrams that constitute the nucleon's electromagnetic current in the quark-diquark model, corresponding to Eq.\,\eqref{FF:Current:QDQ}.} \label{fig:current}
            \end{center}
            \end{figure}

            In the context of Eq.\,\eqref{FF:current}, $T$ is the quark-diquark scattering matrix,
            $G_0 = S\, D$ the product of a dressed quark and diquark propagator, and
            $K = \Gamma \,S\, \conjg{\Gamma}$ the quark-diquark kernel describing the quark exchange.
            The quark-photon and diquark-photon vertices are defined as the gauged inverse propagators:
            $\Gamma^\mu_\text{q} := -\left(S^{-1}\right)^\mu$ and $\Gamma^\mu_\text{dq} := -\left(D^{-1}\right)^\mu$.
            The gauged diquark amplitudes $\Gamma^\mu =: M^\mu$, also referred to as seagull vertices,
            describe the photon coupling to the diquark amplitudes. The ingredients of the current matrix are then
            written as
            \begin{equation}\label{FF:Current:QDQ}
            \begin{split}
                G_0^\mu &= \left( S \,D \right)^\mu = (S\,\Gamma^\mu_\text{q}\, S) \,D + S\, ( D \,\Gamma^\mu_\text{dq} \,D)\,, \\
                K^\mu &= \left( \Gamma \,S \,\conjg{\Gamma} \right)^\mu = M^\mu \,S \,\conjg{\Gamma} + \Gamma \left(S \,\Gamma^\mu_\text{q} \,S\right) \conjg{\Gamma} + \Gamma \,S \,\conjg{M}^\mu\;.
            \end{split}
            \end{equation}
            These diagrams are worked out in detail in App.\,\ref{app:emcurrent} and illustrated in Fig.\,\ref{fig:current}.

            By virtue of Eq.\,\eqref{FF:Current:QDQ}, the electromagnetic current of a baryon
            in a quark-diquark framework is completely specified by identifying the quark-photon vertex,
            the scalar and axial-vector diquark-photon vertices and an ansatz for the seagull terms. These quantities are constrained
            by Ward-Takahashi identities and thereby related to quark and diquark propagators and diquark amplitudes which have already
            been determined previously. The corresponding vertices are collected in Apps.\,\ref{sec:quarkpropagator}, \ref{app:dqphotonvertex} and \ref{app:seagulls}.

\section{Electromagnetic form factors}\label{sec:emff}

    \bigskip
    \fatcol{Sachs form factors.}
            Defining $\tau=Q^2/(4M^2)$, the correspondence between the electric and magnetic Sachs form factors to the Dirac and Pauli form factors
            which appear in \eqref{FF:current1} is given by
            \begin{fshaded1}
              \begin{equation} \label{ff:sachs}
                  \begin{array}{l}
                      G_E = F_1 - \tau F_2 \\[0.1cm]
                      G_M = F_1 + F_2
                  \end{array} \quad \Longleftrightarrow \quad  F_1 = \frac{G_E + \tau\,G_M}{1+\tau}\,, \quad  F_2 = \frac{G_M-G_E}{1+\tau}\,.
              \end{equation}
            \end{fshaded1}

            \noindent
            They can be extracted from the current via
            \begin{equation}
                \frac{1}{2} \, \text{Tr} \big\{ J^\mu P^\mu \big\} = \frac{P^2}{M}\, G_E \,, \qquad
                \frac{1}{4} \, T^{\mu\nu}_{P} \,\text{Tr} \big\{ J^\mu \, i \gamma^\nu \big\} = \tau\, G_M\,.
            \end{equation}
            Unless further specified, the symbol $F(Q^2)$ denotes a generic form factor in the following discussion, either Dirac/Pauli or Sachs.
            The electromagnetic radius $r_F$ corresponding to $F(Q^2)$ is defined as the slope
            at zero momentum transfer via the Taylor expansion
            \begin{align}\label{ff:radii}
                F(Q^2) &= \lambda-\frac{r_F^2}{6} Q^2 + \dots \quad \Longleftrightarrow \quad r_F^2 = -6 \left.\frac{dF}{dQ^2} \right|_{Q^2=0}\,, \\
                \frac{F(Q^2)}{F(0)} &= 1-\frac{r_F^2}{6} Q^2 + \dots \quad \Longleftrightarrow \quad r_F^2 = -\frac{6}{F(0)} \left.\frac{dF}{dQ^2} \right|_{Q^2=0}\,,
            \end{align}
            where $F(Q^2)$ in the first row denotes an 'electric' form factor ($F_1$ or $G_E$) with the electric charge $F(0)=\lambda$,
            in the second row a 'magnetic' form factor ($F_2$, $G_M$) with the (anomalous) magnetic moment
            $F(0) = \kappa$ or $\mu$ such that $G_M(0) = \mu = \lambda + \kappa$. With Eq.\,\eqref{ff:sachs} the correspondence between Sachs and Dirac/Pauli radii is
              \begin{equation}\label{FF:rEr1}
                  r_E^2 = r_1^2 + \frac{3\kappa}{2M^2}\,, \qquad r_M^2 = \frac{r_1^2 + \kappa\, r_2^2}{1+\kappa}\,,
              \end{equation}
             where the second term in $r_E^2$ involving the anomalous magnetic moment is the so-called Foldy term.
             Experimental values for electromagnetic radii and magnetic moments are collected in Table~\ref{tab:FF:exp-values}.

              Form factors are Lorentz invariant, but their interpretation depends on the reference frame.
              Nonrelativistically, the Sachs form factors $G_i(Q^2)$
              are the three-dimensional Fourier transforms of the nucleon's spatial charge and magnetization distributions $\rho_\text{ch}$, $\rho_\text{m}$
              in the Breit frame where the energy of the transferred photon is zero \cite{Sachs:1962zzc}.
              In this case the interpretation of the inferred Sachs radii as charge and magnetization RMS radii is valid:
              \begin{equation}
                  r_E^2 = 4\pi \int dr\, \rho_\text{ch}(r)\,r^4\,, \quad  r_M^2 = 4\pi \int dr \,\rho_\text{m}(r)\,r^4\,, \quad \lambda = 4\pi \int dr\,\rho_\text{ch}(r)\,r^2\,,
              \end{equation}
              where $\lambda$ is the nucleon's electric charge.
              Relativistic boost corrections obscure this interpretation, and the extraction of these densities
              from experimental form factor data becomes model-dependent \cite{Perdrisat:2006hj,Miller:2008jc}.
              Such analyses implicate a positive central charge density of both proton and neutron \cite{Kelly:2002if},
              in agreement with the notion of the neutron being sometimes a proton surrounded by a negatively charged pion cloud.
              On the other hand, the Dirac form factor $F_1(Q^2)$ non-ambiguously corresponds to the nucleon's transverse charge density in the infinite momentum frame \cite{Miller:2007uy},
              and the experimentally observed negativity of $F_1^n(Q^2)$ induces a negative central charge density of a fast-moving neutron.
              It has been argued in Ref.\,\cite{Eichmann:2008ef} that a negative Dirac form factor for the neutron can be explained
              in the context of a quark-diquark model which includes axial-vector diquark degrees of freedom (see also Section~\ref{sec:results}).

            \renewcommand{\arraystretch}{1.2}

            \begin{table}
                \begin{center}

            \begin{equation*}
            \begin{array}{|c|c|c|c|}\hline
               r_{E,M}^{p,n} \; [\text{fm}^{(2)}]   &  \kappa_{p,n,s,v} &   r_{1,2}^{p,n} \; [\text{fm}]    &    \Bigg.\left(r_{1,2}^{v,s}\right)^2\Bigg. \; [\text{fm}^2]    \\[0.2cm] \hline
                                 \begin{array}{ @{\;\;}rcl @{\;\;}}
                                       && \\[-0.4cm]
                                       r_E^p &=& 0.886(15)  \\
                                       \left( r_E^n\right)^2 &=& -0.115(4)  \\
                                       r_M^p &=& 0.855(35) \\
                                       r_M^n &=& 0.873(11)   \\[0.2cm]
                                 \end{array}   &
                                 \begin{array}{ @{\;\;}rcl @{\;\;} }
                                       && \\[-0.4cm]
                                       \kappa_p &=& 1.793  \\
                                       \kappa_n &=& -1.913  \\
                                       \kappa_s &=& -0.120  \\
                                       \kappa_v &=& 3.706 \\[0.2cm]
                                 \end{array}  &
                                 \begin{array}{ @{\;\;}rcl @{\;\;} }
                                       && \\[-0.4cm]
                                       r_1^p &=& 0.82(2)  \\
                                       r_1^n &=& 0.11(2)  \\
                                       r_2^p &=& 0.88(6)  \\
                                       r_2^n &=& 0.88(1)  \\[0.2cm]
                                 \end{array}  &
                                 \begin{array}{ @{\;\;}rcl @{\;\;} }
                                       && \\[-0.4cm]
                                       \left( r_1^s \right)^2 &=& 0.68(3)  \\
                                       \left( r_1^v \right)^2 &=& 0.65(3)  \\
                                       \left( r_2^s \right)^2 &=& 0.78 \pm 2     \\
                                       \left( r_2^v \right)^2 &=& 0.77(6)  \\[0.2cm]
                                 \end{array}  \\ \hline
            \end{array}
            \end{equation*}
                \caption{\backdef Experimental numbers for radii and magnetic moments. The radii in the left column are quoted from Table 1 of Ref.\,\cite{Belushkin:2006qa};
                                  for references, see therein. Magnetic moments are the PDG values \cite{Yao:2006px}.
                                  The remaining quantities are inferred from Eqs.\,\eqref{ff:sachs} and \eqref{FF:iso},
                                  ignoring possible correlations between the statistical errors of the data.  }\label{tab:FF:exp-values}

                \end{center}
            \end{table}

    \bigskip
    \fatcol{Form factor measurement.}
              Electromagnetic form factors are experimentally studied via elastic nucleon-electron scattering.
              Due to the smallness of the electromagnetic fine-structure constant $\alpha=1/137$,
              the Born approximation which describes the scattering process in terms of a single exchanged photon has been commonly employed.
              The scattering off a spin-$1/2$ target with extended structure and an anomalous magnetic moment is then given by the Rosenbluth cross section \cite{Rosenbluth:1950yq}:
              \begin{equation}
                  \frac{d\sigma}{d\Omega} = \left(\frac{d\sigma}{d\Omega}\right)_\text{Mott} \frac{\varepsilon\,G_E^2 + \tau\,G_M^2}{\varepsilon\,(1+\tau)}\,,
              \end{equation}
              where the prefactor represents the Mott cross section of a spin-$0$ point particle,
              $\varepsilon \in [0,1]$ is the virtual photon polarization, and $\theta$ the scattering angle:
              \begin{equation}
                  \left(\frac{d\sigma}{d\Omega}\right)_\text{Mott} = \frac{\alpha^2 E' \cos^2{\theta/2}}{4\,E^3 \sin^4{\theta/2}}\,, \qquad \varepsilon = \frac{1}{1+2\,(1+\tau)\,\tan^2{\theta/2}}\,.
              \end{equation}
              $E$ and $E'$ are the initial and final electron energies.
              The Sachs form factors are extracted from the $\varepsilon$ dependence of the quantity $\varepsilon\,G_E^2 + \tau\,G_M^2$ at fixed $Q^2$
              which entails a reduced sensitivity to $G_E^p$ at large $Q^2$ and $G_M$ at small photon momenta.
              Measurements of the neutron form factors, carried out via electron-deuteron scattering due to the lack of a free neutron target in nature,
              suffer from systematic uncertainties.

              Recent polarization-transfer experiments allow for an increased accuracy
              through a direct extraction of the form factor ratio $G_E/G_M$ via \cite{Arnold:1980zj}
              \begin{equation}
                  \frac{G_E}{G_M} = -\frac{P_t}{P_l} \frac{E+E'}{2M}\,\tan{\theta/2},
              \end{equation}
              where $P_l$ and $P_t$ are the longitudinal and transverse polarization components of the recoil proton, 
              transferred from the longitudinally polarized electron.

              Surprisingly, the corresponding measurements \cite{Jones:1999rz,Gayou:2001qd,Punjabi:2005wq} indicated
              a linear decrease in that ratio with increasing $Q^2$. This
              is incompatible with the result obtained by the Rosenbluth separation technique where the ratio is roughly $\approx 1$ (both are compared in Fig.~\ref{fig:ratio}).
              The discrepancy is currently believed to originate from two-photon exchange corrections
              which have a minimal impact upon the polarization results but
              significantly affect the Rosenbluth cross section \cite{Guichon:2003qm,Arrington:2007ux}.

    \bigskip
    \fatcol{Phenomenological aspects.}
              Dimensional counting rules of perturbative QCD predict the following behavior of the Dirac
              and Pauli form factors at large photon momentum transfer \cite{Brodsky:1974vy}:
              \begin{equation}
                  F_1 \sim 1/Q^4\,, \qquad F_2 \sim 1/Q^6\,, \qquad Q^2 \,F_2/F_1 \sim const.,
              \end{equation}
              where logarithmic corrections \cite{Belitsky:2002kj} have been neglected.
              Correspondingly, the Sachs form factors scale as $G_{E,M} \sim 1/Q^4$, which implies
              that the ratio $G_E/G_M$ becomes constant.
              In this respect, dipole-like parametrizations for the Sachs form factors were found to
              provide a reasonable description of the experimental data:
              \begin{equation}
                  G_i(Q^2) = \frac{g_i(Q^2)}{\left(1+Q^2/\Lambda^2\right)^2}\,,\qquad \Lambda = 0.84\,\text{GeV}\,,
              \end{equation}
              with $g_i(Q^2) = const.$ except for the neutron electric form factor
              where the Galster parametrization \cite{Galster:1971kv} $g_E^n(Q^2) = -\mu_n \tau/(1+5.6\,\tau)$ has been frequently employed to fit the data.

              The deviation from the dipole form, expressed through the dependence of the remainders $g_i(Q^2)$ on the photon momentum,
              is nevertheless sizeable for $Q^2 \gtrsim 2$ GeV$^2$.
              It is especially pronounced in the electric form factor of the proton:
              the polarization-transfer data for the form factor ratio $G_E^p/G_M^p$ (so far only available for $Q^2$ below $6$ GeV$^2$)
              show a linear fall-off in $Q^2$ and even point towards a zero crossing
              at $Q^2 \approx 8$ GeV$^2$, implying the presence of a further small scale $\sim 0.07$ fm.
              The discrepancy between the perturbative prediction and the experimental data has been attributed to the presence of
              non-zero quark orbital angular-momentum content in the proton \cite{Miller:2002qb,Ralston:2003mt,Bloch:2003vn}.
              Possible evidence for an onset of the perturbative scaling can be observed in the $G_M^p$ data above $Q^2 \sim 10$ GeV$^2$ \cite{Perdrisat:2006hj}.

              Pion-cloud effects are expected to play an important role in the form factor structure below $Q^2 \sim 2$ GeV$^2$.
              Such contributions are suppressed at large $Q^2$ where the photon probes the nucleon's quark core.
              Attempts have been made to attribute low-$Q^2$ systematics in the form factor data
              to pionic effects, e.g. through a phenomenological double-dipole fit \cite{Friedrich:2003iz}, in a dispersion-relation approach \cite{Belushkin:2005ds},
              or via implementation in a chiral quark model \cite{Faessler:2005gd}.
              In analogy to the discussion of the pion's charge radius in Chapter~\ref{sec:MESON:results}, pion-loop contributions are believed to
              provide sizeable additions to the nucleon's quark core radii and magnetic moments,
              where the overall strength of these effects can be estimated from chiral perturbation theory.

      \bigskip
      \fatcol{Flavor contributions to form factors.}
              Proton and neutron form factors can be combined to study the flavor dependence of
              the nucleon's charge and magnetization structure.
              Each diagram of Fig.~\ref{fig:current} can be split into terms where the photon either couples to a $u$ or a $d$ quark inside the nucleon.
              The form factors (again generically denoted by $F$) are therefore linear combinations of these $u$- and $d$-quark contributions:
              \begin{equation}\label{FF:flavor-contributions}
              \begin{split}
                  F^p &= 2 q_u \, F^p_u + q_d\,F^p_d = \frac{1}{3} \left(4 F^p_u - F^p_d\right) = \frac{1}{3} \left(4 F^u - F^d\right)\,,   \\
                  F^n &= q_u\,F^n_u + 2q_d\,F^n_d = \frac{2}{3} \left( F^n_u - F^n_d\right) = \frac{2}{3} \left( F^d - F^u\right)\,,
              \end{split}
              \end{equation}
              where because of charge symmetry the $u(d)$ contribution in the proton equals the $d(u)$ contribution in the neutron:
              \begin{equation}\label{FF:flavor-contributions-ud}
                  F^u := F^p_u = F^n_d \,, \qquad F^d := F^p_d = F^n_u\,.
              \end{equation}
              The definition \eqref{FF:flavor-contributions} implies $F^u(0) = F^d(0) = 1$ for the 'charge' form factors $F_1$ and $G_E$ which
              entails
              \begin{equation}
                  \big(r_1^u\big)^2 = \left(r_1^p\right)^2 + \frac{1}{2} \left(r_1^n\right)^2, \qquad
                  \big(r_1^d\big)^2 = \left(r_1^p\right)^2 + 2 \left(r_1^n\right)^2
              \end{equation}
              for the charge radii.
              The flavor contribution to the magnetic form factors $F_2$, $G_M$ is usually separated via $\mu^u = 2\,q_u\,F^p_u(0)$, $\mu_d = q_d\,F^p_d(0)$,
              and charge symmetry implies
              \begin{equation}
                  \mu_p = \mu^u + \mu^d\,, \qquad \mu_n = -2\,\mu^d -\frac{1}{2}\,\mu^u\,.
              \end{equation}

              The contribution from strange quarks, implicit in the seaquark content,
              is missing in Eq.\,\eqref{FF:flavor-contributions} since we are working in isospin-$SU(2)$.
              Nevertheless it has been argued that the present world data are consistent with the proton's strange form factors being zero \cite{Young:2006jc}.
              Under this assumption one may extract the $u$ and $d$ contributions from the experimental quantities.
              Using the values of Table~\ref{tab:FF:exp-values}  yields:
              \begin{equation}
                  r_1^u = 0.82(2) \,\text{fm}\,, \quad
                  r_1^d = 0.83(2) \,\text{fm}\,, \quad
                  \mu^u = 2.44\,, \quad
                  \mu^d = 0.34.
              \end{equation}


\section{Results and discussion}\label{sec:results}

            This section provides results for the nucleon's electromagnetic form factors in the quark-diquark calculation
            where the current is
            constructed from the diagrams in Fig.~\ref{fig:current}.
            Apart from the ingredients which have been determined in previous chapters
            (nucleon amplitude, Section~\ref{sec:qdqbse}; quark propagator, Section~\ref{sec:qcdgreenfunctions}; diquark amplitudes and propagator, Section~\ref{sec:dq} and the respective appendices),
            the calculation involves a quark-photon vertex (App.~\ref{sec:quark-photon-vertex}) and the effective diquark vertices (Apps.~\ref{app:dqphotonvertex} and \ref{app:seagulls}).
            Each vertex satisfies a Ward-Takahashi identity which, in total, ensures
            conservation of the nucleon's charge: $F_1^p(0)=1$, $F_1^n(0)=0$.

 \renewcommand{\arraystretch}{1.5}

\begin{figure}[p]
              \begin{center}
              \includegraphics[scale=0.09]{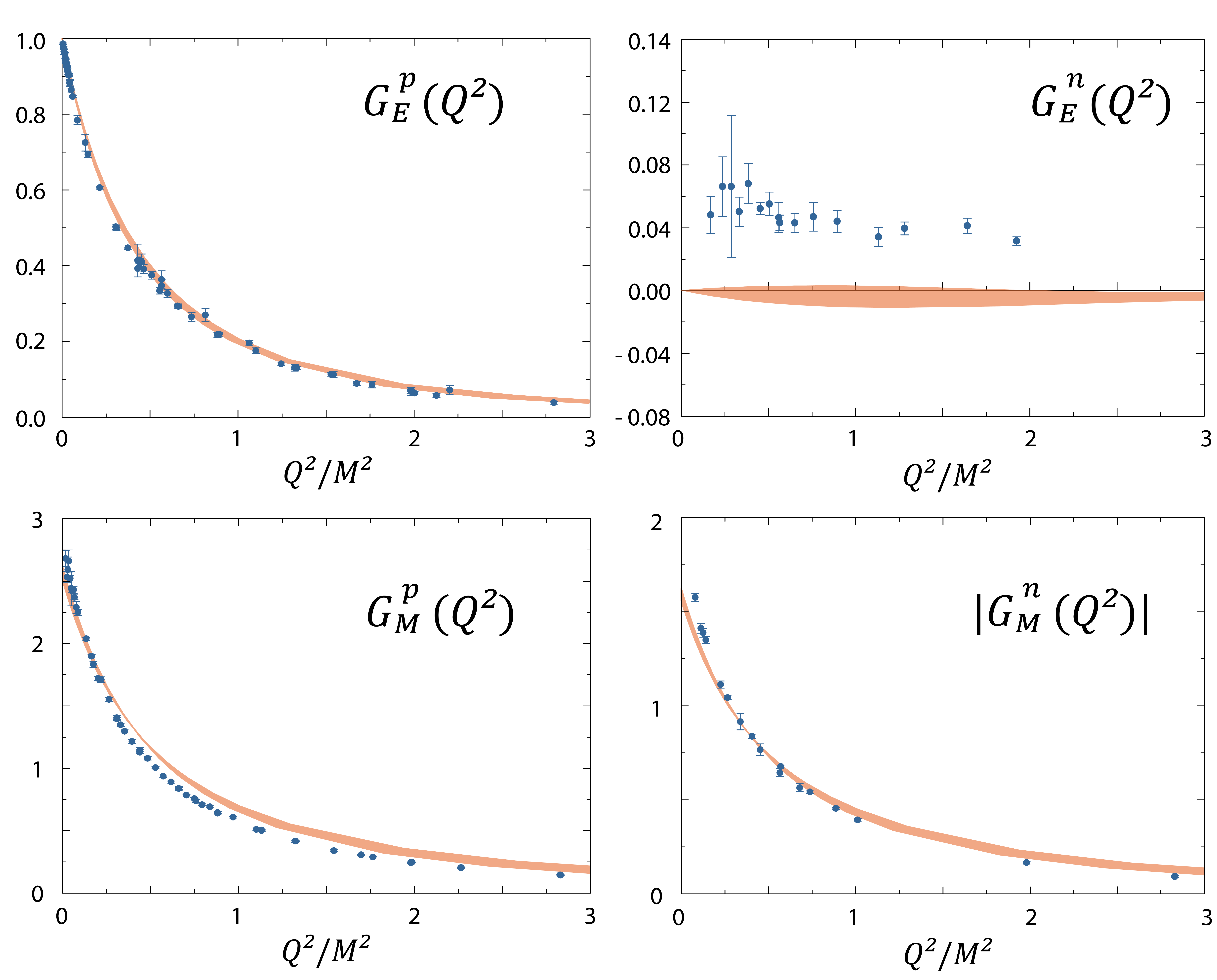}
            \caption[Form factors]{\backdef
                                    Results for the Sachs form factors at the physical point.
                                    The selection of experimental data is based on Ref.\,\cite{Friedrich:2003iz} (data compiled
	                                by P.~Grabmayr).
                                   The bands correspond to a variation of $\omega$ in setup \hyperlink{coupling:c3}{(C3)}.} \label{fig:ff}
              \end{center}
              \end{figure}

      \bigskip
      \fatcol{Form factor contributions.}
              In Fig.\,\ref{fig:ff} the proton's and neutron's electromagnetic Sachs form factors at the physical point are compared to experimental data.
              The main difference between the results presented herein and those of Ref.\,\cite{Eichmann:2007nn} originates from the inclusion
              of the $\rho$-meson pole in the quark-photon vertex, cf. Eq.\,\eqref{vertex:simulate}. In accordance with its effect
              on the pion charge form factor \cite{Maris:1999bh,Eichmann:2008ae}, it reduces the form factors and contributes $~50\%$ to their charge radii.
              Specifically, it cancels the previously positive result for $G_E^n$ to zero (within the considered domain of the $\omega$ parameter).
              Note that the form factors in terms of the underlying quark-photon and diquark-photon vertices are well constrained up to $Q^2 \sim 2$ GeV$^2$;
              the arbitrariness introduced in connection with the transverse seagull term of Eq.\,\eqref{FF:Seagulls:Transverse} becomes important only at larger photon momenta.

              \begin{figure}[p]
              \begin{center}
              \includegraphics[scale=0.33]{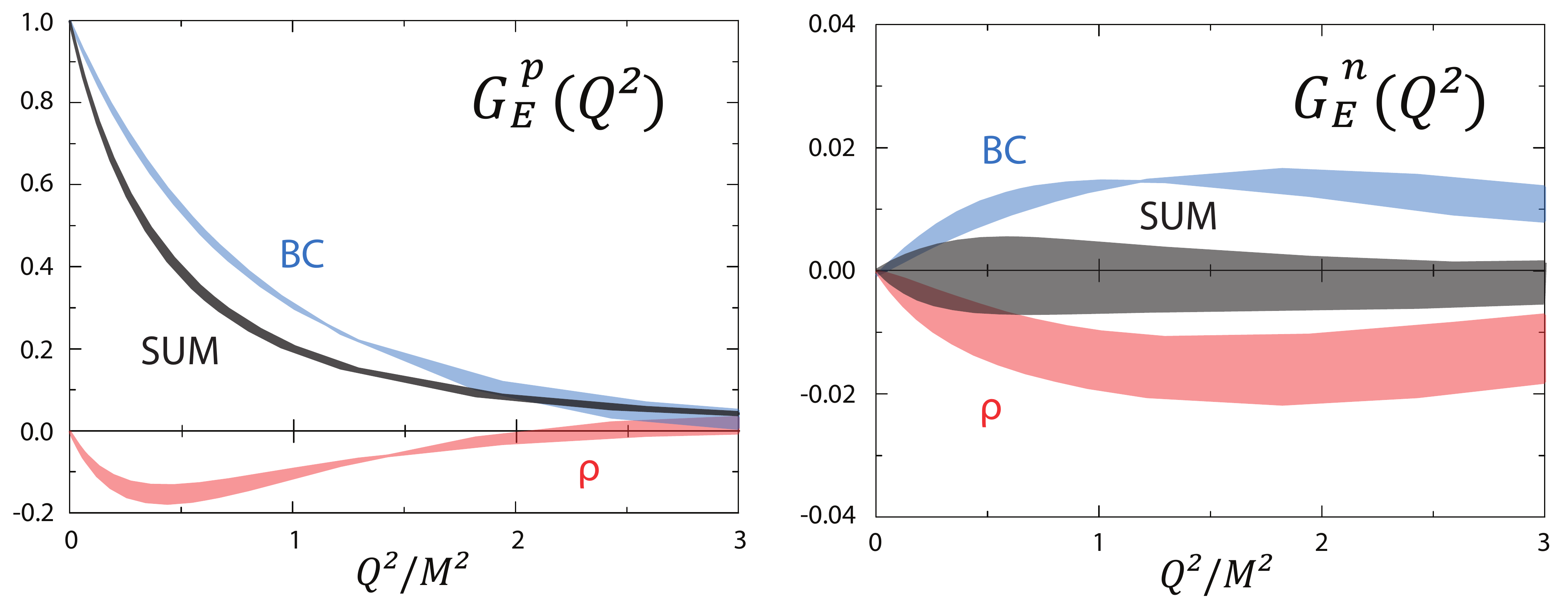}
              \caption[Form factor contributions II]
                                   {\backdef Contributions to $G_E^{p,n}$ from the Ball-Chiu ($\mathsf{BC}$) and vector-meson part ($\mathsf{\rho}$) in the quark-photon vertex.
                                    The overall effect on the magnetic form factors is similar to that in $G_E^p$.} \label{fig:ff-contributions-BC-vs-Rho}
              \end{center}
              \end{figure}

\begin{figure}[tb]
              \begin{center}
              \includegraphics[scale=0.295]{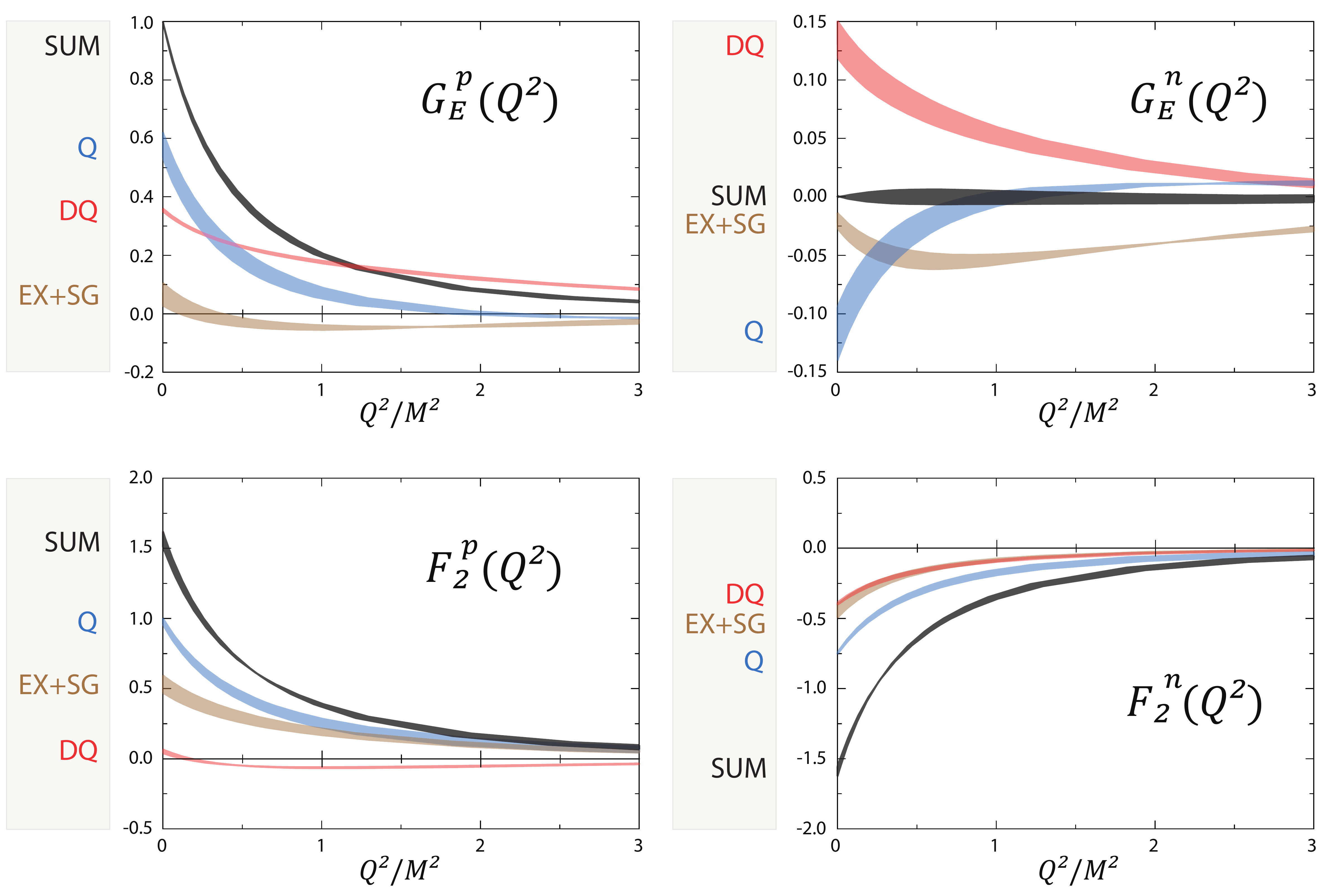}
              \caption[Form factor contributions]
                                   {\backdef Individual contributions to $G_E$ and $F_2$ according to Fig.~\ref{fig:current}. Depicted are the
                                   form factor sum ($\mathsf{SUM}$), photon-quark ($\mathsf{Q}$) and photon-diquark coupling ($\mathsf{DQ}$),
                                   and the combined exchange and seagull terms ($\mathsf{EX+SG}$).
                                   The strength of these contributions in the remaining form factors can be extracted from the plot via
                                   Eqs.\,\eqref{ff:sachs}, \eqref{FF:iso} and \eqref{FF:flavor-contributions}:
                                   $F_1 = G_E + \tau F_2$ and $G_M = F_1+F_2$; for the isoscalar/isovector combinations: $s=p+n$, $v=p-n$;
                                   and the up- and down-quark contributions: $u=p+n/2$, $d=p+2n$.
                                   The $\omega$ variation is identical to Fig.~\ref{fig:ff}.} \label{fig:ff-contributions}
              \end{center}
              \end{figure}

              Fig.~\ref{fig:ff-contributions} depicts the contributions to the form factors $G_E^{p,n}$ and $F_2^{p,n}$ from the quark-photon coupling, the diquark-photon coupling,
              and the exchange and seagull diagrams.
              The strongest contribution at $Q^2=0$ is the direct coupling of the photon to the quark line.
              This feature may change with the $Q^2$ evolution: for instance, the quark-photon contribution to $G_E^p$ exhibits a zero crossing
              at $Q^2/M^2 \approx 2$ which was observed in the model of Ref.\,\cite{Cloet:2008re} as well.
              The diquark contribution to the magnetic form factors (in particular, the scalar--axial-vector transition) provides only a small fraction of the total result.
              The unit charge of the proton results from the canonical normalization of the quark-diquark amplitude.
              Current conservation ensures a vanishing neutron charge in terms of a cancelation of the components in $G_E^n$ at $Q^2=0$.

\begin{figure}[tb]
              \begin{center}
              \includegraphics[scale=0.3]{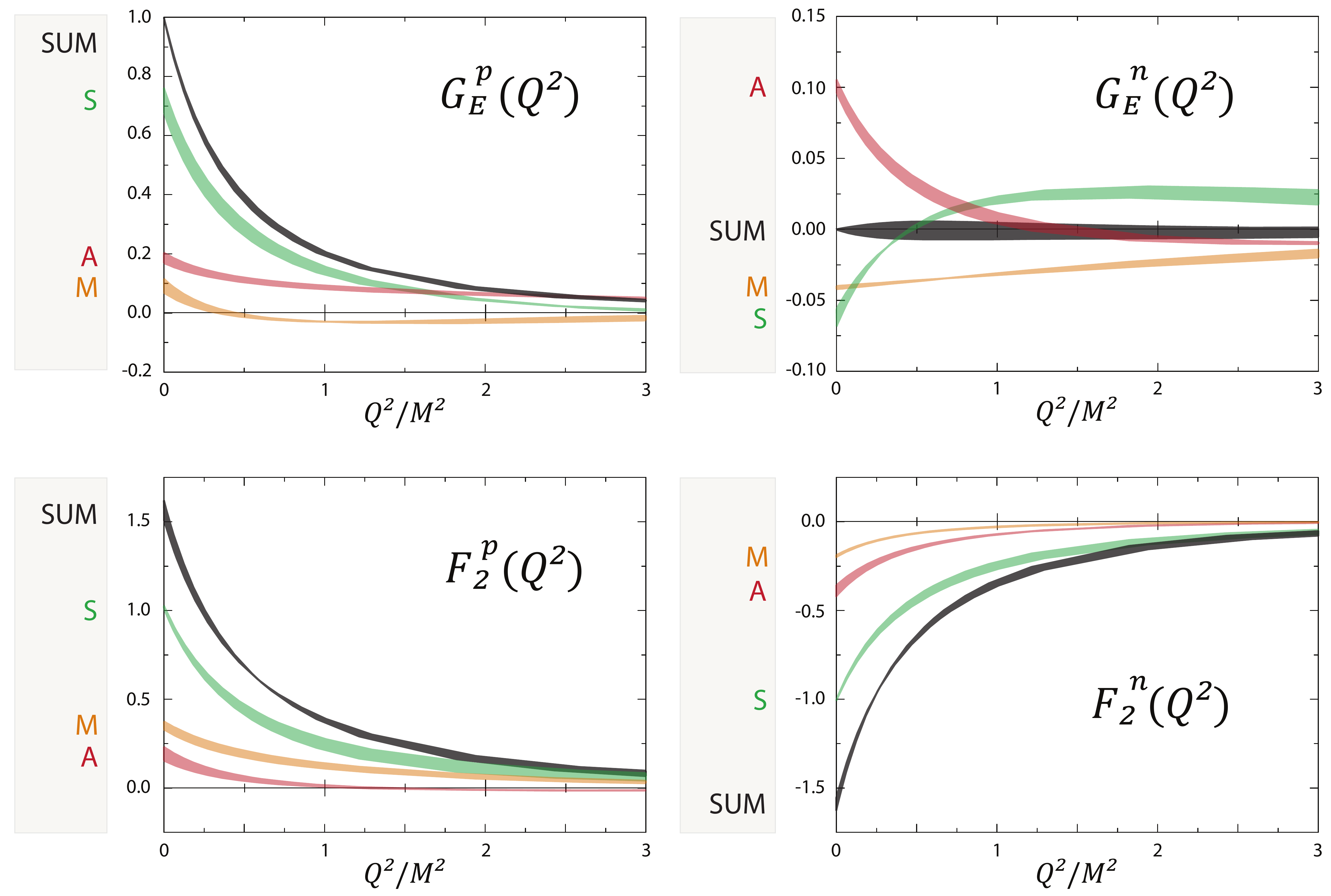}
              \caption[Form factor contributions III]
                                  {\backdef Contributions to $G_E$ and $F_2$ according to the type of incoming and outgoing quark-diquark amplitudes $\Phi$.
                                   The form factor sums ($\mathsf{SUM}$) are decomposed into scalar-scalar ($\mathsf{S}$), axial-vector--axial-vector ($\mathsf{A}$)
                                   and the sum of scalar--axial-vector and axial-vector--scalar contributions ($\mathsf{M}$). } \label{fig:ff-contributions-sa}
              \end{center}
              \end{figure}

              The relative strengths of the scalar--scalar, axialvector-axialvector and scalar--axial-vector components, according
              to the type of the incoming and outgoing nucleon amplitudes,  can be read off from Fig.~\ref{fig:ff-contributions-sa}.
              The dominant contributions are provided by the scalar-diquark contributions.

             \renewcommand{\arraystretch}{1.2}

      \bigskip
      \fatcol{Comparing to experimental and lattice data.}
              Since the nucleon mass as a result of the quark-diquark calculation usually deviates from its experimental value $M_\text{exp}=0.94\,\text{GeV}$,
              some precaution must be taken when comparing form factor results to experimental data.
              There are basically two scales which enter the electromagnetic current and form factors: the photon momentum $Q^2$ and the nucleon mass $M$,
              the latter of which absorbs both the inherent scale $\Lambda_\text{IR}$ and the current-quark mass dependence of the system.
              As form factors are dimensionless they can only depend on the combination $Q^2/M^2$.
              This is accounted for in Figs.\,(\ref{fig:ff}--\ref{fig:ff-contributions-sa})  where the results are plotted as a function of this dimensionless variable.

              The same effect is achieved by an appropriate rescaling of the photon momentum
              \begin{equation}\label{FF:rescaling1}
                  \left(\frac{Q^2}{M^2}\right)_\text{calc} = \frac{\widetilde{Q}^2}{M_\text{exp}^2}\,.
              \end{equation}
              Presenting results as a function of $\widetilde{Q}^2$  removes  the trivial scale dependence induced by the overestimated quark-core result for $M$.
              When investigating the current-mass dependence and comparing to lattice-QCD results,
              the experimental mass in Eq.\,\eqref{FF:rescaling1} should be replaced by that obtained from the lattice: $M_\text{exp} \longrightarrow M_\text{lat}$.
              This affects the comparison of charge radii as well: instead of comparing the dimensionless combinations $r_F^2 \,M^2$ (e.g., as in Ref.\,\cite{Eichmann:2008ef})
              we rescale the radii accordingly,
              \begin{equation}
                  \left(r^2 M^2\right)_\text{calc}  = \widetilde{r}^2 M_\text{lat}^2,
              \end{equation}
              and plot the quantities $\widetilde{r}^2$ instead of $r^2$.
              These issues were not relevant in the context of pion observables $f_\pi$, $r_\pi$ in Chapter~\ref{sec:MESON:results}
              since the mass of the pion is not expected to drastically change upon implementing beyond rainbow-ladder contributions. 

              A second remark concerns the comparison of magnetic moments to lattice-QCD results
              where $F_2$ and $G_M$ are usually expressed in terms of experimental (i.e., fixed instead of current-mass dependent) nuclear magnetons \cite{Gockeler:2003ay}.
              The corresponding quantity which appears in the electromagnetic current \eqref{FF:current1} is $\widetilde{F}_2/(2 M_\text{exp})$.
              Again, our calculated result for $F_2$ (the value in units of 'running' magnetons) is dimensionless, hence we leave it unchanged 
              but remove the current-mass dependence when comparing to lattice data via
              \begin{equation}
                  \left(\frac{F_2}{2 M}\right)_\text{calc} = \frac{\widetilde{F}_2}{2 M_\text{calc}^\text{phys}}\,,
              \end{equation}
              where $M_\text{calc}^\text{phys}$ is the \textit{calculated} nucleon mass at the physical $u/d$ point.

\begin{figure}[p]
              \begin{center}
              \includegraphics[scale=0.09]{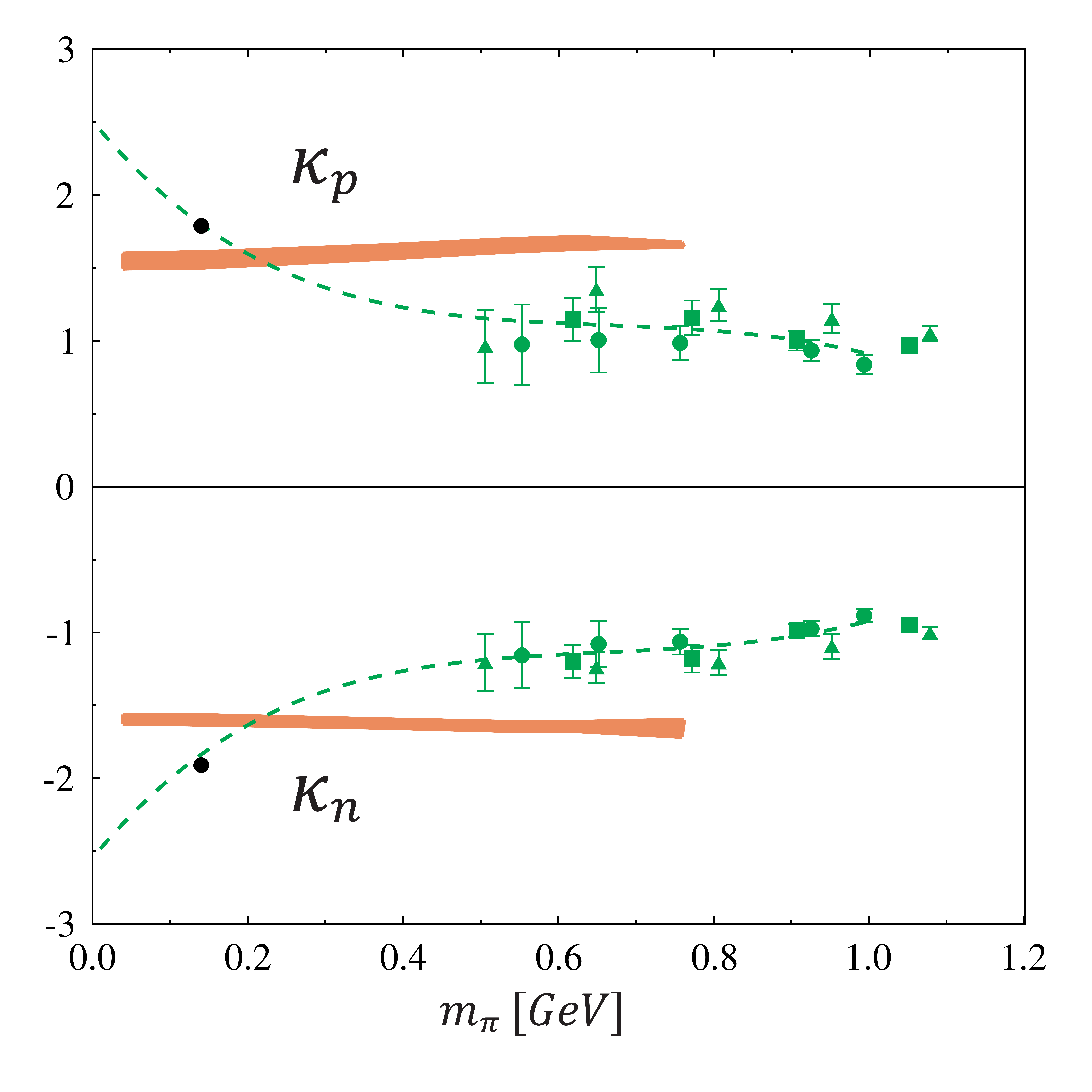}
            \caption[Magnetic moments]{\backdef Anomalous magnetic moments of proton and neutron vs. pion mass.
                                       The results of setup \hyperlink{coupling:c3}{(C3)} are compared to quenched lattice data and
                                       their chiral extrapolations (\textit{dashed curves}) \cite{Gockeler:2003ay}.
                                       Dots denote the experimental values.} \label{fig:magneticmoments}
              \end{center}
              \end{figure}

      \bigskip
      \fatcol{Magnetic moments.}
              The results for the nucleon's magnetic moments at the physical $u/d$ mass are collected in Table~\ref{tab:FF:moments}.
              The current-quark mass dependence of the anomalous magnetic moments $\kappa_p$, $\kappa_n$ is shown in Fig.~\ref{fig:magneticmoments}.
              At large quark masses, where possible pion-cloud effects should be diminished,
              the result still overestimates the lattice data. This suggests further missing corrections beyond pionic contributions.

              In analogy to the discussion of the nucleon mass in Section~\ref{sec:results:nuclenmass}, the pion-cloud effect to magnetic moments can be quantified
              by examining the loop contributions in heavy-baryon chiral effective field theory:
              \begin{equation}
                  \mu_{p,n} (m_\pi^2) = \mu_{p,n}^\text{core} (m_\pi^2,\Lambda) + \mu_{p,n}^\Sigma (m_\pi^2,\Lambda)\,.
              \end{equation}
              Retaining only those loop diagrams which describe a direct coupling of the photon to the intermediate pion,
              where the accompanying baryon is either a $N$ or a $\Delta$,  yields \cite{Young:2004tb,Wang:2007iw}:
              \begin{equation}
                  \mu_{p,n}^\Sigma (m_\pi^2,\Lambda) = \pm \,\frac{\Lambda\,M_N}{3\pi^2 f_\pi^2} \int_0^\infty dx\, \frac{x^4}{\omega^4}
                                                       \left[ g_A^2 \,u_{NN}^2(x) + \frac{C^2}{9}\,\frac{\omega(\omega+2\delta)}{(\omega+\delta)^2}\,u_{N\Delta}^2(x)\right],
              \end{equation}
              where $C=-2\!\cdot\!(0.76)$, and the remaining quantities are explained in the context of Eq.\,\eqref{BARYON:meson-selfenergy}. Using the same input as there,
              i.e. a dipole regulator $u_{NN}(x) = u_{N\Delta}(x)=1/(1+x^2)^2$ with $\Lambda=0.8$ GeV, yields
              at the physical point: $\mu_{p,n}^\Sigma = \pm 0.61$. The inclusion of further meson-loop diagrams can diminish this value \cite{Wang:2007iw}.  

              Since the simplest pion-loop contributions to proton and neutron carry an opposite sign, their
              total cancels in the nucleon's isoscalar and doubles in its isovector combination, defined by
              \begin{equation} \label{FF:iso}
                  F^{s(v)} = F^p \pm F^n\,.
              \end{equation}
              This implies $\kappa_s \approx \kappa_s^\text{core}$, hence isoscalar
              quark-core quantities should be comparable to their experimental or lattice counterparts.
              The experimental isoscalar magnetic moment is small and negative: $\kappa_s=-0.12$. Our result at the $u/d$ current mass is $\kappa_s=-0.03(3)$;
              it is consistent with zero throughout the calculated pion-mass range.

             \renewcommand{\arraystretch}{1.0}

            \begin{table*}[p]
                \begin{center}

                \begin{tabular}{l|lll}
                                                 &$\,\mu_p$    &   $\quad\mu_n$      &   $\quad\kappa_s$          \\[0.1cm] \hline
                                                 &           &                &                        \\[-0.4cm]
                 \hyperlink{coupling:c3}{(C3)}   &$2.56(5)$  &   $-1.58(3)$   &   $-0.03(3)$           \\
                 Exp.                            &$2.79$     &   $-1.91$      &   $-0.12$
                \end{tabular}
                 \caption[]{\backdef
                          Results for proton, neutron and isoscalar magnetic moments in setup \hyperlink{coupling:c3}{(C3)}, compared to experiment. }  \label{tab:FF:moments}

                \end{center}
            \end{table*}

            \begin{table*}[p]
                \begin{center}

                \begin{tabular}{l|llll}
                                                 &   $r_E^p$      &   $\left(r_E^n\right)^2$      &   $r_M^p$    &   $r_M^n$       \\[0.1cm] \hline
                                                 &                &                               &              &          \\[-0.4cm]
                 \hyperlink{coupling:c3}{(C3)}   &   $0.79(2)$    &   $0.00(1)$                   &   $0.73(2)$  &   $0.72(2)$      \\
                 Exp.                            &   $0.89$       &   \!\!\!\!\!$-0.12$                     &   $0.86$     &   $0.87$
                \end{tabular}
                 \caption[]{\backdef
                          Electric and magnetic radii of proton and neutron in setup \hyperlink{coupling:c3}{(C3)} compared to experimental values.
                          All units are fm except $\left(r_E^n\right)^2$ which is fm$^2$. }  \label{tab:FF:radii}

                \end{center}
            \end{table*}

      \bigskip
      \fatcol{Electric and magnetic radii.}
              Table~\ref{tab:FF:radii} shows the results for the charge and magnetic radii together with the experimental values.
              In agreement with the interpretation in terms of an hadronic quark core, and similar to the result for the
              pion charge radius in Fig.~\ref{fig:fpi+rpi}, the radii are sizeably underestimated.

\begin{figure}[p]
              \begin{center}
              \includegraphics[scale=0.108]{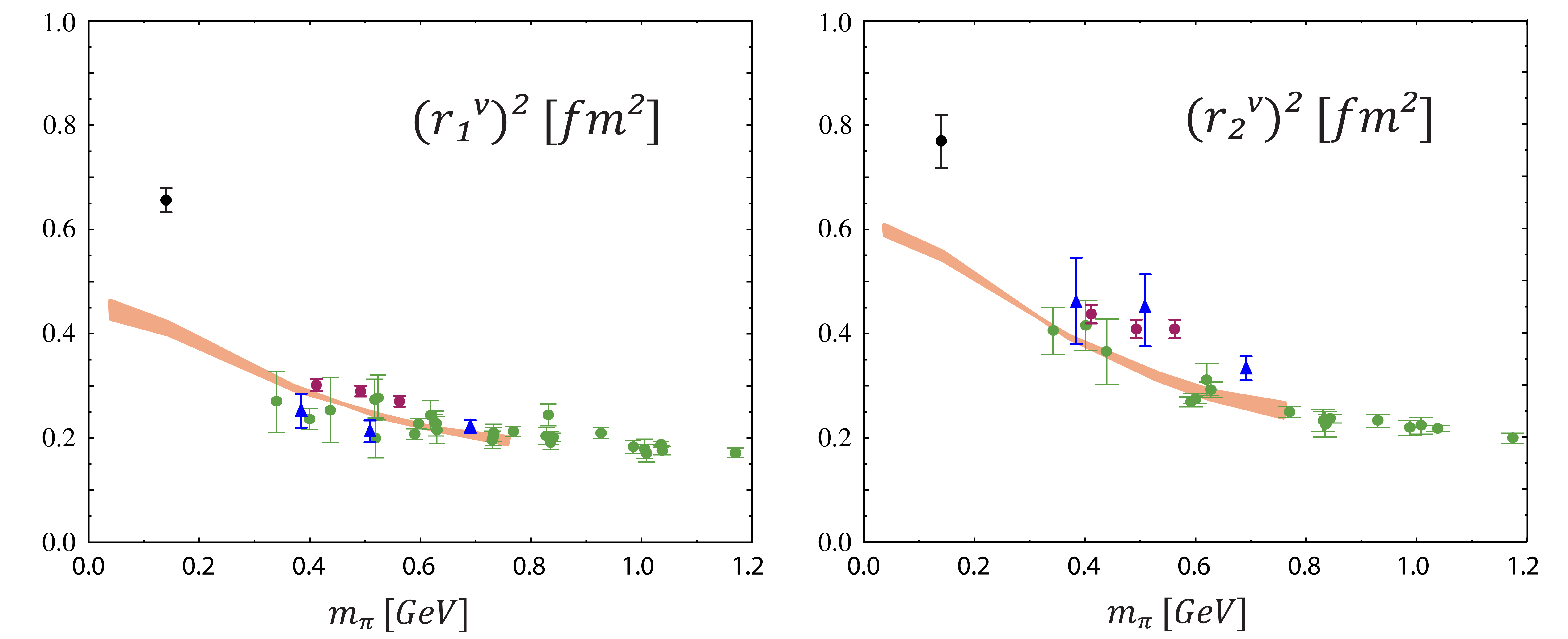}
            \caption[Radii]{\backdef Squared isovector radii corresponding to the Dirac and Pauli form factors $F_{1,2}^v=F_{1,2}^p-F_{1,2}^n$
                                       compared to lattice results \cite{Alexandrou:2006ru,Gockeler:2007ir}.} \label{fig:radii}
              \end{center}
              \end{figure}

              By virtue of Eq.\,\eqref{FF:iso}, the isoscalar and isovector Dirac and Pauli radii read
              \begin{equation}
                  \left(r_1^{s(v)}\right)^2 = \left(r_1^p\right)^2 \pm \left(r_1^n\right)^2\,, \qquad
                  \left(r_2^{s(v)}\right)^2 = \frac{\kappa_p \left(r_2^p\right)^2 \pm \kappa_n \left(r_2^n\right)^2}{\kappa_p\pm\kappa_n}\,.
              \end{equation}
              Because of the smallness of $\kappa_s$ both experimental and calculated values for $r_2^s$ suffer from a large
              statistical uncertainty.
              Lattice studies usually compute isovector form factors where numerically expensive contributions from
              topologically disconnected diagrams cancel. Pion-cloud effects are increased in the isovector channel;
              corresponding lattice results are compared to the quark-diquark calculation in Fig.~\ref{fig:radii} where a mutual agreement at large pion masses is clearly visible.

              \begin{figure}[p]
              \begin{center}
              \includegraphics[scale=0.75]{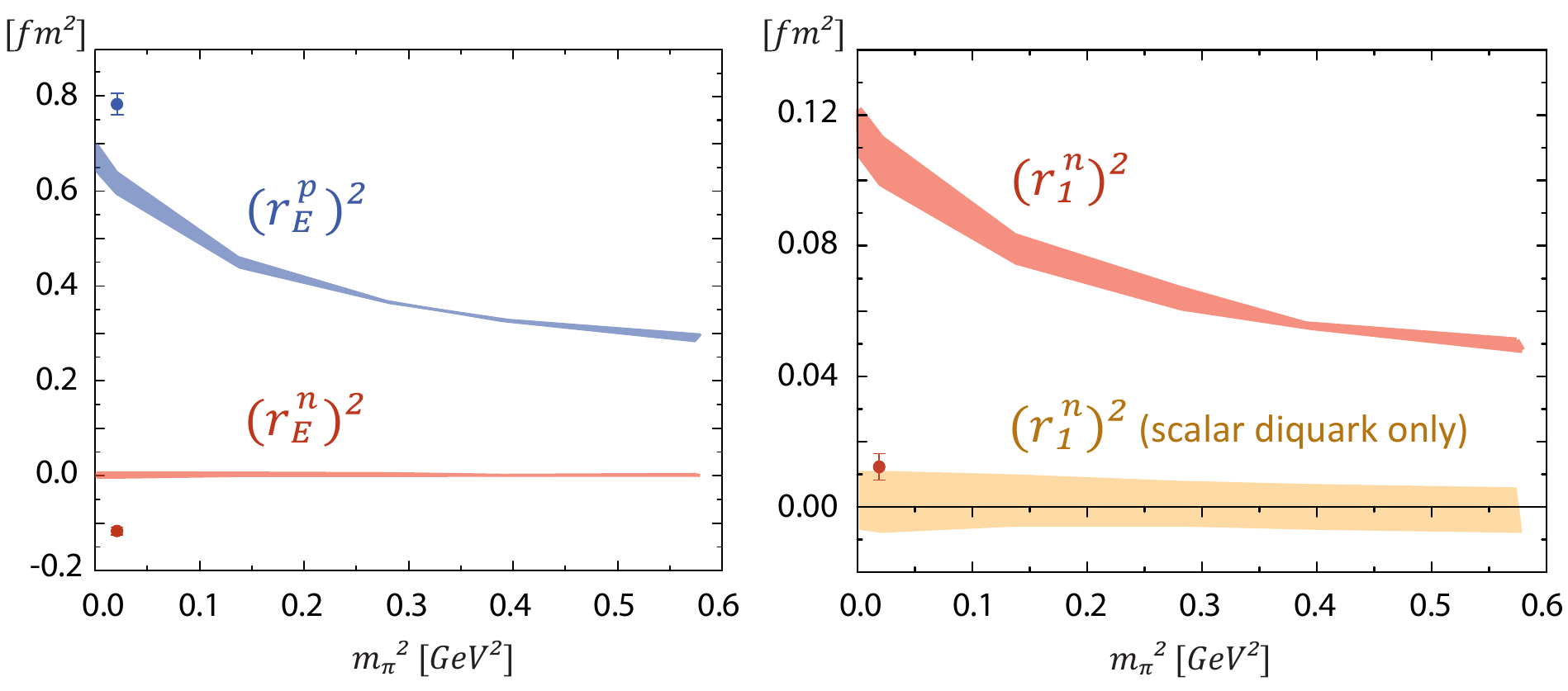}
            \caption[Radii]{\backdef
                            \textit{Left panel:} Electric charge radii for proton and neutron vs. pion mass-squared.
                            \textit{Right panel:} Pion-mass evolution of the neutron's Dirac charge radius.
                                         The full result is compared to the scalar-diquark contribution, cf. Fig.~\ref{fig:ff-contributions-sa}.
                                         Dots with statistical errors denote the experimental values.} \label{fig:radii2}
              \end{center}
              \end{figure}

              In contrast to the magnetic moments, the pionic components of electric and magnetic charge radii $r_E^2$, $r_M^2$
              which correspond to intermediate $N\pi$ and $\Delta\pi$ states  carry opposite signs in the flavor-separated channels, i.e., for the $u$ and $d$-quark contributions \cite{Wang:2008vb}.
              The pion contributions to the charge radii logarithmically diverge in the chiral limit since a massless pion has an infinite Compton wavelength. 
              The quark-diquark result yields $\left(r_E^n\right)^2 \approx 0$ throughout the inspected current-quark mass range and, using Eq.\,\eqref{FF:flavor-contributions}, therefore implies:
              \begin{equation}
                  \big(r_E^u\big)^2 \approx \big(r_E^d\big)^2 \approx \big(r_E^p\big)^2.
              \end{equation}
              A chiral expansion for the electromagnetic radii of proton and neutron similar to \eqref{BARYON:ChiralExpansion} can be found in \cite{Bernard:1995dp}.

              As discussed in Section~\ref{sec:em}, the Dirac form factor $F_1(Q^2)$
              is the Fourier transform of the nucleon's transverse charge distribution in the infinite-momentum frame \cite{Miller:2007uy},
              hence the corresponding Dirac radii feature a direct interpretation in terms of transverse charge radii.
              Experimentally: $F_1^n(Q^2)<0$, and the neutron's Dirac radius is positive: $\big(r_1^n\big)^2>0$.
              With the convention of Eq.\,\eqref{FF:flavor-contributions-ud}, where $r_1^u$ denotes the radius of the up-quark contribution
              in the proton and, via charge symmetry, that of the down-quark in the neutron, Eq.\,\eqref{FF:flavor-contributions} implies
              that the charge radius of the $d$-quark in the neutron is smaller than that of the $u$ quark:
              \begin{equation}
                  \big(r_1^n\big)^2 = \frac{2}{3}\left[\big(r_1^d\big)^2 - \big(r_1^u\big)^2\right] > 0\,.
              \end{equation}
              The extracted transverse charge densities confirm that
              the central charge distribution of a fast-moving neutron (proton) is negative (positive) \cite{Miller:2008jc}.
              This result is at odds with the traditional view of a zero-charge neutron whose $p \,\pi^-$ pion-cloud component
              generates a non-zero charge distribution which has a negative long-range tail but is small and positive at its core.

              The negativity of $F_1^n(Q^2)$ is a natural feature of a quark-diquark model \cite{Eichmann:2008ef,Cloet:2008re}:
              it can be explained by axial-vector $dd$ diquark correlations (corresponding to $\mathsf{s^3}$ in Eq.\,\eqref{dq:flavormatrices}) which,
              other than the scalar-diquark $ud$ and axial-vector $ud$
              contributions, induce a localization of the $d$-quark in the neutron. 
              The quark-diquark model result for the scalar-diquark contribution to $(r_1^n)^2$ at the light quark mass
              is $0.00(1)$ fm$^2$ (Fig.~\ref{fig:radii2}). Adding axial-axial and scalar-axial correlations
              yields $(r_1^n)^2=0.11(1)$ fm$^2$ which  basically cancels the neutron's Foldy term to obtain $r_E^n \approx 0$, cf. Eq.\,\eqref{FF:rEr1}.
              This result is large compared to the experimental value $(r_1^n)^2=0.01$ fm$^2$ and indicative of
              further destructive interference with pion-cloud corrections in the axial-vector diquark channel.

              \begin{figure}[tb]
              \begin{center}
              \includegraphics[scale=0.105]{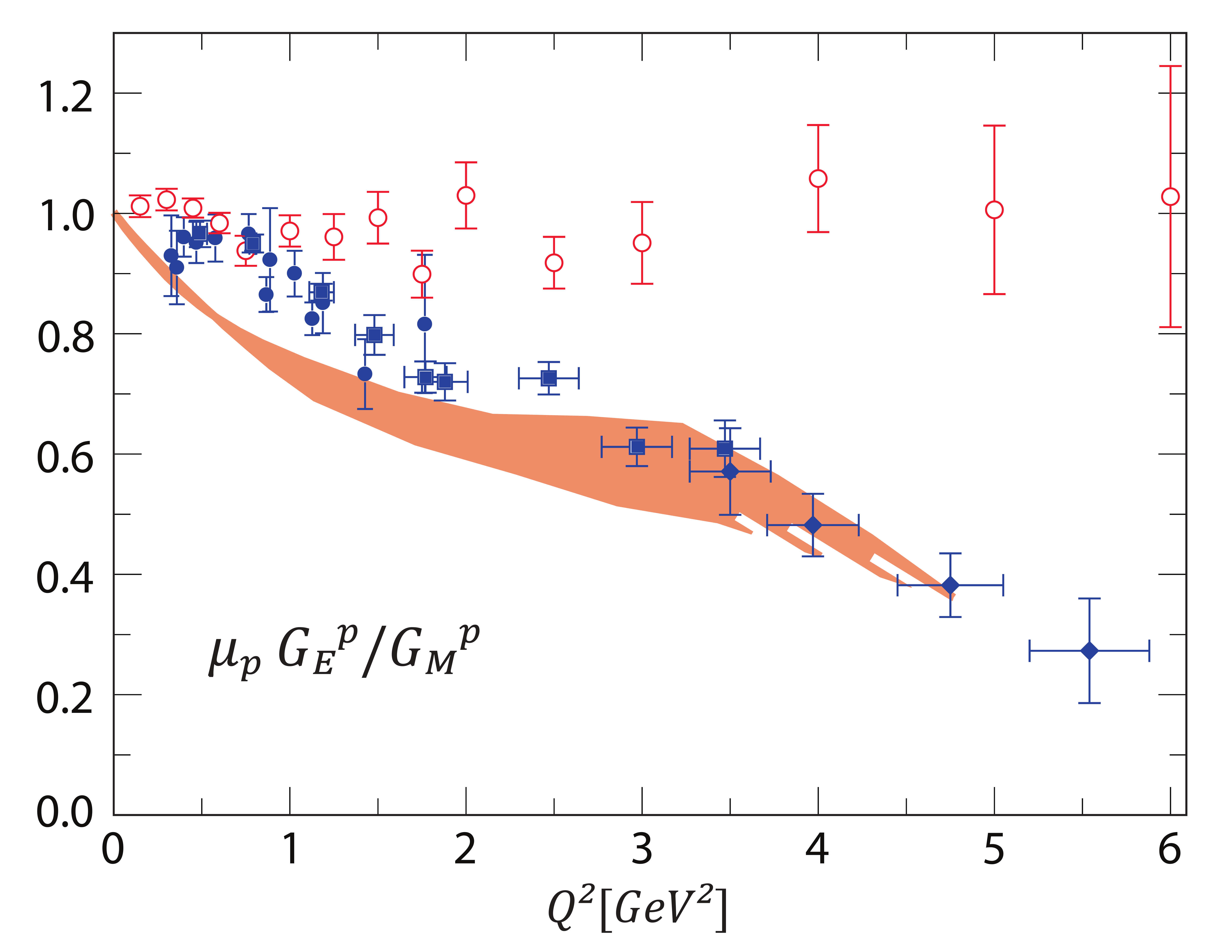}
              \caption[Form factor ratio]{\backdef
                                          $Q^2$-evolution of the proton's form factor ratio $\mu\,G_E/G_M$ in
                                          setup \hyperlink{coupling:c3}{(C3)}, where the band denotes a variation of $\omega$.
                                          The result is compared to experimental data obtained from Rosenbluth-separation \cite{Walker:1993vj}
                                          and polarization-transfer measurements \cite{Jones:1999rz,Gayou:2001qd,Gayou:2001qt}.} \label{fig:ratio}
              \end{center}
              \end{figure}

      \bigskip
      \fatcol{Large-$Q^2$ behavior.}
              At larger momentum transfers $Q^2\gtrsim 2$ GeV$^2$, pion cloud effects must vanish
              since the structure of the nucleon at small distances ($\lesssim 0.15$ fm) is probed.
              The large-$Q^2$ behavior of the form factors therefore reflects genuine properties of the nucleon's quark core.
              Since quark and diquark propagators obtained from the quark DSE or via the T-matrix ansatz necessarily exhibit singularities,
              the applicable $Q^2$ range of the form factor calculation is bounded from above (see App.\,\ref{app:singularities}) and typically limited to few GeV$^2$.
              Access to the large-$Q^2$ domain, where the rainbow-ladder result is expected to become increasingly accurate, necessitates appropriate methods to
              evaluate these Green functions beyond their dominant singularities, i.e., to include the respective residue contributions in the form factor integrals.

              The solution for the proton's form factor ratio $\mu_p G_E^p/G_M^p$ is shown in Fig.\,\ref{fig:ratio} and comparable
              to the quark-diquark model results of Refs.~\cite{Alkofer:2004yf,Cloet:2008re}.
              The small-$Q^2$ structure is a further indication of missing pion-cloud effects.
              A Taylor expansion at small $Q^2$ entails
              \begin{equation}
                  \mu_p \frac{G_E^p}{G_M^p} = 1 - \frac{Q^2}{6}\left( (r_E^p)^2-(r_M^p)^2 \right) + \dots\,.
              \end{equation}
              Experimentally: $r_E^p \approx r_E^n$, whereas the quark-diquark calculation yields $r_E^p > r_E^n$.
              Pionic contributions affect electric and magnetic radii differently \cite{Cloet:2008re}.

              The evolution beyond $Q^2 \gtrsim 2$ GeV$^2$ is mainly dominated by the proton's quark core.
              We note that the phenomenological transverse term introduced in the seagull vertices, Eq.\,\eqref{FF:Seagulls:Transverse},
              has a sizeable impact on the form factor ratio in that region. The parameters \eqref{FF:Seagulls:Transverse:params} optimize
              mutual agreement with the polarization-transfer data for $G_E/G_M$ and those for $G_M^p$ and $G_M^n$.
              Nevertheless a variation of these parameters only minimally affects the small-$Q^2$ region which therefore only depends on the input
              introduced in connection with the effective coupling $\alpha(k^2)$ of Section~\ref{sec:coupling-ansaetze}.
              We note that the effective diquark vertices would naturally no longer appear 
              in a form factor calculation within the three-body approach according to Eq.\,\eqref{FADDEEV:Current}.


\chapter{Conclusions and outlook}\label{sec:conclusions}

             The covariant bound-state formalism, based on the Dyson-Schwinger equations of QCD, provides a versatile tool
             for the calculation of hadron properties. It is formulated within QCD, fully relativistic, and represents a continuum approach;
             it provides access to infrared and ultraviolet momenta; and, with appropriate numerical algorithms
              which have become available in the past years, it covers the full quark mass range from the chiral limit up to arbitrarily large current masses.
              At the present stage, especially in the baryon sector, we are only beginning to explore its potential and possibilities.

             In this thesis a study of pion, $\rho$-meson, nucleon and $\Delta$ properties was presented in such an approach, where
             the inherent link between meson and baryon observables is expressed through a rainbow-ladder truncation in the quark-antiquark and quark-quark channel.
             The effect of dynamical mass generation is manifest, while the truncation at the same time
             preserves the nature of the pion as a Goldstone boson of spontaneously broken chiral symmetry.

             The rainbow-ladder kernel embodies an effective quark-gluon coupling $\alpha(k^2)$
             which represents the only phenomenological input in the calculation of meson and baryon observables.
             Two models for the current-quark mass dependence of the coupling strength were compared:
             the first setup is defined by a current-mass independent strength which has long been known to provide a good description of pseudoscalar and vector-meson properties;
             in the second setup, chiral corrections --- partly owing to pseudoscalar meson-cloud effects --- were anticipated 
             to construct an overestimated quark core for the $\rho$-meson.
             The distinction between the two models arises in the infrared part of the coupling,
             and its impact upon hadronic observables turned out to be
             an approximate additive contribution to quantities with dimension of a squared mass.

             A self-consistent solution of the three-body equation for the nucleon was presented for the first time,
             where the kernel was given by a rainbow-ladder gluon exchange between any two quarks.
             The result was compared to a quark-diquark calculation which is enabled by the occurrence of artificial diquark mass poles
             in the $qq$ scattering matrix induced by the gluon ladder kernel.
             The result for $M_N(m_\pi^2)$ is very similar in both approaches; the quark-diquark model is slightly too attractive, i.e. by $50-100$ MeV.
             The overall properties of an inflated quark core are clearly reproduced in $M_N$ and $M_\Delta$;
             in particular, agreement with lattice data is obtained for all investigated hadron masses when the quark mass becomes large.
             The mass of the $\Delta$ baryon in the quark-diquark calculation appears to be unexpectedly large; it remains to be seen if the three-quark approach
             provides further attraction or, as in the nucleon case, moderate repulsion.

             Results for the nucleon's static electromagnetic properties as a function of the pion mass were presented and
             compared to lattice results and chiral extrapolations.
             Similarly to the pion charge radius, good agreement was found for the nucleon's isovector radii above the strange-quark mass
             whereas missing pion-cloud effects are clearly visible in the chiral region.
             Except for $G_E^n$, the overall consistency of the nucleon's electromagnetic form factors as a function of the photon momentum with experiment
             is quite remarkable.  Pionic effects are missing
             at small $Q^2$, and their absence is amplified in the proton's form factor ratio $G_E/G_M$.
             The implementation of self-consistently obtained quark and diquark propagators with a complex singularity structure leads to an upper limit
             at several GeV$^2$; together with a model dependence of the diquark ingredients at larger photon momentum it
             impedes an unambiguous signal of a possible zero crossing.
             The latter obstacle can be overcome by a form factor calculation within the three-quark approach.

             It is imperative to study the effects of different interactions in the $q\bar{q}$ and $qq$ channels:
             in particular, further admixture of a scalar-scalar interaction, induced by a scalar part in the quark-gluon vertex,
             is expected to yield notable ramifications for the infrared structure of QCD and derived hadron properties;
             the same is true for pionic effects.
             Nevertheless, with a solution of the three-quark equation available,
             the journey beyond rainbow-ladder is now simultaneously and consistently possible in both meson and baryon channels;
             and the covariant bound-state framework as an ab-initio approach evolves one step forward towards an understanding of hadron dynamics.

             The now achieved increase of predictive power finally allows to reach out for more observables which are in the focus of current experimental interest. For instance,
             a calculation of the $\Delta$ mass, electromagnetic form factors and $N$-$\Delta$ transition form factors is a natural application of the three-body approach.
             Further possible future directions are: an extension to the heavy-quark regime, including hadrons with open strange, charm and bottom quantum numbers;
             radial excitations of baryons and a clarification of the role of the Roper resonance; strong and electroweak scattering processes; form factors at large $Q^2$;
             a study of exotic mesons in the bound-state formalism;
             and a description of the hadron's internal momentum and angular momentum structure expressed through parton distribution functions and GPDs.


\begin{appendix}

\chapter[A collection of propagators, vertices and amplitudes]{A collection of propagators, \\ vertices and amplitudes}\label{app:vertices}

            In this appendix we collect the basic Green functions and bound-state amplitudes which appear in the main text.
            We state their general properties and specify the expressions that are used in the numerical computation.
            The quark propagator (\ref{sec:quarkpropagator}) and quark-photon vertex (\ref{sec:quark-photon-vertex}) represent the 'elementary' quantities of the approach.
            Color-singlet mesons and colored diquarks are composite objects;
            their on-shell and effective off-shell properties are summarized in (\ref{app:mesondiquark}) and (\ref{app:mesondiquark-dqprop}), respectively.
            We discuss the bound-state amplitude for the nucleon in the three-body framework (\ref{app:faddeevamp}) and that for $N$ and $\Delta$ in the quark-diquark approach (\ref{app:qdqamp}).
            The diagrams which contribute to the nucleon's electromagnetic current in the quark-diquark model are stated in (\ref{app:emcurrent});
            they involve diquark-photon vertices (\ref{app:dqphotonvertex}) and seagull amplitudes (\ref{app:seagulls}).


\section{Quark propagator}\label{sec:quarkpropagator}

              As a fermionic two-point function which involves one momentum $p$, the
              quark propagator can only depend on the two Dirac structures $\left\{\Slash{p},\,\mathds{1}\right\}$.
              The corresponding dressing functions $\sigma_v$ and $\sigma_s$ may be expressed through
              the quark renormalization function $Z_f$ and the renormaliization-point independent quark mass function $M$:
              \begin{fshaded5}
              \begin{equation}
                  S(p,\mu) =  -i \Slash{p} \,\sigma_v(p^2,\mu^2) + \sigma_s(p^2,\mu^2) = \frac{Z_f(p^2,\mu^2)}{p^2+M(p^2)^2} \left( -i \Slash{p} + M(p^2) \right)\,.
              \end{equation}
              \end{fshaded5}

              \noindent
              Another frequently used notation involves the quantities $A(p^2,\mu^2) = 1/Z_f(p^2,\mu^2)$ and $B(p^2,\mu^2) = M(p^2)/Z_f(p^2,\mu^2)$.
              The inverse propagator reads:
              \begin{equation}
                  S(p,\mu)^{-1} =  A(p^2,\mu^2) \left(i \Slash{p}  + M(p^2) \right).
              \end{equation}

    \bigskip
    \fatcol{Asymptotic behavior and quark condensate.}
            Asymptotically, the DSE solution for the quark mass function reproduces the behavior predicted from perturbation theory (see, e.g., \cite{Roberts:2000aa,Fischer:2003rp}):
            \begin{equation}\label{dse:asymptoticmassf}
                M(p^2) \stackrel{p^2\rightarrow\infty}{\longlongrightarrow} \, \frac{\hat{m}}{\mathcal{F}(p^2)^{\gamma_m}} +
                       \frac{2 \pi^2 \gamma_m}{N_C}  \;\frac{-\langle \bar{q} q\rangle}{\mathcal{F}(p^2)^{1-\gamma_m}\,p^2}
            \end{equation}
            where $\mathcal{F}(p^2)= \frac{1}{2}\ln \left(p^2/\Lambda_{QCD}^2\right)$. The coefficients $\hat{m}$ and $-\langle \bar{q} q \rangle$ defined thereby
            are the renormalization-point independent current mass and chiral condensate. For finite current masses, the second term is suppressed
            by a factor of $p^2$ while in the chiral limit, defined by $\hat{m}=0$, it determines the behavior of the asymptotic mass function.

            The cutoff-dependent bare mass $m_0$ that appears in the bare quark propagator and enters the quark DSE \eqref{dse:qdse} is related to the renormalized mass $m_\mu$ via the mass renormalization constant $Z_m$:
            $m_0(\Lambda^2) = Z_m(\mu^2,\Lambda^2)\,m_\mu$. For a large renormalization point, $M(\mu^2)$ can be identified with $m_\mu$
            which entails $m_0(\Lambda^2) = M(\Lambda^2)$ \cite{Fischer:2003rp}. In the following we will always assume $\mu$ to be large, i.e. $\mu \gg \Lambda_\text{QCD}$;
            in our calculation we use the value $\mu=19$~GeV.

            The renormalization-point-dependent chiral quark condensate is obtained from the trace of the chiral quark propagator:
            \begin{equation}\label{dse:qq}
                -\langle \bar{q} q \rangle_\mu = Z_2(\mu^2,\Lambda^2) \,Z_m(\mu^2,\Lambda^2)\,N_C \int^\Lambda_q\text{Tr}_D \{ S_\text{chiral}(q,\mu) \},
            \end{equation}
            with $Z_m(\mu^2,\Lambda^2) = M(\Lambda^2)/M(\mu^2) $ evaluated at large current masses.
            For a large renormalization point, the current-quark masses and condensates are related via
            \begin{equation}\label{QUARK:M+Cond}
                M(\mu^2) = \frac{\hat{m}}{\mathcal{F}(\mu^2) ^{\gamma_m}}, \quad  -\langle \bar{q} q \rangle_\mu = -\langle \bar{q} q \rangle\, \mathcal{F}(\mu^2) ^{\gamma_m}.
            \end{equation}

    \bigskip
    \fatcol{Solving the quark DSE.}
            The quark DSE \eqref{dse:qdse},
              \begin{equation}
                  S(p,\mu)^{-1} =  Z_2(\mu^2,\Lambda^2) \left( i \Slash{p} + M(\Lambda^2) \right) + \Sigma(p,\mu,\Lambda) \; ,
              \end{equation}
            can be rewritten in terms of two coupled integral equations for the quark propagator's dressing functions $A(p^2,\mu^2)$ and $M(p^2)$:
            \begin{equation}\label{dse:qdse1}
            \begin{split}
                            A(p^2,\mu^2) &=   \quad\quad\quad\;              Z_2(\mu^2,\Lambda^2) + \Sigma_A(p^2,\mu^2,\Lambda^2),     \\
                    M(p^2)\,A(p^2,\mu^2) &=   M(\Lambda^2)\,Z_2(\mu^2,\Lambda^2) + \Sigma_M(p^2,\mu^2,\Lambda^2),         
            \end{split}
            \end{equation}
            where the scalar functions $\Sigma_A$ and $\Sigma_M$ constitute the quark self-energy via
            \begin{equation}
                \Sigma(p,\mu,\Lambda) = i \Slash{p} \, \Sigma_A(p^2,\mu^2,\Lambda^2) + \Sigma_M(p^2,\mu^2,\Lambda^2).
            \end{equation}
            Eqs.\,\eqref{dse:qdse1} can be solved iteratively for chosen values of $Z_2$ and $M(\Lambda^2)$.
            Upon employing a renormalization condition, e.g. $A(\mu^2,\mu^2)=1$,
            and specifying the current mass $M(\mu^2)$ at the renormalization point as an input parameter,
            both $Z_2$ and $M(\Lambda^2)$ are determined together with $A(p^2)$ and $M(p^2)$
            in the course of the iteration via
            \begin{equation}\label{QUARK:Mmu+Z2}
                Z_2(\mu^2,\Lambda^2) = 1 - \Sigma_A(\mu^2,\mu^2,\Lambda^2)\,, \quad M(\Lambda^2) = \frac{M(\mu^2) - \Sigma_M(\mu^2,\mu^2,\Lambda^2)}{Z_2(\mu^2,\Lambda^2)}\,.
            \end{equation}

            Omitting the renormalization-point and cutoff dependence for brevity, the self-energy integral (\ref{dse:qselfenergy},\,\ref{eq:mesons:alpha-i1}) in rainbow truncation reads
            \begin{equation}\label{dse:qdse3}
                \Sigma(p) = \int_q^\Lambda T^{\mu\nu}_k\,g(k^2)\,\gamma^\mu S(q) \,\gamma^\nu,  \quad g(k^2) := Z_2^2  \,\frac{16\pi}{3} \, \frac{\alpha(k^2)}{k^2},
            \end{equation}
            where $k^2 = p^2 + q^2 - 2\,p\cdot q =: p^2 + q^2 - 2\,\sqrt{p^2}\,\sqrt{q^2}\,z$ is the squared gluon momentum and $g(k^2)$ a shorthand notation for the effective coupling.
            The self-energy coefficients become
            \begin{equation}\label{dse:qdse2}
                \Sigma_A(p^2) = \int^\Lambda_q \sigma_v(q^2) \,g(k^2)\,F(p^2,q^2,z), \quad
                \Sigma_M(p^2) = 3\int^\Lambda_q \sigma_s(q^2) \,g(k^2)
            \end{equation}
            and involve the quark dressings $\sigma_v(q^2)$, $\sigma_s(q^2)$ which depend on $A(q^2)$ and $M(q^2)$. The dimensionless quantity $F$ is given by
            \begin{equation}
            \begin{split}
                p^2 \,F(p^2,q^2,z) &= \textstyle{-\frac{1}{4}}\, \text{Tr} \left\{ \Slash{p}\,\gamma^\mu\,\Slash{q}\,\gamma^\nu \right\} T^{\mu\nu}_k \,= \\
                               &= p\cdot q + \frac{2}{k^2}\,(p\cdot k)(q\cdot k)
                                \,=\, 3 \, p\cdot q - \frac{2}{k^2}\left(p^2 \,q^2 - (p\cdot q)^2 \right) \,=  \\
                               &= -k^2 + \frac{p^2+q^2}{2} + \frac{(p^2-q^2)^2}{2k^2}
                                \,=\, p^2 + 3\, p\cdot k + 2 \,\frac{(p\cdot k)^2}{k^2}\,.
            \end{split}
            \end{equation}
            Since the only Lorentz-invariant combinations which appear in the quark self-energy \eqref{dse:qdse2} are $p^2$, $q^2$ and $z$,
            the loop integral becomes two-dimensional (cf. Eq.\,\eqref{hypersphericalintegral}), e.g.:
            \begin{equation*}
                \int_q^\Lambda \sigma_v(q^2)\,g(k^2)\,F(p^2,q^2,z) \sim \int_0^{\Lambda^2} dq^2 \,q^2\, \sigma_v(q^2) \int_{-1}^1 dz\sqrt{1-z^2} \,g(k^2)\,F(p^2,q^2,z).
            \end{equation*}
            For a spacelike external momentum $p^2 \in \mathds{R}_+$, the squared gluon momentum $k^2$ is real and positive as well  and
            the coupled system \eqref{dse:qdse1} can be solved without complications.

    \bigskip
    \fatcol{Quark propagator in the complex plane.}
            If the external quark momentum $p^2$ is complex,
            the argument $k^2$ of the coupling $g(k^2)$ constitutes the interior of a parabola
            \begin{equation}\label{QP:k2-parabola}
                 k^2_P = p^2 + q^2 \pm 2\,\sqrt{p^2}\,\sqrt{q^2} = \left(t\pm i\,|\text{Im}\sqrt{p^2}|\right)^2, \quad t \in \mathds{R}_+,
            \end{equation}
            which passes through the outer point $p^2$ and has its apex at (cf. Fig.\,\ref{fig:dsecomplexplane})
            \begin{equation}
                t=0 \; \Longrightarrow \; k^2_P = -\left(\text{Im}\sqrt{p^2}\right)^2 = \frac{1}{2}\left(\text{Re}\,p^2-|p^2|\right).
            \end{equation}
            If $g(k^2)\,F(p^2,q^2,z)$ exhibits non-analyticities in that $k^2$ domain,
            one must either adjust the size of the parabola \eqref{QP:k2-parabola} by imposing a limit on $p^2$ or resort to refined numerical methods.
            The kinematic singularity in $F$ at $k^2 \rightarrow 0$ induced by the transverse gluon can be compensated by a vanishing $g(k^2 \rightarrow 0)$.
            The infrared behavior of the parametrization \eqref{dse:maristandy} \textit{almost} satisfies that criterion:
            \begin{equation}
                \frac{\alpha(k^2)}{k^2} \stackrel{k^2\rightarrow 0}{\longlongrightarrow} \frac{\pi \gamma_m}{\Lambda_0^2}, \quad \Lambda_0 = 1\,\text{GeV}.
            \end{equation}
            The remainder stems from the ultraviolet term and is relatively small compared to the overall strength of $g(k^2)$.
            It results in small numerical artifacts which are visible in the complex functions $\sigma_v(p^2)$ and $\sigma_s(p^2)$
            upon a straightforward integration.

            The complex conjugate branch points in the logarithmic tail of $\alpha(k^2)$ lead to a theoretical limitation
            \begin{equation}
                -\left(\text{Im}  \sqrt{p^2}\right)^2 > -\Lambda_{QCD}^2 \left(\text{Im}\sqrt{-1\pm i\sqrt{e^2-2}} \right)^2 = -(0.31\,\text{GeV})^2
            \end{equation}
            which is practically unimportant since they are concealed by the large oscillations of the exponential parts in $\alpha(k^2)$.
            The complex conjugate poles of the resulting rainbow-ladder quark propagator are insensitive to these singularities as well.

           \bigskip

            A coupling which does not satisfy $g(k^2\rightarrow 0) \rightarrow 0$ or involves further singularities in the integration domain
            inevitably requires an advanced numerical treatment.
            The same is true for a truncation beyond rainbow-ladder which leads to a more complicated structure of the
            self-energy integrals.
            We shortly discuss two such strategies:

      \begin{itemize}
           \item \fatcol{Complex rays.}
            Upon performing all the integrations except the $q^2$ integral, singular points in $k^2$ lead to branch cuts in the complex $q^2$ plane,
            as illustrated in Fig.\,\ref{fig:dsecomplexplane}. For instance, a pole at $k^2=0$ generates a circular branch cut in the complex $q^2$ plane
	        with an opening at $q^2=p^2$ (dashed line). The original integration contour $q^2 \in \mathds{R}$
            crosses these branch cuts and the resulting numerical artifacts become dominant in a straightforward integration.

            The logarithmic one-loop behavior \eqref{dse:asympcoupling} entails that \textit{any} coupling exhibits singularities at $k^2\neq 0$ as well:
            those will generally lead to more complicated branch cut structures (dash-dotted line) which however still leave the arc $q^2 = r\,e^{i \arg{p^2}}$, $r\in\mathds{R}_+$, unharmed.
            A possible way to avoid all occurring branch cuts is to deform the integration contour $q^2 \in (0,\Lambda^2)$ to a complex arc that passes through the point
            $p^2$ and eventually returns to $\Lambda^2 \in \mathds{R}$ in the far spacelike region. Since $\sigma_v(q^2)$ and $\sigma_s(q^2)$
            must already be known on these complex paths, the complex DSE solution is therefore obtained via iteration of
            \eqref{dse:qdse1} on a family of deformed complex paths in $p^2$.

            This method has been sketched in \cite{Eichmann:2007nn} and is ideally suited if the singularities of the resulting quark propagator which are generated
            during the iteration only appear on the timelike $p^2$ axis.
            Such a singularity structure appears if a Ball-Chiu-like ansatz for the quark-gluon vertex is applied \cite{Alkofer:2003jj}.

            \begin{figure}[tbp]
            \begin{center}
            \includegraphics[scale=0.095]{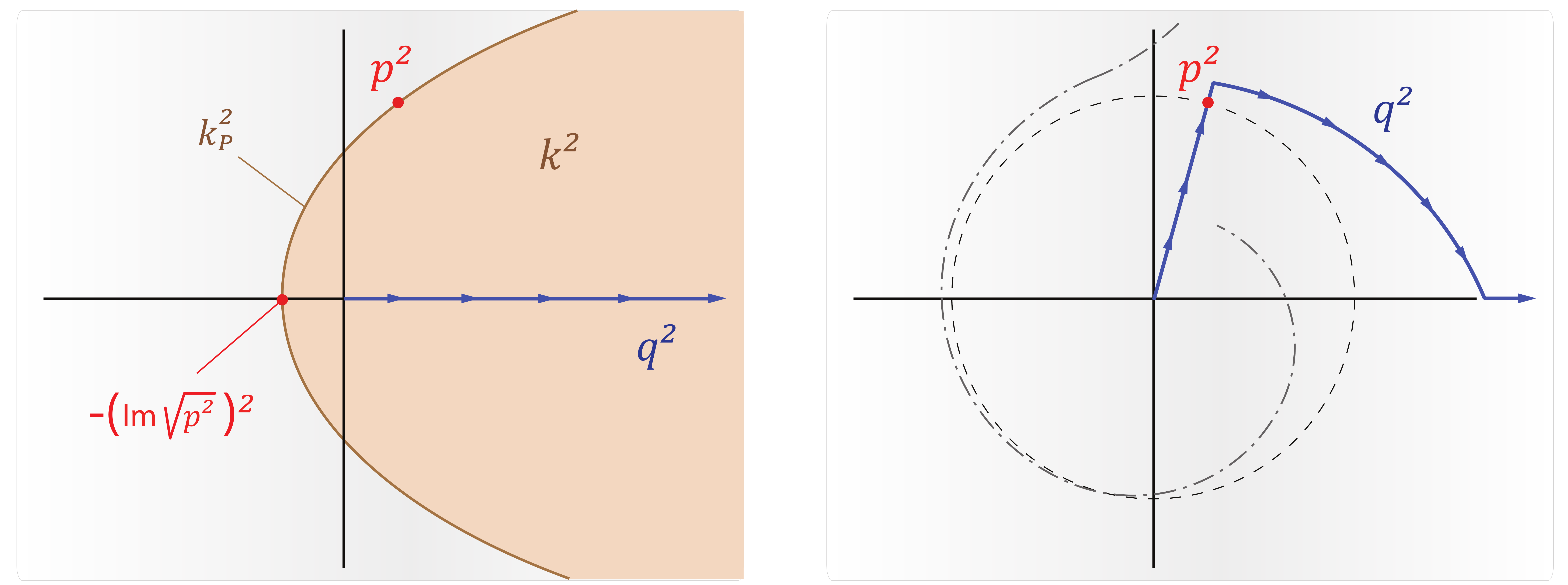}
            \caption[DSE in the complex plane]{\backdef
                                               \textit{Left panel:} Integration domain of quark ($q^2$) and gluon propagator ($k^2$) in the self-energy integral corresponding to a fixed external point $p^2$.
                                               \textit{Right panel:} Branch cuts in the complex $q^2$ plane of the quark propagator and
					                                                 a possible integration path.} \label{fig:dsecomplexplane}
            \end{center}
            \end{figure}

           \item \fatcol{Complex parabolas.}
           A different strategy to avoid singularities in the integration domain has been described in Refs.\,\cite{Fischer:2008sp,Fischer:2009jm,Krassnigg:2009gd}.
           The basic idea is to use the gluon momentum $k$ instead of the quark momentum $q$ as the integration variable\footnote{
           The change of integration variables requires a translationally invariant regularization of the self-energy integral.
           This is not realized by a \textit{hard cutoff} $\Lambda$ which leads to (however small) deviations between results obtained in both methods.
           Implementing a Pauli-Villars regulator $1/(1+k^2/\Lambda^2)$ attached to the gluon propagator suppresses the UV modes and
           restores translational invariance but ignores the perturbative 1-loop behavior of the effective coupling.}.
           As a consequence, the argument $k^2$ of the coupling becomes real and $q^2$ complex;
           hence the only relevant singularities are those of the final quark propagator which are generated during the iteration.
           Now the internal quark momentum $q^2 = p^2 + k^2 + 2\,p\cdot k$, with $k^2 \in \mathds{R}_+$, is bounded by a parabola
           \begin{equation}
               q^2_P = p^2 + k^2 \pm 2\,\sqrt{p^2}\,\sqrt{k^2} = \left(t'\pm i|\text{Im}\sqrt{p^2}|\right)^2, \quad t' \in \mathds{R}_+,
           \end{equation}
           whose position is characterized by the external quark momentum $p^2$.
           The iteration in \eqref{dse:qdse1} takes place on such a parabola, and the Cauchy formula is used to obtain the quark propagator in the interior at each iteration step.

           The method is particularly useful if the resulting quark propagator exhibits complex conjugate poles,
           a feature which is common in rainbow-ladder studies and was also recovered in more general truncations \cite{Fischer:2008sp,Alkofer:2008tt}.
           It furthermore allows for a more accurate determination of the complex-plane propagator at larger quark masses \cite{Krassnigg:2009gd}.

      \end{itemize}

      \noindent
      Analogous procedures can be implemented for evaluating meson and diquark amplitudes in the complex plane of the relative momentum between the contributing quarks.


\section{Quark-photon vertex}\label{sec:quark-photon-vertex}

            The quark-photon vertex is the central ingredient of any hadronic electromagnetic form factor diagram since,
            of all the fundamental degrees of freedom in the QCD Lagrangian, only the quark is electrically charged.
            With the notation $k=(k_+ + k_-)/2$ (relative or Breit momentum) and $Q=k_+ - k_-$ (incoming photon momentum), where $k_+$ and $k_-$ are outgoing and incoming quark momenta,
            the general form of the quark-photon vertex is
                  \begin{align}
                      \Gamma^\mu_\text{(q)}(k,Q) &=  \sum_{i=1}^{12} f^\text{(q)}_i(k^2,\hat{k}\cdot\hat{Q},Q^2) \,\tau_i^\mu(k,Q),
                  \end{align}
            where a possible representation of the Dirac basis elements is given by
                  \begin{equation}\label{QPV:basis}
                         \tau_i^\mu(k,Q)  \in  \left\{ \gamma^\mu, k^\mu, Q^\mu \right\} \times \left\{\mathds{1}, \Slash{Q}, \, \Slash{k}, \, [\Slash{k},\Slash{Q}] \right\}.
                  \end{equation}
            Induced by electromagnetic current conservation, a vector Ward-Takahashi identity constrains the longitudinal contributions $\sim Q^\mu$ of the vertex
            by relating them to the quark propagator:
                  \begin{equation}
                  \begin{split}
                      Q^\mu \,\Gamma^\mu_\text{(q)}(k,Q) &= S^{-1}(k_+)-S^{-1}(k_-)  \\
                                                         &= Q^\mu \Big( i\gamma^\mu\,\Sigma_A +2 k^\mu ( i\Slash{k} \,\Delta_A  + \Delta_B ) \Big).
                  \end{split}
                  \end{equation}
            Here we used the abbreviations
                  \begin{equation}\label{QPV:sigma,delta}
                     \Sigma_F := \frac{F(k_+^2)+F(k_-^2)}{2} , \quad  \Delta_F := \frac{F(k_+^2)-F(k_-^2)}{k_+^2-k_-^2},
                  \end{equation}
            where $A(p^2)$, $B(p^2)$ are the quark propagator's dressing functions, and
            for $Q^2\rightarrow 0$: $\Sigma_F\rightarrow F(k^2)$ and $\Delta_F\rightarrow F'(k^2)$.
            The differential Ward identity for $Q^2\rightarrow 0$ reads
                  \begin{align}\label{QPV:WardIdentity}
                      \Gamma^\mu_\text{(q)}(k,0) &= \frac{d S^{-1}(k)}{d k^\mu}= i\gamma^\mu A(k^2) +2 k^\mu \left( i\Slash{k} \,A'(k^2)+B'(k^2) \right).
                  \end{align}
            Implementing both relations yields the most general expression
            for the quark-photon vertex:
                  \begin{shaded}
                  \begin{equation}\label{vertex:BC}
                      \Gamma^\mu_\text{(q)}(k,Q) =   i\gamma^\mu\,\Sigma_A + 2 k^\mu (i\Slash{k}\, \Delta_A  + \Delta_B) + T^{\mu\nu}_Q \,\Gamma^{\nu}_T(k,Q) \,.
                  \end{equation}
                  \end{shaded}
            \noindent
            The first part is the Ball-Chiu vertex \cite{Ball:1980ay}.
            The transverse contribution $\Gamma^{\mu}_T(k,Q)$ is constructed from the eight basis elements $\sim \gamma^\mu, k^\mu$ and
            must satisfy $\Gamma^{\mu}_T(k,0)=0$ to obey the Ward identity,
            either by a $Q^\mu$ dependence of the basis elements or
            by a vanishing amplitude at $Q^2=0$.

            \bigskip

            We note that, due to the transversality of the photon,
            the purely longitudinal components of the vertex do not contribute to any hadronic matrix elements:
            only the transverse projection of \eqref{vertex:BC} does.
            In this sense the WTI alone provides no constraint on \textit{physics}.
            If, according to the basis \eqref{QPV:basis}, one starts from the general expression
                  \begin{equation}\label{QPV:note1}
                      \Gamma^\mu_\text{(q)}(k,Q) = T^{\mu\nu}_Q \left( i\gamma^\nu\,\Gamma^{(1)}(k,Q) + k^\nu\,\Gamma^{(2)}(k,Q) \right) + Q^\mu\,\Gamma^{(3)}(k,Q),
                  \end{equation}
            then $\Gamma^{(1)}$ and $\Gamma^{(2)}$ determine the physical content, while only the unphysical component $\Gamma^{(3)}$ is constrained by the WTI:
                  \begin{equation}\label{QPV:note2}
                         Q^\mu\,\Gamma^{(3)}(k,Q) = L^{\mu\nu}_Q\,\Big( i\gamma^\nu\,\Sigma_A +2 k^\nu ( i\Slash{k} \,\Delta_A  + \Delta_B ) \Big).
                  \end{equation}
            The transverse and longitudinal parts (i.e., the brackets in Eqs.\,\eqref{QPV:note1} and \eqref{QPV:note2}) must however share the same limit
            at $Q^2\rightarrow 0$ in order to avoid a kinematic singularity. This is ensured by the differential Ward identity which subsequently determines Eq.\,\eqref{vertex:BC}.

            \begin{figure}[tbp]
            \begin{center}
            \includegraphics[scale=0.3]{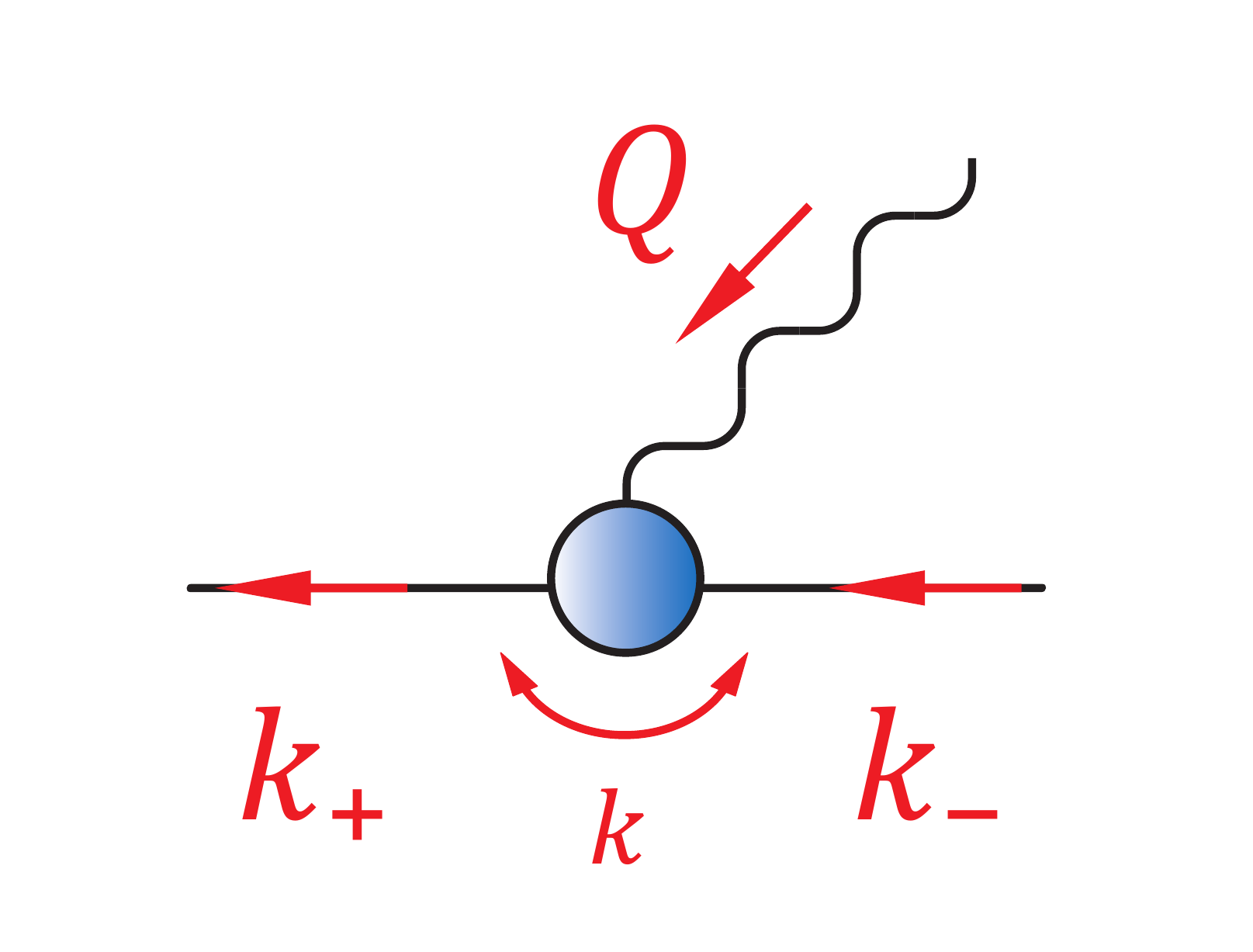}
            \caption[Quark-photon vertex]{\backdef Momentum routing in the quark-photon vertex. } \label{fig:QPV}
            \end{center}
            \end{figure}

    \bigskip
    \fatcol{Modeling the transverse part.}
            The transverse part must vanish for $Q^2=0$ due to current conservation.
            In the perturbative limit $Q^2\rightarrow\infty$ it is $\mathcal{O}(\alpha)$-suppressed compared to the Ball-Chiu construction
            which ensures a bare vertex via $\Sigma_A \rightarrow Z_2$ and $\Delta_{A,B}\rightarrow 0$.
            Several parametrizations for the transverse part have been devised in the literature, amongst which is the
            Curtis-Pennington ansatz that ensures multiplicative renormalizability in the Dyson-Schwinger equations of quenched QED \cite{Curtis:1990zs}:
                \begin{equation}
                    \Gamma^{\mu}_T(k,Q) = i\Delta_A \,\frac{(k\cdot Q) \left( k^2+Q^2/4 \right) }{(k\cdot Q)^2 + \left(\Sigma_{M^2}\right)^2} \,\Big( (k\cdot Q) \, \gamma^\mu - k^\mu\,\Slash{Q} \Big).
                \end{equation}
            A modified ansatz which accomplishes the same constraint in unquenched QED was recently proposed \cite{Kizilersu:2009kg};
            see Ref.\,\cite{privphdwilliams} for a comprehensive overview on the topic.

            A self-consistent solution of the full quark-photon vertex is enabled by its inhomogeneous Bethe-Salpeter equation \cite{Maris:1997hd},
                \begin{equation}\label{QPV:inhomBSE}
                    \Gamma^\mu_\text{(q)}(k,Q)_{\alpha\beta} = Z_2\, i \gamma^\mu + \int_{k'} K(k,k',Q)_{\alpha\gamma,\delta\beta} \left\{S(k_+')\,\Gamma^\mu_\text{(q)}(k',Q)\,S(k_-') \right\}_{\gamma\delta}\,,
                \end{equation}
            which, for consistency with the rainbow-ladder approach, needs to be solved with a gluon-ladder kernel as its input. 
            The resulting vertex \cite{Maris:1999bh} self-consistently generates a vector-meson pole at $Q^2=-m_\rho^2$
            whose contribution significantly increases the charge radii of pseudoscalar and vector mesons \cite{Maris:1999bh,Bhagwat:2006pu}.
            On the domain $-m_\rho^2 < Q^2 < 0.2\,\text{GeV}^2$ it could be described by a Ball-Chiu vertex 
            together with a phenomenological ansatz owing to the $\rho$-meson pole for the transverse part.
            We do not intend to repeat that calculation here and hence adopt a similar ansatz:
            \begin{equation}\label{vertex:simulate}
                \Gamma^{\mu}_T(k,Q) =  - \frac{1}{g_\rho}\, \frac{x}{x+1} \, e^{-g(x) }\, \Gamma^\mu_\text{vc}(k,Q) ,
            \end{equation}
            where $x=Q^2/m_\rho^2$ and $g_\rho=\sqrt{2}\, m_\rho/f_\rho$. 
            Eq.\,\eqref{vertex:simulate} necessitates knowledge of an (unphysical) off-shell $\rho$-meson amplitude $\Gamma^\nu_\text{vc}(k,Q)$.
            On the mass-shell ($Q^2=-m_\rho^2$) it is determined self-consistently from its homogeneous BSE;
            for general $Q^2$ we use the off-shell prescription stated in App.\,\ref{app:mesondiquark-dqprop}.

            To compensate for any arbitrariness from the off-shell part, we introduced the function $e^{-g(x)}$ where the choice $g(x) = (\rho_1+\rho_2 \,x^2)(1+x)$
            optimizes agreement with the results for the quark-photon vertex and the pion charge form factor obtained in Ref.\,\cite{Maris:1999bh} within setup (C1).
            The corresponding value for $r_\pi$ is reproduced if a current-mass dependent parameter $g(0)=\rho_1 = 0.001 + m_\pi^2/(3.72\,\text{GeV}^2)$ is chosen\footnote{
            Note that the different values for $\rho_1$ in Refs.\,\cite{Eichmann:2008ae} and \cite{Eichmann:2008ef} were obtained through
            a slightly different off-shell ansatz for the $\rho$-meson amplitude.}  \cite{Eichmann:2008ef}.
            The remaining parameter
            $\rho_2$ is relevant for the medium--$Q^2$ evolution of the $\rho$-meson part in the vertex and impacts upon the nucleon form factors of Section \ref{sec:results}.
            The value $\rho_2=0.001$, together with the transverse part in the seagull amplitudes (App.\,\ref{app:seagulls}) maximizes agreement with the
            polarization-transfer data for the proton's form factor ratio $G_E(Q^2)/G_M(Q^2)$.

            We repeat that the ansatz \eqref{vertex:simulate} is inspired by low-$Q^2$ phenomenology
            which, at this point, is the only accessible domain in our form factor calculation due to singularity restrictions (cf. App.\,\ref{app:singularities}).
            An exponential suppression of the transverse part clearly disagrees with the large-$Q^2$ analysis,
            and future form factor investigations in that region inevitably
            require an omission of the above ansatz in favor of a self-consistent solution from the inhomogeneous BSE \eqref{QPV:inhomBSE}.


\newpage

\section{Meson and diquark amplitudes}\label{app:mesondiquark}

    \bigskip
    \fatcol{Meson amplitudes.}
            The meson bound-state amplitude has been introduced in connection with its homogeneous BSE \eqref{bse:bse}.
            The Dirac structure of the amplitudes is determined by the Clifford algebra of the gamma matrices.
	        A general fermion-scalar vertex depending on the momenta $q$ and $P$ allows for 4 basis elements
            \begin{equation}\label{bse:scbasisgeneral}
                \tau_i(q,P) \in \left\{ \mathds{1}, \Slash{P}, \, \Slash{q}, \, [\Slash{q},\Slash{P}] \right\},
            \end{equation}
            and a fermion-vector vertex includes 12 structures:
            \begin{equation}\label{bse:vcbasisgeneral}
                \tau_i^\mu(q,P) \in \left\{ \gamma^\mu, q^\mu, P^\mu \right\} \times  \left\{\mathds{1}, \Slash{P}, \, \Slash{q}, \, [\Slash{q},\Slash{P}] \right\}.
            \end{equation}
            The negative parity requirement for pseudoscalar and vector meson amplitudes,
            \begin{equation}
            \begin{split}
                \Gamma(q,P) &= -\gamma^4 \, \Gamma(\Lambda q,\Lambda P)\,\gamma^4 , \\
                \Gamma^\mu(q,P) &=  \gamma^4\, \Lambda^{\mu\nu} \Gamma^\nu (\Lambda q,\Lambda P)\,\gamma^4
            \end{split}
            \end{equation}
            with the parity transformation $\Lambda = \text{diag}(-1,-1,-1,1)$,
            requires the inclusion of a $\gamma^5$ matrix.
            The resulting amplitudes, written with full Dirac, color and flavor dependence, are given by:
            \begin{fshaded}
            \begin{equation}\label{mesamp}
            \begin{split}
                \Gamma(q,P)     &=  \sum_{k=1}^4    f_k^\text{ps}(q^2,z,P^2) \Big\{ i\gamma^5 \,\tau_k(q,P) \Big\}_{\alpha\beta}   \,\otimes\, \frac{\delta_{AB}}{\sqrt{3}}  \,\otimes\,  \mathsf{r^e_{ab}}, \\
                \Gamma^\mu(q,P) &=  \sum_{k=1}^{12} f_k^\text{vc}(q^2,z,P^2) \Big\{ i \tau^\mu_k(q,P) \Big\}_{\alpha\beta}     \,\otimes\, \frac{\delta_{AB}}{\sqrt{3}}  \,\otimes\,  \mathsf{r^e_{ab}}.
            \end{split}
            \end{equation}
            \end{fshaded}

            \noindent
            The dressing functions $f_k(q^2,z,P^2)$ depend on the Lorentz scalars
            $q^2$, $P^2$ and the angular variable $z=\hat{q}\!\cdot\!\hat{P}$.
            Upon solving the meson Bethe-Salpeter equation, they are obtained on the domains $q^2 \in \mathds{R}_+$
            (by implementing the same methods as discussed in App.\,\ref{sec:quarkpropagator} also for $q^2 \in \mathds{C}$), $P^2=-M^2$ (i.\,e., on the mass shell), and $z \in (-1,1)$.
            Greek indices refer to the Dirac structure, and
            the color structure of the meson amplitudes is diagonal ($A,B=1,2,3$).

 \renewcommand{\arraystretch}{1.4}

            The isospin-triplet flavor matrices $\mathsf{r^0}$, $\mathsf{r^\pm}$ corresponding to the states $\pi^0$, $\pi^\pm$ and $\rho^0$, $\rho^\pm$
            are given by
            \begin{equation}\label{meson:flavormatrices}
                \begin{array}{rl}
                \mathsf{r^+} &= \mathsf{ud^\dag} = \frac{1}{2}\left( \sigma_1 + i \sigma_2\right), \\
                \mathsf{r^-} &= \mathsf{du^\dag} = \frac{1}{2}\left( \sigma_1 - i \sigma_2\right),
                \end{array} \qquad
                \begin{array}{rl}
                \mathsf{r^0} &= \frac{1}{\sqrt{2}}\left( \mathsf{uu^\dag}-\mathsf{dd^\dag}\right) = \frac{1}{\sqrt{2}}\,\sigma_3\,,
                \end{array}
            \end{equation}
            where the $\sigma_i$ are the Pauli matrices and $\mathsf{u}=(1,0)$, $\mathsf{d}=(0,1)$.
            They are normalized to unity: $\text{Tr}\{ \mathsf{{r^e}}^\dagger \mathsf{r^{e'}} \}=\delta_{ee'}$.

            \begin{figure}[tbp]
            \begin{center}
            \includegraphics[scale=0.13]{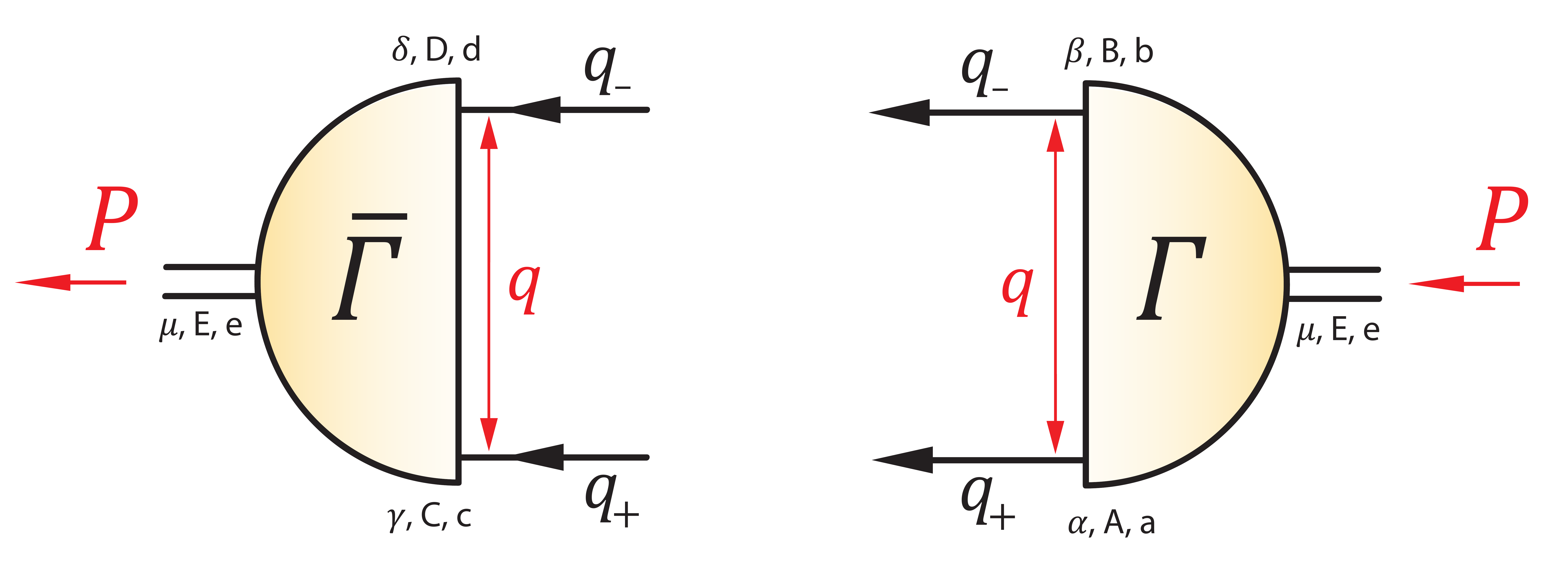}
            \caption[Diquark amplitudes]{\backdef Notational conventions for diquark amplitudes. The quark momenta are $q_\pm = \pm q +P/2$
            and the indices are Dirac/Lorentz, color and flavor indices. } \label{fig:dqamp}
            \end{center}
            \end{figure}

 \newpage

    \bigskip
    \fatcol{Diquark amplitudes.}
            Apart from opposite parity,
            scalar and axial-vector diquark amplitudes exhibit the same Dirac structure as their pseudoscalar and vector meson counterparts.
            The incoming antiquark momentum $-q_-$ is replaced by an outgoing quark momentum $q_-$ which is reflected by the
            charge conjugation matrix $C=\gamma^4 \gamma^2$.
            We denote diquark amplitudes by the same generic symbol $\Gamma$:
            \begin{fshaded}
            \begin{equation}\label{dq:amps}
            \begin{split}
                \Gamma(q,P)     &=  \sum_{k=1}^4    f_k^\text{sc}(q^2,z,P^2) \Big\{ i\gamma^5\,\tau_k(q,P)\,C \Big\}_{\alpha\beta} \,\otimes\,  \frac{\varepsilon_{ABE}}{\sqrt{2}}  \,\otimes\,  \mathsf{s^0_{ab}}\;, \\
                \Gamma^\mu(q,P) &=  \sum_{k=1}^{12} f_k^\text{av}(q^2,z,P^2) \Big\{ i\tau^\mu_k(q,P)\,C \Big\}_{\alpha\beta}   \,\otimes\,  \frac{\varepsilon_{ABE}}{\sqrt{2}}  \,\otimes\,  \mathsf{s^{1,2,3}_{ab}}\;.
            \end{split}
            \end{equation}
            \end{fshaded}

            \noindent
            The isospin singlet and triplet matrices $\mathsf{s^0_{ab}}$ and $\mathsf{s^{1,2,3}_{ab}}$ for the diquarks read
            \begin{equation}\label{dq:flavormatrices}
                \begin{array}{rl}
                \mathsf{s^0} &= \frac{1}{\sqrt{2}}\left(\mathsf{ud^\dag-du^\dag}\right) = \frac{1}{\sqrt{2}}\,i\sigma_2\,,
                \end{array} \qquad
                \begin{array}{rl}
                \mathsf{s^1} &= \mathsf{uu^\dag} = \frac{1}{2}\left(\mathds{1}+\sigma_3\right)\,, \\
                \mathsf{s^2} &= \frac{1}{\sqrt{2}}\left(\mathsf{ud^\dag+du^\dag}\right) = \frac{1}{\sqrt{2}}\,\sigma_1\,, \\
                \mathsf{s^3} &= \mathsf{dd^\dag} = \frac{1}{2}\left(\mathds{1}-\sigma_3\right)
                \end{array}
            \end{equation}
            with the Pauli matrices $\sigma_i$ and $\mathsf{u}=(1,0)$, $\mathsf{d}=(0,1)$.
            They are again normalized via $\text{Tr}\{ \mathsf{{s^e}}^\dagger \mathsf{s^{e'}} \}=\delta_{ee'}$.
            Due to the Pauli principle, diquark amplitudes must be antisymmetric under quark exchange $q_+ \leftrightarrow q_-$,
            \begin{equation}\label{dq:antisymmetry}
                \Gamma(q,P) = -\Gamma^T( \left.-q \right|_{\sigma\leftrightarrow(1-\sigma)} , P)\;,
            \end{equation}
            where the transposition involves all Dirac, color and flavor indices.
            Because of the antisymmetry of the color anti-triplet diquark the combination
            of flavor and spin structure must be symmetric. As a consequence, spin and isospin states coincide for the two-flavor case:
            scalar diquarks correspond to an antisymmetric isospin singlet and axial-vector diquarks to a symmetric isospin
            triplet.

            While the flavor matrices do not give a contribution to the BSE integral \eqref{dq:bse},
            the color factor $\sim\varepsilon_{ABE}$ representing the diquark anti-triplet configuration leads to a prefactor $1/2$
            compared to the meson BSE: diquarks are less bound than mesons.
            The Dirac amplitudes $\sim \mathds{1},\,\gamma^\mu$ in \eqref{bse:scbasisgeneral} and \eqref{bse:vcbasisgeneral}
            are the dominant ones in a solution of the rainbow-ladder BSE
            for the lowest-mass mesons  and diquarks
            and reproduce masses of the full solution within an error of $\lesssim 20 \%$ \cite{Maris:1997tm,Maris:1999nt,Maris:2002yu}.

    \bigskip
    \fatcol{Charge conjugation, C--parity and Pauli principle.}
            The charge-conjugated Dirac amplitudes are defined by
            \begin{equation}\label{bse:conjugation}
            \begin{split}
                \conjg{\Gamma}(q,P) &:= C \,\Gamma^T(-q,-P) \,C^T, \\
                \conjg{\Gamma}^\mu(q,P) &:=  -C \,{\Gamma^\mu}^T(-q,-P) \,C^T\,,
            \end{split}
            \end{equation}
	        where the superscript $T$ denotes matrix transposition\footnote{
            Note that this definition differs in two respects from the somewhat standard convention in the literature. 
            First, the left-hand sides of Eq.\,\eqref{bse:conjugation} are usually denoted by $\conjg{\Gamma}(q,-P)$, whereas
            the above definition is more convenient when using the abbreviation $\conjg{\Gamma}$.
            Second, the conjugated vector-meson amplitude is usually defined without the minus sign. We chose Eq.\,\eqref{bse:conjugation}
            to enable a common definition of charge conjugation %
            for meson and diquark amplitudes, given that the
            dominant amplitudes in both cases carry a factor $i$, i.e.: $i\gamma^5$, $i\gamma^\mu$, $i\gamma^5 C$, $i\gamma^\mu C$.}.
            Pseudoscalar and vector mesons with equal quark masses are $C$-parity eigenstates (e.g.: $J^{PC} = 0^{-+},\,1^{--}$ for $\pi$ and $\rho$) which entails
                \begin{equation}\label{BSE:C-Parity}
                    \conjg{\Gamma}(q,P) = \Gamma(q,-P)\,,\qquad
                    \conjg{\Gamma}^\mu(q,P) = \Gamma^\mu(q,-P)
                \end{equation}
            for the Dirac part of the ground-state amplitudes, assuming equal momentum partitioning. On the other hand, diquarks are subject to the Pauli principle which leads to
                \begin{equation}\label{BSE:Pauli-Principle}
                    \Gamma(q,P) = -\Gamma^T(-q,P)\,,\qquad
                    \Gamma^\mu(q,P) = {\Gamma^\mu}^T(-q,P)\,.
                \end{equation}
            For an appropriate Dirac basis 
            whose elements satisfy Eqs.\,(\ref{BSE:C-Parity}--\ref{BSE:Pauli-Principle}),
            both constraints require the dressing functions to be even in the angular variable $z=\hat{q}\cdot\hat{P}$.
            If a Chebyshev expansion is employed (for details, see App.\,\ref{appendixchebyshev}), only even Chebyshev moments contribute which is ensured
            by the fact that the BSE kernel 
            decouples even and odd Chebyshev moments.
            Conversely, pseudoscalar and vector-meson states with opposite $C-$parity (associated with the 'exotic' quantum numbers $0^{--}$ and $1^{-+}$)
            carry an odd dependence on $z$.

    \bigskip
    \fatcol{Mesons and diquarks on the mass shell.}
            For the actual solution of the meson and diquark BSEs,
            Eqs.\,\eqref{bse:bse} and \eqref{dq:bse}, it is advantageous to construct orthogonalized versions
            of the general Dirac basis elements (\ref{bse:scbasisgeneral},\,\ref{bse:vcbasisgeneral})
            at the respective mass poles $P^2 = -M^2$.
            Suitable choices for the (pseudo--)scalar and (axial--)vector cases are given in Table \ref{bse:basis}.
            Their orthogonality relations read
            \begin{equation}
                \frac{1}{4}\,\text{Tr}\{\tau_i\, \tau_j\} = \delta_{ij}\, a_i(z)\;,\quad
                \frac{1}{4}\,T^{\mu\nu}_P\,\text{Tr}\{\tau_i^{\mu}\, \tau_j^{\nu}\} = \delta_{ij} \,b_i(z)\;,
            \end{equation}
            where
            \begin{equation}
            \begin{split}
                a_1 = a_2 &= \frac{b_1}{3} = -\frac{b_2}{3} = 1\,, \qquad  b_7 = -b_8 = \frac{2}{3}(1-z^2)^2\,,\\
                \frac{a_3}{z^2}  &= \frac{a_4}{4} = -b_3 = \frac{b_4}{z^2} = -\frac{b_5}{2 z^2} = \frac{b_6}{2} = 1-z^2\,. \\
            \end{split}
            \end{equation}
            Contraction of the BSEs with these basis elements and exploiting the orthogonality relations leads to coupled homogeneous
            integral equations for the coefficients $f_k(q^2,z,-M^2)$ which can be solved in any arbitrary frame.
            Exploiting the $O(4)$ symmetry of the problem, a decomposition into Chebyshev polynomials of the second kind is usually employed for numerical convenience (see App.\,\ref{appendixchebyshev}).
	        Typically only a few Chebyshev moments $f_k^n(q^2)$ have to be taken into account
            to match the full solution \cite{Maris:1997tm,Maris:1999nt}.
            This observation will be used in the context of the diquark amplitudes' off-shell continuation
            where the angular dependence is neglected (App.\,\ref{app:mesondiquark-dqprop}).
            Upon introducing an artificial parameter $\lambda(P^2)$ in the homogeneous BSE \eqref{dq:bse},
            the equation becomes an eigenvalue problem where a bound-state solution is obtained for $\lambda(-M^2)=1$.

            \begin{table}
                \begin{center}

                 \renewcommand{\arraystretch}{1.5}
                 \begin{equation*}
                 \begin{array}{|c|c|}\hline
                    J=0   &   J=1 \\ \hline
                                      \begin{array}{ @{\quad}l @{\quad}}
                                          \tau_1 = \mathds{1} \\
                                          \tau_2 = \Slash{\hat{P}} \\
                                          \tau_3 = z\, \Slash{\hat{q}}_T \\
                                          \tau_4 = i\,[ \Slash{\hat{q}}, \Slash{\hat{P}} ]  \\[0.2cm]
                                      \end{array}   &
                                      \begin{array}{ @{\quad}l @{\quad\quad\quad}l @{\quad}}
                                          \tau_1^{\mu} = \gamma^{\mu}                             &   \tau_5^{\mu} = z\,( \gamma^{\mu} \Slash{\hat{q}}_T - \hat{q}^{\mu} )                                           \\
                                          \tau_2^{\mu} = \gamma^{\mu}\Slash{\hat{P}}              &   \tau_6^{\mu} = i\,\hat{q}^{\mu}\Slash{\hat{P}} - \frac{i }{2}\,\gamma^{\mu} [ \Slash{\hat{q}}, \Slash{\hat{P}} ]           \\
                                          \tau_3^{\mu} = i\,\hat{q}^{\mu}                         &   \tau_7^{\mu} = \hat{q}^{\mu}\Slash{\hat{q}}_T - \frac{1}{3}\,\hat{q}^2_T\gamma^{\mu}                                \\
                                          \tau_4^{\mu} = z\,\hat{q}^{\mu} \Slash{\hat{P}}         &   \tau_8^{\mu} = \frac{1}{3}\,\hat{q}^2_T\gamma^{\mu} \Slash{\hat{P}} - \frac{1}{2}\, \hat{q}^{\mu} [ \Slash{\hat{q}}, \Slash{\hat{P}} ]  \\[0.2cm]
                                      \end{array}  \\ \hline
                 \end{array}
                 \end{equation*}
                \caption{\backdef
                         Orthogonal basis elements for pseudoscalar mesons/scalar diquarks ($J=0$)
                         and vector mesons/axialvector diquarks ($J=1$),
                         designed such that all corresponding dressing functions $f_k(q^2,z,-M^2)$ are real and even in $z$ for equal momentum partitioning.
                         $\hat{q}^{\mu}_T=T^{\mu\nu}_P \hat{q}^{\nu}$, where $T^{\mu\nu}_P$ is the transverse projector with respect to $P$.
                         Due to the transversality of on-shell vector mesons (and, by analogy, axialvector diquarks)
                         an additional projector $T^{\mu\nu}_P$ must be attached to the $J=1$ basis elements; hence
                         the four longitudinal basis elements $\sim P^\mu$ do not contribute on the mass shell. }\label{bse:basis}

                \end{center}
            \end{table}

    \bigskip
    \fatcol{Normalization.}
            A bound-state amplitude is normalized by a canonical normalization condition,
	        Eq.\,\eqref{bs:normalization}, which is obtained from evaluating the derivative of Dyson's equation at the mass pole.
            Since a two-body ladder kernel is independent of the total momentum $P$
            only the derivatives of the propagators contribute.
            Defining the quantity
            \begin{equation}\label{bse:norm}
                Q^{(\mu\nu)}(P) :=  \text{Tr}_D \int_q  \left\{ \conjg{\Gamma}^{(\mu)}(q,K) \, S(q_+) \, \Gamma^{(\nu)}(q,K) \, S(-q_-) \right\}_{K^2=-M^2}\,,
            \end{equation}
            where the notation $Q^{(\mu\nu)}$ includes both cases $J=0$ and $J=1$, and $M$ is the pseudoscalar or vector-meson mass,
            the normalization condition reads:
            \begin{equation}\label{bse:normalization}
                \left. \frac{d}{d P^2} \right|_{P^2=-M^2} Q_{(T)}(P^2) = 1\,.
            \end{equation}
            $Q_T=T^{\mu\nu}_P Q^{\mu\nu}/3$ is the transverse component of \eqref{bse:norm} for $J=1$.
            $\text{Tr}_D$ denotes a Dirac trace; the color-flavor trace is $1$ since we use normalized color and flavor matrices.
            An analogous normalization applies to the diquark amplitudes, except for replacing $S(-q_-) \longrightarrow S^T(q_-)$ in \eqref{bse:norm}
            and attaching a symmetrization factor $\nicefrac{1}{2}$ in front of the integral \cite{Ishii:1995bu}.
            In combination with the normalization \eqref{bse:normalization}, the Bethe-Salpeter equation completely determines
            the meson and diquark amplitudes on the mass shell.

    \bigskip
    \fatcol{Electroweak decay constants and pion charge radius.}
            The pseudoscalar and vector-meson leptonic decay constants are defined by \cite{Maris:1999nt}:
            \begin{equation}\label{bse:pdc}
            \begin{split}
               i f_\pi \,m_\pi &=  \sqrt{N_C} \, Z_2 \,\text{Tr}_D
                \int^\Lambda_q \Big[ \gamma^5 \,\Slash{\hat{P}} \,S(q_+) \,\Gamma(q,P) \, S(-q_-) \Big]_{P^2=-m_\pi^2}\,, \\
               i f_\rho \,m_\rho &= \sqrt{N_C} \, Z_2 \,\frac{T_P^{\mu\nu}}{3} \,\text{Tr}_D
                \int^\Lambda_q \Big[ \gamma^\mu  \,S(q_+) \,\Gamma^\nu(q,P) \, S(-q_-)\Big]_{P^2=-m_\rho^2}\,,
            \end{split}
            \end{equation}
            where the prefactor $\sqrt{N_C}$ emerges from the color normalization in \eqref{mesamp}.

            As a consequence of electromagnetic gauge invariance, the pion's electromagnetic current operator which
            is consistent with the rainbow-ladder kernel is the impulse-approximation current (cf. Section~\ref{sec:em}):
            \begin{equation}\label{current:em1}
                J^\mu(Q^2) = \text{Tr}_D \int_q \conjg{\Gamma}(q_f,P_f) \, S(q_+^+) \, \Gamma^\mu_\text{(q)}(q_+,Q) \, S(q_+^-)\,\Gamma(q_i,P_i) \, S(-q_-) \, ,
            \end{equation}
            where $\Gamma^\mu_\text{(q)}$ is the quark-photon vertex of App.~\ref{sec:quark-photon-vertex},
            and the involved momenta depend on the photon momentum $Q$, average total momentum $P=(P_i+P_f)/2$ and loop relative momentum $q$ via
            \begin{equation*}
                 q_{i,f} = q \mp Q/4\,,    \quad   P_{i,f} = P \mp Q/2\,, \quad
                 q_\pm   = \pm q + P/2\,,  \quad   q_+^\pm = q_+ \pm Q/2\,.
            \end{equation*}
            The incoming and outgoing pion momenta are onshell: $P_i^2 = P_f^2 = -m_\pi^2$ and hence $P\cdot Q=0$.
            The most general Poincar\'e-covariant expression of the current is given by (cf. Eq.~\eqref{DQPV:Scalar-Onshell})
            \begin{equation}\label{current:em}
                J^\mu(Q^2) = 2 P^\mu F_\pi(Q^2)\,,
            \end{equation}
            where $F_\pi(Q^2)$ denotes the charged pion's form factor.
            Charge conservation $F(0)=1$ is ensured by the Ward identity \eqref{QPV:WardIdentity}.
            The quark-photon vertex \eqref{vertex:BC}, together with the phenomenological transverse part \eqref{vertex:simulate}
            which incorporates a vector-meson pole, yields for the pion form factor:
            \begin{equation}
                F_\pi(Q^2) =  F_{\pi,\text{BC}}(Q^2) -   F_{\rho\pi\pi}(Q^2) \, \frac{1}{g_\rho} \frac{x}{x+1} \, e^{-g(x) }\,,
            \end{equation}
            i.e., the sum of a Ball-Chiu piece and the $\rho\pi\pi$ triangle diagram whose experimental value
            is $F_{\rho\pi\pi}(Q^2=-m_\rho^2) = g_{\rho\pi\pi} = 6.14$.
            The latter component vanishes for zero photon momentum transfer owing to the transversality of the respective vertex.
            For $Q^2>0$ it reduces the Ball-Chiu contribution to the form factor
            and typically supplies $\sim 50\%$ to the squared pion charge radius \cite{Maris:1999bh,Eichmann:2008ae}, extracted via
            \begin{equation}\label{MESON:pion-charge-radius}
                r_\pi^2 = -6 F'_\pi(0) = r_{\pi,\text{BC}}^2 + \frac{6}{g_\rho m_\rho^2} \, F_{\rho\pi\pi}(0) \,e^{-g(0)}.
            \end{equation}


\section{Off-shell structure of the T-matrix} \label{app:mesondiquark-dqprop}

            Diquarks correlations in the nucleon are off-shell: it is the very requirement that the integration domain of
            scalar and axial-vector diquark propagators in the quark-diquark BSE \eqref{nuc:bse} cannot exceed their pole locations
            without a systematic inclusion of residue contributions.
            The tool to gather information on the off-shell structure of the quark-quark scattering matrix $T^{(2)}$ is the Dyson equation \eqref{BS:dyson}.
            The diquark ansatz \eqref{dq:tmatrixansatz} was introduced as a workaround to avoid its explicit calculation;
            a subsequent solution of the diquark BSE however only determines $T^{(2)}$ on the diquarks' mass shells.
            As will be detailed below, a naive implementation of the on-shell T-matrix for off-shell
            (and in general complex) total diquark momenta $P^2\neq -M^2$, i.e.,  by using on-shell diquark amplitudes and free spin-0 and spin-1 diquark propagators
            \begin{equation}
                D(P^2) =  \frac{1}{P^2+M_\text{sc}^2}, \quad
                D^{\mu\nu}(P) =  \frac{T_P^{\mu\nu}}{P^2+M_\text{av}^2}  +\frac{ L_P^{\mu\nu}}{M_\text{av}^2},
            \end{equation}
             poses several conceptual problems which require closer attention.

            A manifest strategy to bridge the gap between the separable ansatz for $T^{(2)}$ and its  solution from Dyson's equation is to
            consult Eq.\,\eqref{BS:dyson} once again, namely to determine the diquark propagators $D(P^2)$
            upon constructing a sensible analytic continuation of the on-shell amplitudes $\Gamma(q,P)$.
            Thereby the Dirac, color and flavor structure of the separable ansatz is maintained at off-shell momenta
            while certain features of the full self-consistent solution of the T-matrix are implemented as well.

    \bigskip
    \fatcol{Asymptotic behavior I.}
            The asymptotic limit of the kernel $K^{(2)}$ is a gluon ladder exchange. This can be inferred from the kernel's skeleton expansion:
            the ladder diagram is independent of $P^2$ while all higher-order contributions vanish for $P^2 \longrightarrow \infty$ due to the appearance of additional
            quark propagators which behave as $S \longrightarrow 1/\sqrt{P^2}$.
            The Dyson sum \eqref{BS:dysonsum} implies that the gluon ladder kernel \eqref{bse:rlkernel} also dominates the asymptotic behavior of the T-matrix:
            \begin{equation}\label{T:asymptotic-RL}
                 T_{\alpha\gamma, \beta\delta}(p,q,P) \;\; \stackrel{P^2\rightarrow\infty}{\longlongrightarrow} \;\; \sim \frac{\alpha(k^2)}{k^2}
                                                      \left(\frac{\lambda^i}{2}\right)_{\!AC} \! \left(\frac{\lambda^i}{2}\right)_{\!BD}
                                                      (i\gamma^\mu)_{\alpha\gamma} \, T^{\mu\nu}_k \, (i\gamma^\nu)_{\beta\delta}\,.
            \end{equation}
            This is apparently not reproducible through the ansatz \eqref{dq:tmatrixansatz}.
            The implementation of Dyson's equation in the previously described manner
            at best guarantees the correct power-law behavior $\Gamma\,D\,\conjg{\Gamma} \rightarrow const.$ of the T-matrix in the variable $P^2$:
            if the diquark amplitudes are chosen to become constant for large $P^2$ (e.g., via their dominant Dirac structures $\gamma^5 C$, $\gamma^\mu C$),
            the resulting diquark propagators will become constant as well.
            Phrased differently, they pick up finite parts which would appear in a T-matrix \textit{beyond} the diquark ansatz.
            These contributions are suppressed on the mass shells but dominate the ultraviolet region.

    \bigskip
    \fatcol{Offshell ansatz for diquark amplitudes.}
            We start from Eq. \eqref{dq:amps} and for convenience discuss the offshell dependence
            of the diquark amplitudes in terms of the basis elements $\tau_k(q,P)$ and $\tau^\mu_k(q,P)$ alone
            while the dressing functions are left unchanged at their mass-shell values:                             
            \begin{equation}\label{dq:amp-offshell}
            \begin{split}
                \Gamma_\text{sc}(q,P)     &\,=\, \sum_{k=1}^4 f_k^\text{sc}(q^2,z=0,-M_\text{sc}^2) \, i\gamma^5 \tau_k(q,P)\,C,  \\
                \Gamma^\mu_\text{av}(q,P) &\,=\, \sum_{k=1}^8 f_k^\text{av}(q^2,z=0,-M_\text{av}^2) \, i\,\tau^\mu_k (q,P)\,C.
            \end{split}
            \end{equation}
            A solution of the quark-diquark BSE in the nucleon's rest frame requires boosted diquark amplitudes.
            A boost shifts the angular variable $z$ in the diquark amplitudes' dressing functions
            into the complex plane and even outside the convergence radius $|z|<1$ of the Chebyshev expansion which therefore no longer poses a sensible procedure.
            Setting $z=0$ (only) in the dressing functions is a reasonable approximation on the diquark's mass shell and similar to keeping only the zeroth Chebyshev moments.
            .

            The transversality condition for the on-shell axial-vector amplitude need not be included explicitly
            since it is already ensured by the transverse pole in the axial-vector diquark propagator: each diquark amplitude in the
            subsequent calculations appears in
            conjunction with the respective propagator. Likewise, the purely longitudinal off-shell components
            related to $\tau^\mu_{9\dots 12}(q,P)$ are generated by the longitudinal projection of the diquark propagator
            which is suppressed by a factor of $P^2+M_\text{av}^2$ on the mass shell.

            In the following we will make use of the dimensionless diquark momentum variables
            \begin{equation}
                x_\text{sc} := P^2/M_\text{sc}^2, \quad  x_\text{av} := P^2/M_\text{av}^2,
            \end{equation}
            and abbreviate both by the symbol $x$, where the context selects either of the two possibilities.
            Specifically, it entails $x=-1$ on both scalar and axial-vector diquark mass shells.

         \bigskip

            The on-shell basis of Table \ref{bse:basis}, expressed in terms of a normalized diquark momentum $\hat{P}$,
            does not provide a unique analytic continuation to off-shell momenta: 
            \begin{itemize}
            \item Using that basis, all amplitudes would be equally important at $P^2 \rightarrow\infty$ and determined by their strengths on the mass shell.
                  This is not a fundamental problem if one keeps in mind that, in our context,
                  an off-shell diquark amplitude is no meaningful object in itself but merely an auxiliary device for constructing the off-shell T-matrix,
                  especially since the requirement \eqref{T:asymptotic-RL} cannot be met anyway.
                  Nevertheless we choose the dominant diquark amplitudes $\Gamma_\text{sc}(q,P) \sim i\gamma^5$, $\Gamma^\mu_\text{av}(q,P) \sim  i\gamma^\mu$
                  to prevail in the UV such that the diquark becomes 'pointlike' at large diquark momenta.
                  This feature would be explicit in an inhomogeneous BSE solution for a scalar and axial-vector vertex.
                  It is accomplished by suppressing all subleading amplitudes with a factor $g(x)$ which leaves the on-shell value unchanged: $g(-1) = 1$.
                  We use $g(x) = (x+2)^{-\kappa/2}$ and, for the time being, keep $\kappa>0$ as a variable (later we will set $\kappa=2$).
                  For large $\kappa$ the subleading amplitudes are suppressed at spacelike momenta
                  and provide support only in the neighborhood of the mass shell which resembles
                  the case where only the dominant diquark amplitudes are taken into account.
            \item The normalized basis of Table \ref{bse:basis} includes kinematic singularities at $P^2 = 0$ which,
                  if not dealt with, lead to imaginary parts in the spacelike diquark propagator calculated from Eqs.\,\eqref{DQ:prop-equations} below.
                  Contributions $\sim \hat{P}$ and $\sim z$ should behave as $\sqrt{P^2}$ in the vicinity of the origin.
                  For this reason we attach a factor $h(x) \sim \sqrt{x}$ to any occurrence of $\hat{P}^\mu$ or $z=\hat{q}\cdot\hat{P}$.
                  The choice $h(x) = -i\sqrt{x}$ amounts to the replacement of the normalized by an unnormalized basis, i.e. $\hat{P}^\mu \rightarrow P^\mu/(iM)$.
                  It would be sufficient to guarantee the correct on-shell behavior but renders the UV behavior completely arbitrary:
                  some amplitudes would become constant, others would rise with powers of $\sqrt{P^2}$.
                  We therefore choose $h(x) = -i\sqrt{x/(x+2)}$ which entails
                  \begin{equation}
                      \hat{P}^\mu \rightarrow\frac{P^\mu}{iM} \, \sqrt{\frac{M^2}{P^2+2M^2}}
                  \end{equation}
                  and hence $h(-1) = 1$, $h(0)=0$, $h(\infty) = -i$.
            \end{itemize}

            \bigskip
            \noindent
            The final off-shell ansatz for the scalar and axial-vector diquark bases is given in Table \ref{bse:offshellbasis}.
            This is the construction which was used in Refs.\,\cite{Eichmann:2008ef,Nicmorus:2008vb}; it slightly differs from that employed in Ref.\,\cite{Eichmann:2007nn}.
            In combination with the diquark propagator of Eq.\,\eqref{DQ:prop-equations} it provides a prescription for the quark-quark T-matrix
            which is unique on the mass poles and in the ultraviolet whereas its intermediate momentum behavior depends
            on the parameter $\kappa$ inherent in the definition of $g(x)$.
            We note that a further $P^2$-dependent function attached to the full scalar or axial-vector diquark amplitude
            does not change the product $\Gamma\,D\,\bar{\Gamma}$ since it would also appear in the diquark propagator and
            leave the T-matrix itself (and as a consequence, baryonic observables) invariant.

            \begin{table}
                \begin{center}

                \begin{equation*}
                \begin{array}{|c|c|}\hline
                   J=0   &   J=1 \\ \hline
                                     \begin{array}{ @{\quad}l @{\quad}}
                                         \tau_1 = \mathds{1} \\
                                         \tau_2 = g h\,\Slash{\hat{P}} \\
                                         \tau_3 = g h\,z\, \Slash{\hat{q}}_T \\
                                         \tau_4 = g h\,i\,[ \Slash{\hat{q}}, \Slash{\hat{P}} ]  \\[0.2cm]
                                     \end{array}   &
                                     \begin{array}{ @{\quad}l @{\quad\quad}l @{\;}}
                                         \tau_1^{\mu} = \gamma^{\mu}                               &   \tau_5^{\mu} = g h\,z \left\{ \gamma^{\mu} \,\Slash{\hat{q}}_T - \hat{q}^{\mu} \right\}       \\
                                         \tau_2^{\mu} = g h\,\gamma^{\mu}\Slash{\hat{P}}          &   \tau_6^{\mu} = g h\,i \left\{ \hat{q}^{\mu}\Slash{\hat{P}} - \frac{1}{2}\,\gamma^{\mu} [ \Slash{\hat{q}}, \Slash{\hat{P}} ] \right\}          \\
                                         \tau_3^{\mu} = g\,i \hat{q}^{\mu}                        &   \tau_7^{\mu} = g \left\{ \hat{q}^{\mu}\,\Slash{\hat{q}}_T - \frac{1}{3}\,\hat{q}^2_T \,\gamma^{\mu}  \right\}    \\
                                         \tau_4^{\mu} = g h^2\,z\,\hat{q}^{\mu} \Slash{\hat{P}}   &   \tau_8^{\mu} = g h\left\{ \frac{1}{3}\,\hat{q}^2_T\,\gamma^{\mu} \Slash{\hat{P}} - \frac{1}{2}\, \hat{q}^{\mu} [ \Slash{\hat{q}}, \Slash{\hat{P}} ] \right\}\\[0.2cm]
                                     \end{array}  \\ \hline
                \end{array}
                \end{equation*}
                \caption{\backdef Scalar and axial-vector diquark off-shell bases, to be used in conjunction with Eq.\,\eqref{dq:amp-offshell}.
                         The replacements $\Slash{\hat{q}}_T \rightarrow (\Slash{\hat{q}}-h^2 z\, \Slash{\hat{P}})$ and $\hat{q}^2_T \rightarrow (1-h^2 z^2)$ are implicit. }\label{bse:offshellbasis}

                \end{center}
            \end{table}

    \bigskip
    \fatcol{Diquark propagators.}
            With an off-shell ansatz for the diquark amplitudes at hand, the Dyson equation \eqref{BS:dyson} can  be exploited to obtain a
            consistent expression for the diquark propagator. Insertion of the diquark pole ansatz for the T-matrix,
            \begin{equation}
                T = \Gamma\,D\,\conjg{\Gamma} + \Gamma^\mu\,D^{\mu\nu} \conjg{\Gamma}^\nu,
            \end{equation}
            into Dyson's equation yields:
            \begin{equation}
                \Gamma\,D\,\conjg{\Gamma} + \Gamma^\mu D^{\mu\nu} \conjg{\Gamma}^\nu = K + K\,G_0\,\Gamma\,D\,\conjg{\Gamma} + K\,G_0\, \Gamma^\mu D^{\mu\nu} \conjg{\Gamma}^\nu\,.
            \end{equation}
            Successive application of quark propagator pairs $G_0$ and scalar or axial-vector diquark amplitudes from the left and the right
            and closing the loops by integration and tracing  yields (transitions between scalar and axial-vector amplitudes vanish because of their flavor traces):
            \begin{equation}
            \begin{split}
                \big(\conjg{\Gamma}\,G_0\,\Gamma\big) D \big(\conjg{\Gamma}\,G_0\,\Gamma\big) = \;&
                                                        \big(\conjg{\Gamma}\,G_0 K G_0\,\Gamma\big) + \\
                                                        &+\big(\conjg{\Gamma}\,G_0 K G_0\,\Gamma\big) D \big(\conjg{\Gamma}\,G_0\,\Gamma\big), \\[0.1cm]
                \big(\conjg{\Gamma}^\alpha G_0\,\Gamma^\mu\big) D^{\mu\nu} \big(\conjg{\Gamma}^\nu G_0\,\Gamma^\beta\big) = \;&
                                                        \big(\conjg{\Gamma}^\alpha G_0 K G_0\,\Gamma^\beta\big) + \\
                                                         &+\big(\conjg{\Gamma}^\alpha G_0 K G_0\,\Gamma^\mu\big) D^{\mu\nu} \big(\conjg{\Gamma}^\nu G_0\,\Gamma^\beta\big).
            \end{split}
            \end{equation}
            We simplify the notation by introducing the shorthand notation
            \begin{align}\label{DQ:nk}
                n^{(\mu\nu)} &= \frac{1}{M^2} \,\conjg{\Gamma}^{(\mu)}G_0 \Gamma^{(\nu)} = -\frac{1}{2M^2} \,\text{Tr}_D \hspace{-0.1cm}\int\conjg{\Gamma}^{(\mu)}\,S S\,\Gamma^{(\nu)}, \\
                k^{(\mu\nu)} &= \frac{1}{M^2} \,\conjg{\Gamma}^{(\mu)}G_0\,K\,G_0 \Gamma^{(\nu)} = -\frac{1}{2M^2} \,\text{Tr}_D \hspace{-0.1cm}\int\hspace{-0.2cm}\int \conjg{\Gamma}^{(\mu)}\,S S\,K\,S S\,\Gamma^{(\nu)},
            \end{align}
            where we chose the letters $n$ for \textit{normalization} and $k$ for \textit{kernel}:
            $n^{(\mu\nu)}$ is, up to a minus sign, the dimensionless version of the normalization integral $Q(P^2)$ defined in Eq.\,\eqref{bse:norm},
            but now without the $P^2$ dependence of the diquark amplitudes held fixed.
            Upon decomposing each axial-vector quantity $D^{\mu\nu}$, $n^{\mu\nu}$, $k^{\mu\nu}$ into a transverse and longitudinal part
            (e.g. $D^{\mu\nu} = D_T\, T_P^{\mu\nu} + D_L\, L_P^{\mu\nu}$), the above equations become
            \begin{fshaded}
            \begin{equation}\label{DQ:prop-equations}
                D^{-1}   = M_\text{sc}^2 \left( \frac{n^2}{k}-n \right),\;
                D^{-1}_T = M_\text{av}^2 \left( \frac{n_T^2}{k_T}-n_T \right),\;
                D^{-1}_L = M_\text{av}^2 \left( \frac{n_L^2}{k_L}-n_L \right).
            \end{equation}
            \end{fshaded}

            \noindent
            These are the defining relations for the scalar and axial-vector diquark propagators in the quark-diquark model.
            $n$ and $k$ are scalar functions that depend on $x=P^2/M^2$, where $M=M_\text{sc}$ or $M_\text{av}$.

            \begin{figure}[tbp]
            \begin{center}
            \includegraphics[scale=0.16]{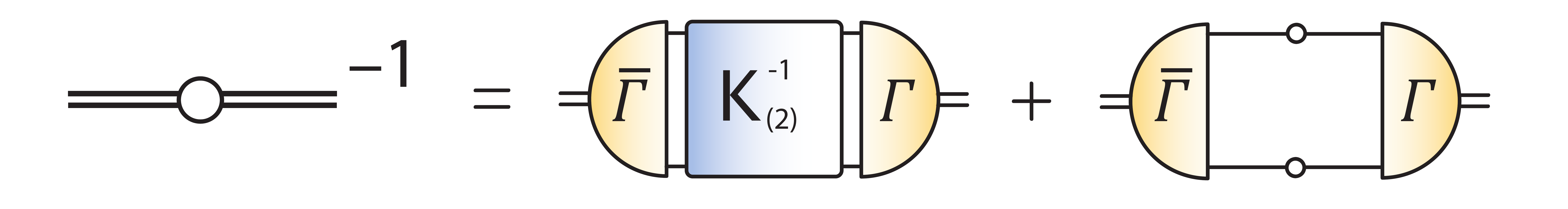}
            \caption[Diquark propagator]{\backdef Defining equation \eqref{DQ:prop-equations} for the diquark propagators.
                                                  The first graph corresponds to $n^2/k$, the second term to $n$.} \label{fig:dqprop}
            \end{center}
            \end{figure}

    \bigskip
    \fatcol{An ansatz for the two-loop integral.}
            $n(x)$ is a one-loop and $k(x)$ a two-loop integral, to be evaluated for $x \in \mathds{C}$ since
            the quark-diquark BSE samples the diquark propagators within a parabolic domain in the complex $P^2$ plane.
            While $n(x)$ can be easily evaluated, we try to circumvent the Monte-Carlo calculation of the quantity $n^2/k$ for general momenta $P^2$ and model it instead by an ansatz,
            a procedure which will also prove useful in App.\,\ref{app:dqphotonvertex} when constructing a diquark-photon vertex.
            What conditions can be imposed upon these functions?
            \begin{itemize}
            \item By construction, the diquark propagators should exhibit timelike poles at $P^2=-M^2$ which enable the derivation of the diquark BSEs and
                  thereby define the diquark masses. Transversality of the axial-vector diquark amplitude on the mass shell is ensured by a transverse pole in the axial-vector propagator.
                  The pole conditions $D^{-1}(P^2=-M_\text{sc}^2) \longrightarrow P^2 + M_\text{sc}^2$ and $D_T^{-1}(P^2=-M_\text{av}^2) \longrightarrow P^2 + M_\text{av}^2$ amount to
                  the relations
                  \begin{equation}
                      n_{(T)}(-1) = k_{(T)}(-1)\,, \quad n'_{(T)}(-1) = 1 + k'_{(T)}(-1).
                  \end{equation}
            \item To avoid kinematic singularities, the transverse and longitudinal parts of the axial-vector propagator should be equal at
                  $P^2=0$: $D_T(0) = D_L(0)$, and for each single term: $n_T(0)=n_L(0)$, $k_T(0)=k_L(0)$.
            \end{itemize}
            Defining $\lambda := n(-1)$ and $\beta:=1+n'(-1)$, one can write
            \begin{equation}
                \frac{n^2}{k}     = \lambda_\text{sc} + \beta_\text{sc} \,F(x), \quad
                \frac{n_T^2}{k_T} = \lambda_\text{av} + \beta_\text{av} \,F_T(x), \quad
                \frac{n_L^2}{k_L} = \lambda_\text{av} + \beta_\text{av} \,F_L(x),
            \end{equation}
            where the 2-loop content has been shifted into the unknown functions $F(x)$.
            The above conditions require $F_{(T)}(-1) = 0$, $F'_{(T)}(-1) = 1$, and $F_T(0)=F_L(0)$.
            A two-loop evaluation of the integral $k^{(\mu\nu)}$ at specific $P^2 \in \mathds{R}$ verifies that these relations,
            together with $F_T(\infty)=F_L(\infty)$, are indeed satisfied.
            The values of $\lambda$ and $\beta$ depend on the number of amplitudes that are taken into account.
            For instance, if only the dominant amplitude is retained: $\beta=0$,
            since by virtue of our off-shell ansatz the diquark amplitudes are then independent of $P^2$
            and $n_{(T)}$ becomes the normalization integral, with $n'_{(T)}=-1$.

            \bigskip

            An ansatz for $F(x)$ can be constructed in the following way.
            Formally, if we had started from the inverse Dyson relation $T^{-1}=K^{-1}-G_0$, we would have arrived at the expression
                  $n^2/k \sim \bar{\Gamma}\,K^{-1}\,\Gamma$. Since the rainbow-ladder kernel is independent of the total momentum $P$ and the
                  dominant diquark amplitudes are as well, the only $P^2$ dependence of that term is introduced by the subleading diquark amplitudes through the functions $g(x)$ and $h(x)$.
            On the mass shell, $\conjg{\Gamma}\,K^{-1}\,\Gamma \sim n^2/k = \lambda$ is real, and so is each single contribution $\sim \conjg{\tau}_i\,K^{-1}\,\tau_j$.
            Hence $F(x)$ can only depend on even powers of $h(x)$, otherwise it would be complex on the positive real axis.
            Consider the scalar off-shell basis given in Table \ref{bse:offshellbasis}: the general form for $F(x)$ is
            \begin{equation}
                F(x) = F_1 + \left(g(x)h(x)\right)^2 \left( F_2 + F_3 h(x)^2 + F_4 h(x)^4 \right).
            \end{equation}
            Exploiting the conditions $F(-1) = 0$ and $F'(-1)=1$, using $g'(-1) = -\kappa/2$ and $h'(-1)=-1$, leads to
            \begin{equation}
            \begin{split}
                F(x) =\;& \frac{1-\left(g(x)\,h(x)\right)^2}{\kappa+2} \left( 1 + 2F_3 + 4F_4 \right) + \\
                     &+\left(g(x)\,h(x)\right)^2 \left\{ F_3 (h^2-1) + F(4) (h^4-1) \right\},
            \end{split}
            \end{equation}
            which for large $x$ becomes
            \begin{equation}
                F(x\rightarrow\infty) = \frac{1 + 2F_3 + 4F_4}{\kappa+2}\,.
            \end{equation}
            The numerical 2-loop result yields $F(x\rightarrow\infty) \approx 1/(\kappa+2)$.
            $F_3$ and $F_4$ only emerge through the transverse projection of $\Slash{\hat{q}}$ in $\tau_3$, so it is conceivable that they are small.
            We therefore choose the approximation:
            \begin{equation}\label{DQ:Fsc}
                F(x) \approx \frac{1-\left(g(x)\,h(x)\right)^2}{\kappa+2} = \frac{1}{\kappa+2}\left( 1 + \frac{x}{(x+2)^{\kappa+1}} \right).
            \end{equation}
            A similar analysis for the axial-vector case (here the Monte-Carlo calculation yields $F(x\rightarrow\infty) \approx 1/\kappa$) leads to
            \begin{equation}\label{DQ:Fav}
                F_T(x) \approx \frac{1-g(x)^2}{\kappa} = \frac{1}{\kappa}\left( 1 - \frac{1}{(x+2)^\kappa} \right)\,.
            \end{equation}
            The longitudinal parts $F_L(x)$ are poorly constrained.
            For simplicity we choose $F_T(x) = F_L(x)$, i.e. $n_T^2/k_T = n_L^2/k_L$.
            Note that this does not implicate $n_T = n_L$, i.e. the axial-vector pole still appears  only in the transverse part of the propagator.

            \pagebreak[4]

            \noindent
            We summarize: the diquark propagators we use are given by
            \begin{fshaded}
            \begin{equation}\label{DQ:prop-final}
            \begin{split}
                D^{-1}(P^2) &= M_\text{sc}^2 \,\big\{ - n(x) +  \lambda_\text{sc} + \beta_\text{sc}\,F_\text{sc}(x) \big\}\\
                \left(D^{-1}\right)^{\mu\nu}(P) &= M_\text{av}^2 \,\big\{ -n^{\mu\nu}(x) + \left( \lambda_\text{av} + \beta_\text{av}\,F_\text{av}(x)\right) \delta^{\mu\nu} \big\}
            \end{split}
            \end{equation}
            \end{fshaded}

            \noindent
            where $x=P^2/M^2$ ($M=M_\text{sc}$ or $M_\text{av}$). $n(x)$ and $n^{\mu\nu}(x)$ are the numerically computed quark-loop integrals of Eq.\,\eqref{DQ:nk} with the
            onshell values $n_{(T)}(-1)=\lambda$ and $n_{(T)}'(-1)=\beta-1$. The functions $F(x)$ are defined in Eqs.\,(\ref{DQ:Fsc}--\ref{DQ:Fav}), and we use the value $\kappa=2$ in our calculation:
            \begin{equation}
                F_\text{sc}(x) = \frac{1}{4}\left( 1 + \frac{x}{(x+2)^3} \right), \; F_\text{av}(x) = \frac{1}{2}\left( 1 - \frac{1}{(x+2)^2}\right).
            \end{equation}
            With Eqs.\,\eqref{DQ:prop-final}, \eqref{dq:amp-offshell} and Table \ref{bse:offshellbasis}, the off-shell behavior of the quark-quark scattering
            matrix is completely determined.

    \bigskip
    \fatcol{Asymptotic behavior II.}    
            By virtue of Eq.\,\eqref{DQ:prop-final}, the asymptotic limits of the inverse diquark propagators are
            \begin{equation}
            \begin{split}
                D^{-1}(P^2\rightarrow\infty) &= M_\text{sc}^2 \,\left\{ \lambda_\text{sc} + \beta_\text{sc}/4 \right\},\\
                \left(D^{-1}\right)^{\mu\nu}(P^2\rightarrow\infty) &= M_\text{av}^2 \,\left\{ \lambda_\text{av} + \beta_\text{av}/2 \right\} \delta^{\mu\nu},
            \end{split}
            \end{equation}
            i.e. they become constant in the ultraviolet since the quark-loop integrals $n$ vanish in the UV due to the $P^2$ suppression of the quark propagators.
            This is the anticipated result which follows from the off-shell ansatz for the amplitudes:
            \begin{equation}
            \begin{split}
                \Gamma_\text{sc}(q,P\rightarrow\infty)     &\,\rightarrow\, f_1^\text{sc}(q^2,0,-M_\text{sc}^2) \, i\gamma^5 C,   \\
                \Gamma^\mu_\text{av}(q,P\rightarrow\infty) &\,\rightarrow\, f_1^\text{av}(q^2,0,-M_\text{av}^2) \, \gamma^\mu C,
            \end{split}
            \end{equation}
            Typical calculated values for $\lambda$, $\beta$ and the amplitude normalizations are:
            \begin{equation}
                \begin{array}{rl} \lambda_\text{sc} &\sim 1, \\ \lambda_\text{av} &\sim 0.5,  \end{array}\qquad
                \begin{array}{rl} \beta_\text{sc} &\sim 1,   \\ \beta_\text{av} &\sim -0.4,   \end{array}\qquad
                \begin{array}{rl} f_1^\text{sc}(0,0,-M_\text{sc}^2) &\sim 20, \\ f_1^\text{av}(0,0,-M_\text{av}^2) &\sim 10.  \end{array}
            \end{equation}
            Consider the propagators constructed from \eqref{DQ:prop-final} in the limit of a single dominant amplitude, i.e. where $\beta=0$:
            \begin{equation}
                D(P^2) = \frac{1}{M_\text{sc}^2} \,\frac{1}{ \lambda_\text{sc} - n(x)  }, \quad
                D^{\mu\nu}(P) = \frac{1}{M_\text{av}^2} \,\left( \frac{T^{\mu\nu}_P}{\lambda_\text{av}-n_T(x) } + \frac{L^{\mu\nu}_P}{\lambda_\text{av}-n_L(x)  } \right)\,.
            \end{equation}
            A simplification which reproduces the perturbative and on-shell behavior (the error in the intermediate $P^2$ region is $\lesssim 30 \%$) reads
            \begin{equation}
                D(P^2) \approx \frac{1}{M_\text{sc}^2} \left( \frac{1}{ \lambda_\text{sc}  } + \frac{1}{x+1} \right)  , \quad
                D^{\mu\nu}(P) \approx \frac{1}{M_\text{av}^2} \,\left( \frac{\delta^{\mu\nu}}{\lambda_\text{av} } + \frac{T^{\mu\nu}_P}{x+1  } \right),
            \end{equation}
            which explicitly demonstrates the appearance of non-resonant contributions in these 'diquark propagators' which serve as a surrogate
            for a more involved structure of the T-matrix. 
            Of course one could have directly extracted a pole contribution from the result \eqref{DQ:prop-final} to arrive at a similar form.


\newpage

\section{Nucleon amplitude} \label{app:faddeevamp}

\begin{figure}[tbp]
                    \begin{center}
                    \includegraphics[scale=0.25]{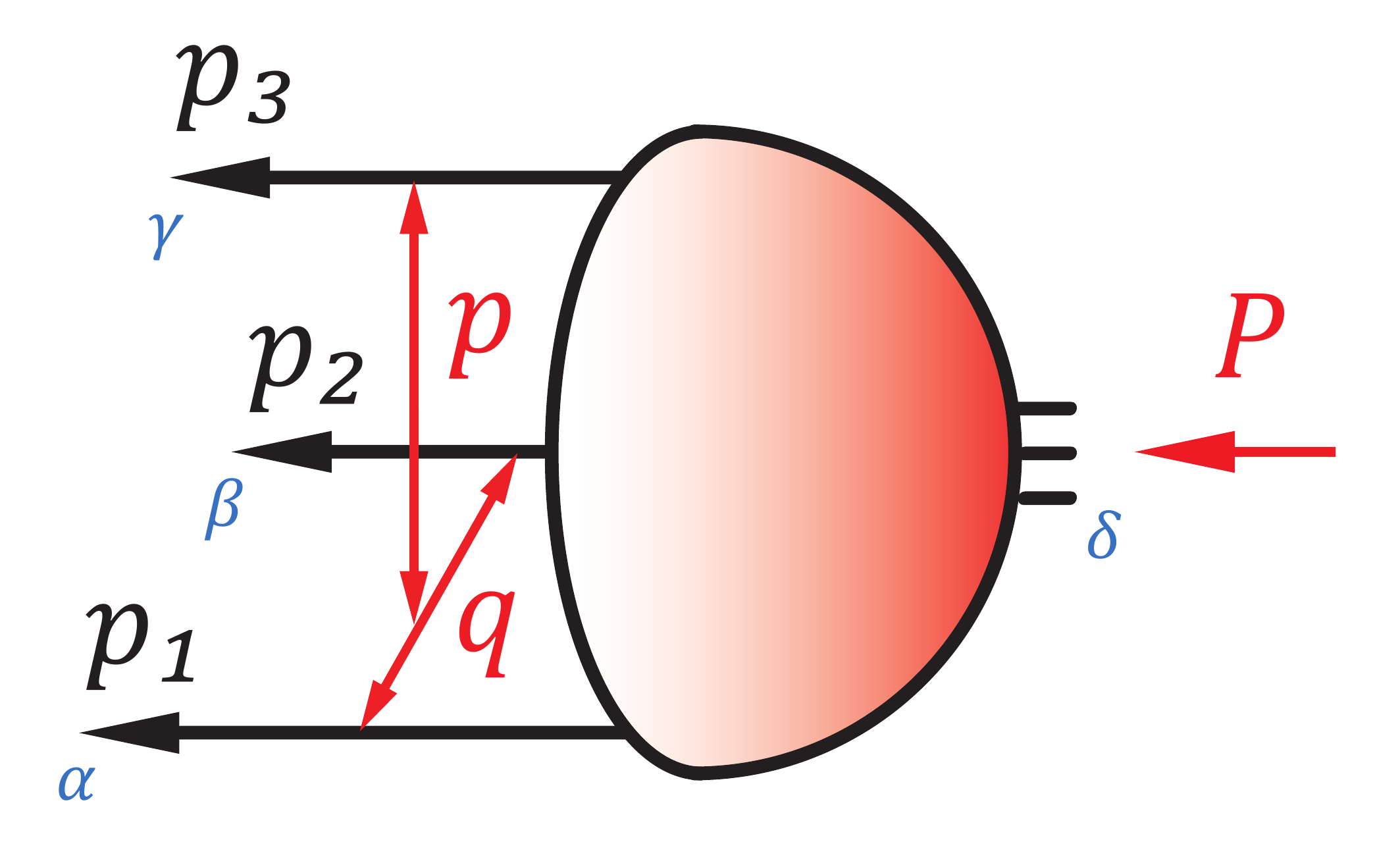}
                    \caption{\backdef Momentum routing in the nucleon's three-body amplitude.}\label{fig:three-body-amp}
                    \end{center}
            \end{figure}

             This section collects supplements to the Faddeev equation of Chapter \ref{chapter:faddeev}.
             The nucleon's three-quark amplitude (depicted in Fig.\,\ref{fig:three-body-amp}) including its full Dirac, flavor and color dependence is given by
             \begin{fshaded2}
              \begin{equation}\label{FE:nucleon_amplitude_full}
                  \Psi(p,q,P) = \Big\{ \Psi(p,q,P)_\mathcal{M_A} \otimes \mathsf{T}_\mathcal{M_A} + \Psi(p,q,P)_\mathcal{M_S}  \otimes \mathsf{T}_\mathcal{M_S} \Big\} \otimes \frac{\varepsilon_{ABC}}{\sqrt{6}}\,.
              \end{equation}
              \end{fshaded2}

              \noindent
              The Dirac amplitudes $\Psi_\mathcal{M_A}$, $\Psi_\mathcal{M_S}$ carry four fermion indices
              and depend on two relative momenta $p$, $q$ and the total onshell nucleon momentum $P$, with $P^2=-M^2$.
              Their decomposition into Dirac basis tensors with Lorentz-invariant coefficient functions is stated in Eq.\,\eqref{faddeev:amp},
              and an orthogonal 64-dimensional basis $\mathsf{X}_{ij}^\pm$ is presented in Table~\ref{tab:faddeev:basis}.
              $\Psi_\mathcal{M_A}$ and $\Psi_\mathcal{M_S}$ are mixed-antisymmetric or mixed-symmetric with respect to the permutation group $\mathbb{S}^3$.
              The Clebsch-Gordan construction of the corresponding flavor tensors reads:
              \begin{equation} \label{FAD:flavor}
              \begin{split}
                  \mathsf{T}_\mathcal{M_A} &= \textstyle\frac{1}{\sqrt{2}}\,i \sigma_2\otimes \mathds{1} = \mathsf{s^0} \otimes (\mathsf{uu^\dag+dd^\dag})\,,\\
                  \mathsf{T}_\mathcal{M_S} &= -\textstyle\frac{1}{\sqrt{6}}\, \vect{\sigma}\,i\sigma_2 \otimes \vect{\sigma} =
                                              \textstyle\sqrt{\frac{2}{3}}\,\mathsf{s^1}\otimes\mathsf{du^\dag} - \textstyle\frac{1}{\sqrt{3}}\,\mathsf{s^2}\otimes(\mathsf{uu^\dag-dd^\dag})
                                                                                     -\textstyle\sqrt{\frac{2}{3}}\,\mathsf{s^3}\otimes \mathsf{ud^\dag}\,,
              \end{split}
              \end{equation}
              where $\sigma_i$ are the Pauli matrices and $\mathsf{s^0}$, $\mathsf{s^{1,2,3}}$ denote the (anti-)symmetric isospin-$0$ and isopin-$1$ quark-quark representations defined in Eq.\,\eqref{dq:flavormatrices}.
             A projection onto the proton's or neutron's flavor state involves a contraction of the rearmost flavor index with either of the two isospin basis states $\mathsf{u}$ or $\mathsf{d}$.

     \bigskip
     \fatcol{Basis transformations.}
              In the Dirac basis of Table~\ref{tab:faddeev:basis} we exchanged the basis elements $\mathsf{S}_{2j}$, $\mathsf{P}_{2j}$ ($j=1\dots 4$)
              with the vector-vector and axialvector-axialvector components $\mathsf{V}_{1j}$, $\mathsf{A}_{1j}$.
              The remaining basis elements of Eqs\,(\ref{faddeev:basisSA}) and (\ref{faddeev:basisAVT}) are linear combinations of the former:
              \begin{equation}\label{faddeev:basis-tf}
                  \begin{array}{rclcrcl}
                         \mathsf{S}_{21}  &=& \pm (\mathsf{P}_{34} -   \mathsf{P}_{43} -   \mathsf{A}_{12} )  &  \quad\quad\quad &
                         \mathsf{V}_{31}  &=&      \mathsf{S}_{13} \pm \mathsf{P}_{31} \mp \mathsf{A}_{13}   \\
                         \mathsf{S}_{22}  &=& \mp (\mathsf{P}_{33}   + \mathsf{P}_{44} -   \mathsf{A}_{11} )  &  \quad\quad\quad &
                         \mathsf{V}_{32}  &=&      \mathsf{S}_{23} \mp \mathsf{P}_{41} +   \mathsf{S}_{14}   \\
                         \mathsf{S}_{23}  &=& \pm (\mathsf{P}_{32}+    \mathsf{P}_{41} -   \mathsf{A}_{14} )  &  \quad\quad\quad &
                         \mathsf{V}_{33}  &=&      \mathsf{S}_{11} \mp \mathsf{P}_{33} \pm \mathsf{A}_{11}   \\
                         \mathsf{S}_{24}  &=& \pm (\mathsf{P}_{42}   - \mathsf{P}_{31}   + \mathsf{A}_{13} )  &  \quad\quad\quad &
                         \mathsf{V}_{34}  &=&      \mathsf{S}_{12} \mp \mathsf{P}_{34} \pm \mathsf{A}_{12}   \\[0.5cm]
                         \mathsf{V}_{21}  &=&      \mathsf{S}_{43}   - \mathsf{S}_{34} \mp \mathsf{P}_{12}   &  \quad\quad\quad &
                         \mathsf{V}_{41}  &=&      \mathsf{S}_{14} \pm \mathsf{P}_{41} \mp \mathsf{A}_{14}   \\
                         \mathsf{V}_{22}  &=&      \mathsf{S}_{33}   + \mathsf{S}_{44} \pm \mathsf{P}_{11}   &  \quad\quad\quad &
                         \mathsf{V}_{42}  &=&      \mathsf{S}_{24} \pm \mathsf{P}_{31}   - \mathsf{S}_{13}   \\
                         \mathsf{V}_{23}  &=&      \mathsf{S}_{32}+    \mathsf{S}_{41} \mp \mathsf{P}_{14}   &  \quad\quad\quad &
                         \mathsf{V}_{43}  &=&      \mathsf{S}_{21} \mp \mathsf{P}_{43} \pm \mathsf{A}_{21}   \\
                         \mathsf{V}_{24}  &=&      \mathsf{S}_{42}   - \mathsf{S}_{31} \pm \mathsf{P}_{13}   &  \quad\quad\quad &
                         \mathsf{V}_{44}  &=&      \mathsf{S}_{11} \mp \mathsf{P}_{44} \pm \mathsf{A}_{11}
                  \end{array}
              \end{equation}
              The positive/negative signs refer to the superscripts $\pm$ which we did not state explicitly. 
              The corresponding relations for $\mathsf{P}$, $\mathsf{A}$ are obtained by interchanging $\mathsf{S}\leftrightarrow \mathsf{P}$, $\mathsf{V}\leftrightarrow \mathsf{A}$,
              and similar dependencies hold for the $\mathsf{T}_{ij}$, e.g.: $\mathsf{T}_{11}^+ = -2\, \mathsf{A}_{11}^+$.
              Eqs.\,\eqref{faddeev:basis-tf}  can be verified by expressing internal products such as $\gamma^\mu_T \,\Slash{p}_T$ through the respective commutator and using
              \begin{equation}
                  \sigma_T^{\mu\nu}\,\Lambda_\pm = \mp \,i\,\varepsilon^{\mu\nu\alpha\beta} \hat{P}^\alpha \,\gamma^\beta_T \,\gamma^5 \Lambda_\pm\,, \qquad
                  \text{with} \quad \sigma_T^{\mu\nu} := -\frac{i}{2} \left[ \gamma^\mu_T, \gamma^\nu_T \right]\,,
              \end{equation}
              together with the $\varepsilon$-tensor identities
              \begin{equation*}
              \begin{array}{rl}
                  \varepsilon^{\mu\alpha\beta\gamma}\,\varepsilon^{\mu\rho\sigma\tau} &= \delta_{\alpha\rho} \left( \delta_{\beta\sigma} \delta_{\gamma\tau} - \delta_{\beta\tau} \delta_{\gamma\sigma}\right) \\
                                                                                      &+ \;\delta_{\alpha\sigma} \left( \delta_{\beta\tau} \delta_{\gamma\rho} - \delta_{\beta\rho} \delta_{\gamma\tau}\right)  \\
                                                                                      &+ \; \delta_{\alpha\tau} \left( \delta_{\beta\rho} \delta_{\gamma\sigma} - \delta_{\beta\sigma} \delta_{\gamma\rho}\right) \,,
              \end{array}\qquad\quad
              \begin{array}{rl}
                   \varepsilon^{\mu\nu\alpha\beta}\,\varepsilon^{\mu\nu\rho\sigma} &= 2 \left( \delta_{\alpha\rho} \delta_{\beta\sigma} - \delta_{\alpha\sigma} \delta_{\beta\rho}\right)\,,\\[-0.1cm]
                   \varepsilon^{\mu\nu\lambda\alpha}\,\varepsilon^{\mu\nu\lambda\beta} &= 6\, \delta_{\alpha\beta}\,.
              \end{array}
              \end{equation*}

    \bigskip
    \fatcol{Explicit implementation.}
            The Faddeev equation is solved via iteration using similar techniques as in the two-body case.
            A 'wave function' $\Phi=SS\,\Psi$ is extracted from the equation $\Psi = KSS \,\Psi$ and evaluated outside the loop integral.
            The subsequent integration $\Psi = K \Phi$ is carried out by calling the wave function $\Phi$ with the loop momenta as its arguments:
                \begin{equation}
                \begin{split}
                    \Phi^{(a)}(p,q,P) &= S(p_b)\,S(p_c)\,\Psi(p,q,P)\,,\\
                    \Psi(p,q,P) &= \lambda(P^2) \sum_{a=1}^3 \int_k K^{(a)}(k)\, \Phi^{(a)}(p^{(a)},q^{(a)},P)\,.
                \end{split}
                \end{equation}
            The index $a=1\dots 3$ denotes the three permutations of the Faddeev kernel, and $\{a,b,c\}$ is an even permutation of $\{1,2,3\}$.
            These steps are repeated for fixed $P^2$ until the eigenvalue $\lambda(P^2)$ converges.
            The value $\lambda(-M^2)=1$ which defines the nucleon mass $M$ is obtained upon processing the above procedure for different $P^2$.

            If the Dirac basis is complete, the wave function can be projected onto the same basis elements as the amplitude (cf. Eq.\,\eqref{faddeev:amp}):
                \begin{align}
                    \Phi^{(a)}_{\alpha\beta\gamma\delta}(p,q,P) &= \sum_{i=1}^{64} \tilde{f}_i^{(a)}(p^2, q^2, \{z\}) \,\tau_i(p,q,P)_{\alpha\beta\gamma\delta}\,,
                \end{align}
            where the $\tau_i$ are given by the orthonormal basis elements $\mathsf{X}^\pm$.
           Using the orthonormality relations \eqref{faddeev:orthogonality} yields coupled equations for the amplitude and wave function dressing functions:
                \begin{equation}\label{faddeev:eqcoupled}
                \begin{split}
                    & f_i(p^2,q^2,\{z\}) =  \sum_j \sum_{a=1}^3 \int_k \mathcal{K}^{(a)}_{ij}(p,q,P,k)\,\tilde{f}^{(a)}_j\Big({p^{(a)}}^2,{q^{(a)}}^2,\{z^{(a)}\}\Big)\,, \\
                    & \tilde{f}_i^{(a)}(p^2,q^2,\{z\}) = \sum_j \mathcal{G}^{(a)}_{ij}(p,q,P)\,f_j\big(p^2,q^2,\{z\}\big)\,,
                \end{split}
                \end{equation}
            where the kernel $\mathcal{K}^{(a)}_{ij}$ and quark propagator matrix $\mathcal{G}^{(a)}_{ij}$ are the matrix elements
                \begin{equation}\label{faddeev:kernel-evaluated}
                \begin{split}
                    \mathcal{K}^{(a)}_{ij}(p,q,P,k) &= \conjg{\tau}_i(p,q,P)_{\beta\alpha,\delta\gamma} \, K^{(a)}_{\alpha\alpha'\beta\beta'\gamma\gamma'}(k) \, \tau_j\big(p^{(a)},q^{(a)},P\big)_{\alpha'\beta',\gamma'\delta}\,, \\
                    \mathcal{G}^{(a)}_{ij}(p,q,P)   &= \conjg{\tau}_i(p,q,P)_{\beta\alpha,\delta\gamma} \, G^{(a)}_{\alpha\alpha'\beta\beta'\gamma\gamma'}(p,q,P) \, \tau_j(p,q,P)_{\alpha'\beta',\gamma'\delta}
                \end{split}
                \end{equation}
            of the quantities
                \begin{equation}
                \begin{split}
                    K^{(3)}_{\alpha\alpha'\beta\beta'\gamma\gamma'}(k) &= K_{\alpha\alpha'\beta\beta'}(k)\,\delta_{\gamma\gamma'}\,, \\
                    G^{(3)}_{\alpha\alpha'\beta\beta'\gamma\gamma'}(p,q,P) &= S_{\alpha\alpha'}(p_1)\,S_{\beta\beta'}(p_2)\,\delta_{\gamma\gamma'}\,,
                \end{split}
                \end{equation}
            and the remaining expressions corresponding to $a=1,2$ are obtained by a cyclic permutation of the index pairs and momentum indices.

            \bigskip
            \fatcol{Numerical aspects.}
            The Faddeev equation \eqref{faddeev:eqcoupled} is equivalent to an iterated (multidimensional) matrix-vector multiplication, where
            in a straightforward implementation the kernel $\mathcal{K}$ and propagator matrix $\mathcal{G}$ would be computed in advance.
            The kernel depends on 9 Lorentz-invariant momentum variables: five $(p^2,q^2,\{z\})$ correspond to the outer momenta $p$, $q$, $P$
            and four $(k^2,\,\hat{k}\cdot\hat{P},\,\hat{k}\cdot\hat{p},\,\hat{k}\cdot\hat{q})$ to the loop momentum $k$. It furthermore involves the amplitude indices $i, j=1 \dots 64$
            and the permutation counter $a=1,2,3$. Choosing 20 grid points for each momentum variable leads to a memory requirement of
            $20^9$ (momenta) $\times\,64^2$ (amplitudes) $\times\,3$ (permutations) $\times\,16$ (double-precision complex number) Byte $\approx 90$ Petabyte,
            which is clearly beyond the capacities of today's computing facilities.

            The uppermost priority in a numerical optimization thus concerns the reduction of memory usage of the kernel $\mathcal{K}_{ij}^{(a)}$.
            This can be accomplished by splitting the kernel in a momentum-independent component which involves all Dirac traces (to be computed in advance)
            and a momentum-dependent part which is evaluated during the iteration process.
            With the momentum alignment \eqref{FADDEEV:momentumalignment}, on whose account
            $\widehat{p_T}$, $\widehat{q_t}$ and $\hat{P}$ become the three orthogonal unit vectors of Eq.\,\eqref{faddeev:orthogonal-unit-vectors},
            the conjugate basis elements $\conjg{\tau}_i$ in Eqs.\,\eqref{faddeev:kernel-evaluated} effectively do not depend on any momentum at all,
            and the momentum dependence is carried by the inner elements $\tau_j(p^{(a)},q^{(a)},P)$.
            One may rewrite the orthonormal basis $\mathsf{X}_{ij}^\pm$ in the following way:
            \begin{equation}
                \mathsf{X}_{ij}^\pm = \left( \begin{array}{rl}
                                              \mathds{1}&\otimes\,\mathds{1} \\
                                              \gamma^5 &\otimes\, \gamma^5 \\
                                              \frac{1}{\sqrt{3}}\,\gamma^\mu_T\gamma^5 &\otimes\, \gamma^\mu_T\gamma^5 \\
                                              \frac{1}{\sqrt{3}}\,\gamma^\mu_T &\otimes\, \gamma^\mu_T
                                              \end{array} \right) \,
                                              T_{ij}(p,q,P) \,
                                              \Big( \Lambda_\pm  \gamma_5 C  \otimes \Lambda_+ \Big), \quad
                                      \left( \begin{array}{lc}
                                              i=1,5,7 \\
                                              i=2,6,8 \\
                                              i=3 \\
                                              i=4
                                              \end{array} \right)
            \end{equation}
            Abbreviating $\widehat{p_T} \rightarrow p$ and $\widehat{q_t} \rightarrow q$,
            the momentum-dependent parts $T_{ij}$ for $i=1\dots 4$ (i.e. the $s=\nicefrac{1}{2}$ covariants) and $i=5\dots 8$ ($s=\nicefrac{3}{2}$) read:
            \begin{equation*}
            \begin{array}{|l|l|l|}\hline
                \quad T_{ij} \;(i\leq 4)\;\;     &   \quad \sqrt{6}\; T_{ij} \;(i=5,6) & \quad \sqrt{2}\;T_{ij} \;(i=7,8)\\ \hline
                                 \begin{array}{ @{\quad}l @{\quad}}
                                        \mathds{1}\otimes \mathds{1} \\
                                        \mathds{1}\otimes \textstyle\frac{1}{2}\, [\, \Slash{p}, \Slash{q} \,] \\
                                        \mathds{1}\otimes \Slash{p}   \\
                                        \mathds{1}\otimes \Slash{q}   \\[0.2cm]
                                 \end{array}   &
                                 \begin{array}{ @{\quad}l @{\;\;}}
                                        3\,\Slash{p} \otimes \Slash{p} - \gamma^\mu_T \otimes \gamma^\mu_T    \\
                                        3\left( \Slash{p} \otimes \Slash{q} -\Slash{q} \otimes \Slash{p} \right) - \gamma^\mu_T \otimes \gamma^\mu_T \,[\, \Slash{p}, \Slash{q} \,] \\
                                        3\,\Slash{p} \otimes \mathds{1} - \gamma^\mu_T \otimes \gamma^\mu_T \, \Slash{p}  \\
                                        3\,\Slash{q} \otimes \mathds{1} - \gamma^\mu_T \otimes \gamma^\mu_T \, \Slash{q}   \\[0.2cm]
                                 \end{array} &
                                 \begin{array}{ @{\quad}l @{\;\;}}
                                        \Slash{p} \otimes \Slash{p} + 2\,\Slash{q} \otimes \Slash{q} -  \gamma^\mu_T \otimes \gamma^\mu_T \\
                                        \Slash{p} \otimes \Slash{q} + \Slash{q} \otimes \Slash{p} \\
                                        \Slash{q} \otimes [\, \Slash{q}, \Slash{p} \,] - \textstyle\frac{1}{2}\,\gamma^\mu_T \otimes [\,\gamma^\mu_T, \, \Slash{p}\,] \\
                                        \Slash{p} \otimes [\, \Slash{p}, \Slash{q} \,] - \textstyle\frac{1}{2}\,\gamma^\mu_T \otimes [\,\gamma^\mu_T ,\, \Slash{q}\,]  \\[0.2cm]
                                 \end{array} \\ \hline
            \end{array}
            \end{equation*}
            This simplifies the extraction of the momentum dependence, e.g. in the nucleon's rest frame via
            \begin{equation}
                T_{i\leq 4,\,2} = \frac{1}{2} \,\varepsilon_{lmn} \left( \mathds{1}\otimes \gamma^m \gamma^n \right) \left(\varepsilon_{lrs} \,p^r q^s\right)\,, \\
            \end{equation}
            and similarly for the remaining ones. Hence the kernel can be written as
            \begin{equation}
                \mathcal{K}^{(a)}_{ij}(p,q,P,k) = \Big[ \mathcal{K}^{(a)\,l}_{ij}(k)\Big]\,\Big[g^l_{ij}(p^{(a)},q^{(a)},P)\Big]\,,
            \end{equation}
            where the momentum dependence carried by the $T_{ij}$ has been shifted into the functions $g^l_{ij}$.
            In the same way one may isolate the kernel's dependence on the gluon momentum $k$:
            \begin{equation}
                \mathcal{K}^{(a)}_{ij}(p,q,P,k) = \bigg[\mathcal{K}^{(a)\,l,\,\mu\nu}_{ij}\bigg]\,\bigg[\frac{4\pi\alpha(k^2)}{k^2}\,T_k^{\mu\nu}\, g^l_{ij}(p^{(a)},q^{(a)},P)\bigg]\,.
            \end{equation}
            This strategy greatly reduces the memory demand to $\lesssim 1$ GB.
            The impact on the run time  due to the above multiplication (which is now processed on the fly)
            is still slightly outweighed by the time consumed to interpolate the dressing functions inside the integral.
            For this reason we drop the dependence on the angular variable $z_0 = \widehat{p_T}\cdot\widehat{q_T}$ which,
            from analogy of the quark-diquark model, is expected to be small.
            In addition we perform an expansion into Chebyshev polynomials of the first kind (see App.\,\ref{appendixchebyshev})
            in the remaining angles $z_1^{(a)}$, $z_2^{(a)}$ that appear in the wave function coefficients inside the integral.
            The resulting run times are accessible by a parallel cluster.


\newpage

\section{Quark-diquark amplitudes for $N$ and $\Delta$} \label{app:qdqamp}

    \fatcol{Nucleon.}
            The matrix-valued quark-diquark amplitudes $\Phi^{(\mu)}_{\alpha\beta}(p,P)$ which were introduced in Sec.\,\ref{sec:qdqbse}
            feature the following Dirac, color and flavor decomposition:
            \begin{fshaded}
            \begin{equation}\label{nuc:amplitudes}
                \begin{split}
                \Phi_{\alpha\beta}^5(p,P) &=   \sum_{k=1}^2 f_k^{N,\text{sc}}(p^2,z) \big\{ \tau_k(p,P) \, \Lambda_+(P) \big\}_{\alpha\beta}
                                               \,\otimes\, \frac{\delta_{AB}}{\sqrt{3}} \,\otimes\, \mathsf{t^0_{ab}}  \\
                \Phi_{\alpha\beta}^\mu(p,P) &= \sum_{k=1}^6 f_k^{N,\text{av}}(p^2,z) \big\{ \tau_k^\mu(p,P) \, \gamma^5 \, \Lambda_+(P) \big\}_{\alpha\beta}
                                               \,\otimes\, \frac{\delta_{AB}}{\sqrt{3}} \,\otimes\, \mathsf{t^{1,2,3}_{ab}}\;  .
                \end{split}
            \end{equation}
            \end{fshaded}

            \noindent
            $p$ is here the relative momentum between quark and diquark and $P$ the total nucleon momentum on the mass shell: $P^2=-M_N^2$.
            The number of basis elements can be inferred from the Clifford algebra.
            The positive parity condition for the full Faddeev amplitude translates into positive parity of the quark-diquark amplitudes.
            This entails a maximum number of four and twelve Dirac basis elements for the scalar and axial-vector quark-diquark amplitudes,
            with a possible basis for both cases given by
            \begin{equation}
                 \{\mathds{1},\,\Slash{p},\,\Slash{P},\,\Slash{p}\,\Slash{P}\}, \quad  \{\gamma^\mu, p^\mu, P^\mu\} \times \{\mathds{1},\,\Slash{p},\,\Slash{P},\,\Slash{p}\,\Slash{P}\}.
            \end{equation}
            The constraint of positive energy for the nucleon, expressed by the positive-energy projector $\Lambda_+(P)=(\mathds{1} + \hat{\Slash{P}})/2$,
            halves the number of basis elements via $\hat{\Slash{P}}\,\Lambda_+ = \Lambda_+$ 
            to 2 Dirac matrices $\{\mathds{1},\Slash{p}\}$ for the scalar quark-diquark amplitude and 6 matrices 
            for the axial-vector one.
            A partial-wave decomposition in terms of quark-diquark total spin and orbital angular momentum eigenstates in the nucleon's
            rest frame \cite{Oettel:1998bk,Alkofer:2005jh} leads to the orthogonal, dimensionless Dirac basis stated in Table~\ref{APP:QDQ:nucbasis}.
            Its orthogonality relations are
            \begin{equation}
                \frac{1}{2} \,\text{Tr}\{\conjg{\tau}_k\, \tau_l\, \Lambda_+\} = (-1)^{k+1} \,\delta_{kl} ,\quad
                \frac{1}{2} \,\text{Tr}\{\conjg{\tau}_k^{\mu}\, \tau_l^{\mu}\,\Lambda_+\} = (-1)^k \,\delta_{kl} .
            \end{equation}
            The dominant amplitudes are those corresponding to the $s$ waves $\tau_1$ and $\tau_1^\mu$, with a further addition
            of $s$ ($\tau_3^\mu$), $p$ ($\tau_2$, $\tau_{2,4,6}^\mu$) and $d$ waves ($\tau_5^\mu$).

            The flavor wave functions of the full nucleon amplitude, constructed from the Clebsch-Gordan prescription, are given in Eq.\,\eqref{FAD:flavor}.
            The flavor matrices $\mathsf{t^e}$ in the quark-diquark amplitudes are their "quark remainders" upon removing the diquark contributions:
            \begin{equation}
            \begin{split}\label{NUC:Isospin}
                \mathsf{t^0} &= \mathsf{uu^\dag+dd^\dag} = \mathds{1} \,,  \\
                \mathsf{t^1} &= \textstyle\sqrt{\frac{2}{3}} \, \mathsf{du^\dag}     = \textstyle\frac{1}{\sqrt{6}} \, (\sigma_1 - i\sigma_2) \,,   \\
                \mathsf{t^2} &= - \textstyle\frac{1}{\sqrt{3}} \, (\mathsf{uu^\dag-dd^\dag})   = -\textstyle\frac{1}{\sqrt{3}}\,\sigma_3  \,,  \\
                \mathsf{t^3} &= -\textstyle\sqrt{\frac{2}{3}}\,\mathsf{ud^\dag}    = - \textstyle\frac{1}{\sqrt{6}}\,(\sigma_1 + i\sigma_2) \,.
            \end{split}
            \end{equation}
            A projection onto the proton and neutron isospinors $\mathsf{u}=(1,0)$ and $\mathsf{d}=(0,1)$ yields the following flavor wave functions:
            \begin{equation}\label{nuc:qdq:flavor}
                \mathsf{p} =  \left( \mathsf{u} \; \Big|  \; \textstyle\sqrt{\frac{2}{3}} \, \mathsf{d}, \, - \textstyle\frac{1}{\sqrt{3}} \, \mathsf{u}, \, 0\right), \quad\quad
                \mathsf{n} =  \left( \mathsf{d} \; \Big|  \; 0,  \, \textstyle\frac{1}{\sqrt{3}} \, \mathsf{d}, \, -\textstyle\sqrt{\frac{2}{3}} \, \mathsf{u}\right).
            \end{equation}

            \begin{table}
                \begin{center}

                \begin{equation*}
                \begin{array}{|c|c|}\hline
                   \Phi^5(p,P)  &   \Phi^\mu(p,P) \\ \hline
                                     \begin{array}{ @{\quad}l @{\quad}}
                                         \tau_1 = \mathds{1} \\
                                         \tau_2 = \Slash{q} \\[0.2cm]
                                     \end{array}   &
                                     \begin{array}{ @{\quad}l @{\quad\quad}l @{\quad\quad}l @{\quad}}
                                         \tau_1^{\mu} = \frac{1}{\sqrt{3}}\,\gamma^\mu_T               &   \tau_3^{\mu} = \hat{P}^\mu           &   \tau_5^{\mu} = \frac{1}{\sqrt{6}}\left( \gamma^\mu_T + 3\, q^\mu \Slash{q}\right)   \\
                                         \tau_2^{\mu} =\frac{1}{\sqrt{3}}\,\gamma^\mu_T \,\Slash{q}    &   \tau_4^{\mu} =\hat{P}^\mu \Slash{q}  &   \tau_6^{\mu} =\frac{1}{\sqrt{6}}\left( \gamma^\mu_T \,\Slash{q} - 3\, q^\mu \right)  \\[0.2cm]
                                     \end{array}  \\ \hline
                \end{array}
                \end{equation*}

                \caption{\backdef Orthogonal basis for the nucleon quark-diquark amplitudes. We abbreviated the normalized, transverse relative momentum by $q^\mu := i\,\widehat{p_T}^\mu$,
                         where $p^\mu_T=T^{\mu\nu}_P p^\nu$ and $T^{\mu\nu}_P$ is the transverse projector
                         with respect to the total nucleon momentum. }\label{APP:QDQ:nucbasis}
                \end{center}
            \end{table}

    \bigskip
    \fatcol{Delta.}
            Compared to the nucleon, only an isospin-$\nicefrac{3}{2}$ diquark can contribute to the spin-$\nicefrac{3}{2}$, isospin-$\nicefrac{3}{2}$ $\Delta$ amplitude
            which excludes the involvement of a scalar diquark.
            Denoting the total $\Delta$ momentum by $P$ with $P^2=-M_\Delta^2$,
            the on-shell quark-diquark amplitude $\Phi^{\mu\nu}$ of Eq.\,(\ref{QDQ:Amplitudes-N-Delta-1}) is decomposed into 8 covariant structures \cite{Oettel:1998bk}:
                        \begin{fshaded1}
                        \begin{equation}\label{deltaamplitdecompos}
                            \Phi_{\alpha\beta}^{\mu\nu}(p,P)\! =\!\! \sum_{k=1}^8 f_k^\Delta(p^2,z) \left\{   \tau_k^{\mu \rho}(p,P)\, \mathds{P}^{\rho \nu}(P) \right\}_{\alpha\beta}
                            \,\otimes\, \frac{\delta_{AB}}{\sqrt{3}} \,\otimes\, \Delta^\mathsf{e}\,,
                        \end{equation}
                        \end{fshaded1}

            \noindent
            where the basis elements include the Rarita-Schwinger projector onto positive-energy and spin-$3/2$ spinors:
                        \begin{equation}
                             \mathds{P}^{\rho \nu}(P) = \Lambda_+(P)\,T^{\rho\alpha}_P \,T^{\beta\nu}_P \left( \delta^{\alpha\beta} -\frac{1}{3} \,\gamma^\alpha \gamma^\beta \right) .
                        \end{equation}
            A general Green function with two fermion legs, two vector legs and two independent momenta
    	    $p$ and $P$ allows for 40 possible Dirac covariants $\tau_{k=1\dots 40}^{\mu\rho}$, constructed from
                \begin{equation}
                     \delta^{\mu\rho} \left\{ \mathds{1}, \,\Slash{p}, \,\Slash{P}, \,\Slash{p}\,\Slash{P} \right\}, \quad
                     \left\{ p^\mu, P^\mu, \gamma^\mu \right\} \times \left\{ p^\rho, P^\rho, \gamma^\rho \right\}  \times \left\{ \mathds{1}, \,\Slash{p}, \,\Slash{P}, \,\Slash{p}\,\Slash{P} \right\}.
                \end{equation}
            The elements $P^\rho$, $\gamma^\rho$, $\Slash{P}$ and $\Slash{p}\,\Slash{P}$ become redundant upon contraction
    	    with the Rarita-Schwinger projector:
            $P^\rho \,\mathds{P}^{\rho \nu} = \gamma^\rho \,\mathds{P}^{\rho \nu} = 0$, $\Slash{P}\,\Lambda_+ = \Lambda_+$.
            This leaves 8 covariants for which a convenient orthogonal set is displayed in Table~\ref{APP:QDQ:deltabasis}.
            The corresponding orthogonality relation reads
                        \begin{equation}
                            \frac{1}{4}\,\textrm{Tr}\left\{ \bar{\sigma}_k^{\mu\rho} \,\sigma_l^{\mu\rho}\right\} = (-1)^{k+1} \,\delta_{kl} ,
                        \end{equation}
            where $\sigma_k^{\mu \nu} (p,P) = \tau_k^{\mu \rho}(p,P)\,\mathds{P}^{\rho \nu}(P)$ and the conjugated amplitude
            is defined as (the superscript ${}^T$ denotes the Dirac transpose):
                        \begin{equation}
                            \conjg{\sigma}_k^{\mu\rho}(p,P) = C\,\sigma_k^{\mu\rho}(-p,-P)^T C^T.
                        \end{equation}
            The basis of Table~\ref{APP:QDQ:deltabasis} again corresponds to a partial-wave decomposition in terms of
            quark-diquark total spin and orbital angular momentum in the $\Delta$ rest frame \cite{Oettel:1998bk,Oettel:2000ig}.
            Since there is only one spherically symmetric $s$-wave component ($\tau_1^{\mu\nu}$),
            the $\Delta$'s deviation from sphericity amounts to an admixture of
            $p$ ($\tau^{\mu\nu}_{2,4,6}$), $d$ ($\tau_{3,5,7}^{\mu\nu}$) and $f$ ($\tau_8^{\mu\nu}$) waves
            which contribute a significant amount of orbital angular momentum to its amplitude.

            \begin{table}
                \begin{center}

                \begin{equation*}
                     \begin{array}{|c|}\hline
                         \Phi^{\mu\nu}(p,P) \\\hline
                     \begin{array}{ @{\quad}l @{\quad\quad}l @{\quad}  }
                         \tau_1^{\mu\nu}  =  \delta^{\mu\nu}                                                                                                  &
                         \tau_2^{\mu\nu}  =  \frac{\scriptstyle 1}{\scriptstyle \sqrt{5}}\left(2\gamma_T^\mu  \,q^\nu -3 \delta^{\mu\nu}\Slash{q}\right)   \\
                         \tau_3^{\mu\nu}  =  -\sqrt{3}  \,\hat{P}^\mu q^\nu \Slash{q}                                                                          &
                         \tau_4^{\mu\nu}  =  \sqrt{3} \,\hat{P}^\mu q^\nu                                             \\
                         \tau_5^{\mu\nu}  =  - \gamma_T^\mu  \,q^\nu \Slash{q}                                          &
                         \tau_6^{\mu\nu}  = -\gamma_T^\mu \,q^\nu                                                      \\
                         \tau_7^{\mu\nu}  =  \gamma_T^\mu \,q^\nu \Slash{q} -\delta^{ \mu \nu}  -  3 \, q^\mu q^\nu   &
                         \tau_8^{\mu\nu}  =  \frac{\scriptstyle 1}{\scriptstyle \sqrt{5}}\!\left( \delta^{\mu \nu}\Slash{q} + \gamma_T^\mu \,q^\nu +   5\, \,q^\mu q^\nu \Slash{q}  \right)  \\[0.2cm]
                     \end{array}  \\ \hline
                     \end{array}
                \end{equation*}

                \caption{\backdef Orthogonal basis for the $\Delta$ quark-diquark amplitudes, with $q^\mu := i \,\widehat{p_T}^\mu$.}\label{APP:QDQ:deltabasis}
                \end{center}
            \end{table}

            In the same way as for the nucleon, the isospin matrices of the $\Delta$ quark-diquark amplitude are constructed via removal of the diquark contributions
            from the full $\Delta$ flavor-amplitude. The equivalent of Eq.\,\eqref{nuc:qdq:flavor} reads:
                \begin{equation}
                \begin{split}
                     &\Delta^{++} = \left(  \mathsf{u} , \, 0 , \, 0 \right),  \quad\quad
                      \Delta^{+}  = \left(  \textstyle\frac{1}{\sqrt{3}} \,\mathsf{d} , \, \textstyle\frac{2}{\sqrt{3}} \,\mathsf{u} , \, 0 \right), \\
                     &\Delta^{0}  = \left( 0 , \, \textstyle\frac{2}{\sqrt{3}} \,\mathsf{d} , \, \textstyle\frac{1}{\sqrt{3}} \,\mathsf{u} \right), \quad\quad
                      \Delta^{-}  = \left( 0 , \, 0 , \, \mathsf{d} \right),
                \end{split}
                \end{equation}
            where the three entries correspond to the axial-vector diquark's three symmetric
    	    isospin-1 states.
            Contraction with the diquark flavor matrices $\mathsf{s^{1,2,3}}$ of Eq.\,\eqref{dq:flavormatrices}
            yields the final flavor tensors corresponding to the four $\Delta$ states.

    \bigskip
    \fatcol{Normalization.}
            The quark-diquark amplitudes for nucleon and $\Delta$ are normalized via the canonical normalization integral of Eq.\,\eqref{bs:normalization}.
            In this context, $\Psi$ is the BSE solution of the quark-diquark amplitude,
            $G_0$ is the product of a dressed quark and diquark propagator, and
            $K$ is the quark-diquark kernel \eqref{nuc:kernel}.
            In contrast to the meson and diquark case,
            the kernel depends on the total nucleon momentum $P$ such that the derivative $dK^{-1}/dP^2$ cannot be omitted.
            The normalization condition for the nucleon is equivalent to the normalization of the proton's electric charge: $G_E^p(0)=1$.


\newpage

\section{Diquark-photon vertex}\label{app:dqphotonvertex}

\begin{figure}[tbp]
            \begin{center}
            \includegraphics[scale=0.3]{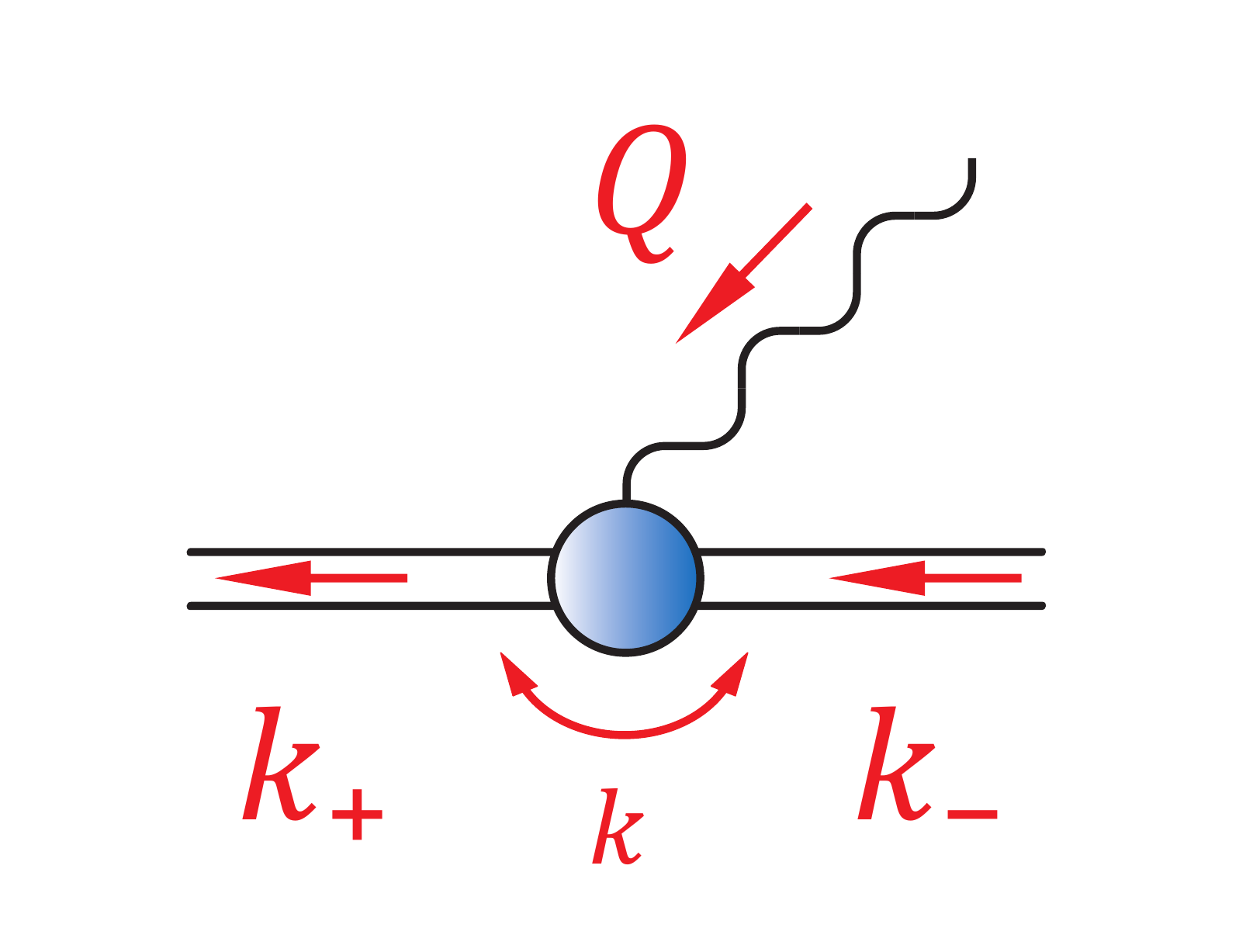}
            \caption[Diquark-photon vertex]{\backdef Momentum routing in the diquark-photon vertex. } \label{fig:DQPV}
            \end{center}
            \end{figure}

     \fatcol{General properties.}
            With the notation $k=(k_++k_-)/2$ and $Q=k_+-k_-$, where $k_+$ and $k_-$ are outgoing and incoming diquark momenta,
            the general form of the scalar and axial-vector diquark-photon vertices is (cf. Fig.\,\ref{fig:DQPV})
                  \begin{align}
                      \Gamma^\mu_\text{(dq)}(k,Q) &= \sum_{i=1}^{2} f^\text{(sc-dq)}_i(k^2,\hat{k}\cdot\hat{Q},Q^2)\, \tau_i^\mu(k,Q)\,, \\
                      \Gamma^{\mu,\alpha\beta}_\text{(dq)}(k,Q) &= \sum_{i=1}^{14} f^\text{(av-dq)}_i(k^2,\hat{k}\cdot\hat{Q},Q^2) \,\tau_i^{\mu,\alpha\beta}(k,Q)\,,
                  \end{align}
            where the set of basis elements is given by
                  \begin{align}\label{DQPV:Basis}
                       \tau_i^\mu(k,Q)                \; \in \;  &\left\{ k^\mu, Q^\mu \right\},   \nonumber \\
                       \tau_i^{\mu,\alpha\beta}(k,Q)  \; \in \;  & \left\{ k^\mu, Q^\mu \right\} \times \{\delta^{\alpha\beta}, k^\alpha k^\beta, \, Q^\alpha Q^\beta, \, k^\alpha Q^\beta, \,Q^\alpha k^\beta \}\,, \\
                                                                 & \;\delta^{\mu\alpha} \times \{k^\beta, Q^\beta \}, \, \delta^{\mu\beta} \times \{k^\alpha, Q^\alpha \} \,.  \nonumber
                  \end{align}
            Both vertices satisfy vector Ward-Takahashi identities which reflect electromagnetic current conservation.
            They constrain the longitudinal contributions $\sim Q^\mu$
            by relating them to the corresponding diquark propagators:
                  \begin{align}
                      Q^\mu \, \Gamma^\mu_\text{(dq)}(k,Q) & = D^{-1}(k_+)-D^{-1}(k_-) = Q^\mu \left\{ 2 k^\mu \Delta_{(D^{-1})} \right\},     \nonumber \\
                      Q^\mu \, \Gamma^{\mu,\alpha\beta}_\text{(dq)}(k,Q) &= D^{-1}_{\alpha\beta}(k_+) - D^{-1}_{\alpha\beta}(k_-) = \label{DQPV:WTI} \\
                                                                         & \hspace{-2cm} = Q^\mu \left\{ 2 k^\mu \left( \Delta_{(D_T^{-1})} \delta^{\alpha\beta}  - \Delta_{(\sigma)} \left( k^\alpha k^\beta + \frac{Q^\alpha Q^\beta}{4} \right) \right)
                                                                                         - \Sigma_{(\sigma)} \left( \delta^{\mu\alpha} k^\beta + \delta^{\mu\beta} k^\alpha \right) \right\}.
                                                                          \nonumber
                  \end{align}
            The average $\Sigma$ and difference quotient $\Delta$ were defined in Eq.\,\eqref{QPV:sigma,delta}.
            $D^{-1}(k^2)$ is the inverse scalar diquark propagator;
            $D_T^{-1}(k^2)$ and $\sigma(k^2):=\left(D^{-1}_T(k^2)-D^{-1}_L(k^2)\right)/k^2$ are the dressing functions
            of the inverse axial-vector diquark propagator:
                  \begin{equation}
                      D^{-1}_{\alpha\beta}(k) = D^{-1}_T(k^2)\,T_k^{\alpha\beta} + D^{-1}_L(k^2)\,L_k^{\alpha\beta} = D^{-1}_T(k^2)\,\delta^{\alpha\beta} - \sigma(k^2)\,k^\alpha k^\beta\,.
                  \end{equation}
            The differential Ward identities read:
                  \begin{equation}
                      \Gamma^\mu_\text{(dq)}(k,0) = \frac{dD^{-1}(k^2)}{dk^\mu}, \quad \Gamma^{\mu,\alpha\beta}_\text{(dq)}(k,0) = \frac{d D^{-1}_{\alpha\beta}(k)}{dk^\mu}\,.
                  \end{equation}

            The final expressions for the vertices upon implementation of the WTIs can be read off from Eq.\,\eqref{DQPV:WTI}:
                  \begin{fshaded4}
                  \begin{align}
                      \Gamma^\mu_\text{(dq)}(k,Q) &= 2 k^\mu \Delta_{(D^{-1})} + T^{\mu\nu}_Q \,\Gamma^\nu_T(k,Q)\,, \label{FF:ScPhotonVertex}\\
                      \Gamma^{\mu,\alpha\beta}_\text{(dq)}(k,Q) &= 2 k^\mu \left( \Delta_{(D_T^{-1})} \delta^{\alpha\beta}  - \Delta_{(\sigma)} \left( k^\alpha k^\beta + \frac{Q^\alpha Q^\beta}{4} \right) \right)  \nonumber \\
                                                                                    &\quad     - \Sigma_{(\sigma)} \left( \delta^{\mu\alpha} k^\beta + \delta^{\mu\beta} k^\alpha \right)  + T^{\mu\nu}_Q \,\Gamma^{\nu,\alpha\beta}_T(k,Q)\,.
                                                                                    \label{FF:AvPhotonVertex}
                  \end{align}
                  \end{fshaded4}

            \noindent
            Similar to the quark-photon vertex of Section \ref{sec:quark-photon-vertex}, the physical properties are encoded in the transverse projection of
            the vertices (\ref{FF:ScPhotonVertex}--\ref{FF:AvPhotonVertex}).
            The transverse parts include the elements of \eqref{DQPV:Basis} in which no $Q^\mu$ is involved, implying
            one further transverse structure $\sim k^\mu$ in the scalar vertex and nine for the axial-vector vertex.
            In the axial-vector case the symmetry requirement
            \begin{equation}
                \Gamma^{\mu,\alpha\beta}_\text{(dq)}(k,Q) = \Gamma^{\mu,\beta\alpha}_\text{(dq)}(k,-Q)
            \end{equation}
            leads to relations between the transverse dressing functions which reduces the number of independent components to six \cite{Salam:1964zk,Eichmann:2007nn}.
            The transverse parts must vanish for $Q^2\rightarrow 0$ and
            do not participate in electromagnetic current conservation. Nevertheless they give contributions to magnetic moments
            and electric quadrupole moments in the case of a spin-$1$ diquark.
            In this respect it is advantageous to add the transverse contribution
            \begin{equation}
            \begin{split}
                    & \frac{1}{2} \, \Sigma_{(\sigma)} \left( \delta^{\mu\alpha} Q^\beta - \delta^{\mu\beta} Q^\alpha \right) + \\
                +\, & \Delta_{(\sigma)} \left\{ k^\mu Q^\alpha Q^\beta - \frac{k\cdot Q}{2} \left( \delta^{\mu\alpha} Q^\beta + \delta^{\mu\beta} Q^\alpha \right) \right\} + \\
                +\, & \Delta_{(\sigma)} \left\{ k^\mu \left( k^\alpha Q^\beta - k^\beta Q^\alpha \right) + k\cdot Q \left( \delta^{\mu\alpha} k^\beta - \delta^{\mu\beta} k^\alpha \right) \right\}
            \end{split}
            \end{equation}
            to Eq.\,\eqref{FF:AvPhotonVertex}. The resulting vertex \cite{Salam:1964zk} may be more conveniently written as
            \begin{fshaded}
            \begin{equation}\label{FF:SalamDelbourgo}
            \begin{split}
                   \Gamma^{\mu,\alpha\beta}_\text{(dq)}(k,Q) = \; & 2 k^\mu \left( \Delta_{(D_T^{-1})} \delta^{\alpha\beta}  - \Delta_{(\sigma)} k_+^\alpha \, k_-^\beta   \right)  \\
                                                                    & - \left( \sigma(k_+^2) \,\delta^{\mu\beta} \, k_+^\alpha + \sigma(k_-^2)\, \delta^{\mu\alpha} \, k_-^\beta \right)
                                                                      + T^{\mu\nu}_Q \,\widetilde{\Gamma}^{\nu,\alpha\beta}_T(k,Q)\,.
            \end{split}
            \end{equation}
            \end{fshaded}

            \noindent
            In the special case of free scalar and axial-vector propagators, i.e.
            \begin{equation}
                D^{-1}(k^2) = k^2 + M_\text{sc}^2, \quad D_T^{-1}(k^2) = k^2 + M_\text{av}^2, \quad D_L^{-1}(k^2) = M_\text{av}^2,
            \end{equation}
            the corresponding vertices satisfy \nocite{Lee:1962vm} 
            \begin{equation}
                \Delta_{(D^{-1})} = \Delta_{(D_T^{-1})} = 1, \quad \sigma(k^2) = \Sigma_{(\sigma)} = 1, \quad \Delta_{(\sigma)} = 0.
            \end{equation}

        \bigskip
        \fatcol{On-shell vertices I.}
            To obtain the electromagnetic current matrix of a spin-$0$ or spin-$1$ particle with mass $M$,
            the respective photon vertex must be taken on its mass shell, where
            \begin{equation}
                k^2 = -M^2 -\frac{Q^2}{4}, \quad   k \cdot Q=0\,,
            \end{equation}
            and hence any transverse projector $T^{\mu\nu}_Q$ acting on $k^\nu$ can be replaced by $\delta^{\mu\nu}$.
            This yields for a scalar diquark:
            \begin{equation} \label{DQPV:Scalar-Onshell}
                J^{\mu}_\text{(dq)}(Q^2) = \Gamma^{\mu}_\text{(dq)}(k,Q) \Big|_{k^2\rightarrow-M_\text{sc}^2-Q^2/4} = 2 k^\mu F_1(Q^2)\,.
            \end{equation}
            The electric form factor $F_1(Q^2)$ is the sum of  $\Delta_{(D^{-1})} = 1$ and the transverse component $\sim k^\mu$ in \eqref{FF:ScPhotonVertex}.
            Electromagnetic current conservation entails
            \begin{equation} \label{FF:DqCurrentConservation}
                Q^\mu J^\mu(Q^2) = 0, \quad J^\mu(0) = 2 k^\mu, \quad F_1(0)=1.
            \end{equation}
            The electromagnetic current of a spin-1 particle is found by applying transverse projectors with respect to outgoing and incoming diquark momentum
            (corresponding to the transverse pole in the propagator) onto the on-shell vertex:
            \begin{equation}
                J^{\mu,\alpha\beta}_\text{(dq)}(Q^2) = T_{k_+}^{\alpha\alpha'} \left\{  \Gamma^{\mu,\alpha'\beta'}_\text{(dq)}(k,Q) \Big|_{k^2\rightarrow-M_\text{av}^2-Q^2/4} \right\} T_{k_-}^{\beta\beta'}\,.
            \end{equation}
            Ignoring all terms which become redundant upon contraction with the transverse projectors, the on-shell diquark-photon vertex reads \cite{Weiss:1993kv,Hawes:1998bz,Alkofer:2004yf}:
            \begin{equation} \label{FF:AvPhotonVertexOnshell}
                \Gamma^{\mu,\alpha\beta}_\text{(dq)}(k,Q) = 2 k^\mu \left( F_1\, \delta^{\alpha\beta} + F_3 \, \frac{Q^\alpha Q^\beta}{2M_\text{av}^2} \right)
                                                             - G_M \left( \delta^{\mu\alpha} Q^\beta - \delta^{\mu\beta} Q^\alpha \right) \,,
            \end{equation}
            with the three $Q^2$-dependent electromagnetic form factors
            $F_1$, $G_M$ and $F_3$.
            Again, $F_1(Q^2)$ is the combination of $\Delta_{(D_T^{-1})} = 1$ and the transverse component $\sim k^\mu \,\delta^{\alpha\beta}$ in
            \eqref{FF:SalamDelbourgo}; and current conservation guarantees the relations of Eq.\,\eqref{FF:DqCurrentConservation}.
            $G_M(Q^2)$ is the magnetic dipole form factor with $G_M(0)=\mu$ being the magnetic moment
            of the axial-vector diquark. $F_3(0)$ is related to an electric quadrupole moment.
            Note that the on-shell projection of Eq.\,\eqref{FF:SalamDelbourgo} \textit{without} its transverse part yields the vertex $2k^\mu\,\delta^{\alpha\beta}$
            while that of Eq.\,\eqref{FF:AvPhotonVertex} already includes a constant contribution to the magnetic moment: $G_M(Q^2) = \sigma(-M_\text{av}^2)$.

            \bigskip

            Eq.\,\eqref{FF:AvPhotonVertexOnshell} is of limited use in the description of a baryon as a bound state of quark and diquark
            since the internal diquark is always off-shell.
            Minimal ans\"atze for the full scalar and axial-vector diquark-photon vertices in the literature
            involve Eqs.\,\eqref{FF:ScPhotonVertex} and \eqref{FF:SalamDelbourgo} which ensure the WTI for arbitrary diquark propagators,
            together with a transverse term of Eq.\,\eqref{FF:AvPhotonVertexOnshell} including the constant diquark's magnetic moment and possibly its electric quadrupole moment \cite{Oettel:2000jj,Alkofer:2004yf,Cloet:2008re}.
            To account for the suppression of the generalized form factors corresponding to $G_M$ and $F_3$ for non-zero photon momentum and off-shell momenta $k$,
            the magnetic moment $\mu$ must be chosen much smaller than its point-like value 2 \cite{Cloet:2008re}.
            This issue is not manifest in our calculation where the diquark-photon vertex is constructed from a different principle, cf. Eq.\,\eqref{ff:quarkloop}.

     \bigskip
     \fatcol{Scalar-axialvector transition.}
            Electromagnetic transitions between scalar and axial-vector diquarks
            appear naturally in the context of nucleon form factors and mediate a spin flip within the diquark correlation that contributes to the Faddeev amplitude.
            These contributions are not constrained by current conservation and hence purely transverse with respect
            to the photon momentum:
                \begin{equation} \label{FF:scaxtransverse}
                    Q^\mu \,\Gamma^{\mu,5\beta}_\text{(dq)} = Q^\mu\,\Gamma^{\mu,\alpha 5}_\text{(dq)} = 0\,,
                \end{equation}
            In analogy to the radiative $\rho\rightarrow\pi\gamma$ decay \cite{Maris:2002mz},
            the transition vertex is described by the Lorentz structure
                \begin{equation} \label{FF:scax}
                    \Gamma^{\mu,5\beta}_\text{(dq)}(k,Q) = \Gamma^{\mu,\beta 5}_\text{(dq)}(-k,Q) = i \varepsilon^{\mu\beta\rho\sigma} \,Q^\rho \,k^\sigma \,\frac{\kappa_{sa}(k^2,\hat{k}\cdot\hat{Q},Q^2)}{M_N}\,.
                \end{equation}
            Since in this case the incoming and outgoing diquarks have different mass shells, the onshell values of the involved momenta are
            \begin{equation}
                k^2 = -\frac{M_\text{sc}^2+M_\text{av}^2}{2}  -\frac{Q^2}{4}, \quad k\cdot Q = \frac{M_\text{av}^2-M_\text{sc}^2}{2}\,.
            \end{equation}

            \begin{figure}[tbp]
            \begin{center}
            \includegraphics[scale=0.15]{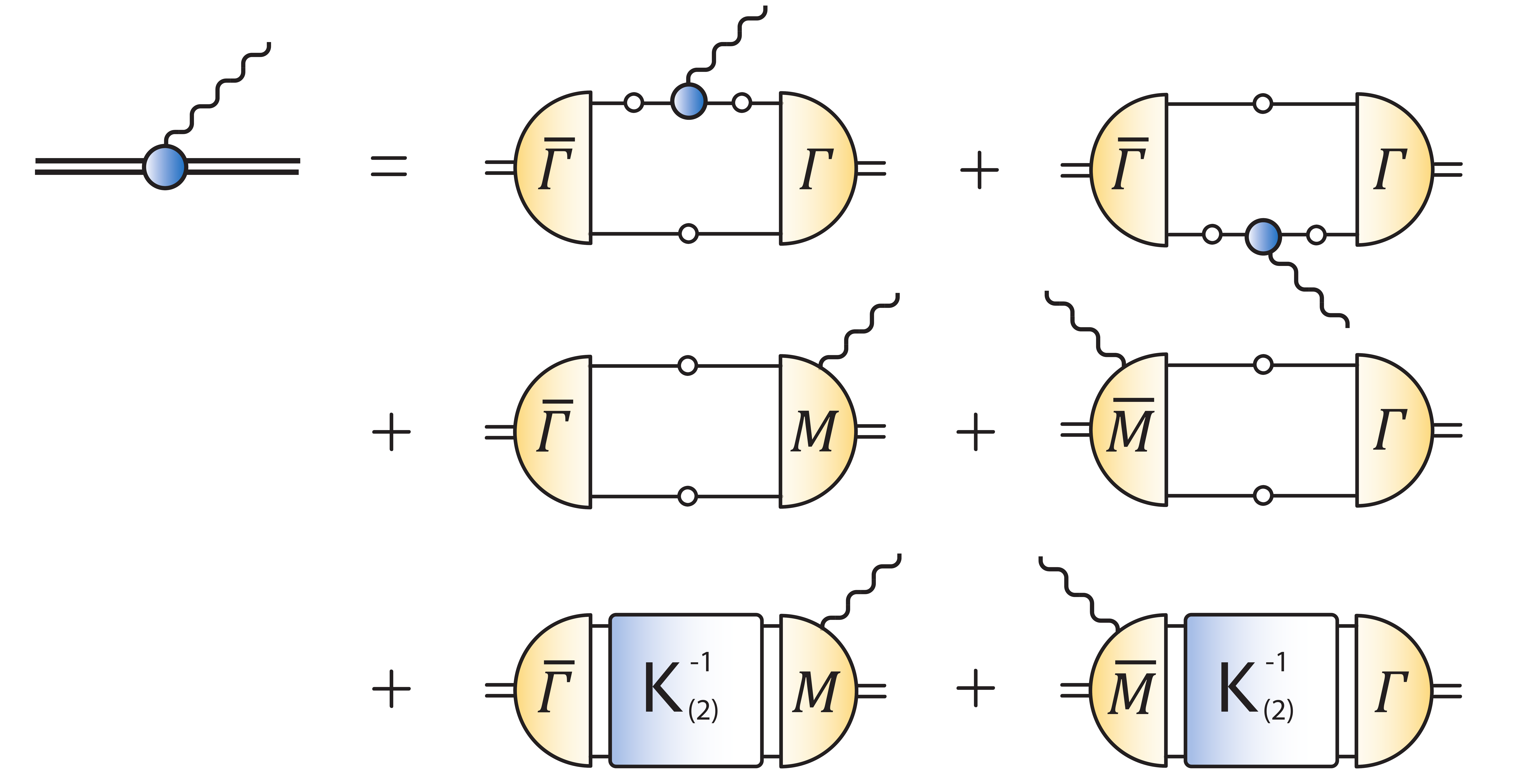}
            \caption[Diquark-photon vertex]{\backdef
                                            The fully resolved diquark-photon vertex of Eq.\,\eqref{ff:quarkloop}.
                                            The first row denotes the impulse approximation \eqref{FF:quarkloop:IMP},
                                            the second row the seagull coupling \eqref{FF:quarkloop:SG},
                                            and the third row the seagull coupling where the inverse rainbow-ladder kernel is involved, Eq.\,\eqref{FF:quarkloop:Kernel}.} \label{fig:dqphotonvertex}
            \end{center}
            \end{figure}

     \bigskip
     \fatcol{Resolving the diquark structure.}
            In the actual calculation we follow an approach explored in Ref.\,\cite{Bloch:1999ke,Oettel:2002wf} and express the diquark-photon vertices in terms of the diquark's substructure,
            i.e. by the "gauged" diquark propagators, cf. Eqs.\,\eqref{DQ:prop-equations}, \eqref{DQ:prop-final} and the discussion in Section~\ref{sec:em}:
                  \begin{fshaded4}
                  \begin{align} \label{ff:quarkloop}
                      \Gamma^{\mu,ab}_\text{(dq)}(k,Q) = \Gamma^{\mu,ab}_\text{IMP}(k,Q) + \Gamma^{\mu,ab}_\text{SG}(k,Q) + \Gamma^{\mu,ab}_\text{K}(k,Q)\,,
                  \end{align}
                  \end{fshaded4}

            \noindent
            where each of the three terms corresponds to a row in Fig.\,\ref{fig:dqphotonvertex}.
            This is the most general form which is consistent with Dyson's equation for the 2-quark T-matrix.
            It satisfies electromagnetic current conservation and provides expressions for the
            hitherto unknown transverse components of the vertices in Eqs.\,(\ref{FF:ScPhotonVertex}--\ref{FF:AvPhotonVertex}).
            The indices $a$, $b$ collect scalar ($a,b=5$) and axial-vector ($a,b=1\dots 4$) quantities as well as
            possible scalar-axialvector transitions.

            The impulse-approximation part in \eqref{ff:quarkloop} expresses the coupling of the photon to the quark propagators within the vertex.
            The two diagrams from the coupling to the upper and the lower quark line are equal, cf. Eq.\,\eqref{DQPV:imp-approx};
            their sum yields:
                  \begin{equation} \label{FF:quarkloop:IMP}
                  \begin{split}
                      \Gamma^{\mu,ab}_\text{IMP}(k,Q) = -\frac{1}{2} \int_q \text{Tr} & \,  \left\{  \conjg{\Gamma}^a(q-\nicefrac{\displaystyle Q}{\textstyle 2},k_+)\,S(q_+)\,\Gamma^b(q,k_-) \right.\times  \\[-0.4cm] 
                                                                                      & \times \left. \left[ S(q_-+Q)\,\Gamma^\mu_\text{(q)}(q_-+\nicefrac{\displaystyle Q}{\textstyle 2},Q)\, S(q_-) \right]^T \right\} .
                  \end{split}
                  \end{equation}
            The second row of Fig.\,\ref{fig:dqphotonvertex} denotes the part of the vertex which owes to the gauged diquark amplitudes,
            i.e. the seagulls $M^{\mu,b}$, $\conjg{M}^{\mu,a}$ which are discussed in App.\,\ref{app:seagulls}:
                  \begin{equation} \label{FF:quarkloop:SG}
                  \begin{split}
                      \Gamma^{\mu,ab}_\text{SG}(k,Q) = -\frac{1}{2} \int_q \text{Tr} & \,  \left\{   \conjg{\Gamma}^a(q,k_+)\,S(q_+')\,M^{\mu,b}(q,k_-,Q)\,S^T(q_-') \right. +\\[-0.4cm]
                                                                                             & \; + \left. \conjg{M}^{\mu,a}(q,k_+,Q)\,S(q_+)\,\Gamma^b(q,k_-)\,S^T(q_-) \right\}.
                  \end{split}
                  \end{equation}
            The internal quark momenta in both integrals are $q_\pm=\pm q+k_-/2$ and $q'_\pm=\pm q+k_+/2$.
            In order to satisfy the Ward-Takahashi identities \eqref{DQPV:WTI} for the full diquark propagators \eqref{DQ:prop-equations},
            the diagrams involving the gauged inverse ladder kernel need to be taken into account as well.
            The photon cannot couple to the rainbow-ladder kernel itself 
            but interacts with the diquark amplitudes, hence its generic form is $\bar{\Gamma}^a\, K^{-1} M^{\mu,b} + \bar{M}^{\mu,a}\,K^{-1} \Gamma^b$.
            Implementation of the ansatz \eqref{DQ:prop-final} for the parts of the diquark propagators which owe to the kernel,
                 \begin{equation}
                 \begin{split}
                     D^{-1}_K(P^2) &= M_\text{sc}^2 \,\big\{ \lambda_\text{sc} + \beta_\text{sc}\,F_\text{sc}(x) \big\},\\
                     \left(D^{-1}_K\right)^{\mu\nu}(P) &= M_\text{av}^2 \,\big\{ \lambda_\text{av} + \beta_\text{av}\,F_\text{av}(x)\big\} \delta^{\mu\nu},
                 \end{split}
                 \end{equation}
            yields with Eqs.\,(\ref{FF:ScPhotonVertex}--\ref{FF:AvPhotonVertex}) and $\sigma_K(k^2)=0$:
                 \begin{equation} \label{FF:quarkloop:Kernel}
                 \begin{split}
                     \Gamma^{\mu,55}_\text{K}(k,Q)          &= 2 k^\mu \,M_\text{sc}^2 \,\beta_\text{sc}\, \Delta_{F_\text{sc}} ,\\
                     \Gamma^{\mu,\alpha\beta}_\text{K}(k,Q) &= 2 k^\mu \,M_\text{av}^2 \,\beta_\text{av}\, \Delta_{F_\text{av}},
                 \end{split}
                 \end{equation}
            where $\Delta$ again denotes the difference quotient defined in \eqref{QPV:sigma,delta}.
            Here we neglected possible further transverse terms and contributions from a scalar-axialvector transition.

        \bigskip
        \fatcol{On-shell vertices II.}
            Only the impulse approximation $\Gamma^{\mu,ab}_\text{IMP}$ contributes on the diquarks' mass shells and hence to the diquarks' form factors
            $F_1$ and (in the axial-vector case) $F_3$, $G_M$.
            Schematically, the full vertex \eqref{ff:quarkloop} is the gauged diquark propagator:
                \begin{equation}
                \begin{split}
                    \Gamma^\mu_\text{(dq)} & \,= \left\{ \conjg{\Gamma} \left(G_0 - K^{-1}\right) \Gamma \right\}^\mu = \\
                                           & \,= \conjg{\Gamma}\,G_0^\mu \,\Gamma + \conjg{\Gamma} \left(G_0 - K^{-1}\right) M^\mu + \conjg{M}^\mu \left(G_0 - K^{-1}\right) \Gamma\,.
                \end{split}
                \end{equation}
            The seagull contributions vanish on the mass shell because of the diquark bound-state equation: $KG_0\Gamma=\Gamma$.
            They are nonetheless necessary to guarantee the Ward-Takahashi identities for general off-shell momenta,
            and they ensure transversality for the
            scalar-axialvector transition vertices
            which can be shown using the relations \eqref{BSE:Pauli-Principle}
            for the Dirac parts of the diquark amplitudes that follow from the general antisymmetry relation \eqref{dq:antisymmetry}.

            At the physical $u/d$ mass, the vertex \eqref{ff:quarkloop} together with the input \hyperlink{coupling:c1}{(C1)} of Section~\ref{sec:coupling-ansaetze} yields the
            on-shell values $\mu_\text{av} = 2.7$ and $\kappa_{sa}=2.3$.
            These are in the ballpark of values used to calculate nucleon electromagnetic form factors \cite{Oettel:2000jj,Alkofer:2004yf}
            in combination with the ans\"atze \eqref{FF:ScPhotonVertex}, \eqref{FF:SalamDelbourgo}, \eqref{FF:AvPhotonVertexOnshell} and \eqref{FF:scax}.

        \bigskip
        \fatcol{Color and flavor.}
            The color and flavor factors in Eq.\,\eqref{ff:quarkloop} have already been worked out.
            The color trace for all contributions is $1$.
            The flavor-charge traces yield
            $e_-=e_+=e_{dq}/2=1/2$ in the scalar and axial-vector case and $e_-=-e_+=1/2$, $e_{dq}=0$ for the transition vertex
            ($e_\pm$ and $e_{dq}$ appear in the definition of the seagulls, cf. App\,\ref{app:seagulls}).
            For instance, using the diquark flavor matrices $\mathsf{s_i}$ ($i=0\dots 3$)
            of Eq.\,\eqref{dq:flavormatrices} entails that the impulse-approximation contribution
            \eqref{FF:quarkloop:IMP} is given by
                  \begin{equation}\label{DQPV:imp-approx}
                      \text{Tr}\left\{ \mathsf{s_i^\dag\,s_j\,Q}\right\} \Lambda_\uparrow^{\mu,ab} +
                      \text{Tr}\left\{ \mathsf{s_i^\dag\,Q\,s_j}\right\} \Lambda_\downarrow^{\mu,ab} =
                      2\,\text{Tr}\left\{ \mathsf{s_i^\dag\,s_j\,Q}\right\} \Lambda_\uparrow^{\mu,ab}\;,
                  \end{equation}
            where $\mathsf{Q}=\text{diag}(q_u,q_d)$ is the quark electric charge matrix, and
            $\Lambda_{\uparrow\downarrow}$ denotes the photon's coupling to the upper and
            lower quark line, respectively.
            In App.\,\ref{app:color-flavor-charge} the diquark charge factors
            \begin{equation}
                2\,\text{Tr}\{\mathsf{s_i^\dag \,s_j \,Q}\} = \left(
                \begin{array}{c|ccc}
                    q_u+q_d    &    0    &    q_d-q_u    &    0    \\ \hline
                    0          &   2q_u  &    0          &    0    \\
                    q_d-q_u    &    0    &    q_u+q_d    &    0    \\
                    0          &    0    &    0          &   2q_d  \\
                \end{array}
                \right)
            \end{equation}
            are explicitly attached to the current matrix diagrams at each occurrence of the diquark-photon vertex.


\newpage

\section{Seagulls} \label{app:seagulls}

            Seagull contributions represent the photon's coupling to the scalar and axial-vector diquark amplitudes.
            They are fermion-scalar-vector and fermion-axialvector-vector four-point functions
             and, together
            with the diquark-photon vertex, reflect the diquark's
            internal substructure.

\begin{figure}[tbp]
            \begin{center}
            \includegraphics[scale=0.15]{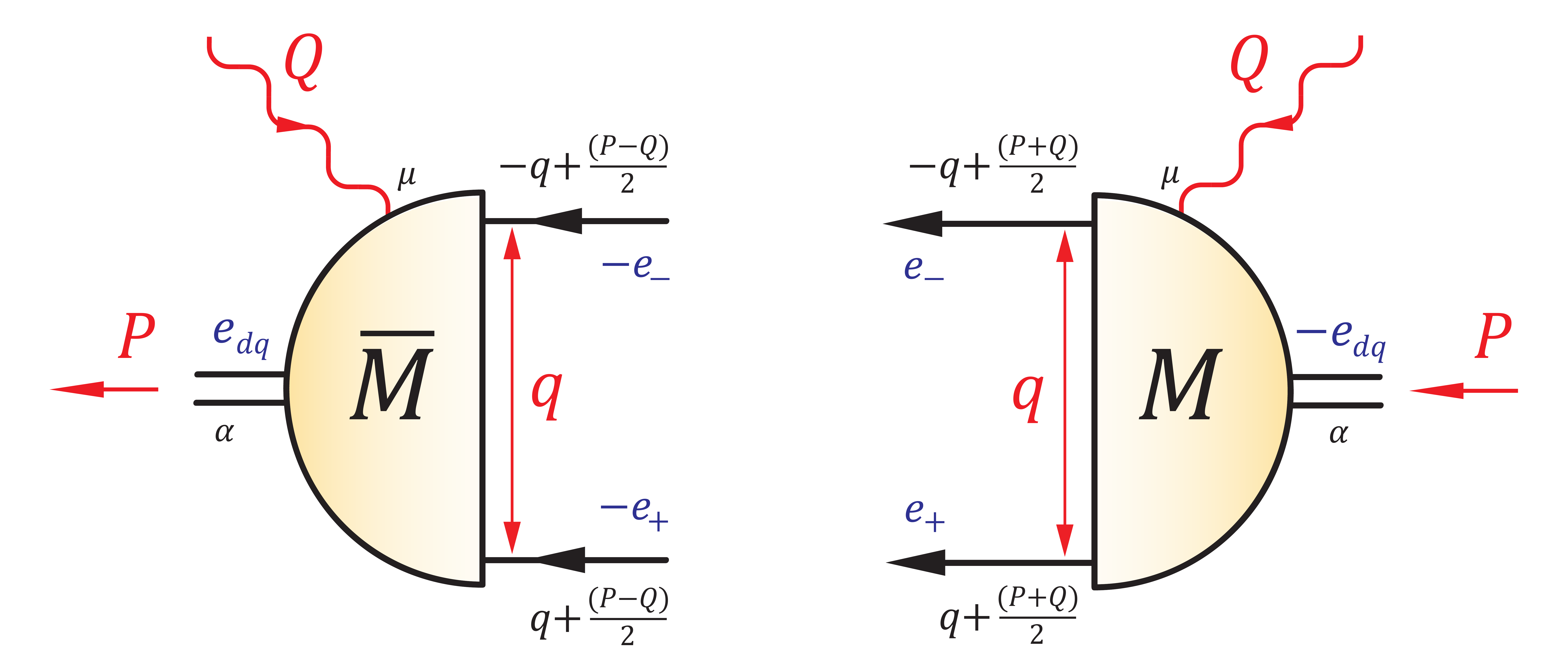}
            \caption[Seagulls]{\backdef Conventions for the seagulls. In accordance with the Ward-Takahashi identity \eqref{ff:seagullwti},
                                         outgoing charges are taken to be positive, incoming charges as negative. } \label{fig:seagulls}
            \end{center}
            \end{figure}

     \bigskip
     \fatcol{Ward-Takahashi identities.}
            The seagull WTIs were derived in \cite{Oettel:1999gc,Wang:1996zu} and read
            \begin{equation}\label{ff:seagullwti}
            \begin{split}
            Q^\mu M^\mu(q,P,Q) \,=\, &+  e_- \, \big\{ \Gamma(q_+,P)-\Gamma(q,P) \big\}  \\
                              &+  e_+ \, \big\{ \Gamma(q_-,P)-\Gamma(q,P) \big\} \\
                              &-  e_{dq}\,  \big\{ \Gamma(q,P_+)-\Gamma(q,P) \big\}\;, \\[0.2cm]
            Q^\mu \conjg{M}^\mu(q,P,Q) \,=\, &-  e_+ \, \big\{ \conjg{\Gamma}(q_+,P)-\conjg{\Gamma}(q,P) \big\} \\
                                    &-  e_- \, \big\{ \conjg{\Gamma}(q_-,P)-\conjg{\Gamma}(q,P) \big\}   \\
                                    &+  e_{dq}\,  \big\{ \conjg{\Gamma}(q,P_-)-\conjg{\Gamma}(q,P) \big\}\;,
            \end{split}
            \end{equation}
            where $q$ is the relative momentum, $P$ is the diquark's total momentum, $Q$ is the photon momentum,
            $q_\pm = q \pm Q/2$, and $P_\pm = P \pm Q$. The charges $e_+$, $e_-$ and $e_{dq}$ correspond to
            quark and diquark legs (cf. Fig.\,\ref{fig:seagulls}).
            For $Q\rightarrow 0$ the WTIs reduce to the differential Ward identities:
            \begin{equation}\label{FF:Seagulls:Ward-id}
            \begin{split}
                     M^\mu(q,P,0) &\,=\,  \left( \frac{e_--e_+}{2} \frac{d}{dq^\mu}-e_{dq}\frac{d}{dP^\mu}\right) \Gamma(q,P)\;, \\
            \conjg{M}^\mu(q,P,0) &\,=\,  \left( \frac{e_--e_+}{2} \frac{d}{dq^\mu}-e_{dq}\frac{d}{dP^\mu}\right) \conjg{\Gamma}(q,P)\;.
            \end{split}
            \end{equation}

     \bigskip
     \fatcol{Dominant diquark amplitudes.}
            In the case where only the dominant diquark amplitudes are retained and their dressing functions $f_1$ (either scalar or axial-vector) depend solely on $q^2$, i.e.
            \begin{equation}
                \Gamma_\text{sc}(q,P) = f^\text{sc}(q^2)\, i\gamma^5 C, \quad\quad  \Gamma_\text{av}^\mu(q,P) = f^\text{av}(q^2)\, i\gamma^\mu C,
            \end{equation}
            the scalar WTI (and the axial-vector WTI accordingly) reduces to
            \begin{equation} \label{SG:WTIsimplified}
            \begin{split}
                Q^\mu M^\mu(q,P,Q) & \, = e_- \, \big\{ \Gamma(q_+,P)-\Gamma(q,P) \big\} +  e_+ \, \big\{ \Gamma(q_-,P)-\Gamma(q,P) \big\} = \\
                                   &    = Q^\mu \left\{ e_- \,(q+Q/4)^\mu \,\Delta_f^+  - e_+  \,(q-Q/4)^\mu\,\Delta_f^- \right\}.
            \end{split}
            \end{equation}
            with the difference quotient
            \begin{equation}
                \Delta_f^\pm := \frac{f(q_\pm^2)-f(q^2)}{q_\pm^2-q^2} = \pm \,\frac{f(q_\pm^2)-f(q^2)}{(q \pm Q/4)\cdot Q}\, .
            \end{equation}

            \newpage

            \noindent
            The expressions for the seagulls which are consistent with the WTI can be immediately read off from Eq.\,\eqref{SG:WTIsimplified},
            cf. Refs.\,\cite{Oettel:1999gc,Oettel:2000jj,Oettel:2002wf,Alkofer:2004yf}:
            \begin{fshaded4}
            \begin{equation}
            \begin{split}
                M^\mu(q,P,Q)          & \, = \left\{ e_- \, (q+Q/4)^\mu \, \Delta_f^+ - e_+ \, (q-Q/4)^\mu \, \Delta_f^- \right\} i\gamma^5 C\,, \\
                M^{\mu,\alpha}(q,P,Q) & \, = \left\{ e_- \, (q+Q/4)^\mu \, \Delta_f^+ - e_+ \, (q-Q/4)^\mu \, \Delta_f^- \right\} i\gamma^\alpha C\,.
            \end{split}
            \end{equation}
            \end{fshaded4}

            \noindent
            The seagulls vanish if the diquark amplitudes are taken to be pointlike, i.e., $f_1=const.$ and all other $f_i=0$.

     \bigskip
     \fatcol{Full diquark substructure.}
            The situation becomes naturally more complicated if the full diquark substructure, i.e. all the remaining Dirac covariants,
            are taken into account. By virtue of the WTI \eqref{ff:seagullwti}, the difference quotients in $\{q_\pm,q\}$ and $\{P_\pm,P\}$ must
            be evaluated in all variables which appear in the diquark amplitude. The resulting expressions are lengthy  yet
            necessary to guarantee the differential Ward identity. A violation of the latter inevitably contaminates the
            small-$Q^2$ structure of the nucleon's electromagnetic form factors and can lead to sizeable deviations
            from the neutron's zero electric charge.

            To simplify the discussion, we transform the diquark basis of the amplitudes \eqref{dq:amp-offshell}
            with dressing functions $f_k(q^2,0,-M^2)$, given in Table~\ref{bse:offshellbasis}, into the general (unnormalized) form 
            \begin{equation*}
            \begin{split}
                \tau_{k=1 \dots 4}(q,P) & =  \left\{ \mathds{1}, \Slash{P}, \, \Slash{q}, \, \Slash{q} \,\Slash{P} \right\},  \\
                \tau_{k=1\dots 8}^\mu(q,P) & =  \gamma^\mu   \left\{\mathds{1}, \Slash{P}, \, \Slash{q}, \, \Slash{q} \,\Slash{P} \right\}, \;
                                                     q^\mu   \left\{\mathds{1}, \Slash{P}, \, \Slash{q}, \, \Slash{q} \,\Slash{P} \right\},
            \end{split}
            \end{equation*}
            such that the diquark amplitudes read:
            \begin{equation} \label{FF:Seagulls:DiquarkAmpRewritten}
            \begin{split}
                \Gamma_\text{sc}(q,P)     &\,=\, \sum_{k=1}^4 f_k^\text{sc}(q^2,q\!\cdot\! P,P^2) \, \tau_k(q,P)\,i\gamma^5 C\,,  \\
                \Gamma^\mu_\text{av}(q,P) &\,=\, \sum_{k=1}^8 f_k^\text{av}(q^2,q\!\cdot\! P,P^2) \, \tau^\mu_k (q,P)\,i\,C\,.
            \end{split}
            \end{equation}
            The new coefficients $f_k^\text{sc}$ and $f_k^\text{av}$ are the respective
            linear combinations of the those in Eq.\,\eqref{dq:amp-offshell} and pick up a dependence on
            the variables $q\cdot P = \sqrt{q^2}\sqrt{P^2} \,z$ and $P^2$.
            In the following we drop the superscripts 'sc' and 'av' for the purpose of brevity.

            In the same way as encountered in previous examples, the strategy to obtain a certain vertex (free of kinematic singularities)
            from its WTI is to write Eq.\,\eqref{ff:seagullwti} as a sum of difference quotients in one variable and extract a factor $Q^\mu$ from the linear terms.
            For example, the first row in the scalar-diquark WTI yields
            \begin{equation*}
            \begin{split}
                 &\left\{  \Gamma(q_+,P) - \Gamma(q,P) \right\} \left(-i \,C^T \gamma^5\right) = \\[0.2cm]
                    & \quad\qquad = \sum_k \Big( f_k(q_+^2,q_+\!\cdot\! P,P^2)\,\tau_k(q_+,P) - f_k(q^2,q\!\cdot\! P,P^2)\,\tau_k(q,P) \Big)  \\
                    & \quad\qquad = \sum_k  \left[\Delta^{(q_+^2)} f_k(q^2,z,P^2)\right] \left(q_+^2-q^2\right)  \tau_k(q_+,P) \,+ \\
                    & \quad\qquad + \sum_k \left[\Delta^{(q_+\cdot P)} f_k(q_+^2,q\!\cdot\! P,P^2)\right] \left(q_+\!\cdot\! P-q\!\cdot\! P\right) \, \tau_k(q_+,P) \,+ \\
                    & \quad\qquad + \sum_k f_k(q^2,z,P^2) \,\Big( \tau_k(q_+,P)-\tau_k(q,P)\Big)
            \end{split}
            \end{equation*}
            where $\Delta^{(\cdot)}$ denotes a difference quotient with respect to the variable $(\cdot)$, e.g.:
            \begin{equation}
                \Delta^{(q_\pm\cdot P)} f(q^2,q\!\cdot\!P,P^2) = \frac{f(q^2,q_\pm\!\cdot\!P,P^2)-f(q^2,q\!\cdot\!P,P^2)}{q_\pm\!\cdot\! P-q\!\cdot\! P}\,.
            \end{equation}
            Expressing the $Q$-linear terms through scalar products with $Q^\mu$,
            \begin{equation}
                \left( q_\pm^2 - q^2 \right) = \pm \, Q\cdot ( q \pm Q/4),  \quad
                \left( q_\pm\!\cdot\! P-q\!\cdot\! P\right) = \pm \, \frac{Q\cdot P}{2},
            \end{equation}
            allows to read off the respective vertex contributions.
            The differences in the basis elements $\tau_k(q_+,P)-\tau_k(q,P)$ only contribute for $k=3,4$.
            Proceeding this way yields the following form for the scalar and axial-vector seagull vertices:
            \begin{fshaded4}
            \begin{equation}\label{FF:Seagulls:Final}
            \begin{split}
                M^\mu(q,P,Q)           &=  \Big( M_1^\mu(q,P,Q) + M_2^\mu(q,P,Q) \Big)  \,i\gamma^5 C\,, \\
                M^{\mu,\alpha}(q,P,Q)  &=  \Big( M_1^{\mu,\alpha}(q,P,Q) + M_2^{\mu,\alpha}(q,P,Q) \Big) \, i C \,,
            \end{split}
            \end{equation}
            \end{fshaded4}

            \noindent
            where the $M_1^{\mu(\alpha)}$ include the difference quotients of the diquarks' dressing functions $f_k$:
            \begin{equation*}
                M_1^{\mu(\alpha)}   =  \sum_{k=1}^{4(8)} \Big\{  e_- (V_+)_k^\mu \,\tau_k^{(\alpha)}(q_+,P)  + e_+ (V_-)_k^\mu \,\tau_k^{(\alpha)}(q_-,P) - e_{dq} \widetilde{V}_k^\mu \,\tau_k^{(\alpha)}(q,P_+) \Big\}.
            \end{equation*}
            The quantities $(V_\pm)_k^\mu$ and $(\widetilde{V}_\pm)_k^\mu$ are free of kinematic singularities and given by
            \begin{align*}
             (V_\pm)_k^\mu  = \,&  \pm \left[ \Delta^{(q_\pm^2)} f_k(q^2,q\!\cdot\!P,P^2) \right] \left(q\pm Q/4\right)^\mu
                                   \pm \left[ \Delta^{(q_\pm \cdot P)} f_k(q_\pm^2,q\!\cdot\!P,P^2) \right] \frac{P^\mu}{2} \,, \\[0.2cm]
             (\widetilde{V}_\pm)_k^\mu  = \,& \pm \,\left[ \Delta^{(P_\pm^2)} f_k(q^2,q\!\cdot\!P,P^2) \right] \left( 2 P \pm Q\right)^\mu
                                              \pm \,\left[ \Delta^{(q\cdot P_\pm)} f_k(q^2,q\!\cdot\!P,P_\pm^2) \right] q^\mu \,.
             \end{align*}
            Their limits for vanishing photon momentum are
            \begin{equation}\label{FF:Seagulls:VZero}
            \begin{split}
             (V_\pm)_k^\mu  \stackrel{Q^2\rightarrow 0}{\longlongrightarrow}  \,&  \pm \frac{1}{2} \left( 2 q^\mu \frac{d f_k}{dq^2} +P^\mu \frac{d f_k}{d (q\!\cdot\! P)} \right) = \pm \,\frac{1}{2} \,\frac{d f_k}{dq^\mu} \,, \\[0.2cm]
             (\widetilde{V}_\pm)_k^\mu  \stackrel{Q^2\rightarrow 0}{\longlongrightarrow} \,& \pm \left( 2 P^\mu \frac{d f_k}{dP^2} + q^\mu\frac{d f_k}{d (q\!\cdot\! P)} \right) = \pm \,\frac{d f_k}{dP^\mu} \,.
             \end{split}
             \end{equation}
            The $M_2^{\mu(\alpha)}$ contain the differences of the basis elements:
            \begin{equation*}
            \begin{split}
                M_2^\mu  &= \frac{e_--e_+}{2} \gamma^\mu (f_3 + f_4 \Slash{P}) - e_{dq} (f_2 + f_4 \Slash{q}) \gamma^\mu\,,  \\
                M_2^{\mu,\alpha}  &= \frac{e_--e_+}{2} \Big\{  \gamma^\alpha \gamma^\mu (f_3 + f_4 \Slash{P})  + \delta^{\mu\alpha} (f_5+f_6 \Slash{P})
                                            + (\delta^{\mu\alpha} \Slash{q} + q^\alpha \gamma^\mu)(f_7+f_8\Slash{P}) \Big\} \\
                                        &   + \frac{e_-+e_+}{4} \,Q^\alpha \gamma^\mu (f_7+f_8\Slash{P})
                                            - e_{dq} \Big\{ \gamma^\alpha(f_2 + f_4 \Slash{q}) +q^\alpha(f_6+f_8\Slash{q}) \Big\} \gamma^\mu\,.
            \end{split}
            \end{equation*}
            Together with \eqref{FF:Seagulls:VZero} this allows for a simple check of the Ward identities \eqref{FF:Seagulls:Ward-id}.
            The final seagull vertices \eqref{FF:Seagulls:Final} are identical to those introduced in Ref.\,\cite{Eichmann:2007nn};
            yet the form \eqref{FF:Seagulls:DiquarkAmpRewritten} allows for a somewhat simplified notation.
            The conjugate seagulls are obtained from
            \begin{align*}
                \conjg{M}^\mu(q,P,Q) &= -C(M^\mu)^T(-q,-P,Q)\,C^{T}\;, \\
                \conjg{M}^{\mu,\alpha}(q,P,Q) &= C (M^{\mu,\alpha})^T(-q,-P,Q)\,C^{T}\;.
            \end{align*}

     \bigskip
     \fatcol{Transverse terms.}
            In the same way as encountered for the quark-photon and diquark-photon vertices,
            further terms transverse to the photon momentum may also contribute to the seagull vertices.
            Such contributions were found to be important for the medium-$Q^2$ structure of electromagnetic form factors \cite{Eichmann:2008ef}:
            ignoring them may cause form factors to rise with increasing photon momentum which is clearly unphysical.
            In the absence of a constraint on the seagulls' transverse structure,
            a minimal ansatz is to introduce a contribution which is proportional to the seagull itself:
            \begin{equation}\label{FF:Seagulls:Transverse}
                M^{\mu(\alpha)}_\text{Tot}  = M^{\mu(\alpha)} + m(x)\,T_Q^{\mu\nu}\,M^{\nu(\alpha)}
                                             = \big( 1-m(x) \big)\, T^{\mu\nu}_Q \,M^{\nu(\alpha)} + L^{\mu\nu}_Q \,M^{\nu(\alpha)},
            \end{equation}
            where $x=Q^2/m_\rho^2$ (the introduction of the $\rho$-meson mass scale will be explained below).
            To avoid a kinematic singularity, the dimensionless function $m(x)$ must vanish at zero photon momentum but is otherwise arbitrary.
            Via electromagnetic current conservation  only the transverse projection of the vertex contributes to observables,
            hence Eq.\,\eqref{FF:Seagulls:Transverse} is physically equivalent to multiplying the full vertex $M^{\mu(\alpha)}$ by a function $1-m(x)$ which is modeled on phenomenological assumptions.

            In the approach presented herein, the diquark-photon vertex is obtained by resolving the diquark's substructure
            which exposes its constituents, i.e. the quark propagator, quark-photon vertex and seagull amplitudes.
            Resolving the seagull structure is not feasible in our context, yet it is a reasonable strategy
            to model the transverse seagull part after that in the quark-photon vertex
            which would remain as fundamental quantity is such an approach.
            Specifically, we use the following parametrization:
            \begin{equation}
                m(x) = -\frac{1}{g_\rho} \,\frac{x^2}{1+x}\,e^{-\rho_3\,(1+x)}\,,
            \end{equation}
            which resembles \eqref{vertex:simulate} and again includes a transverse vector-meson pole with residue $1/g_\rho$.
            The additional factor $x$ in the numerator was implemented to leave the nucleon's static properties
            such as magnetic moments and charge radii unchanged via $m'(0)=0$.

            In connection with the parameter $\rho_2$ which appears in \eqref{vertex:simulate}, the values
            \begin{equation}\label{FF:Seagulls:Transverse:params}
                \rho_2 = 0.001, \quad \rho_3 = 0.075
            \end{equation}
            optimize agreement with the
            polarization-transfer data for the proton's form factor ratio $G_E(Q^2)/G_M(Q^2)$
            and thereby enable a realistic $Q^2$-evolution of the nucleon form factors.
            Nevertheless we emphasize that the implementation of the transverse seagull term \eqref{FF:Seagulls:Transverse} has no noticeable consequences for the
            form factors' small-$Q^2$ behavior, i.e. for photon momenta $Q^2\lesssim 2\,\text{GeV}^2$.


 \renewcommand{\arraystretch}{1.1}

\section[Nucleon electromagnetic current in the quark-diquark model]{Nucleon electromagnetic current in the \\quark-diquark model}\label{app:emcurrent}

            In this appendix we collect the ingredients of the nucleon's electromagnetic current matrix 
            in the quark-diquark model, given by the terms in Eq.\,\eqref{FF:Current:QDQ} and depicted in Fig.\,\ref{fig:current}. They
            depend on the quark-photon vertex $\Gamma^\mu_\text{q}$, the scalar and axial-vector diquark photon vertices $\Gamma^\mu_\text{dq}$,
            and the seagull vertices $M^\mu$ which have been specified
            in Apps.\,\ref{sec:quark-photon-vertex}, \ref{app:dqphotonvertex} and \ref{app:seagulls}.
            The explicit form of the current is given by
            \begin{equation}\label{ff:current2}
                J^\mu_{\alpha\beta}(Q^2) = \int_{p_f}\int_{p_i} \Big[   \conjg{\Phi}^a(p_f,P_f)\, X^{\mu,ab}(p_f,p_i,P_f,P_i)\, \Phi^b(p_i,P_i)  \Big]_{\alpha\beta}\;,
            \end{equation}
            where $P_i$ and $P_f=P_i+Q$ are incoming and outgoing on-shell nucleon momenta. The loop momenta $p_i$ and $p_f$ are arbitrary.
            $\alpha,\beta=1\dots 4$ are quark and $a,b=1\dots 5$ are diquark indices. The quark-diquark amplitudes $\Phi^a$ are the solutions
            of the quark-diquark Bethe-Salpeter equation \eqref{nuc:bse}. The quantity $X^{\mu,ab}$ is given by
            \begin{equation}
            \begin{split}
                X^{\mu,ab} = \,   & \,X^{\mu,ab}_\text{q} \, (2\pi)^4 \delta^4 \left( p_f-p_i-(1-\eta)\,Q \right)   + \\
                             + \, & \,X^{\mu,ab}_\text{dq} \,(2\pi)^4 \delta^4 \left( p_f-p_i+\eta \,Q \right)   \,+\,  X^{\mu,ab}_\text{K}\;,
            \end{split}
            \end{equation}
            and involves the quark and diquark impulse-approximation diagrams (the left column in Fig.\,\ref{fig:current}):
            \begin{align}
                \left( X_\text{q} \right)^{\mu,ab}_{\alpha\beta}  &= \Big[ S(p_+)\,\Gamma^\mu_\text{q}(p_+,p_-)\, S(p_-) \Big]_{\alpha\beta} D^{ab}(p_{d-}) \,, \label{ff:J-q}\\
                \left( X_\text{dq} \right)^{\mu,ab}_{\alpha\beta} &=  S_{\alpha\beta}(p_-) \Big[ D^{aa'}(p_{d+}) \, \Gamma^{\mu,a'b'}_\text{dq}(p_{d+},p_{d-}) \, D^{b'b}(p_{d-}) \Big] \,, \label{ff:J-dq}
            \end{align}
            and a two-loop diagram which represents the gauged quark-diquark kernel (right column of Fig.\,\ref{fig:current}):
            \begin{equation}\label{FF:Diagrams:Kernel}
                \left( X_\text{K} \right)^{\mu,ab}_{\alpha\beta}  = D^{aa'}(p_{d+})\Big[ S(p_+)\,K^{\mu,a'b'}(p_f,p_i,P_f,P_i)\, S(p_-) \Big]_{\alpha\beta}\,D^{b'b}(p_{d-})\,.
            \end{equation}
            The quark-photon coupling $X_\text{q}^{\mu,ab}$ connects scalar ($a,b=5$)
            \textit{or} axial-vector ($a,b=1\dots 4$) amplitudes whereas $X_\text{dq}^{\mu,ab}$ and $X_\text{K}^{\mu,ab}$ additionally allow for scalar--axial-vector transitions.
            The quark and diquark momenta are:
            \begin{align*}
                p_- &= p_i+\eta\,P_i\,,    &   p_{d-} &= -p_i + (1-\eta)\,P_i\;,  \\
                p_+ &= p_f+\eta\,P_f\,,    &   p_{d+} &= -p_f + (1-\eta)\,P_f\;.
            \end{align*}
            The gauged kernel $K^{\mu,ab}$ which appears in \eqref{FF:Diagrams:Kernel} contains the exchange-quark diagram and the seagull contributions:
            \begin{equation}
                K^{\mu,ab} = K^{\mu,ab}_\text{EX} + K^{\mu,ab}_\text{SG} + K^{\mu,ab}_{\overline{\text{SG}}}\;,
            \end{equation}
            with
            \begin{equation}
            \begin{split}
                K_\text{EX}^{\mu,ab}              &= \Gamma_{dq}^b(p_1,p_{d-})\Big[ S(q') \,\Gamma^\mu_\text{q}(q',q)\, S(q) \Big]^T \conjg{\Gamma}^a(p_2,p_{d+})\,,  \\[0.1cm]
                K_\text{SG}^{\mu,ab}              &= M^{\mu,b}(k_1,p_{d-},Q)\, S^T(q')\,\conjg{\Gamma}^a(p_2,p_{d+})\,, \\[0.1cm]
                K_{\overline{\text{SG}}}^{\mu,ab} &= \Gamma_{dq}^b(p_1,p_{d-})\,S^T(q)\,\conjg{M}^{\mu,a}(k_2,p_{d+},Q)\,,
            \end{split}
            \end{equation}
            and momenta:
            \begin{align*}
                q  &= p_{d-}-p_+\;, &
                p_1 &= \frac{p_+-q}{2}\;, &
                p_2 &= \frac{p_--q'}{2}\;, \\
                q' &= p_{d+}-p_-\;, &
                k_1 &= \frac{p_+-q'}{2}\;, &
                k_2 &= \frac{p_--q}{2}\;.
            \end{align*}

    \bigskip
    \fatcol{Breit frame.}
            For explicit calculations we work in the Breit frame where ingoing and outgoing nucleon have opposite $3$-momenta
            and the photon consequentially carries zero energy. With $\tau=Q^2/(4M^2)$, the momenta read:
                 \begin{equation*}\label{FF:Breitframe}
                     P_{i, f} =  i M\left( \begin{array}{c} 0 \\ 0 \\ \pm i \sqrt{\tau} \\ \sqrt{1+\tau} \end{array} \right), \quad
                     Q =  \left( \begin{array}{c} 0 \\ 0 \\ |Q| \\ 0 \end{array} \right), \quad
                     P = \frac{P_f+P_i}{2} = i M\left( \begin{array}{c} 0 \\ 0 \\ 0 \\ \sqrt{1+\tau} \end{array} \right).
                 \end{equation*}


    \bigskip
    \fatcol{Color, flavor and charge coefficients.}\label{app:color-flavor-charge}
             The current matrix diagrams still have to be endued with color and flavor-charge coefficients.
             The color traces for the impulse approximation and exchange/seagull diagrams are given by
             \begin{equation}
                 \frac{\delta_{BA}}{\sqrt{3}}\,\frac{\delta_{AB}}{\sqrt{3}}=1\;, \quad
                 \frac{\delta_{BA}}{\sqrt{3}} \, \frac{\varepsilon_{AED}}{\sqrt{2}}\,\frac{\varepsilon_{CEB}}{\sqrt{2}}\,\frac{\delta_{CD}}{\sqrt{3}}=-1\;,
             \end{equation}
             respectively.
             With the diquark and quark-diquark isospin matrices of Eqs.\,\eqref{dq:flavormatrices} and \eqref{NUC:Isospin}:
             $\mathsf{s_i}$ and $\mathsf{t_i}$, $i=0\dots 3$,
             and the quark charge matrix $\mathsf{Q}=\text{diag}(q_u,q_d)$,
             the flavor-charge matrices for the quark-photon, diquark photon and exchange diagrams read:
             \begin{equation} \label{flavorcharge:1}
                 \sum_{ij} \delta_{ij} \,\mathsf{t_i^\dag} \,\mathsf{Q}\, \mathsf{t_j}\;, \quad
                 \sum_{ij} \mathsf{t_i^\dag} \, \mathsf{t_j} \,2\, \text{Tr}\left\{ \mathsf{s_i^\dag}\,\mathsf{s_j}\,\mathsf{Q} \right\}\; , \quad
                 \sum_{ij} \mathsf{t_i^\dag} \,\mathsf{s_j}\,\mathsf{Q}\, \mathsf{s_i^\dag}\,\mathsf{t_j}\;.
             \end{equation}
             The traces for proton and neutron are obtained by sandwiching these expressions between the isospinors $\mathsf{u}=(1,0)$ or $\mathsf{d}=(0,1)$.
             The index range in $i$, $j$ of the sums in \eqref{flavorcharge:1} depends on the type of the quark-diquark amplitude in the initial and final state:
             for a scalar quark-diquark amplitude in the final state (i.\,e., on the left-hand side): $i=0$, for an axial-vector amplitude: $i=1\dots 3$,
             and likewise for the index $j$ and an incoming amplitude.
             For instance, with the four contributions:
             \begin{align*}
                 &\text{S}\leftarrow\text{S}: &\mathsf{u^\dag}\left(\mathsf{t_0^\dag}\,\mathsf{s_0}\,\mathsf{Q}\, \mathsf{s_0^\dag}\,\mathsf{t_0}\right)\mathsf{u} &= \frac{q_d}{2}\;, \qquad\\
                 &\text{S}\leftarrow\text{A}: &\mathsf{u^\dag}\left(\sum_{j=1}^3\mathsf{t_0^\dag}\,\mathsf{s_j}\,\mathsf{Q}\, \mathsf{s_0^\dag}\,\mathsf{t_j}\right)\mathsf{u} &= -\frac{2q_u+q_d}{2\sqrt{3}}\;, \qquad\\
                 &\text{A}\leftarrow\text{S}: &\mathsf{u^\dag}\left(\sum_{i=1}^3,\mathsf{t_i^\dag}\,\mathsf{s_0}\,\mathsf{Q}\, \mathsf{s_i^\dag}\,\mathsf{t_0}\right)\mathsf{u} &= -\frac{2q_u+q_d}{2\sqrt{3}}\;, \qquad\\
                 &\text{A}\leftarrow\text{A}: &\mathsf{u^\dag}\left(\sum_{i=1}^3\sum_{j=1}^3\mathsf{t_i^\dag}\,\mathsf{s_j}\,\mathsf{Q}\, \mathsf{s_i^\dag}\mathsf{t_j}\right)\mathsf{u} &= -\frac{4q_u-q_d}{6}\;, \qquad
             \end{align*}
             the full exchange contribution to the proton's electromagnetic current (including the color factor) is given by
             \begin{equation}
                 J^\text{EX}_p = -\frac{q_d}{2}\, J^\text{EX}_{SS} + \frac{2q_u+q_d}{2\sqrt{3}}\,
                                  \left( J^\text{EX}_{SA}+J^\text{EX}_{AS}\right) + \frac{4q_u-q_d}{6}\, J^\text{EX}_{AA}\;,
             \end{equation}
             where $J^\text{EX}_{SS}$, $J^\text{EX}_{SA}$, $J^\text{EX}_{AS}$ and $J^\text{EX}_{AA}$ denote the Dirac parts of
             the exchange contributions in Eq.\,\eqref{ff:current2}.
             For the seagull diagrams, the coupling of the photon to both quark lines and the diquark line in the seagull
             vertex (cf. Fig.\,\ref{fig:seagulls}) need to be taken into account; hence all
             occurences of $e_-$, $e_+$ and $e_{dq}$ must be replaced by the combined flavor-charge factors
             \begin{align*}
                  \text{SG:} \qquad &
                               e_- \rightarrow    \sum_{ij} \mathsf{t_i^\dag} \,\mathsf{s_j}\,\mathsf{Q}\, \mathsf{s_i^\dag}\,\mathsf{t_j}\;, \quad
                               e_+ \rightarrow    \sum_{ij} \mathsf{t_i^\dag} \,\mathsf{Q}\,\mathsf{s_j}\, \mathsf{s_i^\dag}\,\mathsf{t_j}\;, \quad \\
                             & e_{dq} \rightarrow \sum_{ij} \mathsf{t_i^\dag} \,\mathsf{s_j}\, \mathsf{s_i^\dag}\,\mathsf{t_j} \;2\, \text{Tr}\left\{ \mathsf{s_j^\dag}\,\mathsf{s_j}\,\mathsf{Q} \right\}, \\
                  \overline{\text{SG:}} \qquad &
                               e_- \rightarrow    \sum_{ij} \mathsf{t_i^\dag} \,\mathsf{s_j}\,\mathsf{Q}\, \mathsf{s_i^\dag}\,\mathsf{t_j}\;, \quad
                               e_+ \rightarrow    \sum_{ij} \mathsf{t_i^\dag} \,\mathsf{s_j}\, \mathsf{s_i^\dag}\,\mathsf{Q}\,\mathsf{t_j}\;, \quad \\
                             & e_{dq} \rightarrow \sum_{ij} \mathsf{t_i^\dag} \,\mathsf{s_j}\, \mathsf{s_i^\dag}\,\mathsf{t_j} \;2\, \text{Tr}\left\{ \mathsf{s_i^\dag}\,\mathsf{s_i}\,\mathsf{Q} \right\} \nonumber
             \end{align*}
             and equipped with an overall color factor $-1$.

\chapter{Utilities}\label{app:utilities}

\section{Euclidean conventions}\label{app:conventions}

    \renewcommand{\arraystretch}{1.2}

            Throughout this thesis we work in Euclidean momentum space with the conventions
            \begin{equation}
                p\cdot q = \sum_{k=1}^4 p_k \, q_k,\quad
                p^2 = p\cdot p,\quad     
                \Slash{p} = p\cdot\gamma,\quad  
                \left\{ \gamma^\mu, \gamma^\nu \right\} = 2\,\delta^{\,\mu\nu},\quad
                \gamma^\mu = \left( \gamma^\mu\right)^\dag.
            \end{equation}
            A vector is spacelike if $p^2 > 0$ and timelike if $p^2<0$. Moreover:
            \begin{equation}
                \sigma^{\mu\nu} = -\frac{i}{2} \left[ \gamma^\mu, \gamma^\nu \right], \quad
                \gamma^5 = -\gamma^1 \gamma^2 \gamma^3 \gamma^4 = -\frac{1}{24} \,\varepsilon^{\mu\nu\rho\sigma} \gamma^\mu \gamma^\nu \gamma^\rho \gamma^\sigma.
            \end{equation}
            The standard representation for the gamma matrices reads:
            \begin{equation}
                \gamma^k  =  \left( \begin{array}{cc} 0 & -i \sigma^k \\ i \sigma^k & 0 \end{array} \right) , \quad
                \gamma^4  =  \left( \begin{array}{c@{\quad}c} \mathds{1} & 0 \\ 0 & \!\!-\mathds{1} \end{array} \right) , \quad
                \gamma^5  =  \left( \begin{array}{c@{\quad}c} 0 & \mathds{1} \\ \mathds{1} & 0 \end{array} \right)\,.
            \end{equation}
            The charge conjugation matrix is defined as
            \begin{equation}
                C = \gamma^4 \gamma^2, \quad C^T = C^\dag = C^{-1} = -C.
            \end{equation}
            We express four-momenta through hyperspherical coordinates:
            \begin{equation}\label{APP:momentum-coordinates}
                p^\mu = \sqrt{p^2} \left( \begin{array}{l} \sqrt{1-z^2}\,\sqrt{1-y^2}\,\sin{\phi} \\
                                                           \sqrt{1-z^2}\,\sqrt{1-y^2}\,\cos{\phi} \\
                                                           \sqrt{1-z^2}\;\;y \\
                                                           \;\; z
                                         \end{array}\right) =
                                   \left( \begin{array}{l} \sin{\psi} \,\sin{\theta} \,\sin{\phi} \\
                                                           \sin{\psi} \,\sin{\theta} \,\cos{\phi} \\
                                                           \sin{\psi} \,\cos{\theta} \\
                                                           \cos{\psi}
                                         \end{array}\right) .
            \end{equation}
            Should $p$ describe the on-shell momentum of a bound state with mass $M$, then: $p^2=-M^2$, and
            in the rest frame: $z=1$. 
            A four-momentum integration reads:
            \begin{equation} \label{hypersphericalintegral}
                \int_p \;\; := \;\;  \int \!\!\frac{d^4 p}{(2\pi)^4} \;\;
                       = \;\; \frac{1}{(2\pi)^4}\,\frac{1}{2} \int_0^{\infty} dp^2 \,p^2 \int_{-1}^1 dz\,\sqrt{1-z^2}  \int_{-1}^1 dy \int_0^{2\pi} d\phi \,.
            \end{equation}

    \section{Angular dependence} \label{appendixchebyshev}

            The Lorentz-invariant coefficients of bound-state amplitudes depend on Euclidean scalar products of the involved momenta.
            Denoting the relative momenta by $p_i$ and the total bound-state momentum with $P$, where $P^2=-M^2$,
            the coefficients $f_k\,(p_i^2,z_i)$ carry a dependence on radial and angular variables
            \begin{equation}
                 p_i^2\,,  \qquad   z_i = \cos{\varphi_i} = \hat{p}_i\cdot\hat{P}\,,\, \hat{p}_1\cdot\hat{p}_2,\, \dots \quad .
            \end{equation}
            In the rest frame of the respective bound state: $p_i^2 \in (0,\infty)$ and $z_i \in (-1,1)$.
            A polynomial expansion in the radial arguments is not practicable as the $p_i^2$ dependence of the amplitudes
            can differ enormously for different amplitudes $f_k$: individual zero crossings might appear, and the specific power laws
            for $p_i^2 \rightarrow 0$, $p_i^2 \rightarrow \infty$ depend on the choice of Dirac covariants (e.g., normalized vs. unnormalized)
            and the inherent IR and UV properties of the physical problem.

            On the other hand, the dependence on the angular variables $z_i$ is usually weak and well-suited for an expansion into orthogonal polynomials.
            We denote these generically by $Y_n(z)$, $n\in \mathds{N}_0$,
            where the (continuous and discretized) orthogonality relations are given by
                \begin{equation}
                    \int_{-1}^1 dz \,\Omega_Y(z) \,Y_m^\ast(z) \,Y_n(z) = \lim_{N\rightarrow\infty} \sum_{k=1}^N \widetilde{\Omega}_Y(z_k) \,Y_m^\ast(z_k) \,Y_n(z_k) = \delta_{mn}
                \end{equation}
            and thus allow for an expansion of a function $f(z)$ via
                \begin{equation}\label{CHEB:expansion}
                      f(z) = \lim_{\conjg{N}\rightarrow\infty} \sum_{n=0}^{\conjg{N}} f_n \,Y_n(z), \qquad
                      f_n = \left\{
                            \begin{array}{rl}
                                & \displaystyle \int_{-1}^1 dz \,\Omega_Y(z) \,Y_n^\ast(z) \,f(z) \\[-0.1cm]
                                & \displaystyle \lim_{N\rightarrow\infty} \sum_{k=1}^{N} \widetilde{\Omega}_Y(z_k) \,Y_n^\ast(z_k) \,f(z_k).
                            \end{array}
                            \right.
                \end{equation}

            \bigskip
            \fatcol{Chebyshev expansion.}
            The common use of Chebyshev polynomials in the bound-state framework is motivated by an approximate $O(4)$ symmetry in a certain type of ladder-exchange Bethe-Salpeter equations \cite{Oettel:2001kd} .
            For $l=0$, i.e., for $s$-wave ground-state wave functions in the context of quark models,
            the hyperspherical harmonics $\mathcal{Y}_{nlm}(\psi,\theta,\phi)$ reduce to Chebyshev polynomials
            of the second kind in the cosine of the azimuthal angle $\psi$.
            The Chebyshev polynomials of the first ($T_n$) and second kind ($U_n$) are given by
                \begin{equation}\label{cheb4}
                \begin{split}
                     T_n(z) &:= \, \frac{1}{2}\left[ \left( z+\sqrt{z^2-1} \right)^n + \left(z-\sqrt{z^2-1} \right)^n \right] \\
                            &\;=\, \cos{(n \arccos{z})} \,=\, 2^{n-1} \prod_{k=1}^n \left[ z-\cos{\left(k-\frac{1}{2}\right)\frac{\pi}{n}} \right], \quad (n \geq 1)  \\
                     U_n(z) &:= \,\frac{\sin{\left[(n+1)\, \arccos{z}\right]}}{\sqrt{1-z^2}}  \,=\, 2^n \prod_{k=1}^n \left[ z-\cos{\frac{k\,\pi}{n+1}} \right], \quad (n \geq 0)
                \end{split}
                \end{equation}
            with $T_0:=\nicefrac{1}{\sqrt{2}}$ to satisfy the orthogonality relation of Eq.\,\eqref{CHEB:expansion} together with \eqref{CHEB:orthogonality}.
            For $n>0$ the Chebyshev polynomials of the second kind are related to the $T_n(z)$ via $T_n(z)=U_n(z)-z \,U_{n-1}(z)$.
            The first few Chebyshev polynomials $(n\geq 0)$ are 
                \begin{equation}
                \begin{split}
                    T_n(z) &= \left\{ \textstyle\frac{1}{\sqrt{2}}, \; z, \; 2 z^2-1, \; 4 z^3-3 z, \; 8 z^4-8 z^2+1, \; \dots \right\}, \\
                    U_n(z) &= \left\{ 1, \; 2z, \; 4 z^2-1, \; 8 z^3-4 z, \; 16 z^4-12 z^2+1, \; \dots \right\}.
                \end{split}
                \end{equation}
            For the explicit expansion in Eq.\,\eqref{CHEB:expansion} one uses $Y_n(z)=i^n\, T_n(z)$ or $Y_n(z)=i^n\,U_n(z)$, for
            each of which the continuous and discrete integral measures are given by
                \begin{equation}\label{CHEB:orthogonality}
                \begin{split}
                    1^\text{st} &: \quad \Omega_T(z) = \frac{2}{\pi}\frac{1}{\sqrt{1-z^2}}\,;\quad \widetilde{\Omega}_T(z_k) = \frac{2}{N}\,, \quad z_k = \cos{\left(k-\frac{1}{2}\right)\frac{\pi}{N}}\,;      \\
                    2^\text{nd} &: \quad \Omega_U(z) = \frac{2}{\pi}\sqrt{1-z^2}\,; \quad  \widetilde{\Omega}_U(z_k) = \frac{2\,(1-z_k^2)}{N+1}\,, \quad z_k = \cos{\frac{k\pi}{N+1}}\,.
                \end{split}
                \end{equation}
            The $z_k$ ($k=1\dots N$) which appear in the discrete versions were chosen to be the roots of the respective polynomials $T_N(z)$ or $U_N(z)$ which,
            for finite $N > \conjg{N}$, optimizes the agreement between the coefficients $f_n$ that appear in the expansion and the projection of \eqref{CHEB:expansion}  \cite{Oettel:2001kd}.

            A Lorentz boost of the bound-state amplitude leads the domain of the angular arguments $z=\hat{p}_i\cdot\hat{P}$ into the complex plane
            whereas the rest-frame solution of the bound-state equation is given only for $z \in (-1,1)$.
            While a Chebyshev expansion allows for a straightforward analytic continuation within the convergence radius $|z|<1$,
            its convergence properties are lost for $|z|>1$ with increasing $\conjg{N}$.
            As a possible workaround one may drop the ideally weak dependence on $z$ altogether, as it was done for the off-shell diquark amplitudes in Eq.\,\eqref{dq:amp-offshell}.
            In the context of nucleon form factors, where the nucleon amplitude is evaluated in the Breit frame,
            one may alternatively introduce a different expansion variable $z_Q$ with $|z_Q|<1$ for each value of the photon momentum $Q^2$ \cite{Cloet:2008re}.
            In general the problem can be overcome by solving the bound-state equation in each boosted Lorentz frame anew
            where the dependence on the complex variable $z$ is substituted into two real arguments $z_{1,2}\in (-1,1)$ \cite{Bhagwat:2006pu}.

    \section{Limitations from the singularity structure}\label{app:singularities}

          \renewcommand{\arraystretch}{1.2}

             For a bound-state equation which is solved in the rest frame, the radial arguments $p_i^2$ of the
             off-shell Green functions which appear in the equation's kernel
             become complex as well. In the following discussion we denote the associated momenta $p_i$ generically by
             $p_{(X)}$, where $X$ is e.g. a quark propagator, a diquark propagator or a diquark amplitude.
             For instance, the arguments of the quark propagators $S(\pm q_\pm)$ in the meson BSE \eqref{bse:bse} read
             \begin{equation}\label{SING:example1}
             \begin{split}
                 p_{(S_1)} &:= q_+ = q + \sigma\,P\,, \\
                 p_{(S_2)} &:= q_- = -q + (1-\sigma)\,P.
             \end{split}
             \end{equation}
             While $q \in \mathds{R}^4$, the square of the on-shell meson momentum is complex: $\sqrt{P^2} = i M$,
             and thus also $p_{(X)}^2$, e.g.
             \begin{equation*}
                p_{(S_1)}^2 = |q|^2 -\sigma^2 M^2 + 2 i\,\sigma M  |q| \,\hat{q}\cdot\hat{P} \,.
             \end{equation*}
             Another example is given by the quark and diquark propagators in the quark-diquark BSE \eqref{nuc:bse}:
             \begin{equation}\label{SING:example2}
             \begin{split}
                 p_{(S_1)} &:= k_q = k + \eta\,P\,, \\
                 p_{(S_2)} &:= q = -p-k + (1-2\,\eta)\,P\,,\\
                 p_{(D)}   &:= k_d= -k + (1-\eta)\,P\,,\\
             \end{split}
             \end{equation}
             where $p,\,k \in \mathds{R}^4$.
             Eqs.\,(\ref{SING:example1}--\ref{SING:example2}) can be summarized as
             \begin{equation} \label{sing:mom}
                 p_{(X)} = R_X + \beta_X(\eta)\,P\,,
             \end{equation}
             where $R_X \in \mathds{R}^4$ denotes the sum of all involved real momenta (e.g. the integration variables)
             and $P$ is the bound-state momentum which satisfies $P^2=-M^2$.
             The coefficients $\beta_X \in \mathds{R}$ depend on an arbitrary momentum partitioning parameter $\eta \in [0,1]$.

             The complex arguments $p_{(X)}^2$ 
             are sampled on a domain
             \begin{equation}
                 p_{(X)}^2 = |R|^2 -\beta^2 M^2 + 2 i \, \beta  M|R|Z\,,   
             \end{equation}
             where we defined $Z:= \hat{R}\cdot \hat{P} \in (-1,1)$.
             This is the interior of a parabola $\left(|R| \pm i \beta\,M\right)^2$
             whose apex $-\beta^2 M^2$ depends on the bound-state mass.
             It is constrained by the occurrence of the nearest singularity in the quantity $X$
             which defines the limiting parabola
             \begin{equation}\label{APP:SING:polemass}
                 p_{(X),\text{max}}^2 = (t \pm i m_X)^2\,, \qquad t \in \mathds{R}_+\,,
             \end{equation}
             where $m_X$ is the respective pole mass (e.g., $m_q$ of Fig.~\ref{fig:MZ} for the complex conjugate poles in the quark propagator,
             $M_\text{sc}$, $M_\text{av}$ for the timelike poles in the diquark propagator and so forth).
             This leads  to a restriction $M<m_X/|\beta_X(\eta)|$, and in total to an upper limit
             for the bound-state mass:
             \begin{equation}\label{sing:cond}
                 M < f(\eta) := \text{min} \, \left\{ \, \frac{m_{X_1}}{|\beta_{X_1}(\eta)|}\,, \, \frac{m_{X_2}}{|\beta_{X_2}(\eta)|}\,, \dots \right\}.
             \end{equation}
             Physics must be independent of $\eta$, as explicitly demonstrated in \cite{Oettel:1998bk,Bhagwat:2006pu}.
             As a consequence one can fix the momentum partitioning $\eta$ to that value $\eta_0$ which maximizes the upper boundary $f(\eta)$
             for the calculable mass $M$. Assuming isospin symmetry with equal current-quark masses, $M_\text{sc} < M_\text{av}$, $M_\text{sc} < 2 m_q$ and
             a singularity-free diquark amplitude, exemplary values are:
             \begin{itemize}
                 \item Meson/diquark BSE: $\eta_0 = \nicefrac{1}{2}$, $M< 2 m_q$\,, \vspace{-0.2cm}
                 \item Quark-diquark BSE: $\eta_0 = m_q/(m_q+M_\text{sc})$, $M< m_q + M_\text{sc}$\,, \vspace{-0.18cm}
                 \item Faddeev equation: $\eta_0 = \nicefrac{1}{3}$, $M< 3 m_q$\,.
             \end{itemize}
             If the bound-state momentum $P$ in Eq.\,\eqref{sing:mom} is replaced by the Breit momentum with $P^2=-(M^2+Q^2/4)$,
             the above considerations can be applied to the nucleon form factor diagrams as well. The result
             \begin{equation}
                 \sqrt{M^2+\frac{Q^2}{4}} < f(\eta)
             \end{equation}
             implies an upper limit for the photon momentum if the nucleon mass $M$ is already known:
             \begin{itemize}
                 \item Quark-diquark BSE: $Q^2 < 4 \left[ (m_q + M_\text{sc})^2 - M^2 \right]$\,, \vspace{-0.2cm}
                 \item Faddeev equation: $Q^2 < 4 \left[ (3\,m_q)^2 - M^2 \right]$\,.
             \end{itemize}
             This is typically in the range of several GeV$^2$ and explains the restrictions encountered in Section~\ref{sec:results}.
             Refined numerical methods, such as an inclusion of pole residues, are necessary to circumvent these limitations and establish a connection with the large-$Q^2$ domain in the form factors.


\end{appendix}


\bibliographystyle{utphys-mod2}
 
\bibliography{literature/MyLiterature6,literature/MissingRefs}


\newpage
\chapter*{Acknowledgements}

             First of all I would like to thank my supervisor Reinhard Alkofer for the opportunity to work with him,
             and for his support, guidance and encouragement throughout the work on this thesis.
             My gratitude goes to Andreas Krassnigg for countless discussions and his invaluable input and support during the past years.
             I am indebted to Diana Nicmorus for her friendship, many useful discussions and, quite generally, being the glue that has bound our group together. 
             I would further like to thank Andreas and Diana for a proof-reading of the thesis.
 
             I am grateful for the opportunity of participating in the PhD program "Hadrons in vacuum, nuclei and stars" at the University of Graz
             which has been a stimulating experience. My thanks go to my fellow students and the staff at the physics department.
 
             I would like to thank Craig Roberts for enabling a pleasant and inspiring
             research stay at the Argonne National Lab.
             I am indebted to Ian Clo\"{e}t, Thomas Klaehn, Ross Young, Bruno El-Bennich,
             Peter Tandy and Wolfgang Bentz for interesting discussions.
             Special thanks go to Ian and Thomas for their help in getting me around during those months.
 
             Last but not least, my thanks go to my parents.
 
             \bigskip

             \fatcol{Financial support.} This thesis was supported by the Austrian Science Fund FWF under
             Grant No.\ W1203 (Doctoral Program ``Hadrons in vacuum, nuclei and stars'') and
             FWF Project P20496-N16.
 
             \bigskip

\end{document}